\documentclass[english,final,numbers]{aipproc}
\layoutstyle{8x11single}

\usepackage{babel}
\usepackage{graphicx}
\usepackage{natbib}

\usepackage{amsmath}
\usepackage{amsfonts}
\usepackage{amssymb}
\usepackage{dcolumn}

\newcolumntype{e}[1]{D{.}{.}{#1}}

\newcommand{\apjl}{Astroph. J. Lett.}
\newcommand{\apj}{Astroph. J.}
\newcommand{\aap}{Astron. Astroph}
\newcommand{\prd}{Phys. Rev. D}
\newcommand{\prl}{Phys. Rev. Lett.}
\newcommand{\MNRAS}{Mon. Not. Royal Astron. Soc.}
\newcommand{\mnras}{Mon. Not. Royal Astron. Soc.}
\newcommand{\araa}{Ann. Rev. Astron. Astroph.}
\newcommand{\nat}{Nature}
\newcommand{\apjs}{Astroph. J. Suppl.}

\begin{document}

\title{The Blackholic energy and the canonical Gamma-Ray Burst\thanks{Part I and Part II of these Lecture notes have been published respectively in \textit{COSMOLOGY AND GRAVITATION: X$^{th}$ Brazilian School of Cosmology and Gravitation; 25$^{th}$ Anniversary (1977-2002)}, M. Novello, S.E. Perez Bergliaffa (eds.), \textit{AIP Conf. Proc.}, \textbf{668}, 16 (2003), see \citet{rubr}, and in \textit{COSMOLOGY AND GRAVITATION: XI$^{th}$ Brazilian School of Cosmology and Gravitation}, M. Novello, S.E. Perez Bergliaffa (eds.), \textit{AIP Conf. Proc.}, \textbf{782}, 42 (2005), see \citet{rubr2}.}}

\author{Remo Ruffini}{address={ICRANet, Piazzale della Repubblica 10, 65122 Pescara, Italy.}, altaddress={Dipartimento di Fisica, Universit\`a di Roma ``La Sapienza'', Piazzale Aldo Moro 5, 00185 Roma, Italy.}}

\author{Maria Grazia Bernardini}{address={ICRANet, Piazzale della Repubblica 10, 65122 Pescara, Italy.}, altaddress={Dipartimento di Fisica, Universit\`a di Roma ``La Sapienza'', Piazzale Aldo Moro 5, 00185 Roma, Italy.}}

\author{Carlo Luciano Bianco}{address={ICRANet, Piazzale della Repubblica 10, 65122 Pescara, Italy.}, altaddress={Dipartimento di Fisica, Universit\`a di Roma ``La Sapienza'', Piazzale Aldo Moro 5, 00185 Roma, Italy.}}

\author{Letizia Caito}{address={ICRANet, Piazzale della Repubblica 10, 65122 Pescara, Italy.}, altaddress={Dipartimento di Fisica, Universit\`a di Roma ``La Sapienza'', Piazzale Aldo Moro 5, 00185 Roma, Italy.}}

\author{Pascal Chardonnet}{address={ICRANet, Piazzale della Repubblica 10, 65122 Pescara, Italy.}, altaddress={Universit\'e de Savoie, LAPTH - LAPP, BP 110, F-74941 Annecy-le-Vieux Cedex, France.}}

\author{Maria Giovanna Dainotti}{address={ICRANet, Piazzale della Repubblica 10, 65122 Pescara, Italy.}, altaddress={Dipartimento di Fisica, Universit\`a di Roma ``La Sapienza'', Piazzale Aldo Moro 5, 00185 Roma, Italy.}}

\author{Federico Fraschetti}{address={ICRANet, Piazzale della Repubblica 10, 65122 Pescara, Italy.}, altaddress={Centre CEA de Saclay (Essonne), Gif-sur-Yvette, 91191 cedex, France.}}

\author{Roberto Guida}{address={ICRANet, Piazzale della Repubblica 10, 65122 Pescara, Italy.}, altaddress={Dipartimento di Fisica, Universit\`a di Roma ``La Sapienza'', Piazzale Aldo Moro 5, 00185 Roma, Italy.}}

\author{Michael Rotondo}{address={ICRANet, Piazzale della Repubblica 10, 65122 Pescara, Italy.}, altaddress={Dipartimento di Fisica, Universit\`a di Roma ``La Sapienza'', Piazzale Aldo Moro 5, 00185 Roma, Italy.}}

\author{Gregory Vereshchagin}{address={ICRANet, Piazzale della Repubblica 10, 65122 Pescara, Italy.}, altaddress={Dipartimento di Fisica, Universit\`a di Roma ``La Sapienza'', Piazzale Aldo Moro 5, 00185 Roma, Italy.}}

\author{Luca Vitagliano}{address={ICRANet, Piazzale della Repubblica 10, 65122 Pescara, Italy.}, altaddress={Dip. di Matematica e Informatica, Universit\`a di Salerno, Via Ponte don Melillo, 84084, Fisciano (SA), Italy.}}

\author{She-Sheng Xue}{address={ICRANet, Piazzale della Repubblica 10, 65122 Pescara, Italy.}}

\begin{abstract}
Gamma-Ray Bursts (GRBs) represent very likely ``the'' most extensive computational, theoretical and observational effort ever carried out successfully in physics and astrophysics. The extensive campaign of observation from space based X-ray and $\gamma$-ray observatory, such as the \emph{Vela}, CGRO, BeppoSAX, HETE-II, INTEGRAL, \emph{Swift}, R-XTE, \emph{Chandra}, XMM satellites, have been matched by complementary observations in the radio wavelength (e.g. by the VLA) and in the optical band (e.g. by VLT, Keck, ROSAT). The net result is unprecedented accuracy in the received data allowing the determination of the energetics, the time variability and the spectral properties of these GRB sources. The very fortunate situation occurs that these data can be confronted with a mature theoretical development. Theoretical interpretation of the above data allows progress in three different frontiers of knowledge: {\bf a)} the ultrarelativistic regimes of a macroscopic source moving at Lorentz gamma factors up to $\sim 400$; {\bf b)} the occurrence of vacuum polarization process verifying some of the yet untested regimes of ultrarelativistic quantum field theories; and {\bf c)} the first evidence for extracting, during the process of gravitational collapse leading to the formation of a black hole, amounts of energies up to $10^{55}$ ergs of blackholic energy --- a new form of energy in physics and astrophysics. We outline how this progress leads to the confirmation of three interpretation paradigms for GRBs proposed in July 2001. Thanks mainly to the observations by \emph{Swift} and the optical observations by VLT, the outcome of this analysis points to the existence of a ``canonical'' GRB, originating from a variety of different initial astrophysical scenarios. The communality of these GRBs appears to be that they all are emitted in the process of formation of a black hole with a negligible value of its angular momentum. The following sequence of events appears to be canonical: the vacuum polarization process in the dyadosphere with the creation of the optically thick self accelerating electron-positron plasma; the engulfment of baryonic mass during the plasma expansion; adiabatic expansion of the optically thick ``fireshell'' of electron-positron-baryon plasma up to the transparency; the interaction of the accelerated baryonic matter with the interstellar medium (ISM). This leads to the canonical GRB composed of a proper GRB (P-GRB), emitted at the moment of transparency, followed by an extended afterglow. The sole parameters in this scenario are the total energy of the dyadosphere $E_{dya}$, the fireshell baryon loading $M_B$ defined by the dimensionless parameter $B\equiv M_Bc^2/E_{dya}$, and the ISM filamentary distribution around the source. In the limit $B\to 0$ the total energy is radiated in the P-GRB with a vanishing contribution in the afterglow. In this limit, the canonical GRBs explain as well the short GRBs. In these lecture notes we systematically outline the main results of our model comparing and contrasting them with the ones in the current literature. In both cases, we have limited ourselves to review already published results in refereed publications. We emphasize as well the role of GRBs in testing yet unexplored grounds in the foundations of general relativity and relativistic field theories.
\end{abstract}



\maketitle

\section{Introduction}

\begin{figure}
\centering 
\includegraphics[width=0.7\hsize,clip]{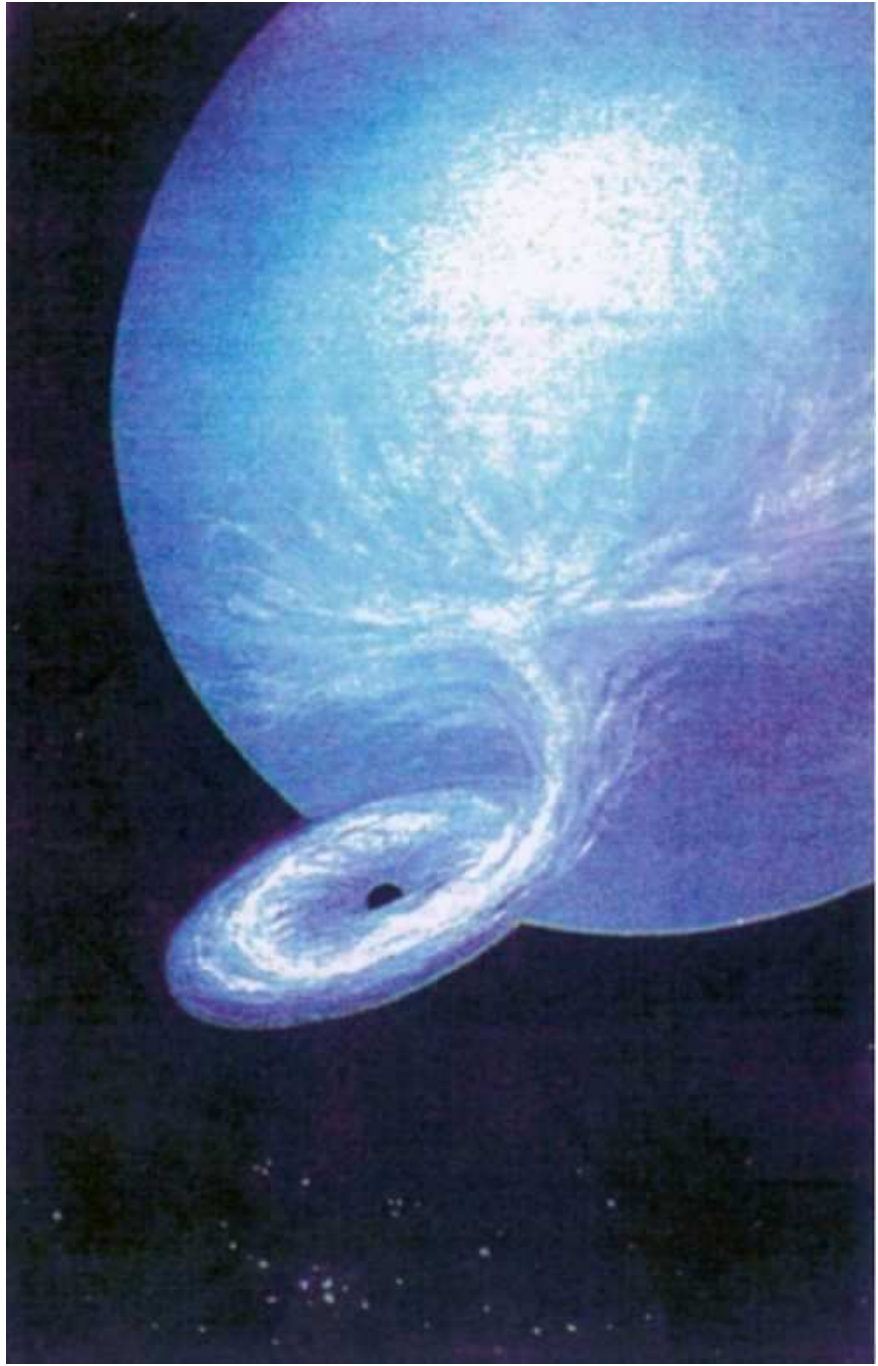}
\caption{Cygnus X-1, here represented in a artist view, offered the possibility of identifying the first black hole in our galaxy (\citet{lr73}). The luminosity $\Phi$ of $10^4$ solar luminosities points to the accretion process into a neutron star or a black hole as the energy source. The absence of pulsation is naturally explained either by a non-magnetized neutron star or a Kerr-Newman black hole, which has necessarily to be axially symmetric. What identifies the black hole unambiguously is that the mass of Cygnus X-1, observed to be larger than $9M_\odot$, exceeds the absolute upper limit of the neutron star mass, estimated at $3.2M_\odot$ by \citet{rr74}.}
\label{CygX1}
\end{figure}

The last century was characterized by three great successes in the field of astrophysics, each one linked to a different energy source:
\begin{enumerate}
\item Jean \citet{p20} and Arthur \citet{e20} were the first to point out, independently, that the nuclear fusion of four hydrogen nuclei into one helium nucleus could explain the energy production in stars. This idea was put on a solid theoretical base by Robert Atkinson and Fritz Houtermans \citep{ah29a,ah29b} using George Gamow's quantum theory of barrier penetration \citep{gh28} further developed by C.F. \citet{vw37,vw38}. The monumental theoretical work by Hans \citet{b39}, and later by \citet{bbfh57}, completed the understanding of the basic role of nuclear energy generated by fusion processes in explaining the energy source of main sequence stars (\citet{msch}).
\item Pulsars, especially NP0532 at the center of the Crab nebula, were discovered by Jocelyn Bell and Tony Hewish \citep{bh67}, and many theorists were actively trying to explain them as rotating neutron stars (see \citet{g68,g69,p68,fw68}). These had already been predicted by George Gamow using Newtonian physics \citep{g38} and by Robert Julius Oppenheimer and students using General Relativity \citep{os38,ov39,os39}. The crucial evidence confirming that pulsars were neutron stars came when their energetics was understood \citep{fw68}. The following relation was established from the observed pulsar period $P$ and its always positive first derivative $dP/dt$:
\begin{equation}
\left(\frac{dE}{dt}\right)_{obs} \simeq
4 \pi^2 \frac{I_{NS}}{P^3}\frac{dP}{dt}\, ,
\label{Pulsar}
\end{equation}
where $\left(\frac{dE}{dt}\right)_{obs}$ is the observed pulsar bolometric luminosity, $I_{NS}$ is its moment of inertia derived from the neutron star theory. This has to be related to the observed pulsar period. This equation not only identifies the role of neutron stars in explaining the nature of pulsars, but clearly indicates the rotational energy of neutron star as the pulsar energy source.
\item The birth of X-ray astronomy thanks to Riccardo Giacconi and his group (see e.g. \citet{gr78}) led to a still different energy source, originating from the accretion of matter onto a star which has undergone a complete gravitational collapse process: a black hole (see e.g. \citet{rw71}). In this case, the energetics is dominated by the radiation emitted in the accretion process of matter around an already formed black hole. Luminosities up to $10^4$ times the solar luminosity, much larger then the ones of pulsars, could be explained by the release of energy in matter accreting in the deep potential well of a black hole (\citet{lr73}). This allowed to probe for the first time the structure of circular orbits around a black hole computed by Ruffini and Wheeler (see e.g. \citet{ll2}). This result was well illustrated by the theoretical interpretation of the observations of Cygnus-X1, obtained by the Uhuru satellite and by the optical and radio telescopes on the ground (see Fig. \ref{CygX1}).
\end{enumerate}
These three results clearly exemplify how the identification of the energy source is the crucial factor in reaching the understanding of any astrophysical or physical phenomenon.

The discovery of Gamma-Ray Bursts (GRBs) may well sign a further decisive progress. GRBs can give in principle the first opportunity to probe and observe a yet different form of energy: the extractable energy of the black hole introduced in 1971 (\citet{cr71}), which we shall refer in the following as the blackholic energy\footnote{This name is the English translation of the Italian words ``energia buconerale'', introduced by Iacopo Ruffini, December 2004, here quoted by his kind permission.}. The blackholic energy, expected to be emitted during the dynamical process of gravitational collapse leading to the formation of the black hole, generates X- and $\gamma$-ray luminosities $10^{21}$ times larger than the solar luminosity, which manifest themselves in the GRB phenomenon. In the very short time they last, GRBs are comparable with the full electromagnetic luminosity of the entire visible universe.

\subsection{The discovery of GRBs by the Vela satellites and the early theoretical works}

We recall how GRBs were detected and studied for the first time using the {\em Vela} satellites, developed for military research to monitor the non-violation of the Limited Test Ban Treaty signed in 1963 (see e.g. \citet{s75}). It was clear from the early data of these satellites, which were put at $150,000$ miles from the surface of Earth, that the GRBs originated neither on the Earth nor in the Solar System. This discovery luckily occurred when the theoretical understanding of gravitationally collapsed objects, as well as the quantum electrodynamics of the vacuum polarization process, had already reached full maturity.

\begin{figure} 
\centering 
\includegraphics[width=0.55\hsize,clip]{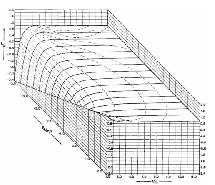}
\includegraphics[width=0.45\hsize,clip]{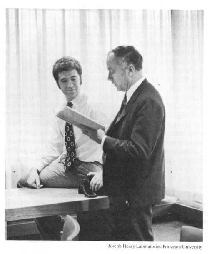}
\caption{The effective potential corresponding to the circular orbits in the equatorial plane of a black hole is given as a function of the angular momentum of the test particle. This digram was originally derived by Ruffini and Wheeler (right picture, reproduced with permission of the Joseph Henry Laboratories). For details see \citet{ll2} and \citet{ReesRufWhe}.}
\label{effp} 
\end{figure}

Three of the most important works in the field of general relativity have certainly been the discovery of the Kerr solution \citep{kerr}, its generalization to the charged case (\citet{newman}) and the formulation by Brandon Carter \citep{carter} of the Hamilton-Jacobi equations for a charged test particle in the metric and electromagnetic field of a Kerr-Newman solution (see e.g. \citet{ll2}). The equations of motion, which are generally second order differential equations, were reduced by Carter to a set of first order differential equations which were then integrated by using an effective potential technique by Ruffini and Wheeler for the Kerr metric (see e.g. \citet{ll2}) and by Ruffini for the Reissner-Nordstr{\o}m geometry (\citet{r70}, see Fig. \ref{effp}).

All the above mathematical results were essential for understanding the new physics of gravitationally collapsed objects and allowed the publication of a very popular article: ``Introducing the black hole'' (\citet{rw71}). In that paper, is was advanced the ansatz that the most general black hole is a solution of the Einstein-Maxwell equations, asymptotically flat and with a regular horizon: the Kerr-Newman solution. Such a solution is characterized only by three parameters: the mass $M$, the charge $Q$ and the angular momentum $L$. This ansatz of the ``black hole uniqueness theorem'' still today after thirty years presents challenges to the mathematical aspects of its complete proof (see e.g. \citet{ckf} and \citet{bcjr}). In addition to the challenges due to the above mathematical difficulties, in the field of physics this ansatz contains most profound consequences. The fact that, among all the possible highly nonlinear terms characterizing the gravitationally collapsed objects, only the ones corresponding solely to the Einstein Maxwell equations survive the formation of the horizon has, indeed, extremely profound physical implications. Any departure from such a minimal configuration either collapses on the horizon or is radiated away during the collapse process. This ansatz is crucial in identifying precisely a standard process of gravitational collapse leading to the formation of the black hole and the emission of GRBs. Indeed, in this specific case, the Born-like nonlinear \citep{b33} terms of the Heisenberg-Euler-Schwinger \citep{he35,s51} Lagrangian are radiated away prior to the formation of the horizon of the black hole (see e.g. \citet{rvx05}). Only the nonlinearity corresponding solely to the classical Einstein-Maxwell theory is left as the outcome of the gravitational collapse process.

The same effective potential technique (see \citet{ll2}), which allowed the analysis of circular orbits around the black hole, was crucial in reaching the equally interesting discovery of the reversible and irreversible transformations of black holes by \citet{cr71}, which in turn led to the mass-energy formula of the black hole:
\begin{equation} 
E_{BH}^2 = M^2c^4 = \left(M_{\rm ir}c^2 + \frac{Q^2}{2\rho_+}\right)^2+\frac{L^2c^2}{\rho_+^2}\, ,
\label{em} 
\end{equation} 
with 
\begin{equation} 
\frac{1}{\rho_+^4}\left(\frac{G^2}{c^8}\right)\left(Q^4+4L^2c^2\right)\leq 1\, , 
\label{s1}
\end{equation} 
where 
\begin{equation} 
S=4\pi\rho_+^2=4\pi(r_+^2+\frac{L^2}{c^2M^2})=16\pi\left(\frac{G^2}{c^4}\right) M^2_{\rm ir}\, ,
\label{sa} 
\end{equation} 
is the horizon surface area, $M_{\rm ir}$ is the irreducible mass, $r_{+}$ is the horizon radius and $\rho_+$ is the quasi-spheroidal cylindrical coordinate of the horizon evaluated at the equatorial plane. Extreme black holes satisfy the equality in Eq.(\ref{s1}).

From Eq.(\ref{em}) follows that the total energy of the black hole $E_{BH}$ can be split into three different parts: rest mass, Coulomb energy and rotational energy. In principle both Coulomb energy and rotational energy can be extracted from the black hole (\citet{cr71}). The maximum extractable rotational energy is 29\% and the maximum extractable Coulomb energy is 50\% of the total energy, as clearly follows from the upper limit for the existence of a black hole, given by Eq.(\ref{s1}). We refer in the following to both these extractable energies as the blackholic energy.

The existence of the black hole and the basic correctness of the circular orbit binding energies has been proven by the observations of Cygnus-X1 (see e.g. \citet{gr78}). However, as already mentioned in binary X-ray sources, the black hole uniquely acts passively by generating the deep potential well in which the accretion process occurs. It has become tantalizing to look for astrophysical objects in order to verify the other fundamental prediction of general relativity that the blackholic energy is the largest energy extractable from any physical object.

We also recall that the feasibility of the blackholic energy extraction has been made possible by the quantum processes of creating, out of classical fields, a plasma of electron-positron pairs in the field of black holes. \citet{he35} clearly evidenced that a static electromagnetic field stronger than the critical value: 
\begin{equation} 
E_c = \frac{m_e^2c^3}{\hbar e} 
\label{ec} 
\end{equation} 
can polarize the vacuum and create electron-positron pairs. As we illustrate in the next sections, the major effort in verifying the correctness of this theoretical prediction has been directed in the analysis of heavy ion collisions (see \citet{rvx05} and references therein). From an order-of-magnitude estimate, it would appear that around a nucleus with a charge: 
\begin{equation} 
Z_c \simeq \frac{\hbar c}{e^2} \simeq 137 
\label{zc} 
\end{equation} 
the electric field can be as strong as the electric field polarizing the vacuum. As we show in the next sections, a more accurate detailed analysis taking into account the bound states levels around a nucleus brings to a value of
\begin{equation} 
Z_c \simeq 173
\label{zc2} 
\end{equation}  
for the nuclear charge leading to the existence of a critical field. From the Heisenberg uncertainty principle it follows that, in order to create a pair, the existence of the critical field should last a time
\begin{equation} 
\Delta t \sim \frac{\hbar}{m_e c^2} \simeq 10^{-18}\, \mathrm{s}\, ,
\label{dt} 
\end{equation} 
which is much longer then the typical confinement time in heavy ion collisions which is 
\begin{equation} 
\Delta t \sim \frac{\hbar}{m_p c^2} \simeq 10^{-21}\, \mathrm{s}\, .
\label{dt2} 
\end{equation} 
This is certainly a reason why no evidence for pair creation in heavy ion collisions has been obtained although remarkable effort has been spent in various accelerators worldwide. Similar experiments involving laser beams meet with analogous difficulties (see e.g. \citet{rvx05} and next sections).

In 1975 \citet{dr75} advanced the alternative idea that the critical field condition given in Eq.(\ref{ec}) could be easily reached, and for a time much larger than the one given by Eq.(\ref{dt}), in the field of a Kerr-Newman black hole in a range of masses $3.2M_\odot \le M_{BH} \le 7.2\times 10^6M_\odot$. In that paper there was generalized to the curved Kerr-Newman geometry the fundamental theoretical framework developed in Minkowski space by \citet{he35} and \citet{s51}. This result was made possible by the work on the structure of the Kerr-Newman spacetime previously done by \citet{carter} and by the remarkable mathematical craftsmanship of Thibault Damour then working with one of us (RR) as a post-doc in Princeton. We give on this topic some additional details in the next sections.

The maximum energy extractable in such a process of creating a vast amount of electron-positron pairs around a black hole is given by:
\begin{equation} 
E_{max} = 1.8\times 10^{54} \left(M_{BH}/M_\odot\right)\, \mathrm{erg}\, \mathrm{.} 
\label{emax} 
\end{equation} 
We concluded in that paper that such a process ``naturally leads to a most simple model for the explanation of the recently discovered $\gamma$-rays bursts''.

At that time, GRBs had not yet been optically identified and nothing was known about their distance and consequently about their energetics. Literally thousands of theories existed in order to explain them and it was impossible to establish a rational dialogue with such an enormous number of alternative theories (see \citet{ruKl}). As we will see, this situation was drastically modified by the observations of BeppoSAX.

\subsection{The role of the CGRO and BeppoSAX satellites and the further theoretical developments}

\begin{figure} 
\centering 
\includegraphics[height=\hsize,clip,angle=90]{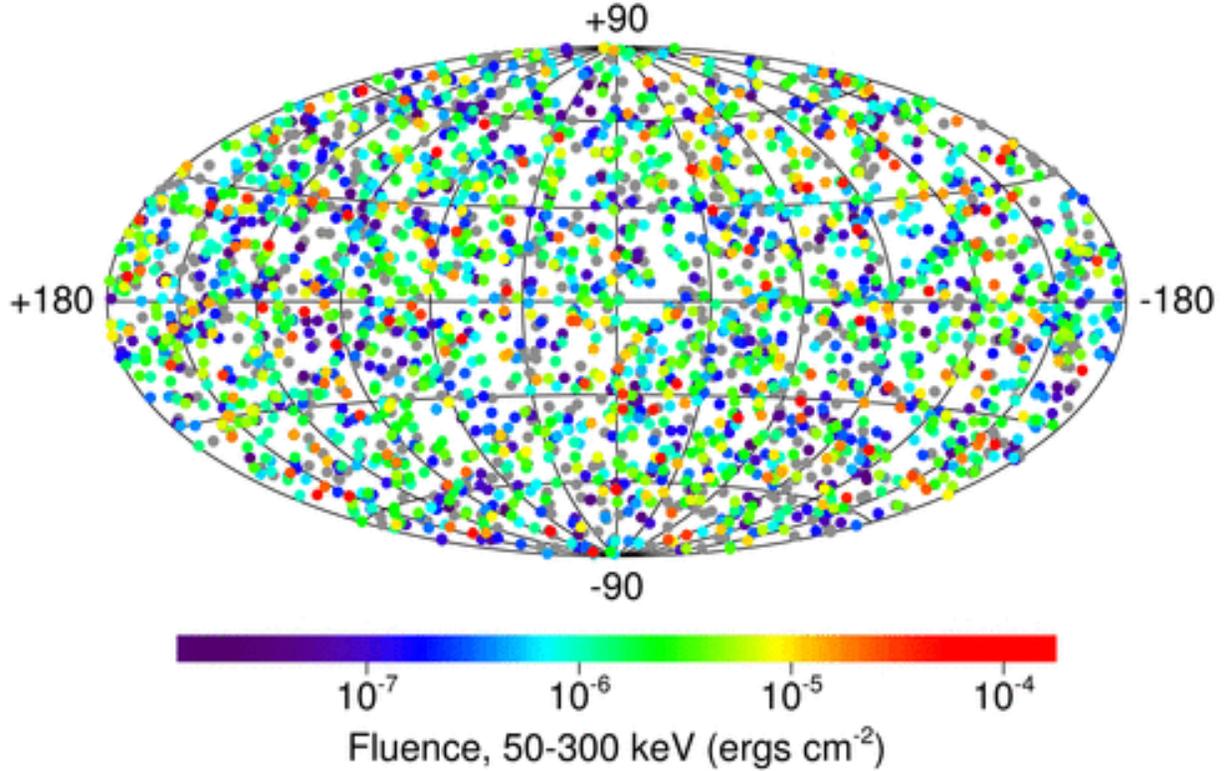} 
\caption{Position in the sky, in galactic coordinates, of 2000 GRB events seen by the CGRO satellite. Their isotropy is evident. Reproduced from BATSE web site by their courtesy.}
\label{batse2k} 
\end{figure} 

\begin{figure} 
\centering 
\includegraphics[width=\hsize,clip]{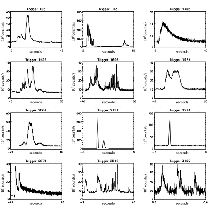} 
\caption{Some GRB light curves observed by the BATSE instrument on board of the CGRO satellite.}
\label{grb_profiles_eng} 
\end{figure}

\begin{figure} 
\centering 
\includegraphics[width=\hsize,clip]{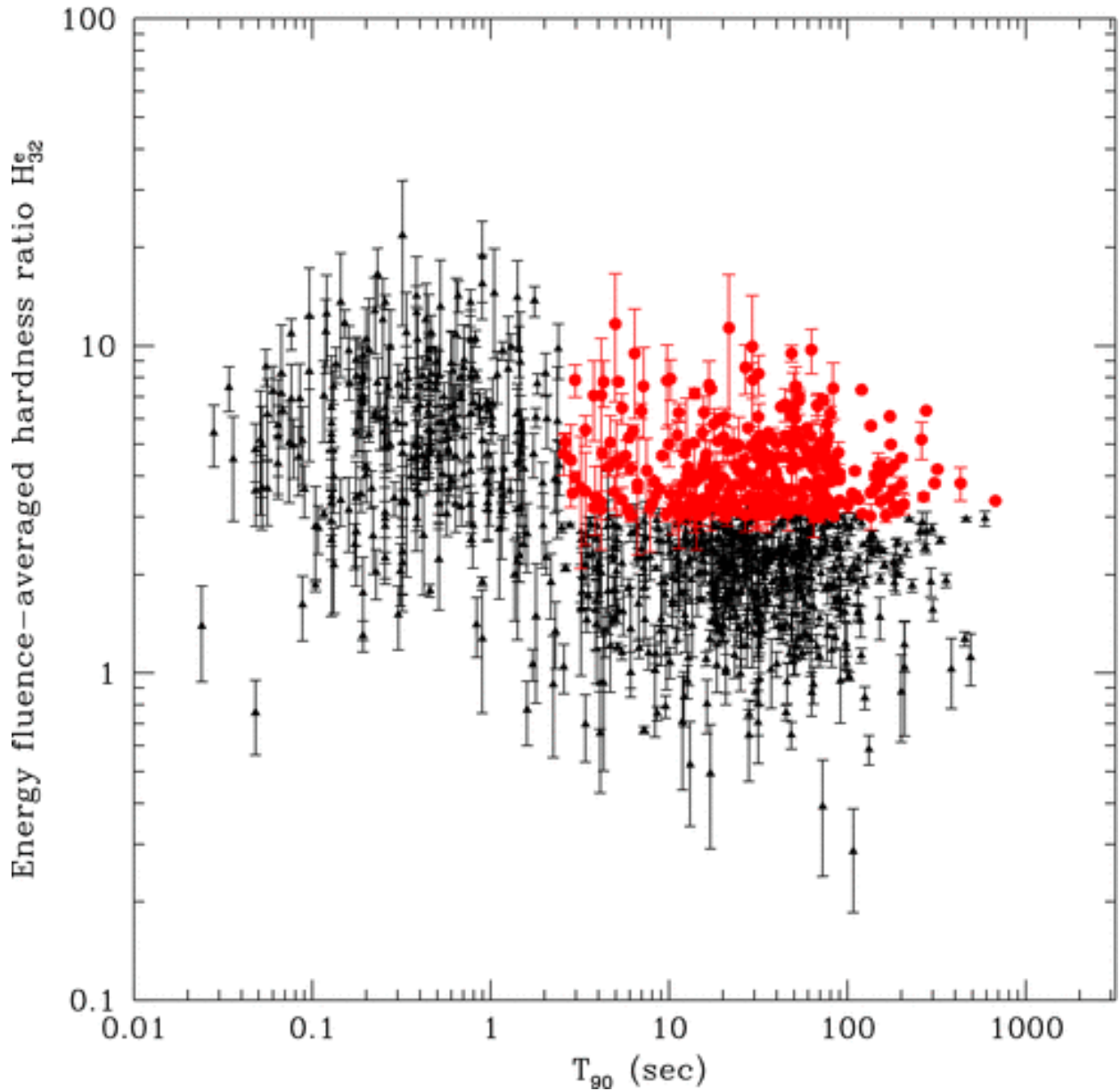} 
\caption{The energy fluence-averaged hardness ratio for short ($T < 1$ s) and long ($T> 1$ s) GRBs are represented. Reproduced, by his kind permission, from \citet{t98} where the details are given.}
\label{tavani} 
\end{figure}

\begin{figure} 
\centering 
\includegraphics[width=\hsize,clip]{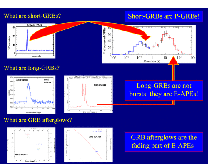} 
\caption{Status of GRB observations following the BATSE and BeppoSAX observations: On the upper right part of the figure are plotted the number of the observed GRBs as a function of their duration. The bimodal distribution corresponding respectively to the short bursts, upper left figure, and the long bursts, middle figure, is quite evident. The afterglow component is represented in the lowest figures. The theoretical goal is to find a coherent astrophysical explanation for all these different phenomena.}
\label{slb} 
\end{figure}

The mystery of GRBs became deeper as the observations of the BATSE instrument on board of the Compton Gamma-Ray Observatory (CGRO) satellite\footnote{see http://cossc.gsfc.nasa.gov/batse/} over $9$ years proved the isotropy of these sources in the sky (See Fig. \ref{batse2k}). In addition to these data, the CGRO satellite gave an unprecedented number of details on the GRB structure, on their spectral properties and time variabilities which have been collected in the fourth BATSE catalog (\citet{batse4b}, see e.g. Fig. \ref{grb_profiles_eng}). Analyzing these BATSE sources it soon became clear (see e.g. \citet{ka93,t98}) the existence of two distinct families of sources: the long bursts, lasting more then one second and softer in spectra, and the short bursts (see Fig. \ref{slb}), harder in spectra (see Fig. \ref{tavani}). We shall return shortly on this topic.

The situation drastically changed with the discovery of the afterglow by the Italian-Dutch satellite BeppoSAX (\citet{ca97}). Such a discovery led to the optical identification of the GRBs by the largest telescopes in the world, including the Hubble Space Telescope, the Keck Telescope in Hawaii and the VLT in Chile, and allowed as well the identification in the radio band of these sources. The outcome of this collaboration between complementary observational technique made possible in 1997 the identification of the distance of these sources from the Earth and of their tremendous energy of the order up to $10^{54}$ erg/second during the burst, which indeed coincides with the theoretical prediction made by \citet{dr75} given in Eq.\eqref{emax}.

The resonance between the X- and gamma ray astronomy from the satellites and the optical and radio astronomy from the ground, had already marked the great success and development of the astrophysics of binary X-ray sources in the seventies (see e.g. \citet{gr78}). This resonance is re-proposed here for GRBs on a much larger scale. The use of much larger satellites, like Chandra and XMM-Newton, and specific space missions, like HETE-II and \emph{Swift}, together with the very lucky circumstance of the coming of age of the development of optical technologies for the telescopes, such as Keck in Hawaii and VLT in Chile, offers today opportunities without precedence in the history of mankind.

Turning now to the theoretical progresses, it is interesting that the idea of using an electron-positron plasma as a basis of a GRB model, introduced in \citet{dr75}, was independently considered years later in a set of papers by \citet{cr78}, \citet{ch81} and \citet{hc83}. However, these authors did not address the issue of the physical origin of their energy source. They reach their conclusions considering the pair creation and annihilation process occurring in the confinement of a large amount of energy in a region of dimension $\sim 10$ km typical of a neutron star. No relation to the physics of black holes nor to the energy extraction process from a black hole was envisaged in their interesting considerations, mainly directed to the study of the creation and consequent evolution of such an electron-positron plasma.

After the discovery of the afterglows and the optical identification of GRBs at cosmological distances, implying exactly the energetics predicted in Eq.(\ref{emax}), we returned to the analysis of the vacuum polarization process around a black hole and precisely identified the region around the black hole in which the vacuum polarization process and the consequent creation of electron-positron pairs occur. We defined this region, using the Greek name for pairs ($\delta\upsilon\acute{\alpha}\varsigma$, $\delta\upsilon\acute{\alpha}\delta o\varsigma$), to be the ``dyadosphere'' of the black hole, bounded  by the black hole horizon and the dyadosphere radius $r_{ds}$ given by (see \citet{rukyoto}, \citet{prx98}):
\begin{equation} 
r_{ds}=\left(\frac{\hbar}{mc}\right)^\frac{1}{2}\left(\frac{GM}{ 
c^2}\right)^\frac{1}{2} \left(\frac{m_{\rm p}}{m}\right)^\frac{1}{2}\left(\frac{e}{q_{\rm p}}\right)^\frac{1}{2}\left(\frac{Q}{\sqrt{G}M}\right)^\frac{1}{2}=1.12\cdot 10^8\sqrt{\mu\xi} \, {\rm cm}, 
\label{rc} 
\end{equation} 
where we have introduced the dimensionless mass and charge parameters $\mu={M_{BH}/M_{\odot}}$, $\xi={Q/(M_{BH}\sqrt{G})}\le 1$. The total energy of the electron positron pairs, $E_{e^\pm}^{tot}$ is equal to the dyadosphere energy $E_{dya}$.

Our GRB model, like all prevailing models in the existing literature (see e.g. \citet{p04,m02,m06} and references therein), is based on the acceleration of an optically thick electron-positron plasma. The mechanism responsible for the origin and the energetics of such a plasma, either in relation to black hole physics or to other physical processes, has often been discussed qualitatively in the GRB scientific literature but never quantitatively with explicit equations. The concept of the dyadosphere (\citet{rukyoto,prx98}) is the only attempt, as far as we know, to do this. It relates such an electron-positron plasma to black hole physics and to the features of the GRB progenitor star, using explicit equations that satisfy the existing physical laws (see e.g. \citet{cr71,rubr2} and references therein, see also \citet{mtw73}). This step is essential if one wishes to identify the physical origin and energetics of GRBs. All the successive evolution of the electron-positron plasma are independent on this step and are indeed common to all prevailing GRB models in the literature. Of course, great differences still exists between the actual treatments of this evolution in the current literature, as we show in the next sections.

Analogies exist between the concept of dyadosphere and the work of \citet{cr78}, as well as marked conceptual differences. In the dyadosphere the created electron-positron pairs are assumed to reach thermal equilibrium and have an essential role in the dynamical acceleration process of GRBs. In \citet{cr78} it is assumed that the created electron-positron pairs do annihilate in a cascade process in a very short bremmsstrahlung time scale: they cannot participate in any way to the dynamical phases of the GRB process. It is interesting that these differences can be checked both theoretically and observationally. It should be possible, in the near future, to evaluate all the cross sections involved by the above annihilation processes and assess by a direct explicit analysis which one of the two above approaches is the correct one. On the other side, such two approaches certainly lead to very different predictions for the GRB structure, especially for the short ones. These predictions will certainly be compared to observations in the near future.

We have already emphasized that the study of GRBs is very likely ``the'' most extensive computational and theoretical investigation ever done in physics and astrophysics. There are at least three different fields of research which underlie the foundation of the theoretical understanding of GRBs. All three, for different reasons, are very difficult.

The first field of research is special relativity. As one of us (RR) always mention to his students in the course of theoretical physics, this field is paradoxically very difficult since it is extremely simple. In approaching special relativistic phenomena the extremely simple and clear procedures expressed by Einstein in his 1905 classic paper \citep{e05} are often ignored. Einstein makes use in his work of very few physical assumptions, an almost elementary mathematical framework and gives constant attention to a proper operational definition of all observable quantities. Those who work on GRBs use at times very intricate, complex and often wrong theoretical approaches lacking the necessary self-consistency. This is well demonstrated in the current literature on GRBs.

The second field of research essential for understanding the energetics of GRBs deals with quantum electrodynamics and the relativistic process of pair creation in overcritical electromagnetic fields as well as in very high density photon gas. This topic is also very difficult but for a quite different conceptual reason: the process of pair creation, expressed in the classic works of Heisenberg-Euler-Schwinger \citep{he35,s51} later developed by many others, is based on a very powerful theoretical framework but has not yet been verified by experimental data. Similarly, the creation of electron-positron pairs from high density and high energy photons lacks still today the needed theoretical description. As we will show in the next sections, there is the tantalizing possibility of observing these phenomena, for the first time, in the astrophysical setting of GRBs on a more grandiose scale.

There is a third field which is essential for the understanding of the GRB phenomenon: general relativity. In this case, contrary to the case of special relativity, the field is indeed very difficult, since it is very difficult both from a conceptual, technical and mathematical point of view. The physical assumptions are indeed complex. The entire concept of geometrization of physics needs a  new conceptual approach to the field. The mathematical complexity of the pseudo-Riemannian geometry contrasts now with the simple structure of the pseudo-Euclidean Minkowski space. The operational definition of the observable quantities has to take into account the intrinsic geometrical properties and also the cosmological settings of the source. With GRBs we have the possibility to follow, from a safe position in an asymptotically flat space at large distance, the formation of a black hole horizon with all the associated relativistic phenomena of light bending and time dilatation. Most important, as we will show in details in the next sections, general relativity in connection with quantum phenomena offers, with the blackholic energy, the explanation of the tremendous GRB energy sources and the possibility to follow in great details the black hole formation.

For these reasons GRBs offer an authentic new frontier in the field of physics and astrophysics. We recall that in the special relativity field, for the first time, we observe phenomena occurring at Lorentz gamma factors of approximately $400$. In the field of relativistic quantum electro-dynamics we see for the first time the interchange between classical fields and high density photon fields with the created quantum matter-antimatter pairs. In  the field of general relativity also for the first time we can test the blackholic energy which is the basic energetic physical variable underlying the entire GRB phenomenon.

The most appealing aspect of this work is that, if indeed these three different fields are treated and approached with the necessary technical and scientific maturity, the model which results has a very large redundancy built-in. The approach requires an unprecedented level of self-consistency. Any departures from the correct theoretical treatment in this very complex system lead to exponential departures from the correct solution and from the correct fit of the observations. 

It is so that, as the model is being properly developed and verified, its solution will have existence and uniqueness. In order to build a theoretical GRB model, we have found necessary to establish clear guidelines by introducing three basic paradigms for the interpretation of GRBs.

\subsection{The first paradigm: The Relative Space-Time Transformation (RSTT) paradigm}

The ongoing dialogue between our work and the one of the workers on GRBs, rests still on some elementary considerations presented by Einstein in his classic article of 1905 \citep{e05}. These considerations are quite general and even precede Einstein's derivation, out of first principles, of the Lorentz transformations. We recall here Einstein's words: ``We might, of course, content ourselves with time values determined by an observer stationed together with the watch at the origin of the co-ordinates, and co-ordinating the corresponding positions of the hands with light signals, given out by every event to be timed, and reaching him through empty space. But this co-ordination has the disadvantage that it is not independent of the standpoint of the observer with the watch or clock, as we know from experience''. 

\begin{figure} 
\centering 
\includegraphics[width=\hsize,clip]{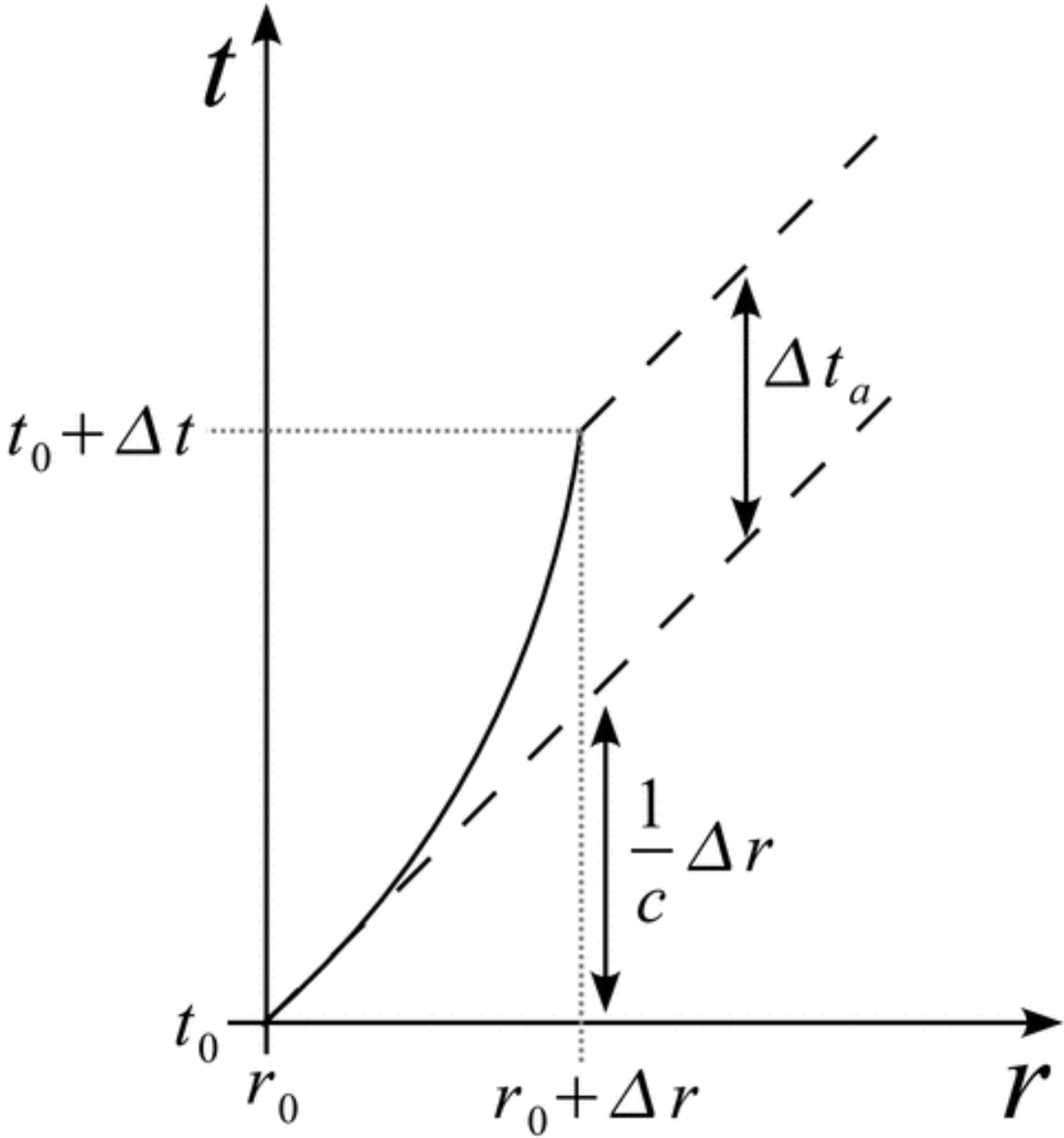} 
\caption{Relation between the arrival time $t_a$ and the laboratory time $t$. Details in \citet{lett1,rubr}.} 
\label{ttasch_new_bn} 
\end{figure} 

The message by Einstein is simply illustrated in Fig. \ref{ttasch_new_bn}. If we consider in an inertial frame a source (solid line) moving with high speed and emitting light signals (dashed lines) along the direction of its motion, a far away observer will measure a delay $\Delta t_a$ between the arrival time of two signals respectively emitted at the origin and after a time interval $\Delta t$ in the laboratory frame, which in our case is the frame where the black hole is at rest. The real velocity of the source is given by: 
\begin{equation} 
v = \frac{\Delta r}{\Delta t} 
\label{v} 
\end{equation} 
and the apparent velocity is given by: 
\begin{equation} 
v_{app} = \frac{\Delta r}{\Delta t_a}\, , 
\label{vapp} 
\end{equation} 
As pointed out by Einstein the adoption of coordinating light signals simply by their arrival time as in Eq.(\ref{vapp}), without an adequate definition of synchronization, is incorrect and leads to unsurmountable difficulties as well as to apparently ``superluminal'' velocities as soon as motions close to the speed of light are considered.

The use of $\Delta t_a$ as a time coordinate, often tacitly adopted by astronomers, should be done, if at all, with proper care. The relation between $\Delta t_a$ and the correct time parameterization in the laboratory frame has to be taken into account:
\begin{equation} 
\Delta t_a = \Delta t - \frac{\Delta r}{c} = \Delta t - 
\frac{1}{c}\int_{t_\circ}^{t_\circ + \Delta t}{v\left(t'\right) dt'}\, . 
\label{tadef} 
\end{equation} 
In other words, the relation between the arrival time and the laboratory time cannot be done without a knowledge of the speed along the \emph{entire} world-line of the source. In the case of GRBs, such a worldline starts at the moment of gravitational collapse. It is of course clear that the parameterization in the laboratory frame has to take into account the cosmological redshift $z$ of the source. We then have, at the detector:
\begin{equation}
\Delta t_a^d = \left(1+z\right) \Delta t_a\, .
\label{taddef}
\end{equation}

In the current GRB literature, Eq.(\ref{tadef}) has been systematically neglected by addressing only the afterglow description neglecting the previous history of the source. Often the integral equation has been approximated by a clearly incorrect instantaneous value: 
\begin{equation} 
\Delta t_a \simeq \frac{\Delta t}{2\gamma^2}\, . 
\label{taapp} 
\end{equation}
The attitude has been adopted to consider separately the afterglow part of the GRB phenomenon, without the knowledge of the entire equation of motion of the source. 

This point of view has reached its most extreme expression in the works reviewed by \citet{p99,p00}, where the so-called ``prompt radiation'', lasting on the order of $10^2$ s, is considered as a burst emitted by the prolonged activity of an ``inner engine''. In these models, generally referred to as the ``internal shock model'', the emission of the afterglow is assumed to follow the ``prompt radiation'' phase (\citet{rm94,px94,sp97,f99,fcrsyn99}).

As we outline in the following sections, such an extreme point of view originates from the inability of obtaining the time scale of the ``prompt radiation'' from a burst structure. These authors consequently appeal to the existence of an ``ad hoc'' inner engine in the GRB source to solve this problem.

We show in the following sections how this difficulty has been overcome in our approach by interpreting the ``prompt radiation'' as an integral part of the afterglow and {\em not} as a burst. This explanation can be reached only through a relativistically correct theoretical description of the entire afterglow (see next sections). Within the framework of special relativity we show that it is not possible to describe a GRB phenomenon by disregarding the knowledge of the entire past worldline of the source. We show that at $10^2$ seconds the emission occurs from a region of dimensions of approximately $10^{16}$ cm, well within the region of activity of the afterglow. This point was not appreciated in the current literature due to the neglect of the apparent superluminal effects implied by the use of the ``pathological'' parametrization of the GRB phenomenon by the arrival time of light signals.

We can now turn to the first paradigm, the relative space-time transformation (RSTT) paradigm (\citet{lett1}) which emphasizes the importance of a global analysis of the GRB phenomenon encompassing both the optically thick and the afterglow phases. Since all the data are received in the detector arrival time it is essential to know the equations of motion of all relativistic phases with $\gamma > 1$ of the GRB sources in order to reconstruct the time coordinate in the laboratory frame, see Eq.\eqref{tadef}. Contrary to other phenomena in nonrelativistic physics or astrophysics, where every phase can be examined separately from the others, in the case of GRBs all the phases are inter-related by their signals received in arrival time $t_a^d$. There is the need, in order to describe the physics of the source, to derive the laboratory time $t$ as a function of the arrival time $t_a^d$ along the entire past worldline of the source using Eq.\eqref{taddef}.

An additional difference, also linked to special relativity, between our treatment and the ones in the current literature relates to the assumption of the existence of scaling laws in the afterglow phase: the power law dependence of the Lorentz gamma factor on the radial coordinate is usually systematically assumed. From the proper use of the relativistic transformations and by the direct numerical and analytic integration of the special relativistic equations of motion we demonstrate (see next sections) that no simple power-law relation can be derived for the equations of motion of the system. This situation is not new for workers in relativistic theories: scaling laws exist in the extreme ultrarelativistic regimes and in the Newtonian ones but not in the intermediate fully relativistic regimes (see e.g. \citet{r70}).

\subsection{The second paradigm: The Interpretation of the Burst Structure (IBS) paradigm}

We turn now to the second paradigm, which is more complex since it deals with all the different phases of the GRB phenomenon. We first address the dynamical phases following the dyadosphere formation.

After the vacuum polarization process around a black hole, one of the topics of the greatest scientific interest is the analysis of the dynamics of the electron-positron plasma formed in the dyadosphere. This issue was addressed by us in a collaboration with Jim Wilson at Livermore. The numerical simulations of this problem were developed at Livermore, while the semi-analytic approach was developed in Rome (see \citet{rswx99,rswx00} and next sections).

The corresponding treatment in the framework of the Cavallo, Rees et al. analysis was performed by \citet{psn93} also using a numerical approach, by \citet{bm95} using an analytic approach and by \citet{mlr93} using a numerical and semi-analytic approach.

Although some analogies exists between these treatments, they are significantly different in the theoretical details and in the final results (see \citet{brvx06} and next sections). Since the final result of the GRB model is extremely sensitive to any departure from the correct treatment, it is indeed very important to detect at every step the appearance of possible fatal errors.

\subsubsection{The optically thick phase of the fireshell}

A conclusion common to all these treatments is that the electron-positron plasma is initially optically thick and expands till transparency reaching very high values of the Lorentz gamma 
factor. A second point, which is common, is the discovery of a new clear feature: the plasma shell expands but the Lorentz contraction is such that its width in the laboratory frame appears to be constant. This self acceleration of the thin shell is the distinguishing factor of GRBs, conceptually very different from the physics of a fireball developed by the inner pressure of an atomic bomb explosion in the Earth's atmosphere. In the case of GRBs the region interior to the shell is inert and with pressure totally negligible: the entire dynamics occurs on the shell itself. For this reason, we refer in the following to the self accelerating shell as the ``fireshell''.

\begin{figure}
\begin{minipage}{\hsize}
\includegraphics[width=\hsize,clip]{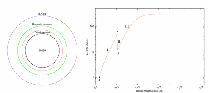}\\
\includegraphics[width=\hsize,clip]{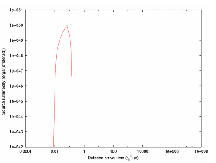}
\end{minipage}
\caption{{\bf Above:} The optically thick phase of the fireshell evolution are qualitatively represented in this diagram. There are clearly recognizable 1) the PEM pulse phase, 2) the impact on the baryonic remnant, 3) the PEMB pulse phase and the final approach to transparency with the emission of the P-GRB. Details in \citet{rubr}. {\rm Below:} The P-GRB emitted at the transparency point at a time of arrival $t_a^d$ which has been computed following the prescriptions of Eq.\eqref{tadef}. Details in \citet{lett2,rubr}.}
\label{cip2}
\end{figure}

There is a major difference between our approach and the ones of Piran, M\'esz\'aros and Rees, in that the dyadosphere is assumed by us to be initially filled uniquely with an electron-positron plasma. Such a plasma expands in substantial agreement with the results presented in the work of \citet{bm95}. In our model the fireshell of electron-positron pairs and photons (PEM pulse, see \citet{rswx99}) evolves and encounters the remnant of the star progenitor of the newly formed black hole. The fireshell is then loaded with baryons. A new fireshell is formed of electron-positron-photons and baryons (PEMB pulse, see \citet{rswx00}) which expands all the way until transparency is reached. At transparency the emitted photons give origin to what we define as the Proper-GRB (P-GRB, see \citet{lett2} and Fig. \ref{cip2}).

In our approach, the baryon loading is measured by a dimensionless quantity
\begin{equation}
B = \frac{M_B c^2}{E_{dya}}\, ,
\label{Bdef}
\end{equation}
which gives direct information about the mass $M_B = N_B m_p$ of the remnant, where $m_p$ is the proton mass. The corresponding treatment done by Piran and collaborators (\citet{sp90,psn93}) and by \citet{mlr93} differs in one important respect: the baryonic loading is assumed to occur since the beginning of the electron-positron pair formation and no relation to the mass of the remnant of the collapsed progenitor star is attributed to it.

A further difference also exists between our description of the rate equation for the electron-positron pairs and the ones by those authors. While our results are comparable with the ones obtained by Piran under the same initial conditions, the set of approximations adopted by \citet{mlr93} appears to be too radical and leads to different results violating energy and momentum conservation (see next sections and \citet{brvx06}).

From our analysis (\citet{rswx00}) it also becomes clear that such expanding dynamical evolution can only occur for values of $B \le 10^{-2}$. This prediction, as we will show shortly in the many GRB sources considered, is very satisfactorily confirmed by observations.

From the value of the $B$ parameter, related to the mass of the remnant, it therefore follows that the collapse to a black hole leading to a GRB is drastically different from the collapse to a neutron star. While in the case of a neutron star collapse a very large amount of matter is expelled, in many instances well above the mass of the neutron star itself, in the case of black holes leading to a GRB only a very small fraction of the initial mass ($\sim 10^{-2}$ or less) is expelled. The collapse to a black hole giving rise to a GRB appears to be much smoother than any collapse process considered until today: almost 99.9\% of the star has to be collapsing simultaneously!

We summarize in Fig. \ref{cip2} the optically thick phase of the fireshell evolution: we start from a given dyadosphere of energy $E_{dya}$; the fireshell self-accelerates outward; an abrupt decrease in the value of the Lorentz gamma factor occurs by the engulfment of the baryonic loading followed by a further self-acceleration until the fireshell becomes transparent.

The photon emission at this transparency point is the P-GRB. An accelerated beam of baryons with an initial Lorentz gamma factor $\gamma_\circ$ starts to interact with the interstellar medium at typical distances from the black hole of $r_\circ \sim 10^{14}$ cm and at a photon arrival time at the detector on the Earth surface of $t_a^d \sim 0.1$ s. These values determine the initial conditions of the afterglow.

We dedicate three sections to outline more closely some of the work we perform and to compare and contrast it with the ones in the current literature. 

In section ``The fireshell in the Livermore code'' we recall the basic hydrodynamics and rate equation for the electron-positron plasma and then we outline the numerical code used to evolve the spherically symmetric general relativistic hydrodynamic equations starting from the dyadosphere. Such a code was not used by us but had already been developed independently for more general astrophysical scenarios by Jim Wilson and Jay Salmonson at the Lawrence Livermore National Laboratory (see \citet{wsm97,wsm98}). In our collaboration, the Livermore code has been used in order to validate the correct choice among a variety of different semi-analytic models developed at the University of Rome ``La Sapienza''.

\begin{figure}
\includegraphics[width=\hsize,clip]{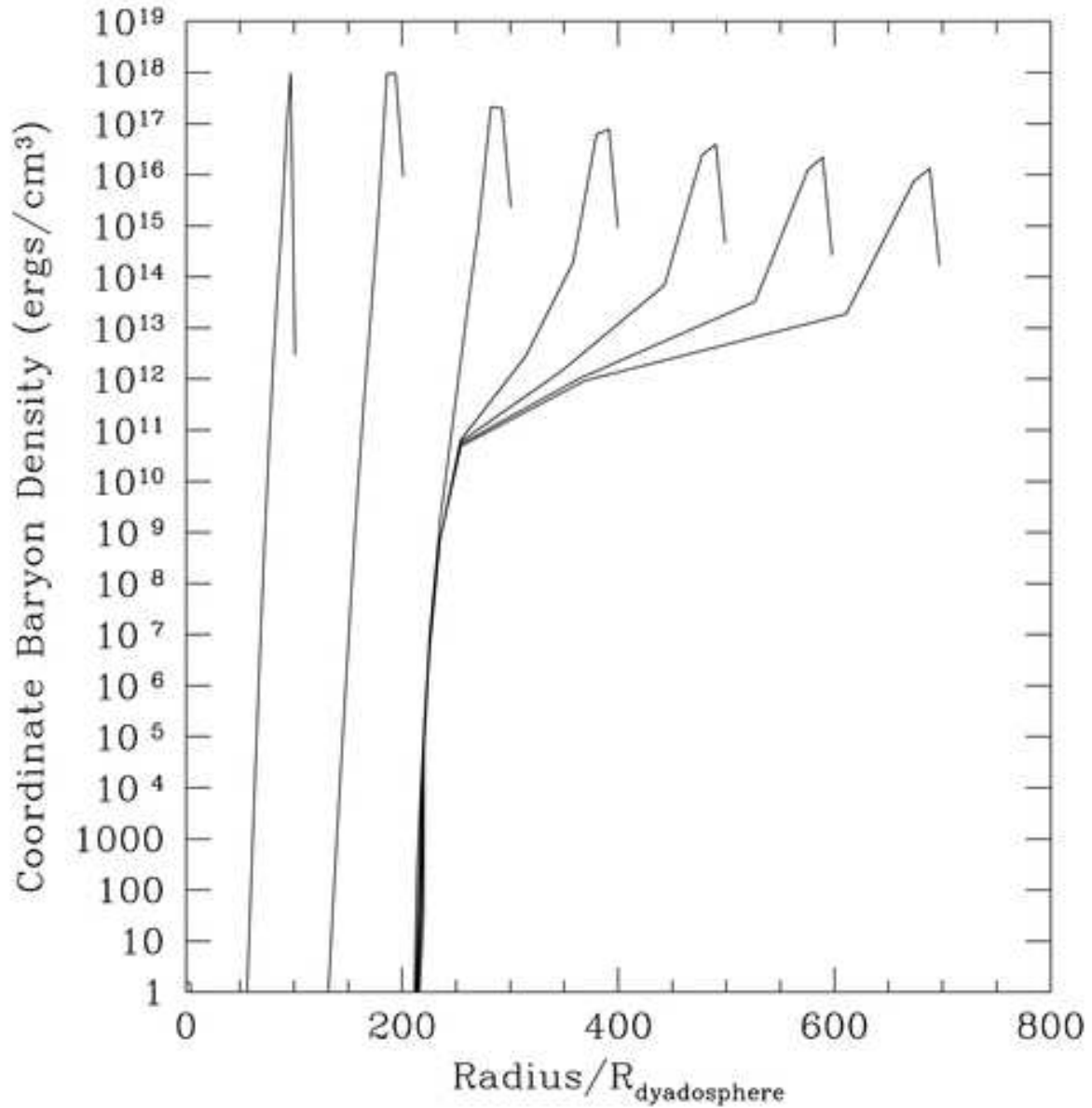}
\caption{A sequence of snapshots of coordinate baryon energy density is shown from the one dimensional hydrodynamic calculations of the Livermore code. The radial coordinate is given in units of dyadosphere radii ($r_{ds}$). At $r \simeq 100 r_{ds}$ there is located a baryonic matter shell corresponding to a baryon loading $B = 1.3\times 10^{-2}$. For this baryon shell mass we see a significant departure from the constant thickness solution for the fireshell dynamics and a clear instability occurs. Details in \citet{rswx00}. As we will see, this result, peculiar of our treatment, will play a major role in the theoretical interpretation of GRBs.}
\label{B10-2}
\end{figure}

In section ``The fireshell in the Rome code'' we first recall the co-variant energy-momentum tensor and the thermodynamic quantities used to describe the electron-positron plasma as well as their expression as functions of Fermi integrals. The thermodynamic equilibrium of the photons and the electron-positron pairs is initially assumed at temperature larger than $e^+e^-$ pairs creation threshold ($T > 1$ MeV). The numerical code implementing entropy and energy conservations as well as the rate equation for the electron-positron pairs is outlined. We recall, as well, the simulation of different geometries assumed for the fireshell and the essential role of the Livermore code in selecting the correct one among these different possibilities for the dynamics of this plasma composed uniquely of electron, positron and photons (PEM pulse). The correct solution resulted to be a very special one: the fireshell is expanding in its comoving frame but its thickness is kept constant in the laboratory frame due to the balancing effect of the Lorentz contraction. We then examine the equations for the engulfment of the baryon loading as well as the further expansion of the fireshell composed by electron, positron, photons and baryon (PEMB pulse) up to the transparency point. We again point out the special role of the Livermore code in validating our results. Quite in addition of this validation procedures, the Livermore code have been essential in evidencing an instability occurring at a critical value of the baryon loading parameter $B=10^{-2}$ (see Fig. \ref{B10-2} and \citet{rswx00}).

In section ``Comparison and contrast of alternative fireshell equations of motion'' we compare our results with the ones in the current literature, in particular with the ones by \citet{sp90,psn93,mlr93,2005ApJ...635..516N}. We indicate a substantial agreement between our results and the early works by Piran and collaborators. The main difference is on the significance of the contribution of the rate equation. Departure from the \citet{mlr93} model are also outlined.

\subsubsection{The afterglow}

\begin{figure} 
\centering 
\includegraphics[width=\hsize,clip]{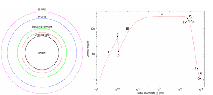} 
\caption{The GRB afterglow phase is here represented together with the optically thick phase (see Fig. \ref{cip2}). The value of the Lorentz gamma factor is here given from the transparency point all the way to the ultrarelativisitc, relativistic and non relativistic regimes. Details in \citet{rubr}.}
\label{cip_tot} 
\end{figure}

After reaching transparency and the emission of the P-GRB, the accelerated baryonic matter (the ABM pulse) interacts with the interstellar medium (ISM) and gives rise to the afterglow (see Fig. \ref{cip_tot}). Also in the descriptions of this last phase many differences exist between our treatment and the other ones in the current literature (see next sections).

We first look to the initial value problem. The initial conditions of the afterglow era are determined at the end of the optically thick era when the P-GRB is emitted. As recalled in the last section, the transparency condition is determined by a time of arrival $t_a^d$, a value of the gamma Lorentz factor $\gamma_\circ$, a value of the radial coordinate $r_\circ$, an amount of baryonic matter $M_B$ which are only functions of the two parameters $E_{dya}$ and $B$ (see Eq.\eqref{Bdef}).

This connection to the optically thick era is missing in the current approach in the literature which attributes the origin of the ``prompt radiation'' to an unspecified inner engine activity (see \citet{p99} and references therein). The initial conditions at the beginning of the afterglow era are obtained by a best fit of the later parts of the afterglow. This approach is quite unsatisfactory since, as we will explicitly show in the next sections, the theoretical treatments currently adopted in the description of the afterglow are not appropriate. The fit which uses an inappropriate theoretical treatment leads necessarily to the wrong conclusions as well as, in turn, to the determination of incorrect initial conditions.

The order of magnitude estimate usually quoted for the characteristic time scale to be expected for a burst emitted by a GRB at the moment of transparency at the end of the optically thick expansion phase is given by $\tau \sim GM/c^3$. For a $10M_\odot$ black hole this will give $\sim 10^{-3}$ s. There are reasons today not to take seriously such an order of magnitude estimate (see next sections and e.g. \citet{rfvx05}). In any case this time is much shorter than the ones typically observed in ``prompt radiation'' of the long bursts, from a few seconds all the way to $10^2$ s. In the current literature (see e.g. \citet{p99} and references therein), in order to explain the ``prompt radiation'' and overcome the above difficulty it has been generally assumed that its origin should be related to a prolonged ``inner engine'' activity preceding the afterglow which is not well identified.

To us this explanation has always appeared logically inconsistent since there remain to be explained not one but two very different mechanisms, independent of each other, of similar and extremely large energetics. This approach has generated an additional very negative result: it has distracted everybody working in the field from the earlier very interesting work on the optically thick phase of GRBs.

\begin{figure}
\centering
\includegraphics[width=\hsize,clip]{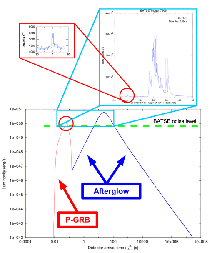}
\caption{The detailed features of GRB 991216 evidenced by our theoretical models are here reproduced. The P-GRB, the ``prompt radiation'' and what is generally called the afterglow. It is clear that the prompt emission observed by BATSE coincides with the extended afterglow peak emission (E-APE) and has been considered as a burst only as a consequence of the high noise threshold in the observations. The small precursor is identified with the P-GRB. Details in \citet{lett2,r02,rubr,rubr2}.}
\label{grb991216}
\end{figure}

The way out of this dichotomy in our model is drastically different: {\bf 1)} indeed the optically thick phase exists, is crucial to the GRB phenomenon and terminates with a burst: the P-GRB; {\bf 2)} the ``prompt radiation'' follows the P-GRB; {\bf 3)} the ``prompt radiation'' is not a burst: it is actually the temporally extended peak emission of the afterglow (E-APE). The observed structures of the prompt radiation can all be traced back to inhomogeneities in the interstellar medium (see Fig. \ref{grb991216} and \citet{r02}).

This approach was first tested on GRB991216. Both the relative intensity and time separation of the P-GRB and the afterglow were duly explained (see Fig. \ref{grb991216}) choosing a total energy of the plasma $E_{e^\pm}^{tot} = E_{dya} = 4.83\times 10^{53}$ erg and a baryon loading $B = 3.0\times 10^{-3}$ (see \citet{lett2,r02,rubr,rubr2}). Similarly, the temporal substructure in the prompt emission was explicitly shown to be related to the ISM inhomogeneities (see next sections).

Following this early analysis, and the subsequent ones on additional sources, it became clear that the ISM structure evidenced by our analysis is quite different from the traditional description in the current literature. Far from considering analogies with shock wave processes developed within fluidodynamic approach, it appears to us that the correct ISM description is a discrete one, composed of uncorrelated overdense ``blobs'' of typical size $\Delta R \sim 10^{14}$ cm widely spaced in underdense and inert regions.

\begin{figure}
\centering
\includegraphics[width=\hsize,clip]{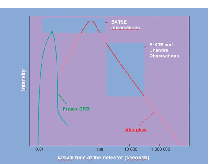}
\caption{Bolometric luminosity of P-GRB and afterglow as a function of the arrival time. Details in \citet{rubr}. Reproduced and adapted from \citet{pls} with the kind permission of the publisher.}
\label{bolum}
\end{figure}

\begin{figure}
\begin{minipage}{\hsize}
\includegraphics[width=0.85\hsize,clip]{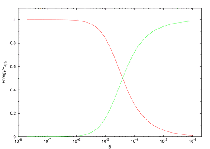}\\
\includegraphics[width=0.85\hsize,clip]{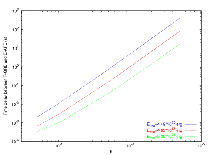}
\end{minipage}
\caption{{\bf Above:} The energy radiated in the P-GRB (the red line) and in the afterglow (the green line), in units of the total energy of the dyadosphere ($E_{dya}$), are plotted as functions of the $B$ parameter. {\bf Below:} The arrival time delay between the P-GRB and the peak of the afterglow is plotted as a function of the $B$ parameter for three selected values of $E_{dya}$.}
\label{bcross}
\end{figure}

We can then formulate the second paradigm, the interpretation of the burst structure (IBS) paradigm (\citet{lett2}), which covers three fundamental issues leading to the unequivocal identification of the canonical GRB structure:\\ 
{\bf a)} the existence of two different components: the P-GRB and the afterglow related by precise equations determining their relative amplitude and temporal sequence (see Fig. \ref{bolum}, \citet{rubr} and next section);\\ 
{\bf b)} what in the literature has been addressed as the ``prompt emission'' and considered as a burst, in our model is not a burst at all --- instead it is just the emission from the peak of the afterglow (see the clear confirmation of this result by the \emph{Swift} data of e.g. GRB 050315 in the next sections and in \citet{050315,Venezia_Orale});\\
{\bf c)} the crucial role of the parameter $B$ in determining the relative amplitude of the P-GRB to the afterglow and discriminating between the short and the long bursts (see Fig. \ref{bcross}). Both short and long bursts arise from the same physical phenomena: the gravitational collapse to a black hole endowed with electromagnetic structure and the formation of its dyadosphere.

The fundamental diagram determining the relative intensity of the P-GRB and the afterglow as a function of the dimensionless parameter $B$ is shown in Fig. \ref{bcross}. The main difference relates to the amount of baryonic matter engulfed by the electron-positron plasma in their optically thick phase prior to transparency. For $B < 10^{-5}$ the intensity of the P-GRB is larger and dominates the afterglow. This corresponds to the short bursts. For $10^{-5} < B \le 10^{-2}$ the afterglow dominates the GRB. For $B > 10^{-2}$ we may observe a third class of ``bursts'', eventually related to a turbulent process occurring prior to transparency (\citet{rswx00}). This third family should be characterized  by  smaller values of the Lorentz gamma factors than in the case of the short or long bursts.

\begin{figure}
\includegraphics[width=\hsize,clip]{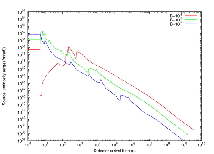}
\caption{The bolometric luminosity of a source with the same total energy and ISM distribution of GRB991216 is here represented for selected values of the $B$ parameter, ranging from $B=10^{-2}$ to $B=10^{-4}$. The actual value for GRB991216 is $B=3.0\times 10^{-3}$. As expected, for smaller values of the $B$ parameter the intensity of the P-GRB increases and the total energy of the afterglow decreases. What is most remarkable is that the luminosity in the early part of the afterglow becomes very spiky and the peak luminosity actually increases.}
\label{Letizia_MultiB}
\end{figure}

Particularly enlightening for the gradual transition to the short bursts as a function of the $B$ parameter is the diagram showing how GRB991216 bolometric light curve would scale changing the sole value of $B$ (see Fig. \ref{Letizia_MultiB}).

\begin{figure}
\includegraphics[width=\hsize,clip]{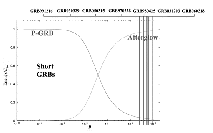}
\caption{Same as Fig. \ref{bcross} with the values determined for selected GRBs. In order to determine the value of the $B$ parameter and the total energy we have performed the complete fit of each source. In particular, we have fitted for each source the observed luminosities in selected energy bands of the entire afterglow including the prompt emission. We have verified that in each source the hard-to-soft spectral evolution is correctly fitted and we have compared the theoretically computed spectral lag with the observations. Where applicable, we have also computed the relative intensity and temporal separation between the P-GRB and the peak of the afterglow and compared these values with teh observed ones. The absence of spectral lag in the P-GRB is automatically verified by our model.}
\label{bcross_sorgenti}
\end{figure}

Moving from these two paradigms, and the prototypical case of GRB 991216, we have extended our analysis to a larger number of sources, such as GRB970228 (\citet{970228}), GRB980425 (\citet{cospar02,Mosca_Orale}), GRB030329 (\citet{030329}), GRB031203 (\citet{031203}), GRB050315 (\citet{050315}), which have led to a confirmation of the validity of our canonical GRB structure (see Fig. \ref{bcross_sorgenti}). In addition, progresses have been made in our theoretical comprehension, which will be presented in the following sections.

In section ``Exact versus approximate solutions in Gamma-Ray Burst afterglows'' we first write the energy and momentum conservation equations for the interaction between the ABM pulse and the ISM in a finite difference formalism. We then express these same equations in a differential formalism to compare our approach with the ones in the current literature. We write the exact analytic solutions of such differential equations both in the fully radiative and in the adiabatic regimes. We then compare and contrast these results with the ones following from the ultra-relativistic approximation widely adopted in the current literature. Such an ultra-relativistic approximation, adopted to apply to GRBs the \citet{bm76} self-similar solution, led to a simple power-law dependence of the Lorentz gamma factor of the baryonic shell on the distance. On the contrary, we show that no constant-index power-law relations between the Lorentz gamma factor and the distance can exist, both in the fully radiative and in the adiabatic regimes. The exact solution is indeed necessary if one wishes to describe properly all the phases of the afterglow including the prompt emission.

In section ``Exact analytic expressions for the equitemporal surfaces in Gamma-Ray Burst afterglows'' we follow the indication by Paul \citet{c39} who pointed out long ago how in all relativistic expansions the crucial geometrical quantities with respect to a physical observer are the ``equitemporal surfaces'' (EQTSs), namely the locus of source points of the signals arriving at the observer at the same time. After recalling the formal definition of the EQTSs, we use the exact analytic solutions of the equations of motion recalled in the previous section to derive the exact analytic expressions of the EQTSs in GRB afterglow both in the fully radiative and adiabatic regimes. We then compare and contrast such exact analytic solutions with the corresponding ones widely adopted in the current literature and computed using the approximate ``ultra-relativistic'' equations of motion discussed in the previous section. We show that the approximate EQTS expressions lead to uncorrect estimates of the size of the ABM pulse when compared to the exact ones. Quite apart from their academic interest, these results are crucial for the interpretation of GRB observations: all the observables come in fact from integrated quantities over the EQTSs, and any minor disagreement in their definition can have extremely drastic consequences on the identification of the true physical processes.

In section ``Exact versus approximate beaming formulas in Gamma-Ray Burst afterglows'' we discuss the possibility that GRBs originate from a beamed emission, one of the most debated issue about the nature of the GRB sources in the current literature after the work by \citet{my94} (see e.g. \citet{p04,m06} and references therein). In particular, on the ground of the theoretical considerations by \citet{sph99}, it was conjectured that, within the framework of a conical jet model, one may find that the gamma-ray energy released in all GRBs is narrowly clustered around $5 \times 10^{50}$ ergs (\citet{fa01}). We have never found in our GRB model any necessity to introduce a beamed emission. Nevertheless, we have considered helpful and appropriate helping the ongoing research by giving the exact analytic expressions of the relations between the detector arrival time $t_a^d$ of the GRB afterglow radiation and the corresponding half-opening angle $\vartheta$ of the expanding source visible area due to the relativistic beaming. We have done this both in the fully radiative and in the adiabatic regimes, using the exact analytic solutions presented in the previous sections. Again, we have compared and contrasted our exact solutions with the approximate ones widely used in the current literature. We have found significant differences, particularly in the fully radiative regime which we consider the relevant one for GRBs, and it goes without saying that any statement on the existence of beaming can only be considered meaningful if using the correct equations.

In the section ``The afterglow bolometric luminosity and the ISM discrete structure'' we derive the expression for the bolometric luminosity of the GRB afterglow and we address the general issue of the possible explanation of the observed substructures in the GRB prompt emission as due to ISM inhomogeneities. On this topic there exist in the literature two extreme points of view: the one by Fenimore and collaborators (see e.g. \citet{fmn96,fcrsyn99,f99}) and Piran and collaborators (see e.g. \citet{sp97,p99,p00,p01}) on one side and the one by Dermer and collaborators (\citet{d98,dbc99,dm99}) on the other. Fenimore and collaborators have emphasized the relevance of a specific signature to be expected in the collision of a relativistic expanding shell with the ISM, what they call a fast rise and exponential decay (FRED) shape. This feature is confirmed by our analysis (see peaks A, B, C in Fig.~\ref{substr_peak}). However they also conclude, sharing the opinion by Piran and collaborators, that the variability observed in GRBs is inconsistent with causally connected variations in a single, symmetric, relativistic shell interacting with the ambient material (``external shocks'') (\citet{fcrsyn99}). In their opinion the solution of the short time variability has to be envisioned within the protracted activity of an unspecified ``inner engine'' (\citet{sp97}); see as well \citet{rm94,pm98,mr01,m02}. On the other hand, Dermer and collaborators, by considering an idealized process occurring at a fixed $\gamma=300$, have reached the opposite conclusions and they purport that GRB light curves are tomographic images of the density distributions of the medium surrounding the sources of GRBs (\citet{dm99}). By applying the exact formulas derived in previous sections, we show that Dermer's conclusions are correct, and we identify that the ``tomography'' purported by \citet{dm99} leads to ISM clouds consistently on the order of $\sim 10^{14}$ cm. Apparent superluminal effects are introduced. In our treatment we have adopted a simple spherically symmetric approximation for the ISM distribution. We show that the agreement of this approximation with the observations is excellent for Lorentz gamma factors $\gamma > 150$ since the relativistic beaming angle introduced in the previous sections provides an effective cut-off to the visible ISM structure. For lower Lorentz gamma factors, a three dimensional description of the ISM would be needed and the corresponding treatment is currently in preparation.

In section ``The theory of the luminosity in fixed energy bands and spectra of the afterglow'', having shown in the previous sections a general agreement between the observed luminosity variability and our treatment of the bolometric luminosity, we have further developed the model in order to explain:\\ 
{\bf a)} the details of the observed luminosity in fixed energy bands, which are the ones actually measured by the detectors on the satellites;\\
{\bf b)} the instantaneous as well as the average spectral distribution in the 
entire afterglow and;\\ 
{\bf c)} the observed hard to soft drift observed in GRB spectra.\\
The fundamental assumption is introduced that the X- and gamma ray radiation during the entire afterglow phase has a thermal spectrum in the co-moving frame. The ratio $\mathcal{R}=A_{eff}/A_{vis}$ between the ``effective emitting area'' $A_{eff}$ of the ABM pulse and its full visible area $A_{vis}$ is introduced. Due to the ISM inhomogeneities, composed of clouds with filamentary structure, the ABM emitting region is in fact far from being homogeneous. We have justified the existence of this thermal emission by considering the ISM filamentary structure and its optical thickness (see \citet{fil}). The theoretical prediction for the observed spectra starting from these premises has been by far the most complex and, in our opinion, the most elegant aspect of the entire GRB model. In order to compute the luminosity in a fixed energy band at a given value of the arrival time it is necessary to perform a convolution over the given EQTS of an infinite number of elementary contributions, each one characterized by a different value of Lorentz and Doppler factors. Therefore, each observed instantaneous spectrum is theoretically predicted to be the result of a convolution of an infinite number of thermal spectra, each one with a different temperature, over the given EQTS and its shape is theoretically predicted to be non-thermal. Moreover, the observed time-integrated spectra depart even more from a thermal shape, being the convolution over the observation time of an infinite number of non-thermal instantaneous spectra. We confirm in this work the qualitative suggestion advanced by \citet{bk99} already in 1999. We then examine the issue of the possible presence or absence of jets in GRBs in the case of GRB 991216. We compare and contrast our theoretically predicted afterglow luminosity in the $2$--$10$ keV band for spherically symmetric versus jetted emission. At these wavelenghts the jetted emission can be excluded and data analysis confirms spherical symmetry. In fact, the actual afterglow luminosity in fixed energy bands, in spherical symmetry, does not have a simple power law dependence on arrival time. This circumstance has been erroneously interpreted, in the usual presentation in the literature, as a broken power-law supporting the existence of jet-like structures in GRBs.

\subsubsection{Theoretical interpretation of luminosity and spectra of selected sources}

Having used GRB 991216 as a prototype, we were constrained by the absence of data in the time range between $\sim 36$ s and $\sim 3500$ s. This same situation was encountered, even more extremely, in all the other sources, like e.g. GRB 970228, GRB 980425, GRB 030329, etc. Fortunately, the launch of the \emph{Swift} mission changed drastically and positively this situation. We could obtain a continuous set of data from the prompt emission to the latest afterglow phases in multiple energy bands. Also the data of INTEGRAL have been important. We obtained for the first time a very good agreement between our theoretical spectral analysis and the observations in the case of GRB 031203 observed by INTEGRAL. We also obtained the first complete analysis of GRB 050315 observed by \emph{Swift}.

In section ``Analysis of GRB 031203'' we show how we are able to predict the whole dynamics of the process which originates the GRB 031203 emission fixing univocally the two free parameters of the model, $E_{dya}$ and $B$. Moreover, it is possible to obtain the exact temporal structure of the prompt emission taking into account the effective ISM filamentary structure. The important point we like to emphasize is that we can get both the luminosity emitted in a fixed energy band and the photon number spectrum starting from the hypothesis that the radiation emitted in the GRB process is thermal in the co-moving frame of the expanding pulse. It has been clearly shown that, after the correct space-time transformations, both the time-resolved and the time-integrated spectra in the observer frame strongly differ from a Planckian distribution and have a power-law shape, although they originate from strongly time-varying thermal spectra in the co-moving frame. We obtain a good agreement of our prediction with the photon number spectrum observed by INTEGRAL and, in addition, we predict a specific hard-to-soft behavior in the instantaneous spectra. Due to the possibility of reaching a precise identification of the emission process in GRB afterglows by the observations of the instantaneous spectra, it is hoped that further missions with larger collecting area and higher time resolving power be conceived and a systematic attention be given to closer-by GRB sources. Despite this GRB is often considered as ``unusual'' (\citet{Wa04,sod}), in our treatment we are able to explain its low gamma-ray luminosity in a natural way, giving a complete interpretation of all its spectral features. In agreement to what has been concluded by \citet{saz}, it appears to us as a under-energetic GRB ($E_{dya}\approx 10^{50}$ erg), well within the range of applicability of our theory, between $10^{48}$ erg for GRB 980425 (\citet{cospar02,Mosca_Orale}) and $10^{54}$ erg for GRB 991216 (\citet{rubr}).

In section ``Analysis of GRB 050315'' we discuss how before the \emph{Swift} data, our model could not be directly fully tested. With GRB 050315, for the first time, we have obtained a good match between the observational data and our predicted intensities, in $5$ energy bands, with continuous light curves near the beginning of the GRB event, including the ``prompt emission'', all the way to the latest phases of the afterglow. This certainly supports our model and opens a new phase of using it to identify the astrophysical scenario underlying the GRB phenomena. In particular:
\begin{enumerate}
\item We have confirmed that the ``prompt emission'' is not necessarily due to the prolonged activity of an ``inner engine'', but corresponds to the emission at the peak of the afterglow.
\item We have a clear theoretical prediction, fully confirmed from the observations, on the total energy emitted in the P-GRB $E_{P-GRB} = 1.98 \times 10^{51}$ erg and on its temporal separation from the peak of the afterglow $\Delta t^d_a = 51$ s. To understand the physics of the inner engine more observational and theoretical attention should be given to the analysis of the P-GRB.
\item We have uniquely identified the basic parameters characterizing the GRB energetics: the total energy of the black hole dyadosphere $E_{dya} = 1.46\times 10^{53}$ erg and the baryon loading parameter $B = 4.55 \times 10^{-3}$.
\item The ``canonical behavior'' in almost all the GRB observed by \emph{Swift}, showing an initial very steep decay followed by a shallow decay and finally a steeper decay, as well as the time structure of the ``prompt emission'' have been related to the fluctuations of the ISM density and of the ${\cal R}$ parameter.
\item The theoretically predicted instantaneous photon number spectrum shows a very clear hard-to-soft behavior continuously and smoothly changing from the ``prompt emission'' all the way to the latest afterglow phases.
\end{enumerate}

After the analysis of the above two sources, only the earliest part of the afterglow we theoretically predicted, which corresponds to a bolometric luminosity monotonically increasing with the photon detector arrival time, preceding the ``prompt emission'', still remains to be checked by direct observations. We hope in the near future to find an intense enough source, observed by the \emph{Swift} satellite, to verify this still untested theoretical prediction.

As a byproduct of the above results, we could explain one of the long lasting unanswered puzzles of GRBs: the light curves in the ``prompt emission'' show very strong temporal substructures, while they are remarkably smooth in the latest afterglow phases. The explanation follows from three factors: 1) the value of the Lorentz $\gamma$ factor, 2) the EQTS structure and 3) the coincidence of the ``prompt emission'' with the peak of the afterglow. For $\gamma \sim 200$, at the peak of the afterglow, the diameter of the EQTS visible area due to relativistic beaming is small compared to the typical size of an ISM cloud. Consequently, any small inhomogeneity in such a cloud produces a marked variation in the GRB light curve. On the other hand, for $\gamma \to 1$, in the latest afterglow phases, the diameter of the EQTS visible area is much bigger than the typical size of an ISM cloud. Therefore, the observed light curve is a superposition of the contribution of many different clouds and inhomogeneities, which produces on average a much smoother light curve (details in \citet{r02,rubr}).

\subsection{The third paradigm: The GRB-Supernova Time Sequence (GSTS) paradigm}

\begin{table}
\caption{see: a) \citet{Mosca_Orale}; b) \citet{031203}; c) \citet{030329}; d) \citet{050315}; e) Bernardini et al. in preparation; f) \citet{rubr2}; g) see \citet{yk}; h) Mazzali, P., private communication at MG11 meeting in Berlin, July 2006; i) evaluated fitting the URCAs with a power law followed by an exponentially decaying part; j) respectively \citet{mh},  \citet{galama98},  \citet{proch}, \citet{greiner}, \citet{kb05}, \citet{inf01}, \citet{bloom}, \citet{piro00}; k) respectively \citet{Ka06}, \citet{Sa06}, XRR is considered in \citet{Ka06}, while XRF as suggested by \citet{Wa04}, \citet{Ka06}, \citet{va05}, \citet{bloom}.}
\label{tab1}
\centering
\begin{tabular}{lcccccrlcc}
\hline
GRB/SN & $E_{e^\pm}^{tot}$ & ${E_{SN}^{bolom}}^g$ & ${E_{SN}^{kin}}^h$ & $E_{URCA}^i$ & $B$ & $\gamma_\circ$ & $z^j$ & $S_X/S_\gamma^k$ &  $\begin{array}{c}\left\langle  n_{ism} \right\rangle\\(\mathrm{\#}/\mathrm{cm}^3)\end{array}$\\
\hline
060218/2006aj & $1.8\times 10^{50}$  & $9.2\times 10^{48}$ & $2.0\times10^{51}$ & $?$ & $1.0\times 10^{-2}$  & $99$ & $0.033$ & $3.54$(XRF) & $1.0$\\ 
980425/1998bw$^a$ & $1.2\times 10^{48}$ & $2.3\times 10^{49}$ &  $1.0\times 10^{52}$ & $3\times 10^{48}$ & $7.7\times 10^{-3}$  & $124$     & $0.0085$ & $0.58$ (XRR) & $2.5\times 10^{-2}$ \\
031203/2003lw$^b$ & $1.8\times 10^{50}$  & $3.1\times 10^{49}$ & $1.5\times10^{52}$ &  $2\times10^{49}$ & $7.4\times 10^{-3}$  & $133$     & $0.105$ & $0.49$(XRR/XRF) & $0.3$\\
030329/2003dh$^c$ & $2.1\times 10^{52}$  & $1.8\times 10^{49}$ & $8.0\times10^{51}$ & $3\times 10^{48}$ & $4.8\times 10^{-3}$  & $206$     & $0.168$   & $0.56$(XRR)     & $1.0$  \\ 
050315$^d$        & $1.5\times 10^{53}$ & & & & $4.5\times 10^{-3}$  & $217$     & $1.949$     & $1.58$(XRF) &   $0.8$\\ 
970228/?$^e$      & $1.4\times 10^{54}$ & & & & $5.0\times 10^{-3}$  & $326$     & $0.695$      & GRB &        $1.0\times 10^{-3}$\\ 
991216$^f$        & $4.8\times 10^{53}$ & & & & $2.7\times 10^{-3}$  & $340$     & $1.0$ & GRB & $3.0$\\ 
\hline
\end{tabular}
\end{table}

Following the classical result of \citet{galama98} who discovered the temporal coincidence of GRB 980425 and SN 1998bw, the association of other nearby GRBs with Type Ib/c SNe has been spectroscopically confirmed (see Tab. \ref{tab1}). The approaches in the current literature have attempted to explain both the SN and the GRB as two aspects of the same astrophysical phenomenon. It is so that GRBs have been assumed to originate from a specially strong SN process, a hypernova or a collapsar (see e.g. \citet{p98,ka98,ia98,wb06} and references therein). Both these possibilities imply very dense and strongly wind-like ISM structure.
 
\begin{figure}
\includegraphics[width=\hsize,clip]{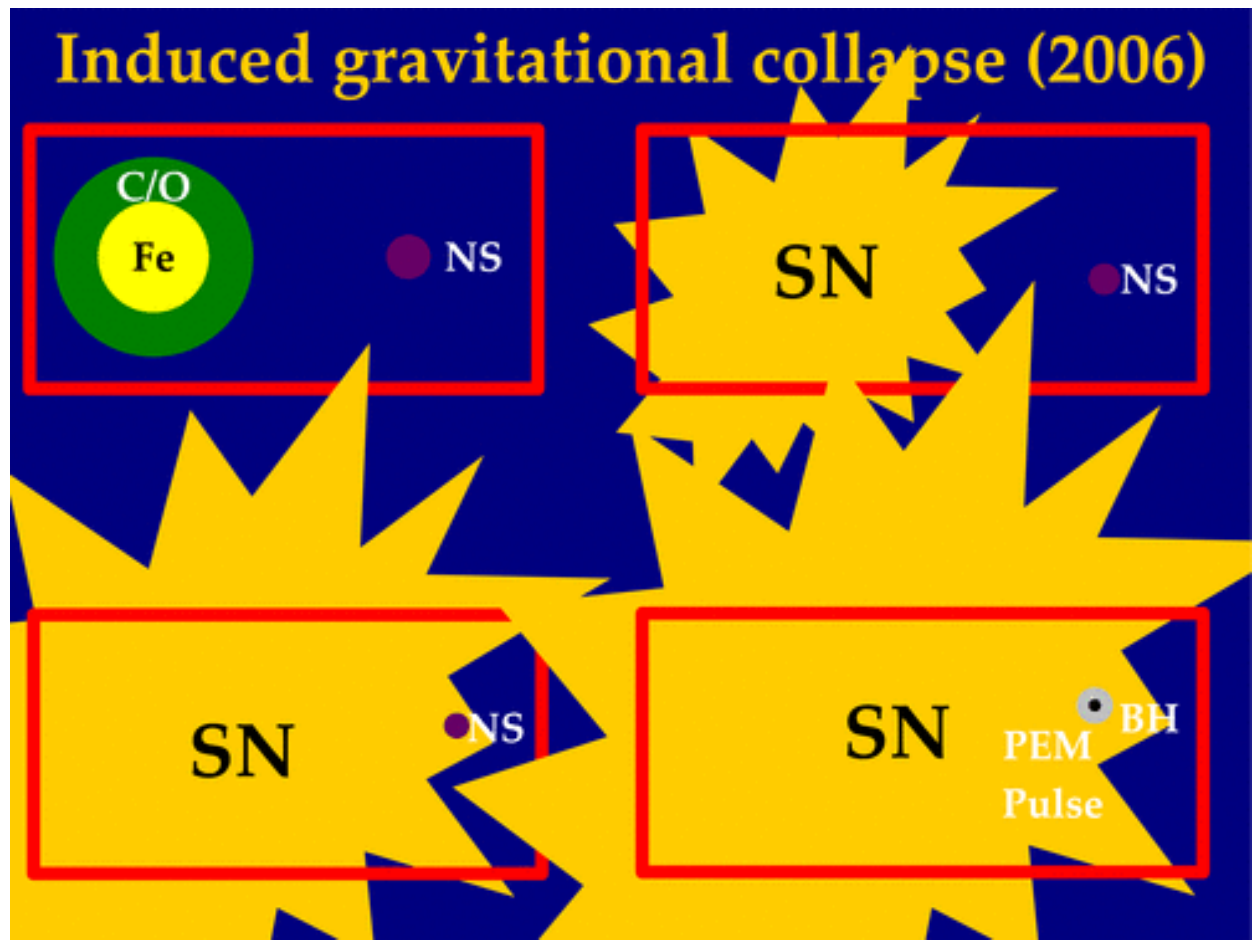}
\caption{A possible process of gravitational collapse to a black hole ``induced'' by the Ib/c SN on a companion neutron star in a close binary system. Details in \citet{mg11}.}
\label{IndColl06}
\end{figure}

In our model we have followed a very different approach. We assumed that the GRB consistently originates from the gravitational collapse to a black hole, embedded in an ISM with average density $\langle n_{ism} \rangle \sim 1$ particle/cm$^3$. The SN follows instead the very complex pattern of the final evolution of a massive star, possibly leading to a neutron star or to a complete explosion but never to a black hole. The temporal coincidence of the two phenomena, the SN explosion and the GRB, have then to be explained by the novel concept of ``induced gravitational collapse'', introduced in \citet{lett3}. We have to recognize that still today we do not have a precise description of how this process of ``induced gravitational collapse'' occurs. At this stage, it is more a framework to be implemented by additional theoretical work and observations. It is so that two different possible scenarios have been outlined. In the first version (\citet{lett3}) we have considered the possibility that the GRBs may have caused the trigger of the SN event. For the occurrence of this scenario, the companion star had to be in a very special phase of its thermonuclear evolution and three different possibilities were considered:
\begin{enumerate}
\item A white dwarf, close to its critical mass. In this case, the GRB may implode the star enough to ignite thermonuclear burning.
\item The GRB enhances in an iron-silicon core the capture of the electrons on the iron nuclei and consequently decreases the Fermi energy of the core, leading to the onset of gravitational instability.
\item The pressure waves of the GRB may trigger a massive and instantaneous nuclear burning process leading to the collapse.
\end{enumerate}
More recently (see \citet{RuffiniTF1,Mosca_Orale}), a quite different possibility has been envisaged: the SN, originating from a very evolved core, undergoes explosion in presence of a companion neutron star with a mass close to its critical one; the SN blast wave may then trigger the collapse of the companion neutron star to a black hole and the emission of the GRB (see Fig. \ref{IndColl06}). It is clear that, in both scenarios, the GRB and the SN occur in a binary system.

There are many reasons to propose this concept of ``induced gravitational collapse'':
\begin{enumerate}
\item The fact that GRBs occur from the gravitational collapse to a black hole.
\item The fact that ISM density for the occurrence of GRBs is inferred from the analysis of the afterglow to be systematically on the order of $1$ particle/cm$^3$ (see Tab. \ref{tab1}). This implies that the process of collapse has occurred in a region of space filled with a very little amount of baryonic matter. The sole significant contribution to the baryonic matter conponent in this process is the one represented by the fireshell baryon loading, which is anyway constrained by the inequality $B \le 10^{-2}$.
\item The fact that the energetics of the GRBs associated with SNe appears to be particularly weak is consistent with the energy originating from the gravitational collapse to the smallest possible black hole: the one with mass $M$ just over the neutron star critical mass.
\end{enumerate}
There are also at work very clearly selection effects among the association between SNe and GRBs:
\begin{enumerate}
\item There is a clear evidence that many type Ib/c SNe exists without an associated GRB (\citet{gdv07}).
\item There is also the opposite case that some GRBs do not show the presence of a SN associated, although they are close enough for the SN to be observed (see e.g. \citet{060614_DellaValle}).
\item There is also the presence in all observed GRB-SN systems of an URCA source, a peculiar late time X-ray emission. These URCA sources have been identified and presented for the first time at the Tenth Marcel Grossmann meeting held in Rio de Janeiro (Brazil) in the Village of Urca, and named consequently. They appears to be one of the most novel issues still to be understood on GRBs. We will return on these aspects in the next sections.
\end{enumerate}

The issue of triggering the gravitational collapse instability induced by the GRB on the progenitor star of the supernova or, vice versa, by the supernova on the progenitor star of the GRB needs accurate timing. The occurrence of new nuclear physics and/or relativistic phenomena is very likely. The general relativistic instability induced on a nearby star by the formation of a black hole needs some very basic new developments.

Only a very preliminary work exists on this subject, by Jim Wilson and his collaborators (see e.g. the paper by \citet{mw}). The reason for the complexity in answering such a question is simply stated: unlike the majority of theoretical work on black holes and binary X-ray sources, which deals mainly with one-body black hole solutions in the Newtonian field of a companion star, we now have to address a many-body problem in general relativity. We are starting in these days to reconsider, in this framework, some classic works by \citet{f21,hr73,m47,p47,p73,Bini1,Bini2} which may lead to a new understanding of general relativistic effects in these many-body systems. This is a welcome effect of GRBs on the conceptual development of general relativity.

In section ``On the GRB-SN association'', after the successful analysis of GRB 991216, GRB 031203 and GRB 050315, we apply our theoretical framework to the analysis of all the other GRBs associated with SNe. We proceed first to GRB 980425; we go then to GRB 030329; finally, we discuss the late time emission of GRB 031203 observed by XMM and Chandra. We summarize the general results of these GRBs associated with SNe and we make some general conclusions on the relations between GRBs, SNe and the URCA sources. We finally present some novel considerations about our third paradigm and the concept of induced gravitational collapse.

\subsection{General relativity, relativistic quantum field theory and GRBs}

We have already seen how the entire physics of the afterglow stands on a very well posed problem of a shell of baryons with an initial very high value of the Lorentz gamma factor ($100 < \gamma_\circ < 400$) interacting with an highly inhomogeneous ISM. The discrete nature of ISM in widely spaced blobs simplifies the problem and, as we have already recalled, these processes are dominated by specific special relativistic effects and not in any way by general relativity. The physics of general relativity and relativistic quantum field theory is contained in the description of the black hole, in the creation of the electron-positron plasma in the dyadosphere, in its dynamics with a finite amount of baryon loading. The only observable effects of this process are at the moment when the fireshell reaches transparency and the P-GRB is emitted. In the limit of $B \to 0$ only the P-GRB emission is observed since the afterglow intensity goes to zero. We recall that all canonical GRBs with $B < 10^{-5}$ correspond with the short GRBs (see Fig. \ref{bcross}, \ref{bcross_sorgenti}).

In our theoretical work on GRBs we have started for simplicity with an already formed black hole. Such a black hole has not to be everlasting! It is used as an approximation to describe the pair production process occurring in the dyadosphere, which lasts for less than $\sim 10^{-2}$ s in the very transient phenomenon of the gravitational collapse. This process, we recall, lasts less than $1$ s (\citet{RuKerr}). This is certainly a good approximation to describe the electron-positron pairs accelerating the baryonic matter giving rise to the afterglow. This allowed also to give the quantitative estimate of the ratio between the afterglow and the P-GRB total energies. However, this treatment is lacking the detailed analysis needed for the description of the fine details of the P-GRBs and, therefore, of the short GRBs. We have also adopted the hypothesis that the electron-positron plasma reaches thermal equilibrium before starting the dynamical phase of expansion and self-acceleration. In recent times, we have given attention to refining our analysis starting from the proofs of the correctness of the above assumption of thermal equilibrium in the electron-positron plasma. We are also exploring the consequences on a deeper understanding of black hole physics made possible by GRB observations and complementary astrophysical phenomenon. Particular attention has been given to the theoretical background for the study of the dynamical formation of the black hole, both from the point of view of relativistic quantum field theory and of general relativity. We focus on the observational consequences on the structure of the P-GRBs. We are also exploring the possibility that Ultra High Energy Cosmic Rays (UHECRs) may be related to the physics of black holes endowed with electromagnetic structure.

We first review results on vacuum polarization and quantum electrodynamics in Minkowski space and then we turn to recent results in general relativity.

In section ``Pair production in Coulomb potential of nuclei and heavy-ion collisions'' we recall the classic theoretical nuclear physics results which have led to the famous $Z_c = 173$ catastrophe. We turn then to the experimental work on heavy-ion collisions and the still ongoing expectation of observing pair production in heavy-ion collisions for $Z > Z_c$.

In the section ``Vacuum polarization in uniform electric field and in Kerr-Newman geometries'' we recall some of the pioneering works by Oscar Klein and Fritz Sauter on pair creation in constant electric fields. We then recall the Heisenberg-Euler-Weisskopf effective theory to describe this phenomenon as well as the classical work of Schwinger on quantum electrodynamics. We then turn to the work by Damour and Ruffini on applying the Schwinger process to the field of a Kerr-Newman geometry.

In the section ``Description of the electron-positron plasma oscillations''
we address the issue of the electron-positron pair creation due to vacuum
polarization process in a uniform electric field and the associated plasma
oscillations regimes for $\mathcal{E}>\mathcal{E}_{\mathrm{c}}$ and $%
\mathcal{E}<\mathcal{E}_{\mathrm{c}}$ ($\mathcal{E}_{\mathrm{c}}\equiv
m_{e}^{2}c^{3}/(e\hbar )$, where $m_{e}$ and $e$ are the electron mass and
charge). Our treatment is based on electro-fluidodynamics approach
consisting of the equation of the continuity, energy-momentum conservation
and the Maxwell equations and is fully consistent with the traditional
Boltzmann-Vlasov framework. For $\mathcal{E}>\mathcal{E}_{\mathrm{c}}$ we
recover previous results about the oscillations of the charges, discuss the
electric field screening and the relaxation of the system to
electron-positron-photon plasma configuration via the process $%
e^{+}e^{-}\rightleftarrows $ $\gamma \gamma $. We evidence the existence of
plasma oscillations also for $\mathcal{E}<\mathcal{E}_{\mathrm{c}}$. We turn then to general relativistic effects. These GRB observations and their consequent theoretical understanding are proposing an authentic renaissance in the field of general relativity. A new set of problematic leads to new understanding of basic issues on the nature of black holes, on the nature of irreducible mass, on the blackholic energy, as well as on the process of black hole formation.

In the section ``On the irreducible mass of the black hole and the role of subcritical and overcritical electric fields'' the dynamics of gravitational collapse is simulated by an exact solution of a thin shell in general relativity. An explicit expression of the irreducible mass of the black hole is derived as a function of the rest mass, of the kinetic energy and the gravitational binding energy of the shell at the horizon. Considerations for the role of an effective ergosphere for undercritical electric field in explaining the origin of UHECRs are outlined as well as the role of overcritical dyadospheres for GRBs.

In the section ``Contributions of GRBs to the black hole theory'' we address the basic issue of the maximum energy extractable during the formation phase of a black hole. From the expression of the irreducible mass derived in the previous section, we show that the maximum energy extractable can never be larger than $50$\% of the initial mass. Some consequences of this result on the Bekenstein and Hawking considerations and on general relativity and thermodynamics are also outlined.

In the section ``On a separatrix in an overcritical collapse'' we have exemplified by an analytic example the gravitational collapse of a thin shell endowed with an electric field. We have only focused on the part relevant for GRBs, namely the case $E > E_c$, where the electron-positron pairs have thermalized and are optically thick leading to the dynamical phase of GRBs. Starting from these initial conditions, we follow the dynamics of the pure electron-positron plasma, without any baryonic contamination. We point out the existence of a separatrix at a distance from the black hole of approximately $4M$, $M$ being the mass of the black hole in geometrical units. For smaller radii, the expanding plasma is captured by the gravitational field of the forming black hole, leading to a clear cut-off in the signal received from far away distance. It is however important to emphasize that these phenomena have been computed only in the case of an electron-positron plasma with zero baryon loading and are therefore relevant uniquely for short GRBs, strictly in the limit $B = 0$.

In the section ``Observational signatures and spectral evolution of short GRBs'' we present theoretical predictions for the spectral, temporal and intensity signatures of the electromagnetic radiation emitted during the process of the gravitational collapse of a stellar core to a black hole, during which electromagnetic field strengths rise over the critical value for $e^+e^-$ pair creation. The last phases of this gravitational collapse are studied, using the result presented in the previous sections, leading to the formation of a black hole with a subcritical electromagnetic field, likely with zero charge, and an outgoing pulse of initially optically thick $e^+e^-$-photon plasma. Such a pulse reaches transparency at Lorentz gamma factors of $10^2$--$10^4$. We find a clear signature in the outgoing electromagnetic signal, drifting from a soft to a hard spectrum, on very precise time-scales and with a very specific intensity modulation. We conclude by making precise predictions for the spectra, the energy fluxes and characteristic time-scales of the radiation for short-bursts. Hopefully new space missions will be planned, with temporal resolution down to fractions of $\mu$s and higher collecting area and spectral resolution than at present, in order to verify the detailed agreement between our model and the observations. It is now clear that if our theoretical predictions will be confirmed, we will have a very powerful tool for cosmological observations: the independent information about luminosity, time-scale and spectrum can uniquely determine the mass, the electromagnetic structure and the distance from the observer of the collapsing core (see e.g. Fig. \ref{ff5} and \citet{rfvx05}). In that case short-bursts, in addition to give a detailed information on all general relativistic and relativistic field theory phenomena occurring in the approach to the horizon, may also become the best example of standard candles in cosmology (\citet{r03tokyo}). We are currently analyzing the introduction of baryonic matter in the optically thick phase of the expansion of the $e^+e^-$ plasma, within this detailed time-varying description of the gravitational collapse, which may affect the structure of the P-GRB (\citet{lett2}) as well as the structure of the long-bursts (\citet{lett2,r02,rubr}).

\begin{figure}
\includegraphics[width=\hsize,clip]{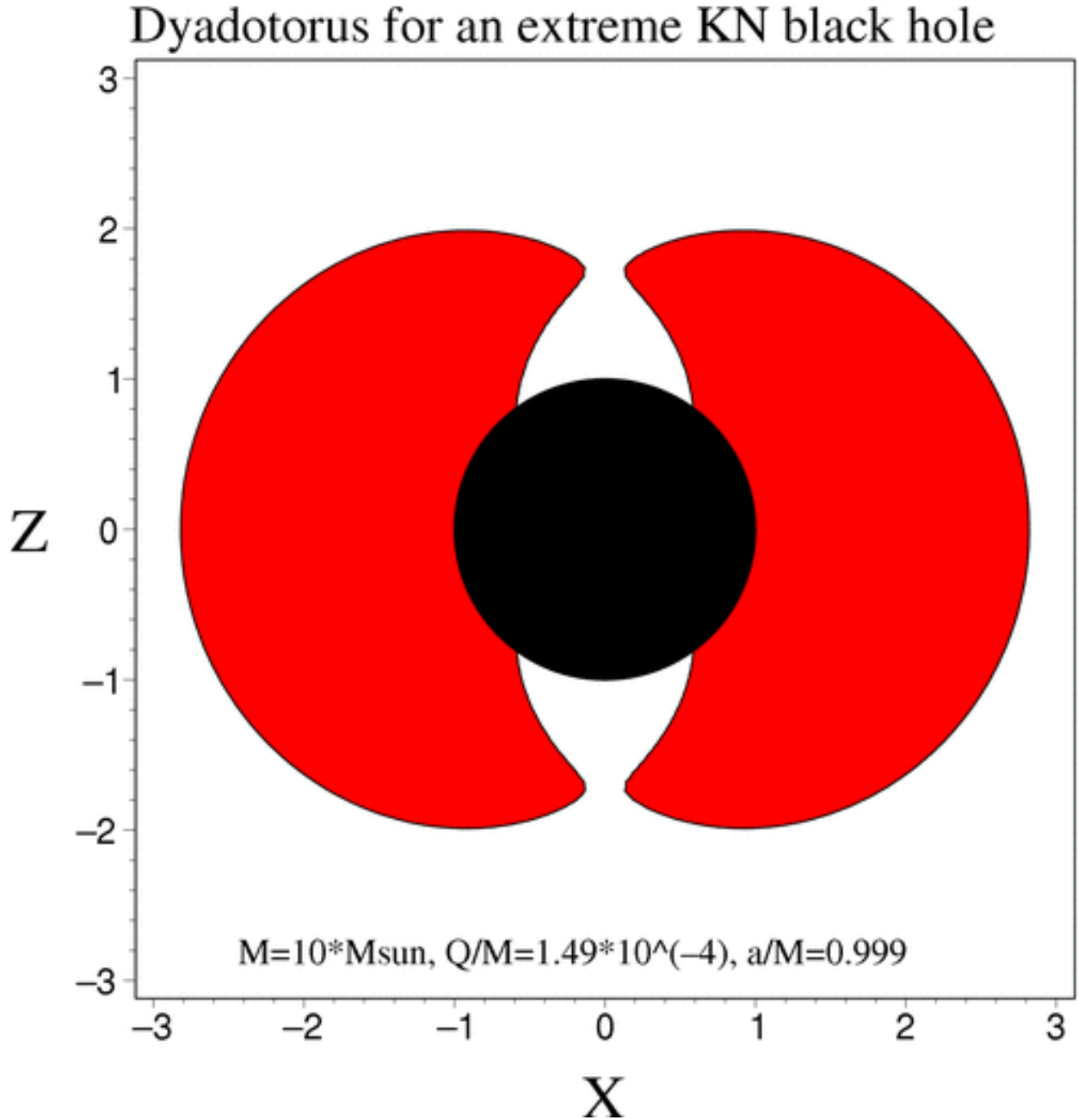}
\caption{The ``Dyado-torus'' is the region outside the horizon of a Kerr-Newman black hole, where the electrodynamical processes generates electron-positron pairs by vacuum polarization processes. Details in \citet{ca07}.}
\label{dyadotorus}
\end{figure}

Finally, in the last section ``Electrodynamics for nuclear matter in bulk'' we present some preliminary results generalizing to the case of a macroscopic core at nuclear density some of the classical works on the very heavy nuclei by Popov and his school and by Greiner and his school. This work clearly points to the possibility to have near the nuclear density core surface an electric field very close to the critical value $E_c$, still fulfilling the overall charge neutrality of the system. The fundamental role of the ultrarelativistic electron gas in obtaining these results is evidenced. From a theoretical point of view, these results follows from some prior works on the relativistic Thomas-Fermi equation in nuclear density matter by Ferreirinho, Ruffini and Stella. The reason we consider these results fundamental for the understanding of the electrodynamics of black hole is that they may represent a very specific example of a system with overall charge neutrality but with a separation between the core and the tail of the electron distribution. This separation leads to electric fields larger than the critical value $E_c$. Such a result appears to be essential in order to give the appropriate initial conditions leading to the electrodynamics of gravitational collapse to a Kerr-Newman black hole and the formation of a dyadosphere.

We have also reported briefly in the meeting on some generalization of the dyadosphere concept introduced in a Reissner-Nordstr{\o}m geometry to the case of a ``dyado-torus'' in a kerr-Newman geometry (see Fig. \ref{dyadotorus}). Indeed, we are interested in a proposal advanced in 2002 by \citet{it02} that the $e^+e^-$ plasma may have a fundamental role as well in the physical process generating jets in the extragalactic radio sources. The concept of dyadosphere originally introduced in Reissner-Nordstr{\o}m black hole in order to create the $e^+e^-$ plasma relevant for GRBs is now being generalized to the process of vacuum polarization originating in a Kerr-Newman black hole due to magneto-hydrodynamical process of energy extraction (see e.g. \citet{pu01} and references therein). The concept of dyado-torus is relevant for both the extraction of rotational and electromagnetic energy from the most general black hole \citet{cr71}. Such research is still ongoing and may be relevant for the analysis of microquasars and active galactic nuclei.

\section{The fireshell in the Livermore code}

\subsection{The hydrodynamics and the rate equations for the plasma of $e^+e^-$-pairs}\label{hydro_pem}

The evolution of the $e^+e^-$-pair plasma generated in the dyadosphere has been treated in \citet{rswx99,rswx00}. We recall here the basic governing equations in the most general case in which the plasma fluid is composed of $e^+e^-$-pairs, photons and baryonic matter. The plasma is described by the stress-energy tensor
\begin{equation}
T^{\mu\nu}=pg^{\mu\nu}+(p+\rho)U^\mu U^\nu\, ,
\label{tensor}
\end{equation}
where $\rho$ and $p$ are respectively the total proper energy density and pressure in the comoving frame of the plasma fluid and $U^\mu$ is its four-velocity, satisfying
\begin{equation}
g_{tt}(U^t)^2+g_{rr}(U^r)^2=-1 ~,
\label{tt}
\end{equation}
where $U^r$ and $U^t$ are the radial and temporal contravariant components of the 4-velocity and 
\begin{equation}
ds^2=g_{tt}(r)dt^2+g_{rr}(r)dr^2+r^2d\theta^2 +r^2\sin^2\theta
d\phi^2 ~,
\label{s}
\end{equation}
where $g_{tt}(r) \equiv - \alpha^2(r)$ and $g_{rr}(r)= \alpha^{-2}(r)$ and $\alpha^{2} = \alpha^{2}\left(r\right) = 1-2M/r+Q^{2}/r^{2}$, where $M$ and $Q$ are the
total energy and charge of the core as measured at infinity.

The conservation law for baryon number can be expressed in terms of the proper baryon number density $n_B$
\begin{eqnarray}
(n_B U^\mu)_{;\mu}&=& g^{-{1\over2}}(g^{1\over2}n_B
U^\nu)_{,\nu}\nonumber\\
&=&(n_BU^t)_{,t}+{1\over r^2}(r^2 n_BU^r)_{,r}=0 ~.
\label{contin}
\end{eqnarray}
The radial component of the energy-momentum conservation law of the plasma fluid reduces to 
\begin{equation}
{\partial p\over\partial r}+{\partial \over\partial t}\left((p+\rho)U^t U_r\right)+{1\over r^2} { \partial
\over \partial r}  \left(r^2(p+\rho)U^r U_r\right)
-{1\over2}(p+\rho)\left[{\partial g_{tt}
 \over\partial r}(U^t)^2+{\partial g_{rr}
 \over\partial r}(U^r)^2\right] =0 ~.
\label{cmom2}
\end{equation}
The component of the energy-momentum conservation law of the plasma fluid equation along a flow line is
\begin{eqnarray}
U_\mu(T^{\mu\nu})_{;\nu}&=&-(\rho U^\nu)_{;\nu}
-p(U^\nu)_{;\nu},\nonumber\\ &=&-g^{-{1\over2}}(g^{1\over2}\rho
U^\nu)_{,\nu} - pg^{-{1\over2}}(g^{1\over2} U^\nu)_{,\nu}\nonumber\\
&=&(\rho U^t)_{,t}+{1\over r^2}(r^2\rho
U^r)_{,r}\nonumber\\
&+&p\left[(U^t)_{,t}+{1\over r^2}(r^2U^r)_{,r}\right]=0 ~.
\label{conse1}
\end{eqnarray}

We define also the total proper internal energy density $\epsilon$ and the baryonic mass density $\rho_B$ in the comoving frame of the plasma fluid,
\begin{equation}
\epsilon \equiv \rho - \rho_B,\hskip0.5cm \rho_B\equiv n_Bmc^2 ~.
\label{cpp}
\end{equation} 

\subsection{The numerical integration}\label{num_int2}

A computer code (\citet{wsm97,wsm98}) has been used to
evolve the spherically symmetric general relativistic hydrodynamic equations starting from the dyadosphere (\citet{rswx99}).

We define the generalized gamma factor $\gamma$ and the radial 3-velocity in the laboratory frame $V^r$
\begin{equation}
\gamma \equiv \sqrt{ 1 + U^r U_r},\hskip0.5cm V^r\equiv {U^r\over U^t}.
\label{asww}
\end{equation}
From Eqs.(\ref{s}, \ref{tt}), we then have
\begin{equation}
(U^t)^2=-{1\over g_{tt}}(1+g_{rr}(U^r)^2)={1\over\alpha^2}\gamma^2.
\label{rr}
\end{equation}
Following Eq.(\ref{cpp}), we also define 
\begin{equation}
E \equiv \epsilon \gamma,\hskip0.5cmD \equiv \rho_B \gamma,
\hskip0.3cm {\rm and}\hskip0.3cm\tilde\rho \equiv \rho\gamma
\label{cp}
\end{equation} 
so that the conservation law of baryon number (\ref{contin}) can then
be written as
\begin{equation}
{\partial D \over \partial t} = - {\alpha \over r^2} {
\partial \over \partial r} ({r^2 \over \alpha} D V^r).
\label{jay1}
\end{equation}
Eq.(\ref{conse1}) then takes the form,
\begin{equation}
{\partial E \over \partial t} = - {\alpha \over r^2} {
\partial \over \partial r} ({r^2 \over \alpha} E V^r) - p
\biggl[ {\partial \gamma \over \partial t} + {\alpha \over r^2}
{\partial \over \partial r} ({ r^2 \over \alpha} \gamma V^r)
\biggr].
\label{jay2}
\end{equation}
Defining the radial momentum density in the laboratory frame
\begin{equation}
S_r\equiv \alpha (p+\rho)U^tU_r = (D + \Gamma E) U_r,  
\label{mstate}
\end{equation}
we can express the radial component of the energy-momentum
conservation law given in Eq.(\ref{cmom2}) by
\begin{eqnarray}
{\partial S_r \over \partial t} &=& - {\alpha \over r^2} { \partial
\over \partial r} ({r^2 \over \alpha} S_r V^r) - \alpha {\partial p
\over \partial r}\nonumber\\ 
&-&{\alpha\over2}(p+\rho)\left[{\partial g_{tt}
\over\partial r}(U^t)^2+{\partial g_{rr} \over\partial
r}(U^r)^2\right]\nonumber\\ 
&=& - {\alpha \over r^2} { \partial \over
\partial r} ({r^2 \over \alpha} S_r V^r) - \alpha {\partial p \over
\partial r}\nonumber\\
&-& \alpha\left({M \over r^2}-{Q^2 \over r^3}\right)
\biggl({D + \Gamma E \over \gamma} \biggr) \biggl[ \left({\gamma \over
\alpha}\right)^2 + {(U^r)^2 \over \alpha^4 } \biggr] ~.
\label{jay3}
\end{eqnarray}

In order to determine the number-density of $e^+e^-$ pairs, we use the pair rate equation. 
We define the $e^+e^-$-pair density in the laboratory frame 
$N_{e^\pm} \equiv\gamma n_{e^\pm}$ and $N_{e^\pm}(T) \equiv\gamma
n_{e^\pm}(T)$, where $n_{e^{\pm}}(T)$ is the total proper number density of pairs in comoving frame at thermodynamic equilibrium with temperature $T$ in the process $e^+ + e^- \rightarrow \gamma + \gamma$ $\left(n_{e^-}(m, T) = n_{\gamma}(T)\right)$, $n_{e^{\pm}}$ is the total proper number density of pairs in comoving frame at a generic time before reaching the equilibrium. We write the rate equation in the form
\begin{equation}
{\partial N_{e^\pm} \over \partial t} = - {\alpha \over r^2} {
\partial \over \partial r} ({r^2 \over \alpha} N_{e^\pm} V^r) +
\overline{\sigma v} (N^2_{e^\pm} (T) - N^2_{e^\pm})/\gamma^2~,
\label{jay:E:ndiff}
\end{equation}
These partial differential equations have been integrated in Livermore starting from the dyadosphere distributions given in Fig. 17 (Right) in \citet{rubr} and assuming as usual ingoing boundary conditions on the horizon of the black hole. A simplified set of ordinary differential equations has been integrated in Rome and the results have been validated by comparison with the ones obtained in Livermore.

\section{The fireshell in the Rome code}

\subsection{Era I: expansion of PEM-pulse}

After the explosion from the dyadosphere a thermal plasma of $e^+e^-$ pairs and photons optically thick with respect to scattering processes begins to expand at ultrarelativistic velocity. In this era the expansion takes place in a region of very low baryonic contamination.

Recalling that the local number density of electron and positron pairs created as a function of radius is given by (\citet{prx98}):
\begin{equation}
n_{e^+e^-}(r) = {Q\over 4\pi r^2\left({\hbar\over
mc}\right)e}\left[1-\left({r\over r^\star}\right)^2\right] ~,
\label{nd}
\end{equation}
the limit on such baryonic contamination, where $\rho_{B_c}$ is the mass-energy density of baryons, is given by 
\begin{equation}
\rho_{B_c}\ll m_pn_{e^+e^-}(r) = 3.2\cdot 10^8\left({r_{ds}\over r}\right)^2\left[1-\left({r\over r_{ds}}\right)^2\right](g/cm^3).
\label{nb} 
\end{equation}
Near the horizon $r\simeq r_+$, this gives
\begin{equation}
\rho_{B_c}\ll m_pn_{e^+e^-}(r) =1.86 \cdot 10^{14}\left({\xi\over\mu}\right)(g/cm^3)\, ,
\label{nb1} 
\end{equation}
and near the radius of the dyadosphere $r_{ds}$:
\begin{equation}
\rho_{B_c}\ll m_pn_{e^+e^-}(r) = 3.2\cdot 10^8\left[1-\left({r\over r_{ds}}\right)^2\right]_{r\rightarrow r_{ds}}(g/cm^3)\, .
\label{nb2} 
\end{equation}
Such conditions can be easily satisfied in the collapse to a black hole, but not necessarily in a collapse to a neutron star. 

\begin{figure}
\centering
\includegraphics[width=9cm,clip]{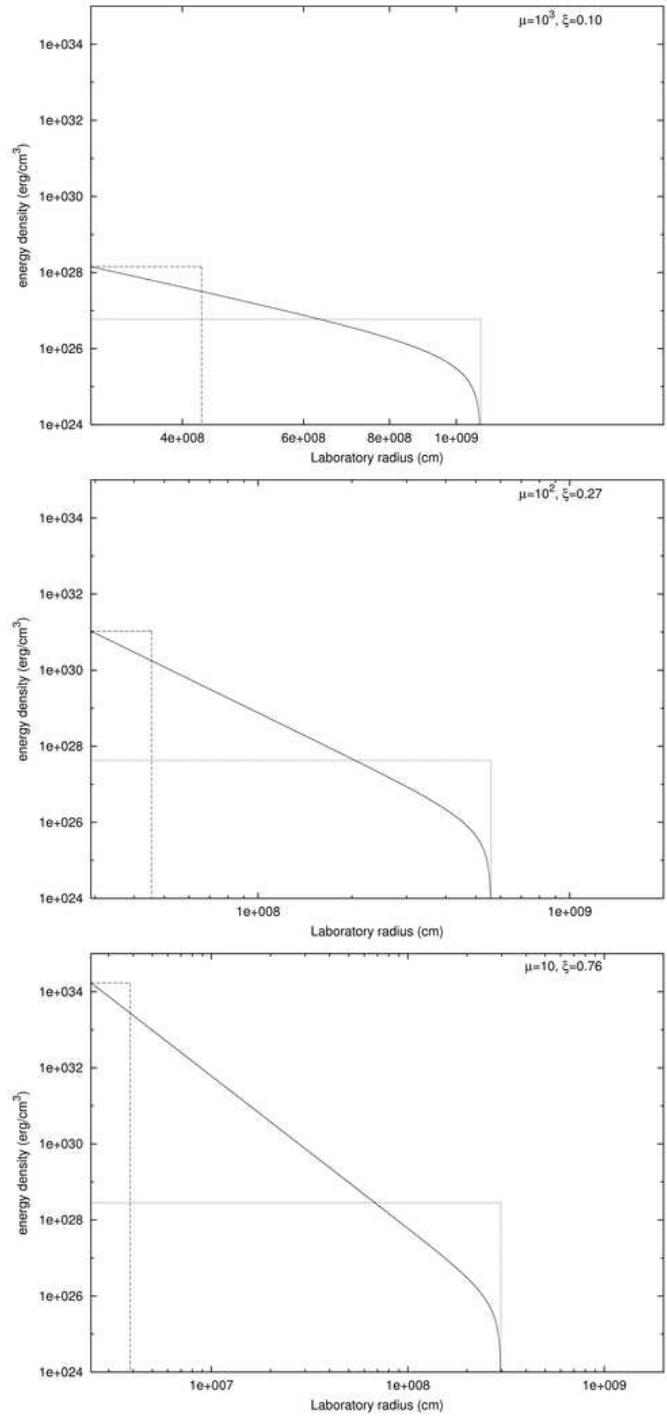}
\caption{{\itshape Three different dyadospheres corresponding to the same value and to different values of the two parameters $\mu$ and $\xi$ are given. The three different configurations are markedly different in their spatial extent as well as in their energy-density distribution (see text)}.}
\label{3dens}
\end{figure}

Consequently we have solved the equations governing a plasma composed solely of $e^+e^-$-pairs and electromagnetic radiation, starting at time zero from the dyadosphere configurations corresponding to constant density in Fig. \ref{3dens}.

The plasma of $e^+e^-$ pairs and photons is described by the covariant energy-momentum tensor $T^{\mu\nu}$ given in Eq.\eqref{tensor}. In general we have $g_{\mu \nu}U^{\mu}U^{\nu} = -1$. For a spherically symmetric motion this reduces to $g_{tt}(U^t)^2 + g_{rr}(U^r)^2 = -1$, where $U^t$ and $U^r$ are respectively temporal and radial controvariant components of 4-velocity $U^{\mu}$.
 
It is assumed that the gravitational interaction with central black hole is negligible with respect to the total energy of PEM-pulse such that a fluid expansion with special relativistic equations can be considered.

Moreover it is assumed that photons remain trapped inside fireball until complete transparency, i.e. the emission of electromagnetic radiation is negligible during the first phases of expansion, being therefore adiabatic (\citet{rswx99}). This assumption is valid until the photon mean free path is negligible with respect to the thickness of pulse.

The thermodynamic quantities used to describe the process are the total proper internal energy density of pulse $\epsilon$, given by $\epsilon = \epsilon_{e^+} + \epsilon_{e^-} +  \epsilon_{\gamma}$, where $\epsilon_{e^+}$ ($\epsilon_{e^-}$) is total proper internal energy density of electrons (positrons) and $\epsilon_{\gamma}$ 
of photons. The proper number density of pairs $n_{e^{\pm}}$, if the system is in thermodynamic equilibrium initially at temperature $T$ of order $T \sim MeV$, enough for $e^+e^-$ pair creation, equals the proper number density of photons $n_{\gamma}$. This is not valid at lower temperature (\citet{brx01}). The pressure is $p={p_{e^+}}+{p_{e^-}}+{p_{\gamma}}$, where $p_{e^{\pm}}$ are electrons and positrons pressures and $p_{\gamma}$ is photons pressure. The system is highly relativistic, so the equation of state $p = {\epsilon} / {3}$ can be considered valid. This equation of state is represented with thermal index $\Gamma$:
\begin{equation}
\Gamma = 1 + { p\over \epsilon} ~.
\label{state}
\end{equation}

\subsubsection{Fermi integrals}    

\begin{figure}
\centering
\includegraphics[width=10cm]{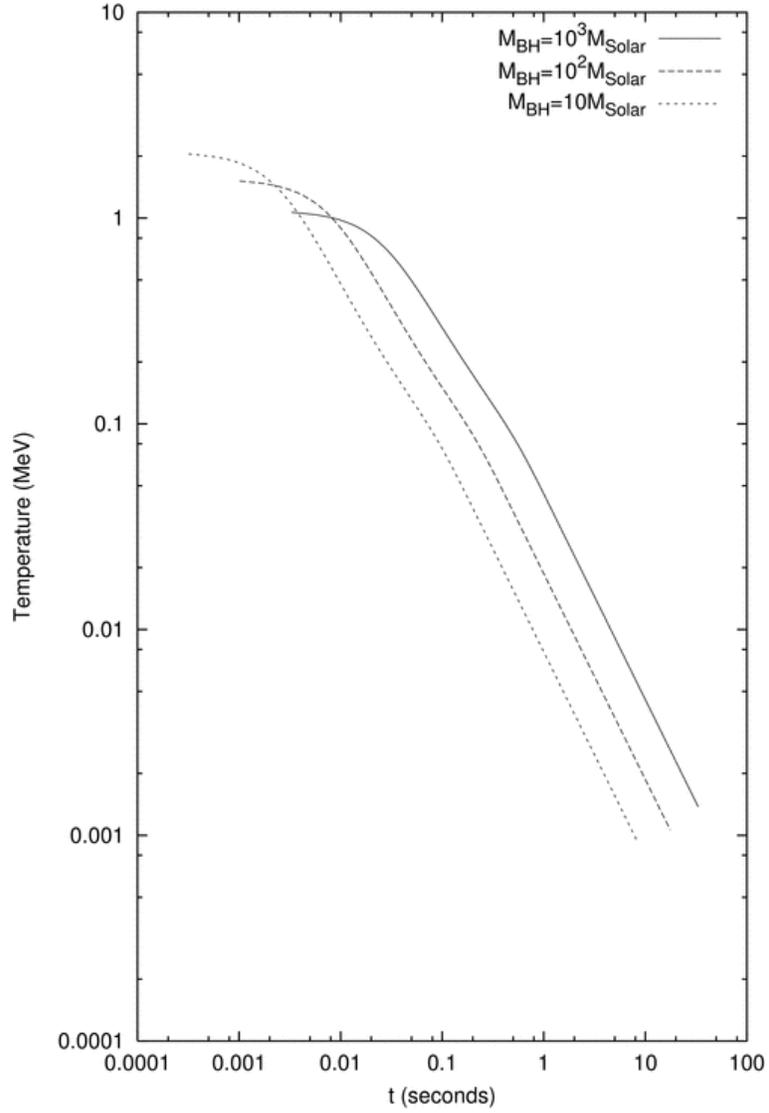}
\caption{{\itshape Temperature in comoving system as a function of emission time for different values of black hole mass $\mu$}.} 
\label{tem2}
\end{figure}

\begin{figure}
\centering
\includegraphics[width=10cm]{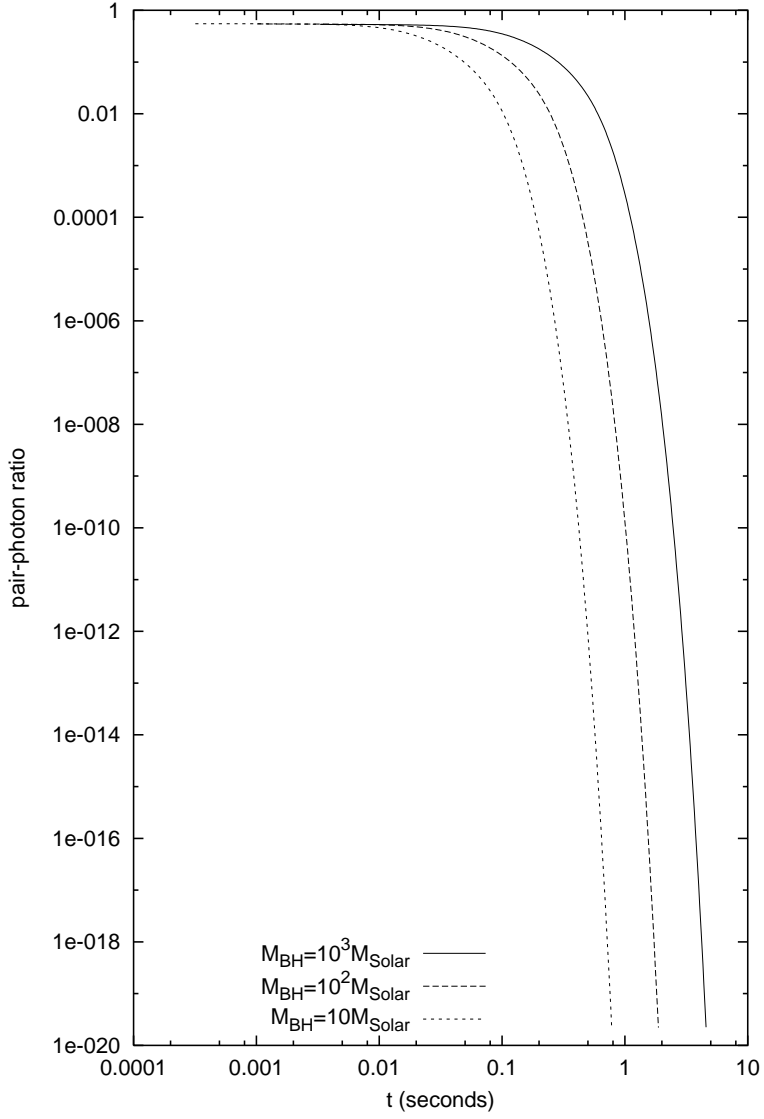}
\caption{{\itshape Ratio between number density of pairs $e^+e^-$ $n_{e^{\pm}}$ and number density of photons ${n_{\gamma}}\left(T\right)$ as a function of emission time for different values of black hole mass $\mu$}.} 
\label{pairs_fig}
\end{figure}

Thermodynamical quantities introduced above are expressed in terms of integrals over Bose distribution for photons and Fermi distribution for $e^+e^-$ pairs with zero chemical potentials $\mu_{\gamma}$ and $\mu_{e^{\pm}}$. We begin from the reaction $e^+ + e^- \rightarrow \gamma + \gamma$. From statistical mechanics it is known that given a thermodynamic system at temperature $T$ kept inside a volume $V$ and made of a number of particle variable $N$, the thermodynamic equilibrium is expressed by the condition that the potential free energy of Helmholtz $F(T, V, N)$ is stationary with respect to $N$ variations:
\begin{equation}
\left({\partial F \over \partial N}\right)_{T,V} = 0 ;
\label{helmh}
\end{equation}
by definition chemical potential $\mu$ is given by
\begin{equation}
\mu = \left({\partial F \over \partial N}\right)_{T,V} ;
\label{potenziale}
\end{equation}
so that for a system made by a photon gas at equilibrium with matter with respect to creation and adsorption processes, we have $\mu_{\gamma}=0$ (\citet{ll5}). We assume the chemical potential of electrons and positrons to be equal to zero: $\mu_{e^-} = 0$, $\mu_{e^+} = 0$. In the following the expressions of thermodynamical quantities as Fermi integrals are listed. The proper number density of electrons (\citet{Weinberg1972}) is given by
\begin{eqnarray}
{n_{e^-}}\left(m, T, \mu_{e^-}\right) & = & \frac{2}{h^3} \int \frac{d^3 p}{e^{\frac{\sqrt{(pc)^2+(mc^2)^2}}{kT}}+1} = \nonumber \\
& = & \frac{8\pi}{h^3} \int_0^{+\infty} \frac{p^2}{e^{\frac{\sqrt{(pc)^2+(mc^2)^2}}{kT}}+1} dp = \nonumber \\ 
& = & \frac{aT^3}{k} {7\over8} {1\over A} \int_0^{+\infty} \frac{z^2}{e^{\sqrt{z^2+(mc^2/kT)^2}}+1} dz,
\label{nel1}
\end{eqnarray}
where $z=pc/kT$, $m$ is the electron mass, $T$[MeV] is the temperature of fireball in comoving frame, $a$ is a constant given by $a= {8\pi^5k^4} / {15h^3c^3} = 1.37 \cdot 10^{26} {erg} / {cm^3 MeV^4}$, $k$ is the Boltzmann constant and $A=(7/4)(\pi^4/15)$ is a numerical constant introduced for convenience.\\
Since the thermodynamic equilibrium is assumed and in all cases considered the initial temperature is larger than $e^+e^-$ pairs creation threshold ($T > 1$~MeV), the proper number density of electrons is roughly equal to that one of photons:
\begin{equation}
n_{e^{\pm}} \sim {n_{e^-}}\left(T\right) \sim {n_{\gamma}}\left(T\right); 
\label{equil_num}
\end{equation}
in these conditions the number of particles is conserved:
\begin{equation}
\left(n_{e^{\pm}}U^{\mu}\right);_{\mu} = 0.
\label{cons_equil}
\end{equation}
Later on, for $T \ll 1$MeV (see Fig. \ref{tem2}), $e^+e^-$ pairs go on in annihilation but can not be created anymore, therefore
\begin{equation}
{n_{\gamma}}\left(T\right) > n_{e^{\pm}} > n_{e^{\pm}}\left(T\right)\, 
\label{equil_num2}
\end{equation}
as shown in Fig. \ref{pairs_fig}.

The total proper internal energy density for photons is given by
\begin{equation}
\epsilon_{\gamma} = \frac{2}{h^3} \int \frac{h\nu}{e^{\frac{h\nu}{kT}}-1} d^3p = aT^4
\label{enpho}
\end{equation}
where $p=h\nu/c$. The total proper internal energy density for electrons is given by:
\begin{eqnarray}
\epsilon_{e^-} & = & \frac{2}{h^3} \int \frac{\sqrt{(pc)^2+(mc^2)^2}}{e^{\frac{\sqrt{(pc)^2+(mc^2)^2}}{kT}}+1} d^3 p = \nonumber \\ 
& = & \frac{8\pi}{h^3} \int_0^{+\infty} \frac{p^2\sqrt{(pc)^2+(mc^2)^2}}{e^{\frac{\sqrt{(pc)^2+(mc^2)^2}}{kT}}+1} dp = \nonumber \\ 
& = & aT^4 {7\over4} {1\over A} \int_0^{+\infty} \frac{z^2\sqrt{z^2+(mc^2/kT)^2}}{e^{\sqrt{z^2+(mc^2/kT)^2}}+1} dz
\label{enel1}
\end{eqnarray}
where $z=pc/kT$ and the integral is computed numerically. Therefore the total proper internal energy density of the PEM-pulse, summing up all the contributions of photons and $e^+e^-$ pairs, is given by
\begin{equation}
\epsilon_{tot} = aT^4 \left[ 1+ {7\over4} {2\over A} \int_0^{+\infty} \frac{z^2\sqrt{z^2+(mc^2/kT)^2}}{e^{\sqrt{z^2+(mc^2/kT)^2}}+1} dz \right]
\label{entot1}
\end{equation}  
where the factor $2$ in front of the integral takes into account of electrons and positrons.\\
About the pressure of the photons it holds
\begin{equation}
p_{\gamma} = {\epsilon_{\gamma} \over 3} = {aT^4 \over 3}; 
\label{ppho}
\end{equation}
and about the pressure of electrons
\begin{eqnarray}
p_{e^-} & = & \frac{2}{3h^3} \int \frac{1}{e^{\frac{\sqrt{(pc)^2+(mc^2)^2}}{kT}} + 1}\cdot \frac{(pc)^2}{\sqrt{(pc)^2+(mc^2)^2}} d^3 p = \nonumber \\ 
& = & \frac{8\pi}{3h^3} \int_0^{+\infty} \frac{p^2}{e^{\frac{\sqrt{(pc)^2+(mc^2)^2}}{kT}} + 1}\cdot \frac{(pc)^2}{\sqrt{(pc)^2+(mc^2)^2}} dp = \nonumber \\ 
& = & {aT^4\over3} {7\over4} {1\over A} \int_0^{+\infty} \frac{z^4}{e^{\sqrt{z^2+(mc^2/kT)^2}} + 1}\cdot \frac{1}{\sqrt{z^2+(mc^2/kT)^2}} dz.
\label{pel}
\end{eqnarray}
Therefore the total pressure of PEM-pulse is given by
\begin{equation}
p_{tot}= {\frac{aT^4}{3}} \left[ 1+ {7\over4} {2\over A} \int_0^{+\infty} \frac{z^4}{e^{\sqrt{z^2+(mc^2/kT)^2}} + 1}\cdot \frac{1}{\sqrt{z^2+(mc^2/kT)^2}} dz \right] .
\label{ptot}
\end{equation}

\subsubsection{Numerical code}    

In the following we recall a zeroth order approximation of the fully relativistic equations of the previous section (\citet{rswx99}):\\
(i) Since we are mainly interested in the expansion of the $e^+e^-$ plasma away from the black hole, we neglect the gravitational interaction.\\
(ii) We describe the expanding plasma by a special relativistic set of equations.\\

In the PEM-pulse phase the expansion in vacuum is described by a set of equation expressing:
\begin{itemize} 
\item entropy conservation, because of the assumption that emission of electromagnetic radiation is negligible up to transparency;
\item energy conservation, because the increase of kinetic energy is compensated by a decrease of total internal energy.
\end{itemize} 
For the expansion of a single shell, the adiabaticity is given by
\begin{equation}
{d\left(V\epsilon\right)}+p{dV}={dE}+p{dV}=0 ~,
\end{equation}
where $V$ is the volume of the shell in the comoving frame and $E = V\epsilon$ is the total proper internal energy of plasma. By using the equation of state Eq.(\ref{state}) we find
\begin{equation}
d{ln{\epsilon}} + \Gamma d{lnV} = 0\,
\label{cons_s_diff}
\end{equation}
and, by integrating, we find
\begin{equation}
\frac{\epsilon_{\circ}}{\epsilon} = \left(\frac{V}{V_{\circ}}\right)^{\Gamma};
\label{scale1}
\end{equation} 
recalling that the volume of the fireball in the comoving frame is given by $V = {\mathcal V} {\gamma}$, where ${\mathcal V}$ is the volume in the laboratory frame, we find
\begin{eqnarray}
{\epsilon_\circ\over \epsilon} &=& 
\left({V\over V_\circ}\right)^\Gamma=
\left({ {\mathcal V}\over  {\mathcal V}_\circ}\right)^\Gamma\left({\gamma
\over \gamma_\circ}\right)^\Gamma.
\label{scale} 
\end{eqnarray}   

The total energy conservation of the shell implies (\citet{rswx99}):
\begin{equation}
(\Gamma\epsilon) {\mathcal V} \gamma^2 = (\Gamma\epsilon_\circ) {\mathcal V}_\circ \gamma_\circ^2;
\label{cons_e2}
\end{equation}
and this gives the evolution for $\gamma$:
\begin{equation}
\gamma = \gamma_\circ\sqrt{{\epsilon_\circ{\mathcal V}_\circ
\over\epsilon{\mathcal V}}} 
\label{gammavar}
\end{equation}
Substituting this expression for $\gamma$ in (\ref{scale}) the final equation for proper internal energy density is found
\begin{equation}
\epsilon = {\epsilon_{\circ}} \left({{\mathcal V}_{\circ} \over {\mathcal V}}\right)^{\frac{\Gamma}{2 - \Gamma}}
\label{scale2}
\end{equation}
The evolution of a plasma of $e^+e^-$ pairs and photons should be treated by relativistic hydrodynamics equations describing the variation of the number of particles in the process. The 4-vector number density of pairs is defined $(n_{e^{\pm}}U^{\mu})$, which in the comoving frame reduces to the 4-vector $(n_{e^{\pm}}, 0, 0, 0)$. The law of number conservation for pairs is
\begin{eqnarray}
\left(n_{e^\pm}U^\mu\right)_{;\mu} &=& \frac{1}{\sqrt {-g}} \left({\sqrt {-g}} n_{e^{\pm}} U^\mu\right)_{,\mu} = \nonumber \\
& = & \left(n_{e^\pm}U^t\right)_{,t}+{1\over r^2}\left(r^2 n_{e^\pm}U^r\right)_{,r} = 0
\label{rate}
\end{eqnarray}
where $g = \parallel g^{\mu \nu} \parallel = -r^4 sin^2\theta $ is the determinant of Reissner-Nordstr{\o}m metric. In the system processes of creation and annihilation of particles occur due to collisions between particles. If the number of particles is conserved, it holds $\left(n_{e^\pm}U^\mu\right)_{;\mu} = 0$; if instead it is not conserved, in the assumptions that only binary collisions between particles occur and in the hypothesis of molecular caos, the Eq.\eqref{rate} becomes
\begin{equation}
\left(n_{e^\pm}U^\mu \right)_{;\mu} = {\overline{\sigma v}} \left[n_{e^-}(T)n_{e^+}(T) - n_{e^-}n_{e^+}\right]
\label{rate1}
\end{equation}
where $\sigma$ is the cross section for the process of creation and annihilation of pairs, given by {}
\begin{equation}
\sigma = {\frac{\pi {r_e}^2}{\alpha_\circ + 1 }} \left[{\frac{{\alpha_\circ}^2 + 4 \alpha_\circ + 1}{{\alpha_\circ}^2 - 1}}\ln\left(\alpha_\circ + \sqrt{{\alpha_\circ}^2 - 1}\right) - {\frac{\alpha_\circ + 3}{\sqrt{{\alpha_\circ}^2 - 1}}}\right],
\label{cross}
\end{equation}
with $\alpha_\circ = \frac{E}{mc^2}$ and $E$ total energy of positrons in the laboratory frame, and $r_e = \frac{e^2}{mc^2}$ the classical radius of electron, $v$ is the sound velocity in the fireball:
\begin{equation}
v = c \sqrt{\frac{p_{tot}}{\epsilon_{tot}}},
\label{vsound}
\end{equation}
and $\overline{\sigma v}$ is the mean value of $\sigma v$; for $\sigma$ we use as a first approximation the Thomson cross section, $\sigma_T = 0.665\cdot 10^{-24} cm^2$; $n_{e^{\pm}}(T)$ is the total proper number density of electrons and positrons in comoving frame at thermodynamic equilibrium in the process $e^+ + e^- \rightarrow \gamma + \gamma$ $\left(n_{e^-}(m, T) = n_{\gamma}(T)\right)$, $n_{e^{\pm}}$ is the total proper number density of electrons and positrons in comoving frame at a generic time before reaching the equilibrium.\\
Using the approximation of special relativity, the 4-velocity is written $U^\mu = (\gamma, \gamma \frac{v}{c})$; Eq.\eqref{rate1} in hybrid form becomes
\begin{equation}
{\partial \left({n_{e^{\pm}} \gamma}\right) \over \partial t} = - {1\over r^2} {\partial \over \partial r} \left(r^2 n_{e^\pm} \gamma V^r\right) + \overline{\sigma v} \left(n^2_{e^\pm} (T) - n^2_{e^\pm}\right) ,
\label{paira}
\end{equation}
valid for electrons and positrons.\\
Now we have a complete set of equations for numerical integration: (\ref{scale2}), (\ref{gammavar}) and (\ref{paira}).

If we now turn from a single shell to a finite distribution of shells, we can introduce the average values of the proper internal energy and pair number densities ($\epsilon, n_{e^\pm}$) for the PEM-pulse, where the average $\gamma$-factor is defined by
\begin{equation}
\gamma={1\over{\mathcal V}}\int_{\mathcal V}\gamma(r) d{\mathcal V},
\label{ga}
\end{equation}
and ${\mathcal V}$ is the total volume of the shell in the laboratory frame (\citet{rswx99}).

\begin{figure}
\centering
\includegraphics[width=15cm]{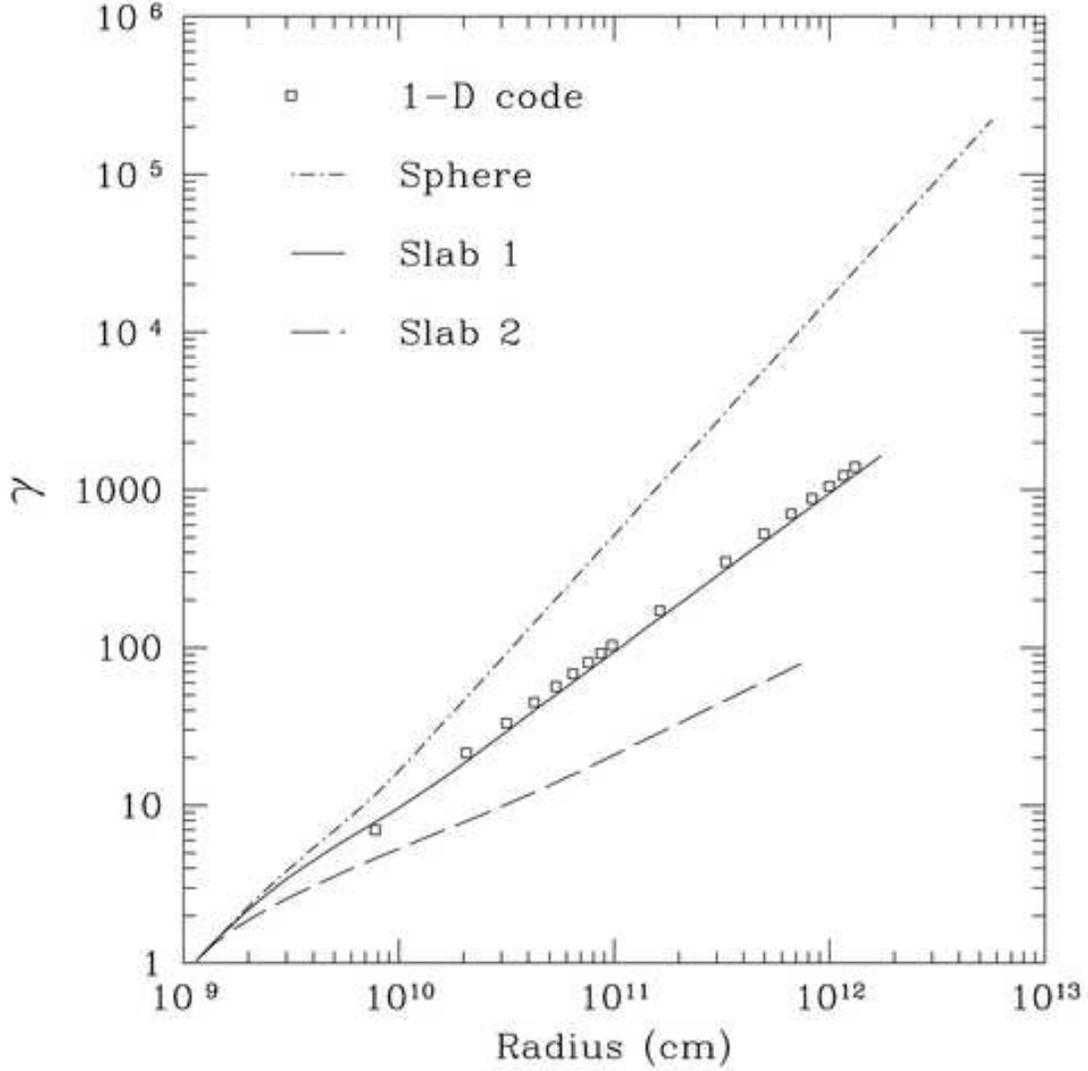}
\caption{{\itshape Lorentz $\gamma$ factor as a function of radial coordinate. Three schemes of expansion of PEM-pulse (see text) are compared with solution of hydrodynamics relativistic equations numerically integrated for a black hole with $\mu = 10^3$ and $\xi = 0.1$. The result is in accordance with the scheme of a fireball with constant thickness in laboratory frame}.}
\label{shells}
\end{figure}

In principle we could have an infinite number of possible schemes to define geometry of the expanding shell. Three different possible schemes have been proposed (\citet{rswx99}):
\begin{itemize}
\item Sphere. An expansion with radial component of 4-velocity proportional to the distance to the black hole $\displaystyle U_r (r) = U \frac{r}{{\mathcal R}(t)}$, where $U$ is the radial component of 4-velocity on the external surface of PEM-pulse (having radius ${\mathcal R}(t)$), the factor $\gamma$ from (\ref{ga}) is 
\begin{equation}
\gamma = \frac{3}{8 U^3}\left[2U\left(1 + U^2\right)^{3 \over 2} - U\left(1 + U^2\right)^{1 \over 2} - ln\left(U + \sqrt{1 + U^2}\right)\right];
\label{gammamean_1sch}
\end{equation}
this distribution corresponds to a uniform and time decreasing density, like in Friedmann model for the universe;
\item Slab 1. An expansion with thickness of fireball constant ${\mathcal D} = r_{ds} - r_+$ in laboratory frame in which the black hole is at rest, with $U_r (r) = U_r = cost$ and $\gamma = \sqrt{1 + {U_r}^2 }$; this distribution does not require an average;
\item Slab 2. An expansion with thickness of fireball constant in comoving frame of PEM-pulse.
\end{itemize} 
The result has been compared with the one of hydrodynamic equation in general relativity (\citet{rswx99}, see Fig. \ref{shells}). Excellent agreement has been found with the scheme in which the thickness of fireball is constant in laboratory frame: what happens is that the thickness in comoving frame increases, but due to the Lorentz contraction, it is kept constant in laboratory frame and equal to ${\mathcal D}=\left(r_{\rm ds}-r_{+}\right)$. In this case $U_r = \sqrt{\gamma^2 - 1}$, where $\gamma$ is computed by conservation equations.\\
A similar situation occurs for the temperature of PEM-pulse. In the comoving frame the temperature decreases as $T'\sim{R^{-1}}$, in accordance with results in literature (\citet{p99}). Since $\gamma$ monotonically increases as $\gamma\sim{R}$ (\citet{lett1}), in laboratory frame $T=\gamma{T'}\sim{constant}$ (\citet{rswx00}); photons are blue-shifted in laboratory frame in such away that, at least in the first phase, the temperature measured by an observer at infinity is constant. The numerical value of the temperature of equilibrium at each instant is found by imposing the equivalence, within a certain precision, of (\ref{entot1}) numerically computed and (\ref{scale2}).

Even if the PEM-pulse is optically thick in the expansion before transparency, photons located at a distance from the external surface less their mean free path can escape and reach the observer at infinity. The mean free path in the comoving frame is given by
\begin{equation}
L_{\gamma} = \frac{1}{\sigma{n_{e^+e^-}}}  \sim  10^{-6}cm 
\label{lambda2}
\end{equation}
while in laboratory frame is given by $\lambda = {L_{\gamma} / \gamma} \sim 10^{-8}$cm.
However the luminosity emitted at this stage is negligible, since the ratio between $\lambda$ and the thickness of the fireball ${\mathcal D}$ in the laboratory frame (with ${\mathcal D}= (r_{ds}- r_+) \sim 10^{9}$cm) is of the order of $\lambda / {\mathcal D} \simeq 10^{-17}$.

\subsection{Era II: interaction of the PEM pulse with remnant}\label{era2}\

The PEM pulse expands initially in a region of very low baryonic contamination created  by the process of gravitational collapse. As it moves outside the baryonic remnant of the progenitor star is swept up.
The existence of such a remnant is necessary in order to guarantee the overall charge neutrality of the system: the collapsing core has the opposite charge of the remnant and the system as a whole is clearly neutral. The number of extra charges in the baryonic remnant negligibly affects the overall charge neutrality of the PEM pulse.

The baryonic matter remnant is assumed to be distributed well outside the dyadosphere in a shell of thickness $\Delta$ between an inner radius $r_{\rm in}$ and an outer radius $r_{\rm out}=r_{\rm in}+\Delta$ at a distance from the black hole not so big that the PEM pulse expanding in vacuum has not yet reached transparency and not so small that the system will reach enoughly high value of Lorentz $\gamma$ in order to not be stopped in the collision (see Fig. \ref{cip2}). For example we choose
\begin{equation}
r_{\rm in}=100r_{\rm ds},\hskip 0.5cm \Delta = 10r_{\rm ds}.
\label{bshell_1}
\end{equation}
The total baryonic mass $M_B=N_Bm_p$ is assumed to be a fraction of the dyadosphere initial total energy $(E_{\rm dya})$. The total baryon-number $N_B$ is then expressed as a function of the dimensionless parameter $B$ given by Eq.\eqref{Bdef}. We shall see below the role of $B$ in the determination of the features of the GRBs. We already saw the sense in which $B$ and $E_{dya}$ can be considered to be the only two free parameters of the black hole theory for the entire GRB family, the so called ``long bursts''. For the so called ``short bursts'' the black hole theory depends on the two other parameters $\mu$, $\xi$, since in that case $B=0$ since most of the energy, unless the whole energy, in the pulse is emitted at transparency. 
The baryon number density $n^\circ_B$ is assumed to be a constant
\begin{equation}
n^\circ_B={N_B\over V_B},\hskip0.5cm \bar\rho^\circ_B=m_pn^\circ_B c^2.
\label{bnd}
\end{equation}
 
As the PEM pulse reaches the region $r_{\rm in}<r<r_{\rm out}$, it interacts with the baryonic matter which is assumed to be at rest. In our model we make the following assumptions to describe this interaction: 
\begin{itemize}
\item the PEM pulse does not change its geometry during the interaction;
\item the collision between the PEM pulse and the baryonic matter is assumed to be inelastic,
\item the baryonic matter reaches thermal equilibrium with the photons and pairs of the PEM pulse.
\end{itemize}
These assumptions are valid if: (i) the total energy of the PEM pulse is much larger than the total mass-energy of baryonic matter $M_B$, $10^{-8}<B \le 10^{-2}$ (see Fig. \ref{B10-2}), (ii) the ratio of the comoving number density  of pairs and baryons at the moment of collision $n_{e^+e^-}/n^\circ_B$ is very high (e.g., $10^6 <n_{e^+e^-}/ n^\circ_B <10^{12}$) and (iii)  the PEM pulse has a large value of the gamma factor ($\gamma > 100$).

In the collision between the PEM pulse and the baryonic matter at $r_{\rm out}>r>r_{\rm in}$ , we impose total conservation of energy and momentum. We consider the collision process between two radii $r_2,r_1$ satisfying 
$r_{\rm out}>r_2>r_1>r_{\rm in}$ and $r_2-r_1\ll \Delta$. The amount of baryonic mass acquired by the PEM pulse is
\begin{equation}
\Delta M = {M_B\over V_B}{4\pi\over3}(r_2^3-r_1^3) ,
\label{mcc_2}
\end{equation}
where $M_B/ V_B$ is the mean-density of baryonic matter at rest in the laboratory frame.

As for energy density of dyadosphere, here also we choose a simplification for the energy density: in fact during the passage of the shell a deposition of material on the external surface of the fireball creates; however we neglected this effect and assumed that this material after collision diffuses instantaneously in the pulse with a constant density:
\begin{equation}
n'_B={N'_B\over V},
\label{nB}
\end{equation}   
where $N'_B$ is the number of particle of the remnant shell swept up by the pulse and $V$ is the comoving volume of the fireball.

The conservation of total energy leads to the estimate of the corresponding quantities before (with ``$\circ$'') and after such a collision  
\begin{equation}
(\Gamma\epsilon_\circ + \bar\rho^\circ_B)\gamma_\circ^2{\mathcal V}_\circ + \Delta M = (\Gamma\epsilon + \bar\rho_B + {\Delta M\over V} + \Gamma\Delta\epsilon)\gamma^2{\mathcal V},
\label{ecc_2}
\end{equation}
where $\Delta\epsilon$ is the corresponding increase of internal energy due to the collision. Similarly the momentum-conservation gives
\begin{equation}
(\Gamma\epsilon_\circ + \bar\rho^\circ_B)\gamma_\circ U^\circ_r{\mathcal V}_\circ = (\Gamma\epsilon + \bar\rho_B + {\Delta M\over V} + 
\Gamma\Delta\epsilon)\gamma U_r{\mathcal V},
\label{pcc_2}
\end{equation}
where the radial component of the four-velocity of the PEM pulse is $U^\circ_r=\sqrt{\gamma_\circ^2-1}$ and $\Gamma$ is the thermal index. 
We then find 
\begin{eqnarray}
\Delta\epsilon & = & {1\over\Gamma}\left[(\Gamma\epsilon_\circ + \bar\rho^\circ_B) {\gamma_\circ U^\circ_r{\mathcal V}_\circ \over \gamma U_r{\mathcal V}} - (\Gamma\epsilon + \bar\rho_B + {\Delta M\over V})\right],\label{heat_2}\\
\gamma & = & {a\over\sqrt{a^2-1}},\hskip0.5cm a\equiv {\gamma_\circ  \over  
U^\circ_r}+ {\Delta M\over (\Gamma\epsilon_\circ + \bar\rho^\circ_B)\gamma_\circ U^\circ_r{\mathcal V}_\circ}.
\label{dgamma_2}
\end{eqnarray}
These equations determine the gamma factor $\gamma$ and the internal energy density $\epsilon=\epsilon_\circ +\Delta\epsilon$ in the capture process of baryonic matter by the PEM pulse.

The effect of the collision of the PEM pulse with the remnant leads to the following consequences:
\begin{itemize}
	\item a reheating of the plasma in the comoving frame but not in the laboratory frame; an increase of the number of 					$e^+e^-$ pairs and of free electrons originated from the ionization of those atoms remained in the baryonic 						remnant; correspondingly this gives an overall increase of the opacity of the pulse; 
	\item the more the amount of baryonic matter swept up, the more internal energy of the PEMB pulse is converted in 						kinetic energy of baryons.
\end{itemize}

By describing the interaction of PEM pulse with remnant as completely inelastic collision of two particles, one can compute by the energy-momentum conservation equation the decrease of Lorentz $\gamma$ and the increase of internal energy as function of $B$ parameter and also the ultrarelativistic approximation ($\gamma_\circ \rightarrow \infty$):
\begin{enumerate}
	\item an abrupt decrease of the gamma factor given by
				\begin{displaymath}
				\gamma_{coll} = \gamma_\circ \frac{1+B}{\sqrt{ {\gamma_\circ}^2 \left(2B+B^2 \right) +1}} 															\stackrel{\rightarrow}{\gamma_\circ \rightarrow \infty}  \frac{B+1}{\sqrt{B^2+2B}}\, ,
				\label{gamma_circ}
				\end{displaymath}
				where $\gamma_\circ$ is the gamma factor of the PEM pulse before the collision,
	\item an increase of the internal energy in the comoving frame $E_{coll}$ developed in the collision given by
				\begin{displaymath}
				\frac{E_{coll}}{E_{dya}} =  \frac{\sqrt{ {\gamma_\circ}^2 \left(2B+B^2 \right) +1}}{\gamma_\circ} - 										\left(\frac{1}{\gamma_\circ} + B \right) \stackrel{\rightarrow}{\gamma_\circ \rightarrow \infty} 	-B+\sqrt{B^2+2B} \, ,
				\label{E_int/E}
				\end{displaymath}
\end{enumerate}
This approximation applies when the final gamma factor at the end of the PEM pulse era is larger than $\gamma_{coll}$, right panel in Fig. \ref{cip2}.

In this phase of expansion, another thermodynamic quantity has not been considered: the chemical potential $\mu$ of the electrons from ionization of baryonic remnant. We remind that the total proper number density of electrons of ionization is given by
\begin{equation}
{n^b_{e^-}}(m, T, \mu) = \frac{aT^3}{k} {7\over8} {1\over A} \int_0^{+\infty} \frac{z^2}{e^{\sqrt{z^2+(mc^2/kT)^2} + \frac{\mu}{kT}}+1} dz\,
\label{num_pote1}
\end{equation}
four equations are imposed to find a formula useful for numerical computation: the first one is the thermodynamical equilibrium of fireball, or
\begin{equation}
n_{e^{\pm}} ({T_{\circ}}) = n_{\gamma} ({T_{\circ}}) ;
\label{equil}
\end{equation} 
the second one is
\begin{equation}
n^b_{e^-} = \bar Z n_B
\label{neb}
\end{equation} 
where $1/2<\bar Z<1$, with $\bar Z=1$ for hydrogen atoms and $\bar Z=1/2$ for baryonic matter in general; the third one derives from the definition of $B$, and states a relation between the two densities $n_B$ and $n_{e^\pm}$: from definition of $B$, we have
\begin{equation}
\frac{N_B}{N_{e^{\pm}}(T_{\circ})} = B \frac{E_{dya}}{m_pc^2} \frac{1}{N_{e^{\pm}}(T_{\circ})} = 10^b 
\label{def_di_b_1}
\end{equation}
where $T_{\circ}$ is the initial temperature of fireball and $b$ is a parameter ($b < 0$) defined by (\ref{def_di_b_1}); so if $V_{\circ}$ is the initial volume of dyadosphere and $w$ the initial volume of the baryonic shell 
\begin{equation}
n^\circ_B = 10^b n_{e^\pm}(T_{\circ}) \frac {V_{\circ}}{w} ;  
\label{def_di_b}
\end{equation}
finally the fourth one is the conservation law of baryonic matter
\begin{equation}
(n^b_{e^-}U^\mu)_{;\mu} = 0.
\label{fourth}
\end{equation}
Therefore the chemical potential $\mu$ is numerically determined at a certain time of expansion if the initial temperature $T_{\circ}$ of fireball and the initial volume of baryonic shell $w$ are known and, at that time, the volume $V$, the temperature $T$ and the Lorentz factor $\gamma$ of the fireball, the volume of the baryonic shell swept up $vb$ and the ratio $\displaystyle {n^b_{e^-}(T) \over n^b_{e^-}}$:
\begin{equation}
2 \zeta (3) {\bar Z  10^b {n^b_{e^-}(T) \over n^b_{e^-}} {{{T_0}^3 w} \over {T^3 V \gamma}} \left({vb \over w}\right)} =
 \int_0^{+\infty} \frac{z^2}{e^{\sqrt{z^2+(mc^2/kT)^2} + \frac{\mu}{kT}}+1} dz
\label{pot_chi_fin}
\end{equation}
where the factor in brackets $\left({vb \over w}\right)$ must be considered only for $r > r_{\rm out}$, while the proportionality factor is the function zeta of Riemann $\zeta (x)$ for computation of $n_{\gamma}$, with $\zeta (3) = 1.202$.\\ 
Therefore the equations for this phase are (\ref{paira}), (\ref{nB}), (\ref{heat_2}), (\ref{dgamma_2}), and (\ref{pot_chi_fin}).

\subsection{Era III: expansion of PEMB pulse}

After the engulfment of the baryonic matter of the remnant the plasma formed of $e^+e^-$-pairs, electromagnetic radiation and baryonic matter expands again as a sharp pulse, namely the PEMB pulse. The calculation is continued as the plasma fluid expands,
cools and the $e^+e^-$ pairs recombine until it becomes optically
thin:
\begin{equation} 
\int_R dr(n_{e^\pm}+\bar
Zn_B)\sigma_T\simeq O(1),
\label{thin_1}
\end{equation}
where $\sigma_T =0.665\cdot 10^{-24}
{\rm cm^2}$ is the Thomson cross-section and the integration is over the radial interval of the PEMB pulse in the
comoving frame. 
In order to study the PEMB pulse expansion the validity of the slab approximation adopted for the PEM pulse phase has to be verified; otherwise the full hydrodynamics relativistic equations should be integrated. The PEMB pulse evolution firstly has been simulated by integrating the general relativistic hydrodynamical equations with the Livermore codes, for a total energy in the dyadosphere of $3.1\times 10^{54}$ erg and a baryonic shell of thickness $\Delta =10 r_{\rm ds}$ at rest at a radius of $100 r_{\rm ds}$ and $B \simeq 1.3\cdot 10^{-4}$. 

\begin{figure}
\centering
\includegraphics[width=15cm]{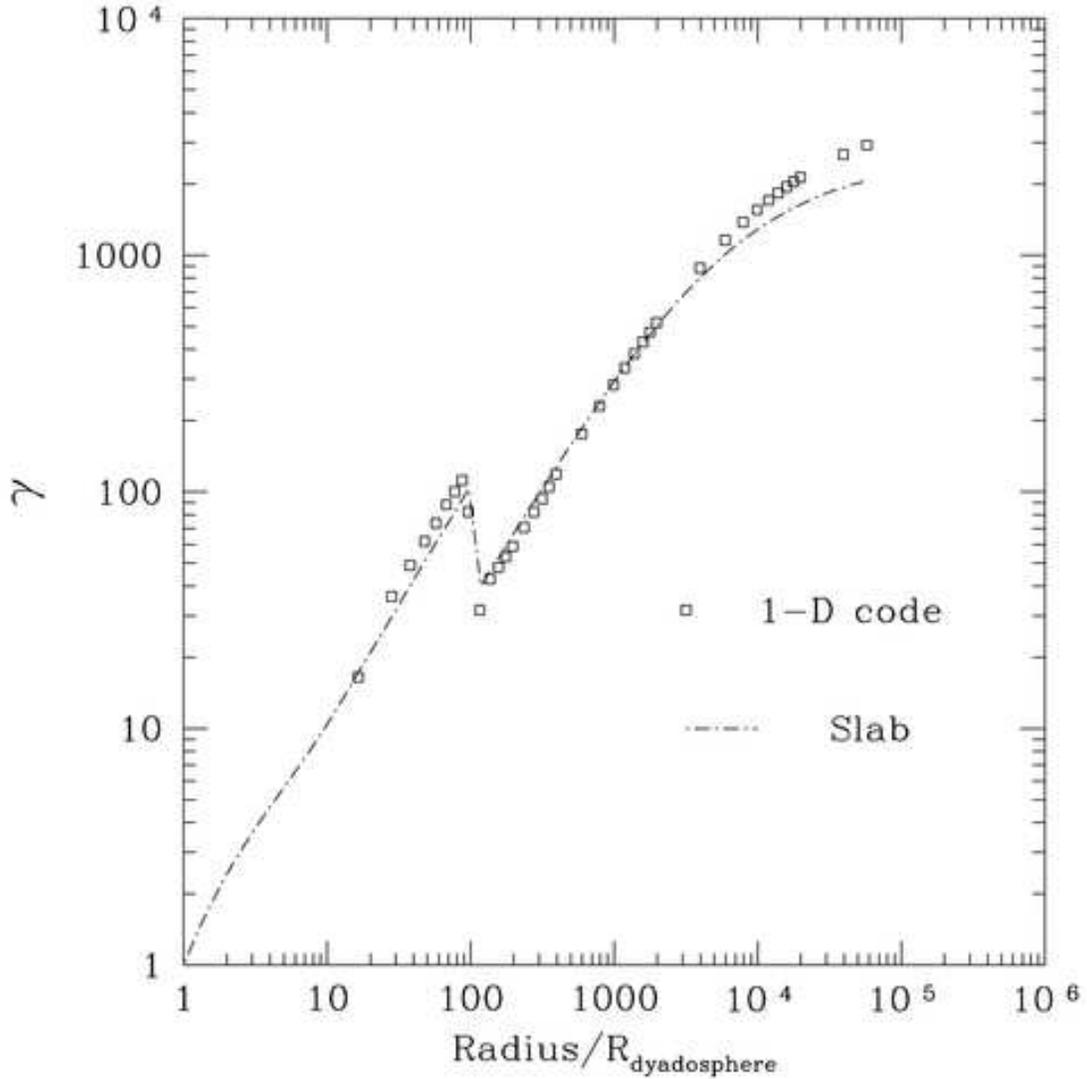}
\caption{{\itshape Lorentz $\gamma$ factor as a function of radial coordinate from the PEMB-pulse simulation is compared with the $\gamma$ factor as solution of hydrodynamics relativistic equations numerically integrated (open squares) for $E_{dya} = 3.1\times 10^{54}$erg  and $B=1.3\times 10^{-4}$, $r_{\rm in}=100r_{\rm ds}$ and $\Delta = 10r_{\rm ds}$. The result is in accordance with the scheme of a fireball with constant thickness in laboratory frame which is valid up to $B= 10^{-2}$}.}
\label{shells_2}
\end{figure}

In analogy with the special relativistic treatment for the PEM pulse, presented above (see also \citet{rswx99}), for the adiabatic expansion of the PEMB pulse in the constant-slab approximation described by the Rome codes the following hydrodynamical equations with $\bar\rho_B\not=0$ has been found
\begin{eqnarray}
{n_B^\circ\over n_B}&=& { V\over  V_\circ}={ {\mathcal V}\gamma
\over {\mathcal V}_\circ\gamma_\circ},
\label{be'}\\
{\epsilon_\circ\over \epsilon} &=& 
\left({V\over V_\circ}\right)^\Gamma=
\left({ {\mathcal V}\over  {\mathcal V}_\circ}\right)^\Gamma\left({\gamma
\over \gamma_\circ}\right)^\Gamma,
\label{scale1'}\\
\gamma &=&\gamma_\circ\sqrt{{(\Gamma\epsilon_\circ+\bar\rho^\circ_B){\mathcal V}_\circ
\over(\Gamma\epsilon+\bar\rho_B) {\mathcal V}}},
\label{result1'}\\
{\partial \over \partial t}(N_{e^\pm}) &=& -N_{e^\pm}{1\over{\mathcal V}}{\partial {\mathcal V}\over \partial t}+\overline{\sigma v}{1\over\gamma^2}  (N^2_{e^\pm} (T) - N^2_{e^\pm}).
\label{paira'_2}
\end{eqnarray}
In these equations ($r>r_{\rm out}$) the comoving baryonic mass and number densities are $\bar\rho_B=M_B/V$ and $n_B=N_B/V$, where $V$ is the comoving volume of the PEMB pulse.

The result is shown in Fig. \ref{shells_2} (\citet{rswx00}) where the bulk gamma factor as computed from the Rome and Livermore codes are compared and very good agreement has been found. This validates the constant-thickness approximation in the case of the PEMB pulse as well. On this basis we easily estimate a variety of physical quantities for an entire range of values of $B$.

\begin{figure}
\centering
\includegraphics[width=0.45\hsize,clip]{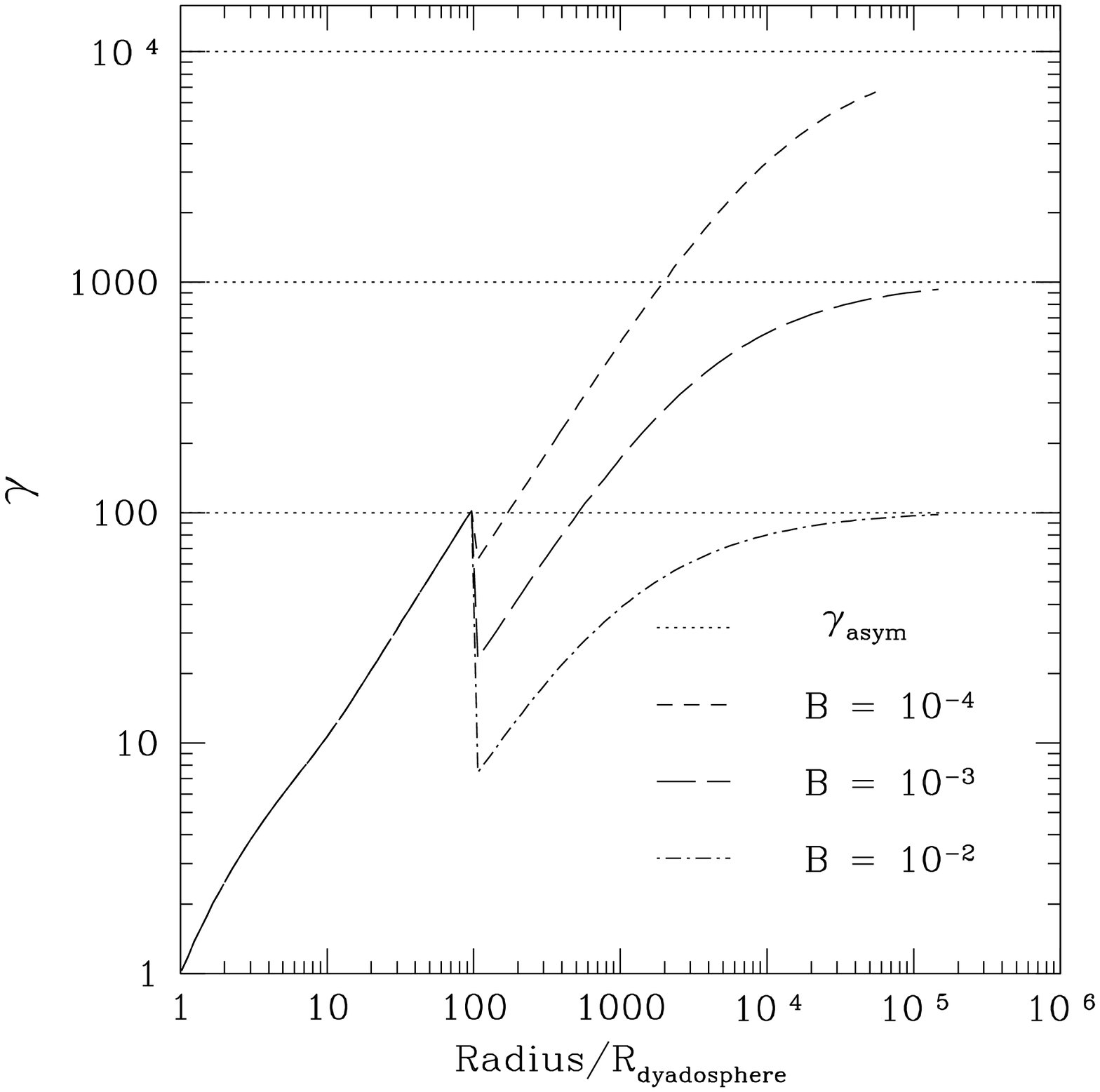}
\includegraphics[width=0.55\hsize,clip]{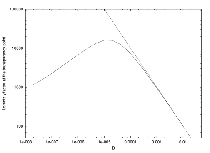}
\caption{{\itshape {\bf Left)} The gamma factors are given as functions of the radius in units of the dyadosphere radius for selected values of $B$ for the typical case $E_{dya}=3.1\times 10^{54}$ erg. The asymptotic values $\gamma_{\rm asym} = E_{\rm dya}/(M_Bc^2)=10^4,10^3,10^2$ are also plotted. The collision of the PEM pulse with the baryonic remnant occurs at $r/r_{ds}=100$ where the jump occurs.{\bf Right)} The $\gamma$ factor (the solid line) at the transparency point is plotted as a function of the $B$ parameter. The asymptotic value (the dashed line) $E_{\rm dya}/ (M_Bc^2)$ is also plotted}.}
\label{3gamma}
\end{figure}

For the same black hole different cases have been considered (\citet{rswx00}). The results of the integration show that for the first parameter range the PEMB pulse propagates as a sharp pulse of constant thickness in the laboratory frame, but already for $B\simeq 1.3\cdot 10^{-2}$ the expansion of the PEMB pulse becomes much more complex, turbulence phenomena can not be neglected any more and the constant-thickness approximation ceases to be valid.

It is also interesting to evaluate the final value of the gamma factor of the PEMB pulse when the transparency condition given by Eq.(\ref{thin_1}) is reached as a function of $B$, see Fig. \ref{3gamma}. For a given black hole, there is a {\em maximum} value of the gamma factor at transparency. By further increasing the value of $B$ the entire $E_{dya}$ is transferred into the kinetic energy of the baryons (see also \citet{rswx00}).

In Fig. \ref{3gamma}-Left we plot the gamma factor of the PEMB pulse as a function of radial distance for different amounts of baryonic matter. The diagram extends to values of the radial coordinate at which the transparency condition given by Eq.(\ref{thin_1}) is reached. The ``asymptotic'' gamma factor
\begin{equation}
\gamma_{\rm asym}\equiv {E_{\rm dya}\over M_B c^2}
\label{asymp}
\end{equation}
is also shown for each curve. The closer the gamma value approaches the ``asymptotic'' value (\ref{asymp}) at transparency, the smaller the intensity of the radiation emitted in the burst and the larger the amount of kinetic energy left in the baryonic matter (see Fig. \ref{3gamma}-Right).

\subsection{The approach to transparency: the thermodynamical quantities}

As the condition of transparency expressed by Eq.(\ref{thin_1}) is reached the {\em injector phase} terminates. The electromagnetic energy of the PEMB pulse is released in the form of free-streaming photons --- the P-GRB. The remaining energy of the PEMB pulse is released as an ABM pulse.

We now proceed to the analysis of the approach to the transparency condition. It is then necessary to turn from the pure dynamical description of the PEMB pulse described in the previous sections to the relevant thermodynamic parameters. Also such a description at the time of transparency needs the knowledge of the thermodynamical parameters in all previous eras of the GRB. 

\begin{figure}
\includegraphics[width=0.4\hsize,clip]{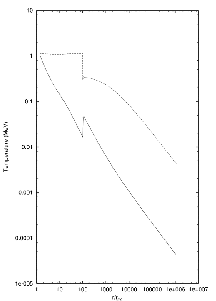}
\includegraphics[width=0.4\hsize,clip]{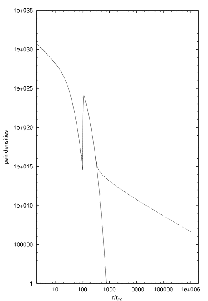}
\caption{{\bf Left)} The temperature of the plasma in the comoving frame $T'$(MeV) (the solid line) and in the laboratory frame $\gamma T'$ (the dashed line) are plotted as functions of the radius in the unit of the dyadosphere radius $r_{\rm ds}$. {\bf Right)} The number densities $n_{e^+e^-}(T)$ (the solid line) computed by the Fermi integral and $n_{e^+e^-}$ (the dashed line) computed by the rate equation (see section~\ref{hydro_pem}) are plotted as functions of the radius. $T'\ll m_ec^2$, two curves strongly divergent due to $e^+e^-$-pairs frozen out of the thermal equilibrium. The peak at $r\simeq100r_{\rm ds}$ is due to the internal energy developed in the collision.}
\label{pair}
\end{figure}

As above we shall consider as  a typical case an EMBH of $E_{dya}=3.1\times 10^{54}$ erg and $B=10^{-2}$. One of the key thermodynamical parameters is represented by the temperature of the PEM and PEMB pulses. It is given as a function of the radius both in the comoving and in the laboratory frames in Fig.~\ref{pair}. 
Before the collision the PEM pulse expands keeping its temperature in the laboratory frame constant while its temperature in the comoving frame falls (see \citet{rswx99}). In fact we have:
\begin{equation}
{d(\epsilon\gamma^2{\cal V})\over dt}=0,
\label{tc}
\end{equation}
where the baryon mass-density is $\rho_B=0$ and the thermal energy-density of photons and $e^+e^-$-pairs is $\epsilon=\sigma_B T^4(1+f_{e^+e^-})$, $\sigma_B$ is the Boltzmann constant and $f_{e^+e^-}$ is the Fermi-integral for $e^+$ and $e^-$. This leads to
\begin{equation}
\epsilon\gamma^2{\cal V}=E_{\rm dya},\hskip0.3cm T^4\gamma^2{\cal V}={\rm const.}
\label{econ}
\end{equation}
Since $e^+$ and $e^-$ in the PEM pulse are extremely relativistic, we have the equation of state $p\simeq\epsilon/3$ and the thermal index (\ref{state}) $\Gamma\simeq 4/3$ in the evolution of PEM pulse. Eq.(\ref{econ}) is thus equivalent to
\begin{equation}
T^3\gamma {\cal V}\simeq {\rm const.}
\label{encon}
\end{equation}
These two equations (\ref{tc}) and (\ref{encon}) result in the constancy of the laboratory temperature $T\gamma$ in the evolution of the PEM pulse.

It is interesting to note that Eqs.(\ref{econ}) and (\ref{encon}) hold as well in
the cross-over region where $T\sim m_ec^2$ and $e^+e^-$ annihilation takes place.  
In fact from the conservation of entropy it follows that asymptotically we have
\begin{equation}
      \frac{(V T^3)_{T<m_ec^2}}{(V T^3)_{T>m_ec^2}}  =\frac{11}{4}\ ,
\label{reheat}
\end{equation}
exactly for the same reasons and physics scenario discussed in the cosmological framework by Weinberg (see e.g. Eq.~(15.6.37) of \citet{Weinberg1972}). 
The same considerations when
repeated for the conservation of the total energy 
$\epsilon\gamma V=\epsilon\gamma^2{\cal V}$
following from Eq.~(\ref{tc}) then lead to
\begin{equation}
      \frac{(V T^4 \gamma)_{T<m_ec^2}}{(V T^4 \gamma)_{T>m_ec^2}}  
             =\frac{11}{4}\ .
\end{equation}
The ratio of these last two quantities gives asymptotically 
\begin{equation}
      T_\circ= (T \gamma)_{T>m_ec^2}= (T \gamma)_{T<m_ec^2},
\label{rt}
\end{equation}
where $T_\circ$ is the initial average temperature of the dyadosphere at rest.

During the collision of the PEM pulse with the remnant we have an increase in the number density of $e^+e^-$ pairs (see Fig.~\ref{pair}). This transition corresponds to an {\em increase} of the temperature in the comoving frame and a {\em decrease} of the temperature in the laboratory frame as a direct effect of the dropping of the gamma factor (see Fig.~\ref{3gamma}). 

\begin{figure}
\includegraphics[width=0.4\hsize,clip]{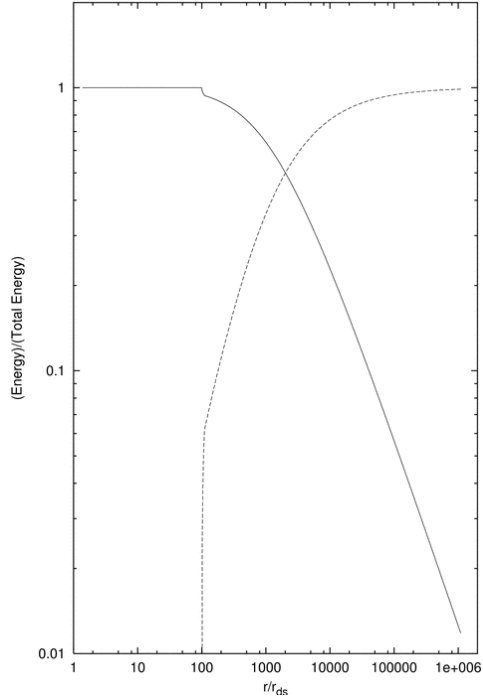}
\caption{The energy of the non baryonic components of the PEMB pulse (the solid line) and the kinetic energy of the baryonic matter (the dashed line) in unit of the total energy are plotted as functions of the radius in the unit of the dyadosphere radius $r_{\rm ds}$.}
\label{intkin}
\end{figure}

After the collision we have the further acceleration of the PEMB pulse (see Fig.~\ref{3gamma}). The temperature now decreases both in the laboratory and the comoving frame (see Fig.~\ref{pair}). Before the collision the total energy of the $e^+e^-$ pairs and the photons is constant and equal to $E_{\rm dya}$. After the collision
\begin{equation}
E_{\rm dya}=E_{\rm Baryons}+E_{e^+e^-}+E_{\rm photons},
\label{Etotal}
\end{equation}
which includes both the total energy $E_{e^+e^-}+E_{\rm photons}$ of the nonbaryonic components
and the kinetic energy $E_{\rm Baryons}$ of the baryonic matter
\begin{equation}
E_{\rm Baryons}=\bar\rho_B V(\gamma -1).
\label{kinetic}
\end{equation}
In Fig.~\ref{intkin} we plot both the total energy $E_{e^+e^-}+E_{\rm photons}$ of the nonbaryonic components and the kinetic energy $E_{\rm Baryons}$ of the baryonic matter as functions of the radius for the typical case $E_{\rm dya}=3.1\times 10^{54}$ erg and $B=10^{-2}$. Further details are given in \citet{rswx00}.

\section{Comparison and contrast of alternative fireshell equations of motion}

We compare and contrast the different approaches to the optically thick
adiabatic phase of GRB all the way to the transparency. Special attention is
given to the role of the rate equation to be self consistently solved with the
relativistic hydrodynamic equations. The works by \citet{sp90,psn93,mlr93} and by \citet{rswx99,rswx00} are compared and contrasted. The role of the baryonic loading in these three treatments is
pointed out. We also discuss recent paper by \citet{2005ApJ...635..516N}.

\subsection{Energy-momentum principle}

The basis of description for relativistic fireballs is the energy-momentum
principle. It allows to obtain relativistic hydrodynamical equations, or
equations of motion for the fireball; energy and momentum conservation
equations which are used extensively to describe interaction of relativistic
baryons of the fireball with the interstellar gas; and also boundary conditions
which are used to understand shock waves propagation in the decelerating
baryons and in the outer medium. Consider energy-momentum conservation in the
most general form\footnote{Greek indices denote four-dimensional components
and run from 0 to 3 while Latin indices run from 1 to 3. The general
relativistic effects are neglected, which is a good approximation, but we left
the general definition of the energy-momentum conservation to take into
account the most general coordinate system.}:%

\begin{align}
T^{\mu\nu}{}_{;\nu}=\frac{\partial(\sqrt{-g}\,T^{\mu\nu})}{\partial x^{\nu}%
}+\sqrt{-g}\,\Gamma^{\mu}_{\nu\lambda} T^{\nu\lambda}=0, \label{ce}%
\end{align}
where $\Gamma^{\mu}_{\nu\lambda}$ are Cristoffel symbols and $g$ is
determinant of the metric tensor. Integrating over the whole three-dimensional
volume we obtain
\begin{align}
\int_{V} T^{\mu\nu}{}_{;\nu} dV=0. \label{cev}%
\end{align}
Integrating over the whole four-dimensional volume and applying divergence
theorem we get (\citet{Taub})
\begin{align}
\int_{t}\int_{V} T^{\mu\nu}{}_{;\nu} dV dt=\oint_{V} T^{\mu\nu} \lambda_{\nu
}dV=0, \label{ceom}%
\end{align}
where $\lambda_{\alpha}$ are covariant components of the outward drawn normal
to the three-dimensional hypersurface (volume $V$).

Define the momentum four-vector $P^{\mu}$:
\[
P^{\mu}=\int_{V}T^{0\mu}dV.
\]
From Eq.(\ref{ce}) and Eq.(\ref{cev}) in Minkowski metric (when $\Gamma_{\nu
\lambda}^{\mu}=0$) we see that
\begin{align}
\frac{dP^{0}}{dt}  &  =\int_{V}\frac{\partial T^{00}}{\partial t}dV=-\int
_{V}\frac{\partial T^{i0}}{\partial x^{i}}dV=-\oint_{S}T^{i0}dS_{i},\\
\frac{dP^{j}}{dt}  &  =\int_{V}\frac{\partial T^{0j}}{\partial t}dV=-\int
_{V}\frac{\partial T^{ij}}{\partial x^{i}}dV=-\oint_{S}T^{ij}dS_{i},
\end{align}
so, if the energy and momentum fluxes through the surface $S$ bounding
the considered volume $V$ are absent, the energy and momentum are constants during
system evolution. Supposing this is the case we arrive to the conservation of
energy and momentum:
\[
P^{\mu}=\mathrm{const}.
\]
This equation is important to describe interaction of the baryons left from
the fireball with the interstellar gas. Assume the energy-momentum tensor in
the form of the ideal fluid
\[
T^{\mu\nu}=p\,g^{\mu\nu}+\omega\,U^{\mu}U^{\nu},
\]
where $\omega=\epsilon+p$ is proper entalpy, $p$ is proper pressure and
$\epsilon$ is proper energy densities. Now suppose spherical
symmetry\footnote{The only nonvanishing components of the energy-momentum
tensor are $T^{00},\,T^{01,},\,T^{10},\,T^{11},\,T^{22},\,T^{33}.$}, which is
usually done for fireballs description. Using spherical coordinates with the
interval $ds^2$ given in Eq.\eqref{s} with $\alpha(r) = 1$, we rewrite (\ref{ce}):
\begin{align}
\frac{\partial T^{00}}{\partial t}+\frac{1}{r^{2}}\frac{\partial}{\partial
r}\left(  r^{2}T^{01}\right)   &  =0,\\
\frac{\partial T^{10}}{\partial t}+\frac{1}{r^{2}}\frac{\partial}{\partial
r}\left(  r^{2}T^{11}\right)  -r\left(  T^{33}+T^{44}\sin^{2}\theta\right)
&  =0,
\end{align}
arriving to equations of motion for relativistic fireballs
(\citet{psn93,mlr93,rswx99,bm76,bm95}):
\begin{align}
\frac{\partial(\gamma^{2}\omega)}{\partial t}-\frac{\partial p}{\partial
t}+\frac{1}{r^{2}}\frac{\partial}{\partial r}\left(  r^{2}\gamma^{2}%
u\omega\right)   &  =0,\label{conseq1}\\
\frac{\partial(\gamma^{2}u\omega)}{\partial t}+\frac{1}{r^{2}}\frac{\partial
}{\partial r}\left[  r^{2}(\gamma^{2}-1)\omega\right]  +\frac{\partial
p}{\partial r}  &  =0, \label{conseq2}%
\end{align}
where the four-velocity and the relativistic gamma factor are defined as
follows:
\[
U^{\mu}=(\gamma,\gamma u,0,0),\quad\quad\gamma\equiv(1-u^{2})^{-1/2},
\]
the radial velocity $u$ is measured in units of speed of light $u=v/c$.

Now suppose that there is a discontinuity on the fluid flow. Suppose the
three-dimensional volume is a spherical shell and choose the coordinate system
where the discontinuity is at rest so that in (\ref{ceom}) for normal vectors
to the discontinuity hypersurface $\lambda_{\alpha}$ we have
\begin{align}
\lambda_{\alpha}\lambda^{\alpha}=1, \quad\quad\lambda_{0}=0.
\end{align}
Let the radius of the shell $R_{s}$ be very large and shell thickness $\Delta$
be very small. With $R_{s}\rightarrow\infty$ and $\Delta\rightarrow0$ from
(\ref{ceom}) we arrive to
\begin{align}
\left[  T^{\alpha i} \right]  =0,
\end{align}
where the brackets mean that the quantity inside is the same on both sides of
the discontinuity surface. This equation together with continuity conditon for
particle density flux $[n U^{i}]=0$ was used by \citet{Taub} to obtain relativistic Rankine-Hugoniot equations.
Such equations govern shock waves dynamics which are supposed to appear
during collision of the baryonic material left from the fireball with the ISM
(\citet{bm76}). The origin of the afterglow could be connected to
the conversion of kinetic energy into radiative energy in these shocks
(\citet{1992MNRAS.258P..41R,1992ApJ...395L..83N,1994ApJ...422..248K,p99}).
However, our scenario differs from that, namely we suppose that fully
radiative condition during this interaction is satisfied. Our model allows to
explain the afterglow phenomenon without consideration of shocks as sources of radiation (\citet{rubr,rubr2}).

\subsection{Quasi-analytic model of GRBs}

The first detailed models for relativistic fireballs were suggested in the
beginning of nineties (\citet{sp90,psn93,mlr93}).
Independent calculations performed in \citet{rswx99} and
\citet{rswx00} give precise understanding and we first describe our
approach, mentioning the main differences with the existing literature.

First of all, the source of energy, which was obscure in previous models, is
supposed to be the energy extraction process from the black hole within dyadosphere model (\citet{rubr}). The second difference is that initially not
photons but pairs are created by overcritical electric field and
these pairs produce photons later. The resulting plasma, referred to as
pair-electro-magnetic (PEM) pulse expands initially into vacuum surrounding
the black hole reaching very soon relativistic velocities. Then collision with
the baryonic remnant takes place and the PEM pulse becomes
pair-electro-magnetic-baryonic (PEMB) pulse (see \citet{rubr}
for details). This difference is not large, since it was shown that the final
gamma factor does not depend on the distance to the baryonic remnant and
parameters of the black hole. The only crucial parameters are again the energy
of dyadosphere, or simply $E_{0}$, and the baryon loading $B$ given in Eq.\eqref{Bdef} which,
in the current literature, is usually defined as:%
\begin{equation}
\eta = B^{-1}.
\end{equation}

The exact model is based on numerical integration of relativistic
energy-momentum conservation equations (\ref{conseq1},\ref{conseq2}) together
with the baryonic number conservation equation\footnote{Instead of
(\ref{conseq1}) the projection on the flow line $U_{\mu}T^{\mu\nu}{}_{;\nu}=0$
is used in \citet{rswx99} and \citet{rswx00}.}
\begin{align}
(n_{B} U^{\mu})_{;\mu}=0. \label{conseq3}%
\end{align}
However, the most important distinct point from all previous models is that
the \emph{rate equation} for electron-positron pairs is added to the model and
integrated simultaneously in order to reach self-consistency.

Here we concentrate on the simple quasi-analytical treatment presented in
\citet{rswx99,rswx00} (see also
\citet{rubr}). The PEMB pulse is supposed to contain finite
number of shells each with flat density profile. The dynamics is governed by
Eqs.\eqref{be'}--\eqref{paira'_2}.

For an infinitesimal expansion of the coordinate volume from $\mathcal{V}_{0}
$ to $\mathcal{V}$ in the coordinate time interval $t-t_{0}$ one can
discretize the last differential equation for numerical computations.

The most importants outcomes from analysis performed in
\citet{rswx00} are the following:
\begin{itemize}
\item[-] the appropriate model for geometry of expanding fireball (PEM-pulse)
is given by the constant width approximation (this conclusion is achieved by
comparing results obtained using (\ref{conseq1},\ref{conseq2}) and simplified
treatment described above),
\item[-] there is a bound on parameter $B$ which comes from violation of
constant width approximation, $B\leq10^{-2}$ ($\eta\geq10^{2}$).
\end{itemize}
The last conclusion is crucial since it shows that there is a critical loading
of baryons. When their presence produce a turbulence in the outflow from the
fireball, its motion becomes very complicated and the fireball evolution does
not lead in general to the GRB.

Exactly because of this reason, the optically thick fireball never reaches
such large radius as $r_{b}=r_{0}\eta^{2}$ (discussed in
\citet{mlr93}, see below) since to do this, the
baryonic fraction should overcome the critical value $B_{c}=10^{-2}$. For
larger values of $B_{c}$ the theory reviewed here does not apply. This means
in particular, that all conclusions in \citet{mlr93} obtained for
$r>r_{b}$ are invalid. In fact, for $B<B_{c}$ the gamma factor even does not
reach saturation.

\begin{figure}
\centering
\includegraphics[width=0.6\hsize]{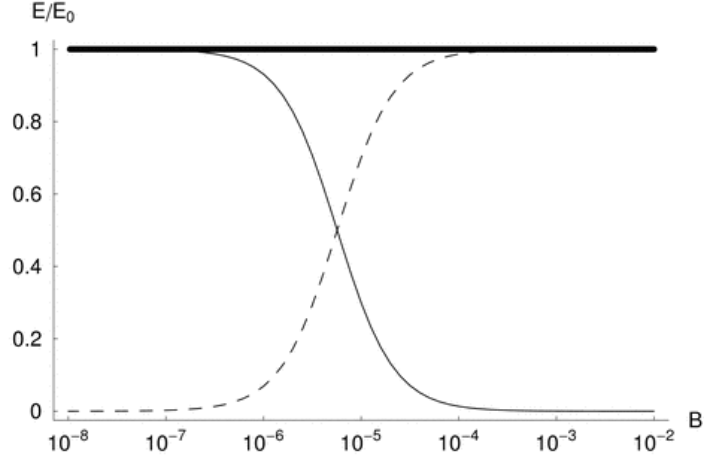}
\caption{Relative energy release in
the form of photons emitted at transparency point $E_{\gamma}/E_{0}$ (solid
line) and kinetic energy of the plasma $E_{k}/E_{0}$ (dashed line) of the
baryons in terms of initial energy of the fireball depending on parameter
$B=\eta^{-1}$ obtained on the basis of quasi-analytic model. Thick line
denotes the total energy of the system in terms of initial energy $E_{0}$.}%
\label{diag}%
\end{figure}

\begin{figure}
\centering
\includegraphics[width=0.6\hsize]{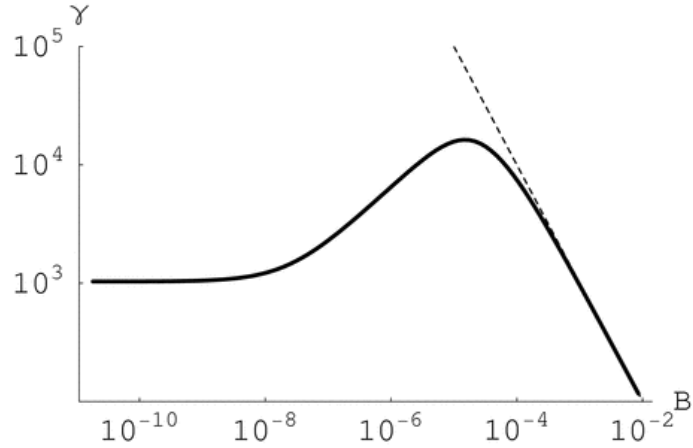}
\caption{Relativistic gamma factor of
the fireball when it reaches trasparency depending on the value of parameter
$B$. Dashed line gives asymptotic value $\gamma=B^{-1}$.}%
\label{gammab}%
\end{figure}

The fundamental result coming from this model are the diagrams presented at
fig. \ref{diag} and \ref{gammab}. The first one shows basically which portion
of initial energy is emitted in the form of gamma rays $E_{\gamma}$ when the
fireball reaches transparency condition $\tau\simeq1$ and how much energy gets
converted into the kinetic form of the baryons $E_{k}$ left after pairs
annihilation and photons escape. The second one gives the value of gamma factor at the moment when the systems reaches transparency.

The energy conservation holds, namely
\begin{align}
E_{0}=E_{\gamma}+E_{k}, \label{enRcon}%
\end{align}
Clearly when the baryons abundance is low most energy is emitted when the
fireball gets transparent. It is remarkable that almost all initial energy is
converted into kinetic energy of baryons already in the region of validity of
constant thickness approximation $B \le 10^{-2}$, so the region $10^{-8}%
<B \le 10^{-2}$ is the most interesting from this point of view.

\subsection{Shemi and Piran model}

In this section we discuss the model, proposed by \citet{sp90}.
This quantitative model gives rather good general
picture of relativistic fireballs.

Shemi and Piran found that the temperature at which the fireball becomes
optically thin is determined as%
\begin{equation}
\mathcal{T}_{esc}=\min(\mathcal{T}_{g},\mathcal{T}_{p}),
\end{equation}
where $\mathcal{T}_{g}$ and $\mathcal{T}_{p}$\ is the temperature when it
reaches transparency with respect to gas (plasma) or pairs:%
\begin{align}
\mathcal{T}_{g}^{2}  &  \simeq\frac{45}{8\pi^{3}}\frac{m_{p}}{m_{e}}\frac
{1}{\alpha^{2} g_{0}^{\frac{1}{3}}}\frac{1}{\mathcal{T}_{0}^{2}\mathcal{R}%
0}\eta,\label{eqn_sp_Tg}\\
\mathcal{T}_{p}  &  \simeq0.032,
\end{align}
where $m_{p},m_{e},$ are proton and electron masses, $g_{0}=\frac{11}{4},$
$\alpha=\frac{1}{137},$ dimensionless temperature $\mathcal{T}$ and radius
$\mathcal{R}$ of the fireball are measured in units of $\frac{m_{e}c^{2}}{k}$
and $\lambda_{e}\equiv\frac{\hbar}{m_{e}c}$ correspondingly, and the subscript
"0" denotes initial values. The temperature at transparency point in the case
when plasma admixture is unimportant is nearly a constant for a range of
parameters of interest and it nearly equals%
\begin{equation}
T_{p}=15\,\,\mathrm{keV}. \label{eqn_Tp}%
\end{equation}
Adiabatic expansion of the fireball implies:%
\begin{equation}
\frac{\mathcal{E}}{\mathcal{E}_{0}}=\frac{\mathcal{T}}{\mathcal{T}_{0}}%
=\frac{\mathcal{R}_{0}}{\mathcal{R}} , \label{eqn_sp_adi}%
\end{equation}
where $\mathcal{E}=\frac{E}{m_{e}c^{2}}$ is a radiative energy. From the
energy conservation (\ref{ce}), supposing the fluid to be pressureless and
its energy density profile to be constant we have in the coordinate frame:%
\begin{align}
\int T^{00}d\mathcal{V}=\gamma^{2}\rho\mathcal{V}=\gamma\rho V=\gamma
E_{tot}=\mathrm{const}.
\end{align}

Supposing at initial moment $\gamma_{0}=1$ and remembering that $E_{tot}=E+M
c^{2}$ we arrive to the following fundamental expression of relativistic gamma
factor $\gamma$ at transparency point:%
\begin{equation}
\gamma=\frac{\mathcal{E}_{0}+\mathcal{M}c^{2}}{\mathcal{E}+\mathcal{M}c^{2}}=
\frac{\eta+1}{(\frac{\mathcal{T}_{esc}}{\mathcal{T}_{0}})\eta+1},
\label{eqn_sp_gamma}%
\end{equation}
where $\mathcal{M}=\frac{M}{m_{e}}.$

\begin{figure}
\centering
\includegraphics[width=0.6\hsize]{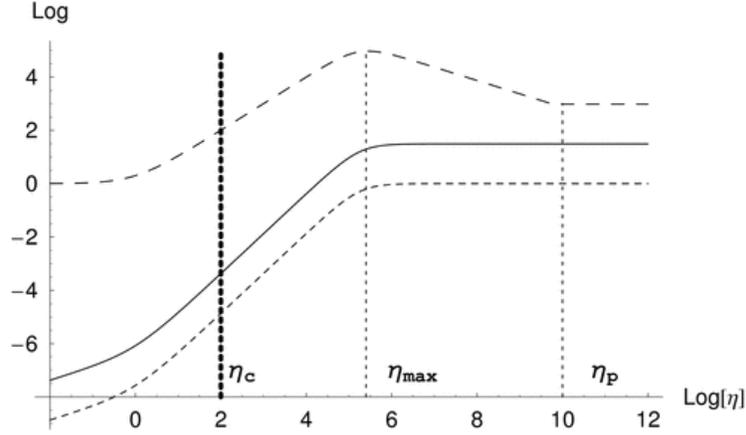} \caption{The relativistic gamma factor
(upper dashed line), the observed temperature (solid line), and the ratio of
observed energy to the initial energy of the fireball (lower dashed line) as a
function of $\eta$ (see \citet{sp90}). The values of parameters
are the same as in the cited paper. Thick dashed line denotes the limiting
value $\eta_{c}$. The values of $\eta$ when gamma factor reaches maximum and
gets constant are also shown.}%
\label{fig_sp}%
\end{figure}

One can use this relation to get such important characteristics of the GRB as
observed temperature and observed energy. In fact, they can be expressed as follows:%
\begin{align}
\mathcal{T}_{obs}  &  = \gamma\mathcal{T}_{esc},\\
\mathcal{E}_{obs}  &  = \mathcal{E}_{0}\frac{\mathcal{T}_{obs}}{\mathcal{T}%
_{esc}}.
\end{align}
These results are presented at fig. \ref{fig_sp}. In the limit of small $\eta$
we have $\gamma=(1+\eta)$, while, for very large $\eta$ the value of gamma
factor at transparency point is $\gamma=\mathcal{T}_{0}/\mathcal{T}_{esc}$,
and it has a maximum at intermediate values of $\eta$. We donote by dashed
thick line the limiting value of $\eta$ parameter $\eta_{c}\equiv B_{c}^{-1}$.
For $\eta<\eta_{c}$ the approximations used to construct the model do not
hold. It is clear that because of the presence of bound $\eta_{c}$ the value
$\gamma=\eta$ can be reached only as asymptotic one. In effect, the value
$\eta_{c}$ cuts the region where saturation of the gamma factor happens before
the moment when the fireball becomes transparent.

It was found that for relatively large $\eta\geq10^{5}$ the photons emitted
when the fireball becomes transparent carry most of the initial energy.
However, since the observed temperature in GRBs is smaller than initial
temperature of the fireball, one may suppose that a large part of initial
energy is converted to kinetic energy of the plasma.

\subsection{Shemi, Piran and Narayan model}

\citet{psn93} present a generalization of
this model to arbitrary initial density profile of the fireball. These authors
performed numerical integrations of coupled energy-momentum relativistic
conservation equations (\ref{conseq1},\ref{conseq2}) and baryon number
conservation equation (\ref{conseq3}). They were mainly interested in the
evolution of the observed temperature, gamma factor and other quantities with
the radius. Their study results in the number of important conclusions, namely:
\begin{itemize}
\item[-] the expanding fireball has two basic phases: a radiation dominated
phase and a matter-dominated phase. In the former, the gamma factor grows
linearly as the radius of the fireball: $\gamma\varpropto r$, while in the
latter the gamma factor reaches asymptotic value $\gamma\simeq\eta+1$.
\item[-] the numerical solutions are reproduced with a good accuracy by
frozen-pulse approximation, when the pulse width is given by initial radius of
the fireball.
\end{itemize}

The last conclusion is important, since the volume $V$ of the fireball can be
calculated as%
\begin{align}
V=4\pi R^{2} \Delta, \label{volume}%
\end{align}
where $\Delta\simeq R_{0}$ is the width of the leading shell with consant
energy density profile, $R$ is the radius of the fireball.

They also present the following scaling solution:%
\begin{align}
R  &  = R_{0}\left(  \frac{\gamma_{0}}{\gamma}D^{3}\right)  ^{1/2},\\
\frac{1}{D}  &  \equiv\frac{\gamma_{0}}{\gamma}+\frac{3\gamma_{0}}{4\gamma
\eta}-\frac{3}{4\eta},
\end{align}
where subscript "0" denotes some initial time when $\gamma\gtrsim$ few, which
can be inverted to give $\gamma(R)$.

\subsection{M\'{e}sz\'{a}ros, Laguna and Rees model}

The next step in developing this model was made in \citet{mlr93}.
In order to reconcile the model with observations, these authors proposed a
generalization to anisotropic (jet) case. Nevertheless, their analytic results
apply to the case of homogeneous isotropic fireballs and we will follow their
analytical isotropic model in this section.

Starting from the same point as Shemi and Piran, consider (\ref{eqn_sp_adi})
and (\ref{eqn_sp_gamma}). The analytic part of the paper describes the geometry of the fireball, the gamma factor behavior and the final energy balance between
radiation and kinetic energy. Magnetic field effects are also considered, but
we are not interested in this part here.

Three basic regimes are found in \citet{mlr93} for evolution of
the fireball. In two first regimes there is a correspondence between the
analysis presented in the paper and results of \citet{psn93}, so the
constant thickness approximation holds. It is claimed in
\citet{mlr93} that when the radius of the fireball reaches very
large values such as $R_{b}=R_{0}\eta^{2}$ the noticeable departure from
constant width of the fireball occurs. However, it is important to note, that
the fireball becomes transparent much earlier and this effect never becomes
important (see above).

The crucial quantity presented in the paper is $\Gamma_{m}$ -- the maximum
possible bulk Lorentz factor achievable for a given initial radiation energy
$E_{0}$ deposited within a given initial radius $R_{0}$:%
\begin{align}
\Gamma_{m}  &  \equiv\eta_{m}=\left(  \tau_{0}\eta\right)  ^{1/3}=\left(
\Sigma_{0}\kappa\eta\right)  ^{1/3},\label{eqn_mlr_Gm}\\
\Sigma_{0}  &  =\frac{M}{4\pi R_{0}^{2}}, \quad\quad\kappa=\frac{\sigma_{T}%
}{m_{p}},
\end{align}
where $\Sigma_{0}$ is initial baryon (plasma) mass surface density.

All subsequent calculations in the paper by \citet{mlr93} involves
this quantity. It is evident from (\ref{eqn_mlr_Gm}) that the linear
dependence between the gamma factor $\Gamma$ and parameter $\eta$ is assumed.
However, this is certainly not true as can be seen from fig. \ref{fig_sp}. We
will come back to this point in the following section.

Another important quantity is given in this paper, namely%
\begin{equation}
\Gamma_{p}=\frac{T_{0}}{T_{p}}.
\end{equation}
This is just the asymptotic behavior of the gamma factor at fig.
\ref{fig_sp} for very large $\eta$. Using it, the authors calculate the value
of $\eta$ parameter above which the pairs dominated regime occurs:%
\begin{equation}
\eta_{p}=\frac{\Gamma_{m}^{3}}{\Gamma_{p}^{2}}.
\end{equation}
This means that above $\eta_{p}$ the presence of baryons in the fireball is
insufficient to keep it optically thick after pairs are annihilated and
almost all initial energy deposited in the fireball is emitted immediately.

\begin{figure}
\centering
\includegraphics[width=0.6\hsize]{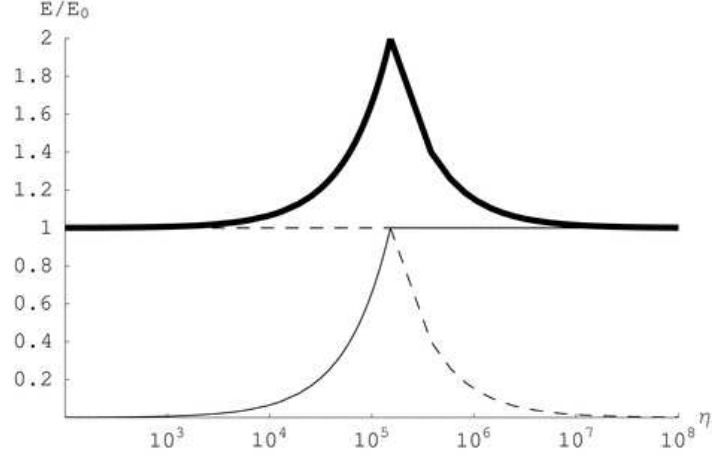} \caption{The ratios of radiation
and kinetic energy to the initial energy of the fireball predicted by
M\'{e}sz\'{a}ros, Laguna and Rees model. Thick line denotes the total energy
of the system in terms of initial energy. Energy conservation does not hold.}%
\label{lmr_diag}%
\end{figure}

The estimate of the final radiation to kinetic energy ratio made in
\citet{mlr93} is incorrect, because kinetic and radiation
energies do not sum up to initial energy of the fireball thus violating energy
conservation (\ref{enRcon}). This is illustrated at fig. \ref{lmr_diag}. The
correct analytic diagram is presented instead in fig. \ref{fig_sp_diag}.

\subsection{Approximate results}

All models for isotropic fireballs are based on the following points:

\begin{enumerate}
\item Flat space-time,
\item Relativistic energy-momentum principle,
\item baryonic number conservation.
\end{enumerate}

Anthough the dyadosphere model starts with Reissner-Nordstr{\o}m geometry, the numerical
code is written for the case of flat space-time simply because curved
space-time effects becomes insignificant soon after the fireball reachs
relativistic expansion velocities. The presence of rate equation in the model
by \citet{rswx99,rswx00} has a deep physical
ground and its lack in the other treatments means they are incomplete.
Indeed, the number density of pairs influences the speed of expansion
of the fireball. However, in this section we neglect the rate equation and
discuss the common points between all considered models.

First of all, let us come back to fig. \ref{fig_sp}. For almost all values of
$\eta$ parameter the gamma factor is determined by gas (i.e. plasma or
baryons) admixture according to (\ref{eqn_sp_Tg}), consider this case below.
For given initial energy and radius this temperature depends only on $\eta$, so one can write:%
\begin{equation}
\gamma=\frac{\eta+1}{(\frac{\mathcal{T}_{g}}{\mathcal{T}_{0}})\eta+1}%
=\frac{\eta+1}{a \eta^{\frac{3}{2}}+1}, \label{eqn_an}%
\end{equation}
where%
\begin{equation}
a =2.1\cdot10^{3}\mathcal{T}_{0}^{-2}\mathcal{R}_{0}^{-0.5}.
\end{equation}
From this formula we can get immediately the two asymptotic regimes, namely:%
\begin{equation}
\gamma=\left\{
\begin{array}
[c]{cc}%
\eta+1, & \eta<\eta_{\max},\\
\frac{1}{a \sqrt{\eta}}, & \eta>\eta_{\max}%
\end{array}
\right.  . \label{eqn_g_as}%
\end{equation}
Notice that the constant $a$ is an extremely small number, so that after
obtaining precise value of $\eta_{\max}$ by equating to zero the derivative of
function (\ref{eqn_an}) one can expands the result in Taylor series and get in
the lowest order in $a$, that:%
\begin{align}
\eta_{\max}  &  \simeq\left(  \frac{2}{a}\right)  ^{\frac{2}{3}}-2,\\
\gamma_{\max}  &  \equiv\gamma(\eta_{\max})\simeq\frac{1}{3}\left[  1+\left(
\frac{2}{a}\right)  ^{\frac{2}{3}}\right]  .
\end{align}
In particular, in the case shown in fig. \ref{fig_sp} one has $\eta_{\max
}=2.8\cdot10^{5}$, $\gamma_{\max}=9.3\cdot10^{4}$ while according to
(\ref{eqn_mlr_Gm}) $\Gamma_{m}=\eta_{\max}=1.75\cdot10^{5}$. Clearly, our
result is much more accurate. Actually, the value $\Gamma_{m}$ in
(\ref{eqn_mlr_Gm}) is obtained from equating asymptotes in (\ref{eqn_g_as})
and there exists the following relation:%
\begin{align}
a=(\tau_{0}\eta)^{-1/2}.
\end{align}

Now we are ready to explain why the observed temperature (and consequently the
observed energy) does not depend on $\eta$ in the region $\eta_{max}<\eta
<\eta_{p}$. From the second line in (\ref{eqn_g_as}) it follows that the gamma
factor in this region behaves as $\gamma\propto\eta^{-1/2}$, while
$\mathcal{T}_{esc}\propto\eta^{1/2}$. These two exactly compensate each other
leading to independence of the observed quantities on $\eta$ in this region.
This remains the same for $\eta>\eta_{p}$ also, since here $\mathcal{T}%
_{esc}=\mathcal{T}_{p}=$const and from (\ref{eqn_an}) $\gamma=$const.

\subsection{Significance of the rate equation}

The rate equation describes the number densities evolution for electrons and
positrons. In analytic models it is supposed that pairs are annihilated
instantly when transparency condition is fulfilled. Moreover, the dynamics of
expansion is influenced by the electron-positron energy density as can be seen
from Eqs.\eqref{be'}--\eqref{paira'_2}. Therefore, it is important to make clear
whether neglect of the rate equation is a crude approximation or not.

\begin{figure}
\centering
\includegraphics[width=0.6\hsize]{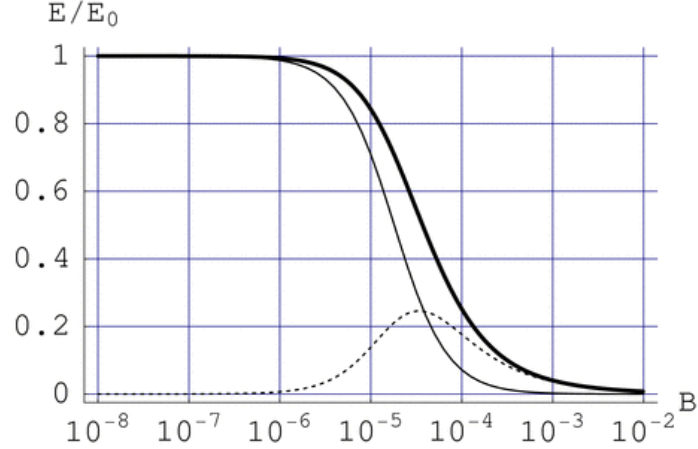} \caption{Relative energy release in
the form of photons emitted at transparency point $E_{obs}/E_{0}$ of the GRB
in terms of initial energy of the fireball depending on parameter $B=\eta
^{-1}$. Thick line represents numerical results and it is the same as in fig.
\ref{diag}. Normal line shows results for the analytic model of \citet{sp90}.
Dashed line shows the difference between
exact numerical and approximate analytical results.}%
\label{fig_sp_diag}%
\end{figure}

Using eq. (\ref{eqn_sp_gamma}) one can obtain analytic dependence of the
energy emitted at transparency point on parameter $B$ and we compare it at
fig. \ref{fig_sp_diag}.

We also show the difference between numerical results based on integration of
Eqs.\eqref{be'}--\eqref{paira'_2} and analytic results from Shemi and Piran
model. The values of parameters are: $\mu=10^{3}$ and $\xi=0.1$ (which
correspond to $E_{0}=2.87\cdot10^{54}\,$ergs and $R_{0}=1.08\cdot10^{9}\,$cm).
One can see that this difference peaks at intermediate values of $B$. The
crucial deviations however appear for large $B$, where analytical predictions
for observed energy are about two orders of magnitude smaller than the
numerical ones. This is due to the difference in predictions of the radius of
the fireball at transparency moment. In fact, the analytical model
overestimates this value at about two orders of magnitude for $B=10^{-2}$. So
for large $B$ with correct treatment of pairs dynamics the fireball gets
transparent at \emph{earlier} moments comparing to the analytical treatment.

\begin{figure}
\centering
\includegraphics[width=0.6\hsize]{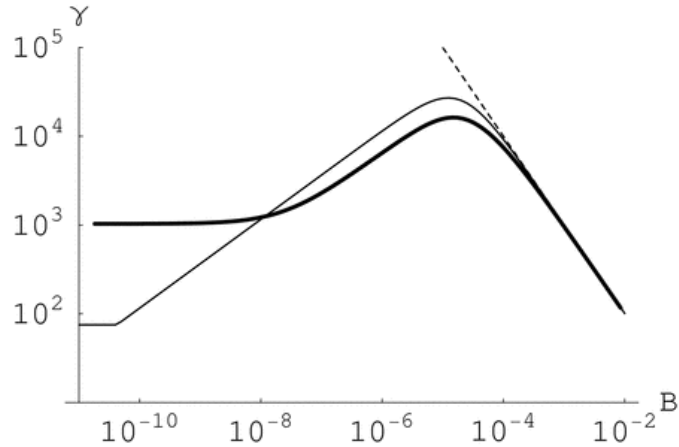} \caption{Relativistic gamma factor
when transparency is reached. The thick line denotes exact numerical results,
the normal line corresponds to analytical estimate from Shemi and Piran model,
the dotted line denotes the asymptotic value $\gamma=B^{-1}$. The dashed line
shows results of Nakar, Piran and Sari.}%
\label{gamma}%
\end{figure}

At the same time, the difference between numerical and analytical results for
gamma factor is significant for small $B$ as illustrated at fig. \ref{gamma}.
While both results coincide for $B>10^{-4}$ there is a constant difference for
the range of values $10^{-8}<B<10^{-4}$ and asymptotic constant values for the
gamma factor are also different. Besides, this asymptotic behavior takes place
for larger values of $B$ in disagreement with analytical expectations. Thus
the acceleration of the fireball for small $B$ is larger if one accounts for
pairs dynamics.

It is clear that the error coming from neglect of the rate equation is
significant. This implies that simple analytic model of Shemi and Piran gives
only qualititive picture of the fireball evolution and in order to get correct
description of the fireball one cannot neglect the rate equation.

Moreover, the difference between exact numerical model
(\citet{rswx99,rswx00}) and approximate
analytical model by \citet{sp90} becomes apparent in various
physical aspects, namely in predictions of the radius of the shell when it
reaches transparency, the gamma factor at transparency and the ratio between
the energy released in the form of photons and the one converted into kinetic
form. The last point is crucial. It is assumed in the literature that the
whole initial energy of the fireball gets converted into kinetic energy of the
shell during adiabatic expansion. Indeed, taking typical value of parameter
$B$ as $10^{-3}$ we find that according to Shemi and Piran model we have only
$0.2\%$ of initial energy left in the form of photons. However, exact
numerical computations (\citet{rswx99,rswx00}) give $3.7\%$ for the energy of photons radiated when the fireball reaches
transparency, which is a significant value and it cannot be neglected.

\begin{table}
\centering
\begin{tabular}
[c]{||c|c|c|c|c||}\hline
& {\small Ruffini et al.} & {\small Shemi, Piran} & {\small Piran, et al.} &
{\small M\'{e}sz\'{a}ros, et al.}\\\hline
{\small conservation:} &  &  &  & \\
{\small energy-momentum,} & {\small yes} & {\small yes} & {\small yes} &
{\small yes}\\
{\small baryon number,} & {\small yes} & {\small do not consider} &
{\small yes} & {\small yes}\\\hline
{\small rate equation} & {\small yes} & {\small no} & {\small no} &
{\small no}\\\hline
{\small constant width} & {\small justify} & {\small do not consider} &
{\small justify} & {\small in part}\\
{\small approximation} &  &  &  & \\\hline
{\small model for }$\gamma(r)$ & {\small num./analyt.} & {\small no} &
{\small num./analyt.} & {\small num./analyt.}\\\hline
{\small model for }$\gamma(\eta)$ & {\small num.} & {\small analyt.} &
{\small do not consider} & {\small analytic}\\\hline
\end{tabular}
\caption{Comparison of different models for fireballs.}%
\label{mod}%
\end{table}

To summurize the above discussion, we present the result of this survey in the
Table \ref{mod}. It is important to notice again that comparing to simplified
analytic treatment, accounting for the rate of change of electron-positron
pairs densities gives quantitatively different results on the ratio of kinetic
versus photon energies produced in the GRB and the gamma factor at
transparency moment, which in turn leads to different afterglow properties.
Therefore, although analytical models presented above agree and give correct qualitative description of the fireball,
one should use the numerical approach in order to
compare the theory and observations.

\subsection{Nakar, Piran and Sari revision}

Recently revision of the fireball model was made by \citet{2005ApJ...635..516N}. These authors presented new diagram for final
Lorentz gamma factor and for energy budget of the fireball. Their work was
motivated by observation of giant flares with the following afterglow
spreading up to radio region with thermal spectrum. They concluded that the
fireball have to be loaded by either baryons or magnetic field, and cannot be
only pure $e^{\pm},\gamma$ plasma in order to have $10^{-3}$ of the total
energy radiated in the giant flare.

In analogy with cosmology, authors define the number density of pairs which
survives because expansion rate becomes larger than annihilation
rate\footnote{This effect is accounted for automatically in our approach where
rate equations for pairs include expansion term.}, which gives the condition
\begin{align}
n_{\pm}\approx\frac{1}{\sigma_{T} R_{0}}.
\end{align}
Then, recalling (\ref{eqn_sp_adi}), if we want to estimate number of pairs it
turn out to be
\begin{align}
N_{\pm}=\frac{4\pi R_{0} ct}{\sigma_{T}}\,\left(  \frac{T_{0}}{T_{\pm}%
}\right)  ^{2},
\end{align}
where we identify $\Delta=ct$ in (\ref{volume}). In \citet{2005ApJ...635..516N}
the authors obtained third power of the ratio of temperatures which influences
all their subsequent results.

Obtaining the conclusion that the afterglow cannot be obtained as the result
of interaction of $e^{\pm},\gamma$ plasma with ISM authors turn to baryonic
loading consideration. They attempt to define critical values of the loading
parameter $\eta$ finding in general 4 such values\footnote{The last value
$\eta_{4}$ corresponds to the case of heavy loading where spreading of the
expanding shell is observed, and is not considered here.}, in particular:
\begin{align}
\eta_{1}=\frac{E_{0}\sigma_{T}}{4\pi R_{0} ct m_{e} c^{2}}\left(  \frac
{T_{\pm}}{T_{0}}\right)  ^{3},\\
\eta_{2}=\frac{E_{0}\sigma_{T}}{4\pi R_{0} ct m_{p} c^{2}}\left(  \frac
{T_{\pm}}{T_{0}}\right)  ^{3},\\
\eta_{3}=\left(  \frac{E_{0}\sigma_{T}}{4\pi R_{0} ct m_{p} c^{2}}\right)
^{1/4}.
\end{align}

We recall that the first two quantities are based on the formula for $N_{\pm}$
and should contain factors $\left(  \frac{T_{\pm}}{T_{0}}\right)  ^{2}$ instead.

The first `critical' value, $\eta_{1}$, comes from the condition $N_{p}
m_{p}=N_{\pm}m_{e}$, where $N_{p}\equiv\frac{E_{0}}{m_{p} c^{2}\eta}$ is just
the number of protons in plasma admixture. It does not correspond to any
critical change in the physics of the phenomena; for instance, it cannot be
interpreted as equality of masses (equal inertia) of pairs and baryons since
the former is given mainly by their total energy $E_{\pm}$, while the latter
by their rest mass $N_{p} m_{p}$. This value is however close to the one
defined above $\eta_{1}\approx\eta_{p}$.

The second `critical' value, $\eta_{2}$, corresponds to the condition
$N_{p}=N_{\pm}$, namely equality of numbers of protons and pairs. It is also
incorrectly interpreted as equal contribution to the Thompson scattering. In
fact, cross-section for the Thompson scattering for protons contains additions
factor $\left(  \frac{m_{e}}{m_{p}}\right)  ^{2}$ with respect to the usual
formula for electrons.

Definition of the third `critical' value, $\eta_{3}$, is not clear, but
important is its vicinity to the critical value $\eta_{c}$ quoted above.

On the basis of adiabatic conditions (\ref{eqn_sp_adi}) authors present the
new diagram for the final gamma factor and energy budget of the pair-baryonic
plasma at transparency. In fact, this diagram, shown in our fig.\ref{gamma} by
dashed curve for parameter $B$, is very similar to the one, obtained by
\citet{gw98}, who considered
hydrodynamics of relativistic $e^{\pm},\gamma$ winds. That problem is very
different from ours, because of different boundary conditions\footnote{Note,
that \citet{gw98} also use rate equations
desribing decoupling plasma from photons.}. In particular, in the wind energy
conservation Eq.(\ref{ce}) does not hold; the reason is that constant energy
(mass) supply takes place parametrized in \citet{gw98} by $\dot
E$ ($\dot M$). In that paper, in fact, authors present the diagram for
asymptotic value of the Lorentz gamma factor depending on the ratio
$\frac{\dot E}{\dot M}$ which is very different from the quantity $\eta$.

Surprisingly, the foundamental result about the presence of maximum in the
diagram for gamma factor on fig.\ref{gamma} which was found by the same
authors previously in \citet{sp90} (see fig.\ref{fig_sp}) that
comes from the energy conservation (\ref{eqn_sp_gamma}) is ignored in
\citet{2005ApJ...635..516N}. It can be understood in the following way. For
small loading (small $B$) the more baryons are present in the plasma the
larger becomes the number density of corresponding electrons, the larger
optical depth is. Therefore, transparency is reached later, which gives larger
gamma factor at transparency. From the other hand, for heavy baryon loading
(relatively large $B$) the more baryons are present, the more inertia has the
plasma, and by energy conservation, the less final gamma factor has to be.

To conclude, we compared existing isotropic models of GRBs, so called fireball
models. It is shown that the crucial difference between our approach and other
models in the literature is the presence of the rate equation which accounts
for electron-positron pairs densities evolution during expansion of the
fireball. This results in quantitative difference between predictions of our
quasi-analytic model and analytic models in the literature. Considering its
significance we conclude that in order to compare theory and observations it
is necessary to take into account rate equation together with energy and mass
conservation conditions.

Another important difference is the presence of bound on baryonic loading
parameter $B_{c}=10^{-2}$ which comes from violation of constant thickness
approximation used in our quasi-analytic model. The same bound should be
present in all analytic models in the literature. As a consequence, the
broadening of the relativistic shell resulting from the fireball never happens
before it reaches transparency. Besides, the gamma factor does not reach
saturation and the value $\gamma=\eta=B^{-1}$ is only asymptotic one for
$B \le 10^{-2}$.

\section{Exact versus approximate solutions in Gamma-Ray Burst afterglows}

The consensus has been reached that the afterglow emission originates from a relativistic thin shell of baryonic matter propagating in the ISM and that its description can be obtained from the relativistic conservation laws of energy and momentum. In both our approach and in the other ones in the current literature (see e.g. \citet{p99,cd99,rubr,PowerLaws}) such conservations laws are used. The main difference is that in the current literature an ultra-relativistic approximation, following the \citet{bm76} self-similar solution, is widely adopted while we use the exact solution of the equations of motion. We first express such equations in a finite difference formulation, and we later express them in a differential formulation which will be most useful in comparing and contrasting our exact solutions with the ones in the current literature.

\subsection{The equations of the afterglow dynamics - Finite difference formulation}

In analogy and by extension of the results obtained for the PEM and PEMB pulse cases, we also assume that the expansion of the ABM pulse through the ISM occurs keeping its width constant in the laboratory frame, although the results are quite insensitive to this assumption. Then we assume that this interaction can be represented by a sequence of inelastic collisions of the expanding ABM pulse with a large number of thin and cold ISM spherical shells at rest in the laboratory frame. Each of these swept up shells of thickness $\Delta r$ has a mass $\Delta M_{\rm ism}$ and is assumed to be located between two radial distances $r_1$ and $r_2$ (where $r_2-r_1 = \Delta r \ll r_1$) in the laboratory frame. These collisions create an internal energy $\Delta E_{\rm int}$.

We indicate by $\Delta\epsilon$ the increase in the proper internal energy density due to the collision with a single shell and by $\rho_B$ the proper energy density of the swept up baryonic matter. This includes the baryonic matter composing the remnant, already swept up in the PEMB pulse formation, and the baryonic matter from the ISM swept up by the ABM pulse:
\begin{equation}
\rho_B=\frac{\left(M_B+M_{\rm ism}\right)c^2}{V}.
\label{rhob}
\end{equation}
Here $V$ is the ABM pulse volume in the comoving frame, $M_B$ is the mass of the baryonic remnant and $M_{\rm ism}$ is the ISM mass swept up from the transparency point through the $r$ in the laboratory frame:
\begin{equation}
M_{\rm ism}=m_pn_{\rm ism}{4\pi\over3}\left(r^3-{r_\circ}^3\right)\, ,
\label{dgm1}
\end{equation}
where $m_p$ the proton mass and $n_{\rm ism}$ the number density of the ISM in the laboratory frame.

The energy conservation law in the laboratory frame at a generic step of the collision process is given by
\begin{equation}
\rho_{B_1} {\gamma_1}^2{\cal V}_1 + \Delta M_{\rm ism} c^2 = \left(\rho_{B_1}\frac{V_1}{V_2} + {{\Delta M_{\rm ism} c^2}\over V_2} + \Delta\epsilon \right){\gamma_2}^2{\cal V}_2,
\label{ecc}
\end{equation}
where the quantities with the index ``$1$'' are calculated before the collision of the ABM pulse with an elementary shell of thickness $\Delta r$ and the quantities with ``$2$'' after the collision, $\gamma$ is the gamma factor and ${\cal V}$ the volume of the ABM pulse in the laboratory frame so that $V=\gamma {\cal V}$. The momentum conservation law in the laboratory frame is given by
\begin{equation}
\rho_{B_1} \gamma_1 U_{r_1} {\cal V}_1 = \left(\rho_{B_1}\frac{V_1}{V_2} + {\Delta M_{\rm ism} c^2\over V_2} + 
\Delta\epsilon \right)\gamma_2 U_{r_2} {\cal V}_2,
\label{pcc}
\end{equation}
where $U_r=\sqrt{{\gamma}^2 - 1}$ is the radial covariant component of the four-velocity vector (see \citet{rswx99,rswx00}). We thus obtain
\begin{eqnarray}
\Delta\epsilon & = & \rho_{B_1} {\gamma_1 U_{r_1} {\cal V}_1 \over \gamma_2 U_{r_2} {\cal V}_2} - \left(\rho_{B_1}\frac{V_1}{V_2} + {\Delta M_{\rm ism} c^2\over V_2} \right) ,\label{heat}\\
\gamma_2 & = & {a\over\sqrt{a^2-1}},\hskip0.5cm a\equiv {\gamma_1  \over  
U_{r_1}}+ {\Delta M_{\rm ism} c^2\over \rho_{B_1} \gamma_1 U_{r_1} {\cal V}_1}.
\label{dgamma}
\end{eqnarray}

We can use for $\Delta \varepsilon$ the following expression
\begin{equation}
\Delta \varepsilon = \frac{E_{{\rm int}_2}}{V_2}-\frac{E_{{\rm int}_1}}{V_1} = \frac{E_{{\rm int}_1}+\Delta E_{\rm int}}{V_2}-\frac{E_{{\rm int}_1}}{V_1} = \frac{\Delta E_{\rm int}}{V_2}
\label{deltaexp}
\end{equation}
because we have assumed a ``fully radiative regime'' and so $E_{{\rm int}_1}=0$.
Substituting Eq.(\ref{dgamma}) in Eq.(\ref{heat}) and applying Eq.(\ref{deltaexp}), we obtain:
\begin{equation}
 \Delta E_{\rm int}  = \rho_{B_1} {V_1}\sqrt {1 + 2\gamma_1 \frac{{\Delta M_{\rm ism} c^2 }}{{\rho_{B_1} V_1 }} + \left( {\frac{{\Delta M_{\rm ism} c^2 }}{{\rho_{B_1} V_1 }}} \right)^2 }
  - \rho_{B_1}{V_1} \left( 1 + \frac{{\Delta M_{\rm ism} c^2 }}{\rho_{B_1} V_1} \right)\, , \label{heat2}
\end{equation}
\begin{equation}
 \gamma_2  = \frac{{\gamma_1  + \frac{{\Delta M_{\rm ism} c^2 }}{{\rho_{B_1} V_1 }}}}{{\sqrt {1 + 2\gamma_1 \frac{{\Delta M_{\rm ism} c^2 }}{{\rho_{B_1} V_1 }} + \left( {\frac{{\Delta M_{\rm ism} c^2 }}{{\rho_{B_1} V_1 }}} \right)^2 } }}\, . \label{dgamma2} 
\end{equation}

\subsection{The equations of the afterglow dynamics - differential formulation}

Under the limit:
\begin{equation}
\frac{\Delta M_{\rm ism} c^2}{\rho_{B_1} V_1 } \ll 1\, ,
\label{limite}
\end{equation}
and performing the following substitutions:
\begin{equation}
\Delta E_{\rm int} \to d E_{\rm int}\, , \quad \gamma_2-\gamma_1 \to d\gamma\, , \quad \Delta M_{\rm ism} \to d M_{\rm ism}\, ,
\label{substitutions}
\end{equation}
Eqs.(\ref{heat2},\ref{dgamma2}) are equivalent to:
\begin{subequations}\label{Taub_Eq} 
\begin{eqnarray} 
dE_{\mathrm{int}} &=& \left(\gamma - 1\right) dM_{\mathrm{ism}} c^2 
\label{Eint}\, ,\\ 
d\gamma &=& - \textstyle\frac{{\gamma}^2 - 1}{M} dM_{\mathrm{ism}}\, , 
\label{gammadecel}\\ 
dM &=& 
\textstyle\frac{1-\varepsilon}{c^2}dE_{\mathrm{int}}+dM_\mathrm{ism}\, 
,\label{dm}\\ 
dM_\mathrm{ism} &=& 4\pi m_p n_\mathrm{ism} r^2 dr \, , \label{dmism} 
\end{eqnarray} 
\end{subequations} 
where, we recall, $E_{\mathrm{int}}$, $\gamma$ and $M$ are respectively the internal energy, the Lorentz factor and the mass-energy of the expanding pulse, $n_\mathrm{ism}$ is the ISM number density which is assumed to be constant, $m_p$ is the proton mass, $\varepsilon$ is the emitted fraction of the energy developed in the collision with the ISM and $M_\mathrm{ism}$ is the amount of ISM mass swept up within the radius $r$: $M_\mathrm{ism}=(4/3)\pi(r^3-{r_\circ}^3)m_pn_\mathrm{ism}$, where $r_\circ$ is the starting radius of the baryonic shell.

\subsection{The exact analytic solutions}

In both our work and in the current literature (see \citet{p99,cd99,rubr,PowerLaws}) a first integral of these equations has been found, leading to expressions for the Lorentz gamma factor as a function of the radial coordinate. In the ``fully radiative condition'' (i.e. $\varepsilon = 1$) we have: 
\begin{equation} 
\gamma=\frac{1+\left(M_\mathrm{ism}/M_B\right)\left(1+\gamma_\circ^{-1}\right)\left[1+\left(1/2\right)\left(M_\mathrm{ism}/M_B\right)\right]}{\gamma_\circ^{-1}+\left(M_\mathrm{ism}/M_B\right)\left(1+\gamma_\circ^{-1}
\right)\left[1+\left(1/2\right)\left(M_\mathrm{ism}/M_B\right)\right]}\, , 
\label{gamma_rad} 
\end{equation} 
while in the ``fully adiabatic condition'' (i.e. $\varepsilon = 0$) we have:
\begin{equation} 
\gamma^2=\frac{\gamma_\circ^2+2\gamma_\circ\left(M_\mathrm{ism}/M_B\right) 
+\left(M_\mathrm{ism}/M_B\right)^2}{1+2\gamma_\circ\left(M_\mathrm{ism}/M_B\right)+\left(M_\mathrm{ism}/M_B\right)^2}\, , 
\label{gamma_ad} 
\end{equation} 
where $\gamma_\circ$ is the initial value of the Lorentz gamma factor of the accelerated baryons at the beginning of the afterglow phase.

A major difference between our treatment and the ones in the current literature is that we have integrated the above equations analytically. Thus we obtained the explicit analytic form of the equations of motion for the expanding shell in the afterglow for a constant ISM density. For the fully radiative case we have explicitly integrated the differential equation for $r\left(t\right)$ in Eq.\eqref{gamma_rad}, recalling that $\gamma^{-2}=1-\left[dr/\left(cdt\right)\right]^2$, where $t$ is the time in the laboratory reference frame. The new explicit analytic solution of the equations of motion we have obtained for the relativistic shell in the entire range from the ultra-relativistic to the non-relativistic regimes is (\citet{EQTS_ApJL2}):
\begin{equation} 
\begin{split} 
& t = \tfrac{M_B  - m_i^\circ}{2c\sqrt C }\left( {r - r_\circ } \right) 
+ \tfrac{{r_\circ \sqrt C }}{{12cm_i^\circ A^2 }} \ln \left\{ 
{\tfrac{{\left[ {A + \left(r/r_\circ\right)} \right]^3 \left(A^3  + 
1\right)}}{{\left[A^3  + \left( r/r_\circ \right)^3\right] \left( {A + 1} 
\right)^3}}} \right\} - \tfrac{m_i^\circ r_\circ }{8c\sqrt C }\\ 
& + t_\circ + \tfrac{m_i^\circ r_\circ }{8c\sqrt C } \left( 
{\tfrac{r}{{r_\circ }}} \right)^4 + \tfrac{{r_\circ \sqrt{3C}}}{{6 c 
m_i^\circ A^2 }} \left[\arctan \tfrac{{2\left(r/r_\circ\right) - 
A}}{{A\sqrt 3 }} - \arctan \tfrac{{2 - A}}{{A\sqrt 3 }}\right] 
\end{split} 
\label{analsol} 
\end{equation} 
where $A=\sqrt[3]{\left(M_B-m_i^\circ\right)/m_i^\circ}$, $C={M_B}^2(\gamma_\circ-1)/(\gamma_\circ +1)$ and $m_i^\circ = \left(4/3\right)\pi m_p n_{\mathrm{ism}} r_\circ^3$.

Correspondingly, in the adiabatic case we have (\citet{EQTS_ApJL2}): 
\begin{equation} 
t = \left(\gamma_\circ-\tfrac{m_i^\circ}{M_B}\right)\tfrac{r-r_\circ}{c\sqrt{\gamma_\circ^2-1}} 
+ \tfrac{m_i^\circ}{4M_Br_\circ^3}\tfrac{r^4-r_\circ^4}{c\sqrt{\gamma_\circ^2-1}} 
+ t_\circ\, . 
\label{analsol_ad} 
\end{equation} 

\subsection{Approximations adopted in the current literature}

We turn now to the comparison of the exact solutions given in Eqs.\eqref{analsol} with the approximations used in the current literature. We show that such an approximation holds only in a very limited range of the physical and astrophysical parameters and in an asymptotic regime which is reached only for a very short time, if any, and that therefore it cannot be used for modeling GRBs. Following \citet{bm76}, a so-called ``ultrarelativistic'' approximation $\gamma_\circ \gg \gamma \gg 1$ has been widely adopted by many authors to solve Eqs.\eqref{Taub_Eq} (see e.g. \citet{s97,s98,w97,rm98,gps99,pm98c,pm99,cd99,p99,gw99,vpkw00,m02} and references therein). This leads to simple constant-index power-law relations:
\begin{equation}
\gamma\propto r^{-a}\, ,
\label{gr0}
\end{equation}
with $a=3$ in the fully radiative case and $a=3/2$ in the fully adiabatic case.

We address now the issue of establishing the domain of applicability of the simplified Eq.\eqref{gr0} used in the current literature both in the fully radiative and adiabatic cases.

\subsubsection{The fully radiative case}\label{fr}

We first consider the fully radiative case. If we assume:
\begin{equation}
1/\left(\gamma_\circ+1\right) \ll M_\mathrm{ism}/M_B \ll \gamma_\circ/\left(\gamma_\circ+1\right) < 1\, ,
\label{app_FB}
\end{equation}
we have that in the numerator of Eq.\eqref{gamma_rad} the linear term in $M_\mathrm{ism}/M_B$ is negligible with respect to $1$ and the quadratic term is {\em a fortiori} negligible, while in the denominator the linear term in $M_\mathrm{ism}/M_B$ is the leading one. Eq.\eqref{gamma_rad} then becomes:
\begin{equation}
\gamma\simeq\left[\gamma_\circ/\left(\gamma_\circ+1\right)\right] M_B/M_\mathrm{ism}\, .
\label{gamma_rad_2}
\end{equation}
If we multiply the terms of Eq.\eqref{app_FB} by $(\gamma_\circ+1)/\gamma_\circ$, we obtain $1/\gamma_\circ\ll(M_\mathrm{ism}/M_B)[(\gamma_\circ+1)/\gamma_\circ]\ll 1$, which is equivalent to $\gamma_\circ\gg[\gamma_\circ/(\gamma_\circ+1)](M_B/M_\mathrm{ism})\gg 1$, or, using Eq.\eqref{gamma_rad_2}, to:
\begin{equation}
\gamma_\circ\gg\gamma\gg 1\, ,
\label{cond_rad}
\end{equation}
which is indeed the inequality adopted in the ``ultrarelativistic'' approximation in the current literature. If we further assume $r^3 \gg r_\circ^3$, Eq.\eqref{gamma_rad_2} can be approximated by a simple constant-index power-law as in Eq.\eqref{gr0}:
\begin{equation}
\gamma \simeq \left[\gamma_\circ/\left(\gamma_\circ+1\right)\right] M_B/\left[\left(4/3\right)\pi n_\mathrm{ism}m_pr^3\right]\, \propto\, r^{-3}\, .
\label{gamma_rad_3}
\end{equation}

We turn now to the range of applicability of these approximations, consistently with the inequalities given in Eq.\eqref{app_FB}. It then becomes manifest that these inequalities can only be enforced in a finite range of $M_\mathrm{ism}/M_B$. The lower limit (LL) and the upper limit (UL) of such range can be conservatively estimated:
\begin{subequations}\label{LL-UL}
\begin{equation}
\left(\tfrac{M_\mathrm{ism}}{M_B}\right)_{LL}=10^2\tfrac{1}{\gamma_\circ+1}\, ,\quad \left(\tfrac{M_\mathrm{ism}}{M_B}\right)_{UL}=10^{-2}\tfrac{\gamma_\circ}{\gamma_\circ+1}\, .
\label{LL}
\end{equation}
The allowed range of variability, if it exists, is then given by:
\begin{equation}
\left(\tfrac{M_\mathrm{ism}}{M_B}\right)_{UL}-\left(\tfrac{M_\mathrm{ism}}{M_B}\right)_{LL}=10^{-2}\tfrac{\gamma_\circ-10^4}{\gamma_\circ+1}>0\, .
\label{delta}
\end{equation}
\end{subequations}
A necessary condition for the applicability of the above approximations is therefore:
\begin{equation}
\gamma_\circ > 10^4\, .
\label{g0}
\end{equation}
It is important to emphasize that Eq.\eqref{g0} is only a {\em necessary} condition for the applicability of the approximate Eq.\eqref{gamma_rad_3}, but it is not {\em sufficient}: Eq.\eqref{gamma_rad_3} in fact can be applied only in the very limited range of $r$ values whose upper and lower limits are given in Eq.\eqref{LL}. See for explicit examples section \ref{exampl} below.

\subsubsection{The adiabatic case}\label{ad}

We now turn to the adiabatic case. If we assume:
\begin{equation}
1/\left(2\gamma_\circ\right)\ll M_\mathrm{ism}/M_B \ll \gamma_\circ/2\, ,
\label{app_FB_ad}
\end{equation}
we have that in the numerator of Eq.\eqref{gamma_ad} all terms are negligible with respect to $\gamma_\circ^2$, while in the denominator the leading term is the linear one in $M_\mathrm{ism}/M_B$. Eq.\eqref{gamma_ad} then becomes:
\begin{equation}
\gamma\simeq\sqrt{\left(\gamma_\circ/2\right) M_B/M_\mathrm{ism}}\, .
\label{gamma_ad_2}
\end{equation}
If we multiply the terms of Eq.\eqref{app_FB_ad} by $2/\gamma_\circ$, we obtain $1/\gamma_\circ^2\ll(2/\gamma_\circ)(M_\mathrm{ism}/M_B)\ll 1$, which is equivalent to $\gamma_\circ^2\gg(\gamma_\circ/2)(M_B/M_\mathrm{ism})\gg 1$, or, using Eq.\eqref{gamma_ad_2}, to:
\begin{equation}
\gamma_\circ^2\gg\gamma^2\gg 1\, .
\label{cond_ad}
\end{equation}
If we now further assume $r^3 \gg r_\circ^3$, Eq.\eqref{gamma_ad_2} can be approximated by a simple constant-index power-law as in Eq.\eqref{gr0}:
\begin{equation}
\gamma \simeq \sqrt{\left(\gamma_\circ/2\right) M_B/\left[\left(4/3\right)\pi n_\mathrm{ism}m_pr^3\right]}\, \propto\, r^{-3/2}\, .
\label{gamma_ad_3}
\end{equation}

We turn now to the range of applicability of these approximations, consistently with the inequalities given in Eq.\eqref{app_FB_ad}. It then becomes manifest that these inequalities can only be enforced in a finite range of $M_\mathrm{ism}/M_B$. The lower limit (LL) and the upper limit (UL) of such range can be conservatively estimated:
\begin{subequations}\label{LL-UL_ad}
\begin{equation}
\left(\tfrac{M_\mathrm{ism}}{M_B}\right)_{LL}=10^2\tfrac{1}{2\gamma_\circ}\, ,\quad \left(\tfrac{M_\mathrm{ism}}{M_B}\right)_{UL}=10^{-2}\tfrac{\gamma_\circ}{2}\, .
\label{LL_ad}
\end{equation}
The allowed range of variability, if it exists, is then given by:
\begin{equation}
\left(\tfrac{M_\mathrm{ism}}{M_B}\right)_{UL}-\left(\tfrac{M_\mathrm{ism}}{M_B}\right)_{LL}=10^{-2}\tfrac{\gamma_\circ^2-10^4}{2\gamma_\circ}>0\, .
\label{delta_ad}
\end{equation}
\end{subequations}
A necessary condition for the applicability of the above approximations is therefore:
\begin{equation}
\gamma_\circ > 10^2\, .
\label{g0_ad}
\end{equation}
Again, it is important to emphasize that Eq.\eqref{g0_ad} is only a {\em necessary} condition for the applicability of the approximate Eq.\eqref{gamma_ad_3}, but it is not {\em sufficient}: Eq.\eqref{gamma_ad_3} in fact can be applied only in the very limited range of $r$ values whose upper and lower limits are given in Eq.\eqref{LL_ad}. See for explicit examples section \ref{exampl} below.

\subsection{A specific example}\label{exampl}

\begin{figure}
\centering
\includegraphics[width=\hsize,clip]{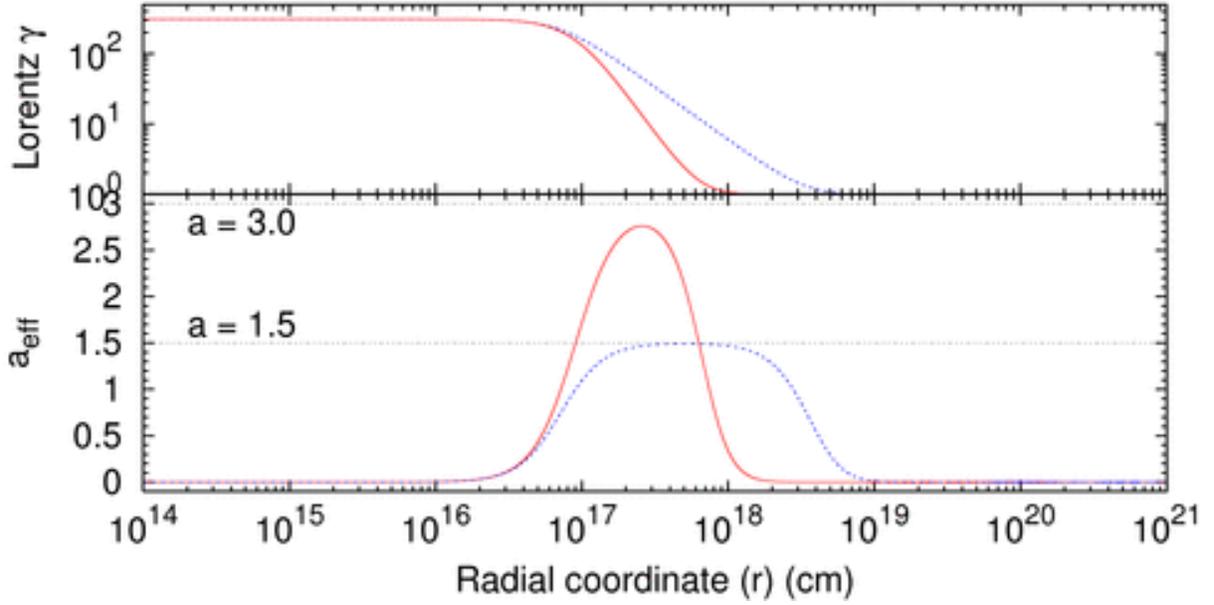}
\caption{In the upper panel, the analytic behavior of the Lorentz $\gamma$ factor during the afterglow era is plotted versus the radial coordinate of the expanding thin baryonic shell in the fully radiative case of GRB 991216 (solid red line) and in the adiabatic case starting from the same initial conditions (dotted blue line). In the lower panel are plotted the corresponding values of the ``effective'' power-law index $a_{eff}$ (see Eq.\eqref{eff_a}), which is clearly not constant, is highly varying and systematically lower than the constant values $3$ and $3/2$ purported in the current literature (horizontal dotted black lines). Details in \citet{PowerLaws}.}
\label{2gamma}
\end{figure}

Having obtained the analytic expression of the Lorentz gamma factor for the fully radiative case in Eq.\eqref{gamma_rad}, we illustrate in Fig. \ref{2gamma} the corresponding gamma factor as a function of the radial coordinate in the afterglow phase for GRB 991216 (see \citet{rubr} and references therein). We have also represented the corresponding solution which can be obtained in the adiabatic case, using Eq.\eqref{gamma_ad}, starting from the same initial conditions. It is clear that in both cases there is not a simple power-law relation like Eq.\eqref{gr0} with a constant index $a$. We can at most define an ``instantaneous'' value $a_{eff}$ for an ``effective'' power-law behavior:
\begin{equation}
a_{eff} = - \frac{d\ln\gamma}{d\ln r}\, .
\label{eff_a}
\end{equation}
Such an ``effective'' power-law index of the exact solution smoothly varies from $0$ to a maximum value which is always smaller than $3$ or $3/2$, in the fully radiative and adiabatic cases respectively, and finally decreases back to $0$ (see Fig. \ref{2gamma}). We see in particular, from Fig. \ref{2gamma}, how in the fully radiative case the power-law index is consistently smaller than $3$, and in the adiabatic case $a_{eff} = 3/2$ is approached only for a small interval of the radial coordinate corresponding to the latest parts of the afterglow with a Lorentz gamma factor of the order of $10$. In this case of GRB 991216 we have, in fact, $\gamma_\circ = 310.13$ and neither Eq.\eqref{cond_rad} nor Eq.\eqref{cond_ad} can be satisfied for any value of $r$. Therefore, neither in the fully radiative nor in the adiabatic case the constant-index power-law expression in Eq.\eqref{gr0} can be applied.

\begin{figure}
\begin{minipage}{\hsize}
\includegraphics[width=0.5\hsize,clip]{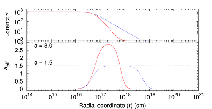}
\includegraphics[width=0.5\hsize,clip]{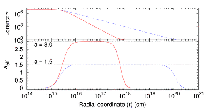}\\
\includegraphics[width=0.5\hsize,clip]{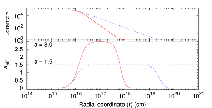}
\includegraphics[width=0.5\hsize,clip]{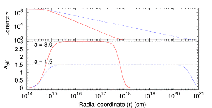}
\end{minipage}
\caption{In these four diagrams we reproduce the same quantities plotted in Fig. \ref{2gamma} for four higher values of $\gamma_\circ$. The upper (lower) left diagram corresponds to $\gamma_\circ = 10^3$ ($\gamma_\circ = 10^5$). The upper (lower) right diagram corresponds to $\gamma_\circ = 10^7$ ($\gamma_\circ = 10^9$). It is manifest how asymptotically, by increasing the value of $\gamma_\circ$, the values $a = 3$ and $a = 3/2$ (horizontal black dotted lines) are reached, but only in a limited range of the radial co-ordinate and anyway for values of $\gamma_\circ$ much larger than the ones actually observed in GRBs. Details in \citet{PowerLaws}.}
\label{multigamma}
\end{figure}

For clarity, we have integrated in Fig. \ref{multigamma} an ideal GRB afterglow with the initial conditions as GRB 991216 for selected higher values of the initial Lorentz gamma factor: $\gamma_\circ = 10^3, 10^5, 10^7, 10^9$. For $\gamma_\circ = 10^3$, we then see that, again, in the fully radiative condition $a_{eff} = 3$ is never reached and in the adiabatic case $a_{eff} \simeq 3/2$ only in the region where $10 < \gamma < 50$. Similarly, for $\gamma_\circ = 10^5$, in the fully radiative case $a_{eff} \simeq 3$ is only reached around the point $\gamma = 10^2$, and in the adiabatic case $a_{eff} \simeq 3/2$ for $10 < \gamma < 10^2$, although the non-power-law behavior still remains in the early and latest afterglow phases corresponding to the $\gamma \equiv \gamma_\circ$ and $\gamma \to 1$ regimes. The same conclusion can be reached for the remaining cases $\gamma_\circ = 10^7$ and $\gamma_\circ = 10^9$.

We like to emphasize that the early part of the afterglow, where $\gamma \equiv \gamma_\circ$, which cannot be described by the constant-index power-law approximation, do indeed corresponds to the rising part of the afterglow bolometric luminosity and to its peak, which is reached as soon as the Lorentz gamma factor starts to decrease. We have shown (see e.g. \citet{lett2,rubr,rubr2} and references therein) how the correct identifications of the raising part and the peak of the afterglow are indeed crucial for the explanation of the observed ``prompt radiation''. Similarly, the power-law cannot be applied during the entire approach to the newtonian regime, which corresponds to some of the actual observations occurring in the latest afterglow phases.

\section{Exact analytic expressions for the equitemporal surfaces in Gamma-Ray Burst afterglows}

\subsection{The definition of the EQTSs}

\begin{figure}
\centering
\includegraphics[width=\hsize,clip]{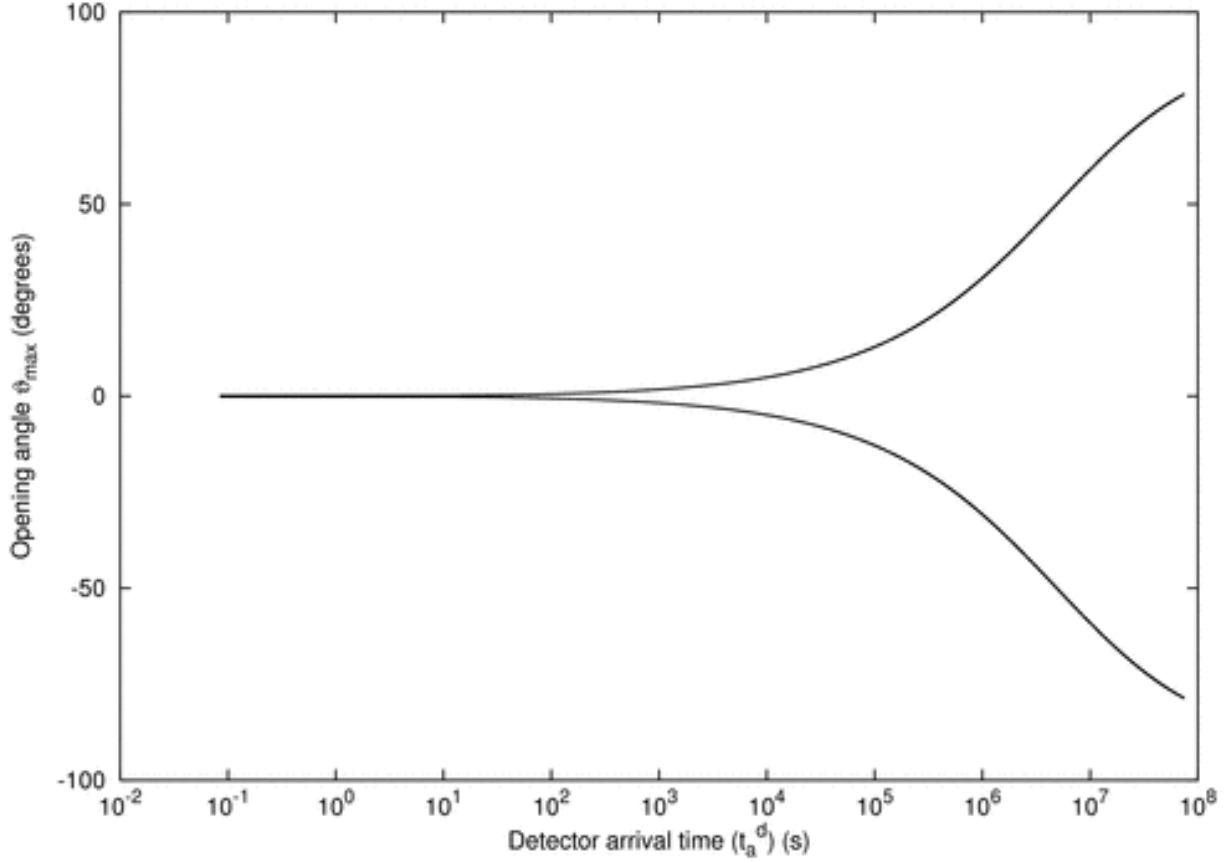}
\caption{Not all values of $\vartheta$ are allowed. Only photons emitted at an angle such that $\cos\vartheta \ge \left(v/c\right)$ can be viewed by the observer. Thus the maximum allowed $\vartheta$ value $\vartheta_{max}$ corresponds to $\cos\vartheta_{max} = (v/c)$. In this figure we show $\vartheta_{max}$ (i.e. the angular amplitude of the visible area of the ABM pulse) in degrees as a function of the arrival time at the detector for the photons emitted along the line of sight (see text). In the earliest GRB phases $v\sim c$ and so $\vartheta_{max}\sim 0$. On the other hand, in the latest phases of the afterglow the ABM pulse velocity decreases and $\vartheta_{max}$ tends to the maximum possible value, i.e. $90^\circ$. Details in \citet{r02,rubr}}
\label{openang}
\end{figure}

\begin{figure}
\centering
\includegraphics[width=\hsize,clip]{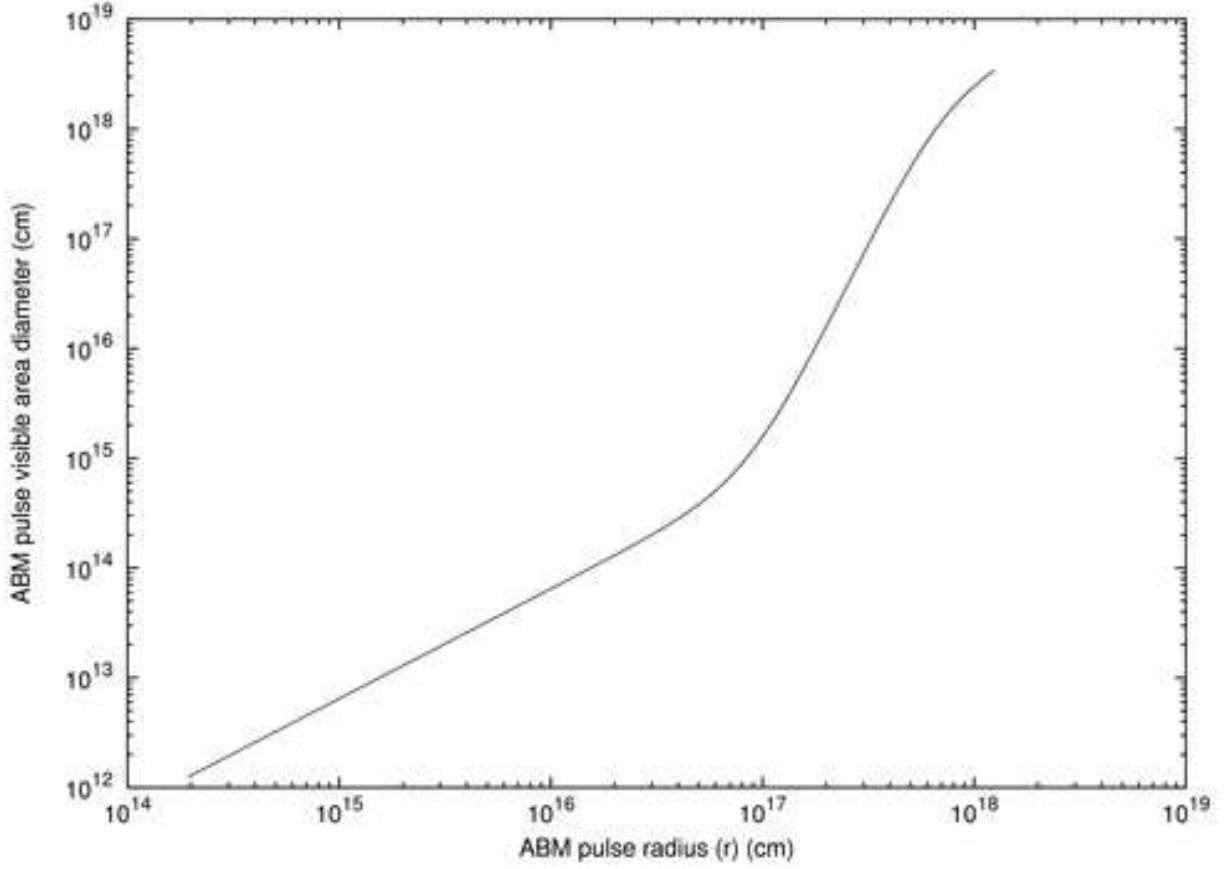}
\caption{The diameter of the visible area is represented as a function of the ABM pulse radius. In the earliest expansion phases ($\gamma\sim 310$) $\vartheta_{max}$ is very small (see left pane and Fig. \ref{opening}), so the visible area is just a small fraction of the total ABM pulse surface. On the other hand, in the final expansion phases $\vartheta_{max} \to 90^\circ$ and almost all the ABM pulse surface becomes visible. Details in \citet{r02,rubr}}
\label{opensrad}
\end{figure}

\begin{figure}
\centering
\includegraphics[width=0.35\hsize,clip]{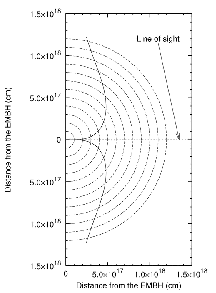}
\includegraphics[width=0.65\hsize,clip]{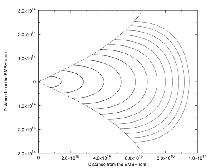}
\caption{{\bf Left:} This figure shows the temporal evolution of the visible area of the ABM pulse. The dashed half-circles are the expanding ABM pulse at radii corresponding to different laboratory times. The black curve marks the boundary of the visible region. The black hole is located at position (0,0) in this plot. Again, in the earliest GRB phases the visible region is squeezed along the line of sight, while in the final part of the afterglow phase almost all the emitted photons reach the observer. This time evolution of the visible area is crucial to the explanation of the GRB temporal structure. Details in \citet{r02,rubr}. {\bf Right:} Due to the extremely high and extremely varying Lorentz gamma factor, photons reaching the detector on the Earth at the same arrival time are actually emitted at very different times and positions. We represent here the surfaces of photon emission corresponding to selected values of the photon arrival time at the detector: the {\em equitemporal surfaces} (EQTS). Such surfaces differ from the ellipsoids described by Rees in the context of the expanding radio sources with typical Lorentz factor $\gamma\sim 4$ and constant. In fact, in GRB~991216 the Lorentz gamma factor ranges from $310$ to $1$. The EQTSs represented here (solid lines) correspond respectively to values of the arrival time ranging from $5\, s$ (the smallest surface on the left of the plot) to $60\, s$ (the largest one on the right). Each surface differs from the previous one by $5\, s$. To each EQTS contributes emission processes occurring at different values of the Lorentz gamma factor. The dashed lines are the boundaries of the  visible area of the ABM pulse and the black hole is located at position $(0,0)$ in this plot. Note the different scales on the two axes, indicating the very high EQTS ``effective eccentricity''. The time interval from $5\, {\rm s}$ to $60\, {\rm s}$ has been chosen to encompass the E-APE emission, ranging from $\gamma=308.8$ to $\gamma=56.84$. Details in \citet{r02,rubr}.}
\label{opening}
\end{figure}

For a relativistically expanding spherically symmetric source the ``equitemporal surfaces'' (EQTSs) are surfaces of revolution about the line of sight. The general expression for their profile, in the form $\vartheta = \vartheta(r)$, corresponding to an arrival time $t_a$ of the photons at the detector, can be obtained from (see e.g. \citet{rubr,EQTS_ApJL,EQTS_ApJL2} and Figs. \ref{openang}--\ref{opening}):
\begin{equation} 
ct_a = ct\left(r\right) - r\cos \vartheta  + r^\star\, , 
\label{ta_g} 
\end{equation} 
where $r^\star$ is the initial size of the expanding source, $\vartheta$ is the angle between the radial expansion velocity of a point on its surface and the line of sight, and $t = t(r)$ is its equation of motion, expressed in the laboratory frame, obtained by the integration of Eqs.(\ref{Taub_Eq}). From the definition of the Lorentz gamma factor $\gamma^{-2}=1-(dr/cdt)^2$, we have in fact:
\begin{equation} 
ct\left(r\right)=\int_0^r\left[1-\gamma^{-2}\left(r'\right)\right]^{-1/2}dr'\, , 
\label{tdir} 
\end{equation} 
where $\gamma(r)$ comes from the integration of Eqs.(\ref{Taub_Eq}). 

It is appropriate to underline a basic difference between the apparent superluminal velocity orthogonal to the line of sight, $v^\bot \simeq \gamma v$, and the apparent superluminal velocity along the line of sight, $v^\parallel \simeq \gamma^2 v$. In the case of GRBs, this last one is the most relevant: for a Lorentz gamma factor $\gamma \simeq 300$ we have $v^\parallel \simeq 10^5 c$. This is self-consistently verified in the structure of the ``prompt radiation'' of GRBs, see e.g. \citet{r02}.

\subsection{The analytic expressions for the EQTSes}

\subsubsection{The fully radiative case}

The analytic expression for the EQTS in the fully radiative regime can then be obtained substituting $t(r)$ from Eq.(\ref{analsol}) in Eq.(\ref{ta_g}). We obtain (\citet{EQTS_ApJL2}):
\begin{equation}
\begin{split}
&\cos\vartheta=\frac{M_B  - m_i^\circ}{2r\sqrt{C}}\left( {r - r_\circ } \right) +\frac{m_i^\circ r_\circ }{8r\sqrt{C}}\left[ {\left( {\frac{r}{{r_\circ }}} \right)^4  - 1} \right] \\[6pt]
&+\frac{{r_\circ \sqrt{C} }}{{12rm_i^\circ A^2 }} \ln \left\{ {\frac{{\left[ {A + \left(r/r_\circ\right)} \right]^3 \left(A^3  + 1\right)}}{{\left[A^3  + \left( r/r_\circ \right)^3\right] \left( {A + 1} \right)^3}}} \right\} +\frac{ct_\circ}{r}-\frac{ct_a}{r} \\[6pt] & + \frac{r^\star}{r} +\frac{{r_\circ \sqrt{3C} }}{{6rm_i^\circ A^2 }} \left[ \arctan \frac{{2\left(r/r_\circ\right) - A}}{{A\sqrt{3} }} - \arctan \frac{{2 - A}}{{A\sqrt{3} }}\right]\, ,
\end{split}
\label{eqts_g_dopo}
\end{equation}
where $A$, $C$ and $m_i^\circ$ are the same as in Eq.(\ref{analsol}).

\subsubsection{The adiabatic case}

The analytic expression for the EQTS in the adiabatic regime can then be obtained substituting $t(r)$ from Eq.(\ref{analsol_ad}) in Eq.(\ref{ta_g}). We obtain (\citet{EQTS_ApJL2}):
\begin{equation}
\begin{split}
\cos\vartheta & = \frac{m_i^\circ}{4M_B\sqrt{\gamma_\circ^2-1}}\left[\left(\frac{r}{r_\circ}\right)^3  - \frac{r_\circ}{r}\right] + \frac{ct_\circ}{r} \\[6pt] & - \frac{ct_a}{r} + \frac{r^\star}{r} - \frac{\gamma_\circ-\left(m_i^\circ/M_B\right)}{\sqrt{\gamma_\circ^2-1}}\left[\frac{r_\circ}{r} - 1\right]\, .
\end{split}
\label{eqts_g_dopo_ad}
\end{equation}

\subsubsection{Comparison between the two cases}

\begin{figure}
\begin{minipage}{\hsize}
\includegraphics[width=0.85\hsize,clip]{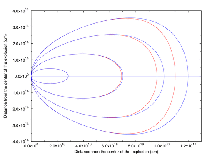}\\
\includegraphics[width=0.85\hsize,clip]{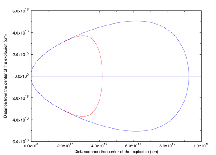}
\end{minipage}
\caption{Comparison between EQTSs in the adiabatic regime (blue lines) and in the fully radiative regime (red lines). The upper plot shows the EQTSs for $t_a=5$ s, $t_a=15$ s, $t_a=30$ s and $t_a=45$ s, respectively from the inner to the outer one. The lower plot shows the EQTS at an arrival time of 2 days.  Details in \citet{EQTS_ApJL2}.}
\label{eqts_comp}
\end{figure}

The two EQTSs are represented at selected values of the arrival time $t_a$ in Fig.~\ref{eqts_comp}, where the illustrative case of GRB~991216 has been used as a prototype. The initial conditions at the beginning of the afterglow era are in this case given by $\gamma_\circ = 310$, $r_\circ = 1.94 \times 10^{14}$ cm, $t_\circ = 6.48 \times 10^{3}$ s, $r^\star = 2.35 \times 10^8$ cm (see \citet{lett1,lett2,r02,rubr,EQTS_ApJL2}).

\subsection{Approximations adopted in the current literature}\label{approx}

In the current literature two different treatments of the EQTSs exist: one by \citet{pm98c} and one by \citet{s98} later applied also by \citet{gps99} (see also \citet{p99,p00,vpkw00} and references therein).

In both these treatments, instead of the more precise dynamical equations given in Eqs.(\ref{gamma_ad},\ref{gamma_rad}), the simplified formula, based on the ``ultrarelativistic'' approximation, given in Eq.\eqref{gr0} has been used. A critical analysis comparing and contrasting our exact solutions with Eq.(\ref{gr0}) has been presented in the previous section and in \citet{PowerLaws}. As a further approximation, instead of the exact Eq.(\ref{tdir}), they both use the following expansion at first order in $\gamma^{-2}$:
\begin{equation}
ct\left(r\right) = \int_0^r \left[1+\frac{1}{2\gamma^2\left(r'\right)}\right] dr'\, .
\label{tdir_app}
\end{equation}
Correspondingly, instead of the exact Eq.(\ref{analsol_ad}) and Eq.(\ref{analsol}), they find:
\begin{subequations}\label{t_app_pm98c_s98}
\begin{eqnarray}
t\left(r\right) & = & \frac{r}{c}\left[1+\frac{1}{2\left(2a+1\right)\gamma^2\left(r\right)}\right]\, ,\\
t\left(r\right) & = & \frac{r}{c}\left[1+\frac{1}{16\gamma^2\left(r\right)}\right]\, .
\end{eqnarray}
\end{subequations}
The first expression has been given by \citet{pm98c} and applies both in the adiabatic ($a=3/2$) and in the fully radiative ($a=3$) cases (see their Eq.(2)). The second one has been given by \citet{s98} in the adiabatic case (see his Eq.(2)). Note that the first expression, in the case $a=3/2$, does not coincide with the second one: \citet{s98} uses a Lorentz gamma factor $\Gamma$ of a shock front propagating in the expanding pulse, with $\Gamma = \sqrt{2} \gamma$.

Instead of the exact Eqs.(\ref{ta_g}), \citet{pm98c} and \citet{s98} both uses the following equation:
\begin{equation}
ct_a = ct\left(r\right) - r\cos \vartheta\, ,
\label{ta_g_app}
\end{equation}
where the initial size $r^\star$ has been neglected. The following approximate expressions for the EQTSs have been then presented:
\begin{subequations}\label{eqts_app}
\begin{eqnarray}
\vartheta & = & 2\arcsin\left[\frac{1}{2\gamma_\circ}\sqrt{\frac{2\gamma_\circ^2ct_a}{r}-\frac{1}{2a+1}\left(\frac{r}{r_\circ}\right)^{2a}}\right]\, ,\\
\cos\vartheta & = & 1-\frac{1}{16\gamma_L^2}\left[\left(\frac{r}{r_L}\right)^{-1}-\left(\frac{r}{r_L}\right)^{3}\right]\, .
\end{eqnarray}
\end{subequations}
The first expression has been given by \citet{pm98c} and applies both in the adiabatic ($a=3/2$) and in the fully radiative ($a=3$) cases (see their Eq.(3)). The second expression, where $\gamma_L \equiv \gamma(\vartheta=0)$ over the given EQTS and $r_L=16\gamma_L^2ct_a$, has been given by \citet{s98} in the adiabatic case (see his Eq.(5)).

\begin{figure}
\begin{minipage}{\hsize}
\includegraphics[width=0.85\hsize,clip]{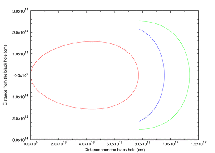}\\
\includegraphics[width=0.85\hsize,clip]{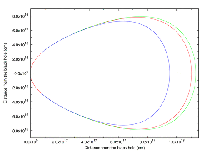}
\end{minipage}
\caption{Comparison between the EQTSs computed using the approximate formulas given by \citet{pm98c} (blue line) and by \citet{s98,gps99} (green line) in the fully adiabatic case ($a=3/2$ in Eqs.(\ref{eqts_app})) and the corresponding ones computed using the exact analytic expression given in Eq.(\ref{eqts_g_dopo_ad}) (red line). The difference between the dashed line and the dotted line is due to the factor $\sqrt{2}$ in the Lorentz $\gamma$ factor adopted by Sari (see text). The upper (lower) panel corresponds to $t_a^d=35$ s ($t_a^d=4$ day). The approximate curves are not drawn entirely because Eqs.(\ref{eqts_app}) are declared to be valid only where $\gamma < 2/3 \gamma_\circ$. Details in \citet{EQTS_ApJL,EQTS_ApJL2}.}
\label{eqts_comp_ad}
\end{figure}

\begin{figure}
\begin{minipage}{\hsize}
\includegraphics[width=0.85\hsize,clip]{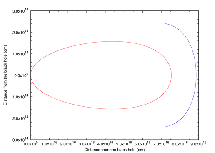}\\
\includegraphics[width=0.85\hsize,clip]{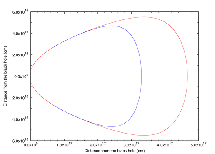}
\end{minipage}
\caption{Comparison between the EQTSs computed using the approximate formulas given by \citet{pm98c} (blue line) in the fully radiative case ($a=3$ in the first of Eqs.(\ref{eqts_app})) and the corresponding ones computed using the exact analytic expression given in Eq.(\ref{eqts_g_dopo}) (red line). The upper (lower) panel corresponds to $t_a^d=35$ s ($t_a^d=4$ day). Details in \citet{EQTS_ApJL,EQTS_ApJL2}.}
\label{eqts_comp_rad}
\end{figure}

Without entering into the relative merit of such differing approaches, we show in Figs.~\ref{eqts_comp_ad}--\ref{eqts_comp_rad} that both of them lead to results different from the ones computed with the exact solutions. The consequences of using the approximate formula given in Eq.(\ref{gr0}) to compute the expression $t \equiv t(r)$, instead of the exact solution of Eqs.(\ref{Taub_Eq}), are clearly shown in Figs.~\ref{eqts_comp_ad}--\ref{eqts_comp_rad}. The EQTSs represented in these figures are computed at selected values of the detector arrival time both in the early ($\sim 35$ s) and in the late ($\sim 4$ day) phases of the afterglow. Both the fully radiative and fully adiabatic cases are examined. Note the approximate expression of the EQTS can only be defined for $\gamma < \gamma_d$ and $r > r_d$ (see \citet{EQTS_ApJL2}). Consequently, at $t_a^d=35$ s the approximate EQTSs are represented by arcs, markedly different from the exact solution (see the upper panels of Figs.~\ref{eqts_comp_ad}--\ref{eqts_comp_rad}). The same conclusion is found for the EQTS at $t_a^d=4$ days, where marked differences are found both for the fully radiative and adiabatic regimes (see the lower panels of Figs.~\ref{eqts_comp_ad}--\ref{eqts_comp_rad}).

\section{Exact versus approximate beaming formulas in Gamma-Ray Burst afterglows}

Using the exact solutions introduced in the previous sections, we here introduce the exact analytic expressions of the relations between the detector arrival time $t_a^d$ of the GRB afterglow radiation and the corresponding half-opening angle $\vartheta$ of the expanding source visible area due to the relativistic beaming (see e.g. \citet{rubr}). Such visible area must be computed not over the spherical surface of the shell, but over the EQuiTemporal Surface (EQTS) of detector arrival time $t_a^d$, i.e. over the surface locus of points which are source of the radiation reaching the observer at the same arrival time $t_a^d$ (see \citet{EQTS_ApJL,EQTS_ApJL2} for details). The exact analytic expressions for the EQTSs in GRB afterglows, which have been presented in Eqs.(\ref{eqts_g_dopo_ad})--(\ref{eqts_g_dopo}) and in \citet{EQTS_ApJL2}, are therefore crucial in our present derivation. This approach clearly differs from the ones in the current literature, which usually neglect the contributions of the radiation emitted from the entire EQTS.

The analytic relations between $t_a^d$ and $\vartheta$ presented in this section allow to compute, assuming that the expanding shell is not spherically symmetric but is confined into a narrow jet with half-opening angle $\vartheta_\circ$, the value $(t_a^d)_{jet}$ of the detector arrival time at which we start to ``see'' the sides of the jet. A corresponding ``break'' in the observed light curve should occur later than $(t_a^d)_{jet}$ (see e.g. \citet{sph99}). In the current literature, $(t_a^d)_{jet}$ is usually defined as the detector arrival time at which $\gamma \sim 1/\vartheta_\circ$, where $\gamma$ is the Lorentz factor of the expanding shell (see e.g. \citet{sph99} and also our Eq.(\ref{Sist}) below). In our formulation we do not consider effects of lateral spreadings of the jet.

In the current literature, in the case of adiabatic regime, different approximate power-law relations between $(t_a^d)_{jet}$ and $\vartheta_\circ$ have been presented, in contrast to each other (see e.g. \citet{sph99,pm99,p06}). We show here that in four specific cases of GRBs, encompassing more than $5$ orders of magnitude in energy and more than $2$ orders of magnitude in ISM density, both the one by \citet{pm99} and the one by \citet{sph99} overestimate the exact analytic result. A third relation just presented by \citet{p06} slightly underestimate the exact analytic result. We also present an empirical fit of the numerical solutions of the exact equations for the adiabatic regime, compared and contrasted with the three above approximate relations. In the fully radiative regime, and therefore in the general case, no simple power-law relation of the kind found in the adiabatic regime can be established and the general approach we have outlined has to be followed.

Although evidence for spherically symmetric emission in GRBs is emerging from observations (\citet{sa06}) and from theoretical argumentations (\citet{Spectr1,cospar04}), it is appropriate to develop here an exact theoretical treatment of the relation between $(t_a^d)_{jet}$ and $\vartheta_\circ$. This will allow to make an assessment on the existence and, in the positive case, on the extent of beaming in GRBs, which in turn is going to be essential for establishing their correct energetics.

\subsection{Analytic formulas for the beaming angle}

The boundary of the visible region of a relativistic thin and uniform shell expanding in the ISM is defined by (see e.g. \citet{rubr} and references therein):
\begin{equation}
\cos\vartheta = \frac{v}{c}\, ,
\label{bound}
\end{equation}
where $\vartheta$ is the angle between the line of sight and the radial expansion velocity of a point on the shell surface, $v$ is the velocity of the expanding shell and $c$ is the speed of light. To find the value of the half-opening beaming angle $\vartheta_\circ$ corresponding to an observed arrival time $(t_a^d)_{jet}$, this equation must be solved together with the equation describing the EQTS of arrival time $(t_a^d)_{jet}$ (\citet{EQTS_ApJL2}). In other words, we must solve the following system:
\begin{equation}
\left\{
\begin{array}{rcl}
\cos\vartheta_\circ & = & \frac{v\left(r\right)}{c}\\
\cos\vartheta_\circ & = & \cos\left\{\left.\vartheta \left[r;(t_a^d)_{jet}\right]\right|_{EQTS\left[(t_a^d)_{jet}\right]}\right\}
\end{array}
\right.\, .
\label{Sist}
\end{equation}
It should be noted that, in the limit $\vartheta_\circ \to 0$ and $v \to c$, this definition of $(t_a^d)_{jet}$ is equivalent to the one usually adopted in the current literature (see above).

\subsubsection{The fully radiative regime}

In this case, the analytic solution of the equations of motion gives (see Eq.(\ref{gamma_rad}) and \citet{EQTS_ApJL2,PowerLaws}):
\begin{equation}
\frac{v}{c} = \frac{\sqrt{\left(1-\gamma_\circ^{-2}\right)\left[1+\left(M_{ism}/M_B\right)+\left(M_{ism}/M_B\right)^2\right]}}{1+\left(M_{ism}/M_B\right)\left(1+\gamma_\circ^{-1}\right)\left[1+\textstyle\frac{1}{2}\left(M_{ism}/M_B\right)\right]}\, .
\label{vRad}
\end{equation}
Using the analytic expression for the EQTS given in Eq.(\ref{eqts_g_dopo}) and in \citet{EQTS_ApJL2}, Eq.(\ref{Sist}) takes the form (\citet{beaming}):
\begin{equation}
\left\{
\begin{array}{rcl}
\cos\vartheta_\circ & = & \frac{\sqrt{\left(1-\gamma_\circ^{-2}\right)\left[1+\left(M_{ism}/M_B\right)+\left(M_{ism}/M_B\right)^2\right]}}{1+\left(M_{ism}/M_B\right)\left(1+\gamma_\circ^{-1}\right)\left[1+\textstyle\frac{1}{2}\left(M_{ism}/M_B\right)\right]}\\[18pt]
\cos\vartheta_\circ & = & \frac{M_B  - m_i^\circ}{2r\sqrt{C}}\left( {r - r_\circ } \right) +\frac{m_i^\circ r_\circ }{8r\sqrt{C}}\left[ {\left( {\frac{r}{{r_\circ }}} \right)^4  - 1} \right] \\[6pt]
& + & \frac{{r_\circ \sqrt{C} }}{{12rm_i^\circ A^2 }} \ln \left\{ {\frac{{\left[ {A + \left(r/r_\circ\right)} \right]^3 \left(A^3  + 1\right)}}{{\left[A^3  + \left( r/r_\circ \right)^3\right] \left( {A + 1} \right)^3}}} \right\} \\[6pt]
& + & \frac{ct_\circ}{r} - \frac{c(t_a^d)_{jet}}{r\left(1+z\right)} + \frac{r^\star}{r} \\[6pt]
& + & \frac{{r_\circ \sqrt{3C} }}{{6rm_i^\circ A^2 }} \left[ \arctan \frac{{2\left(r/r_\circ\right) - A}}{{A\sqrt{3} }} - \arctan \frac{{2 - A}}{{A\sqrt{3} }}\right]
\end{array}
\right.
\label{SistRad}
\end{equation}
where $t_\circ$ is the value of the time $t$ at the beginning of the afterglow phase, $m_i^\circ=(4/3)\pi m_p n_{\mathrm{ism}} r_\circ^3$, $r^\star$ is the initial size of the expanding source, $A=[(M_B-m_i^\circ)/m_i^\circ]^{1/3}$, $C={M_B}^2(\gamma_\circ-1)/(\gamma_\circ +1)$ and $z$ is the cosmological redshift of the source.

\subsubsection{The adiabatic regime}

In this case, the analytic solution of the equations of motion gives (see Eq.(\ref{gamma_ad}) and \citet{EQTS_ApJL2,PowerLaws}):
\begin{equation}
\frac{v}{c} = \sqrt{\gamma_\circ^2-1}\left(\gamma_\circ+\frac{M_{ism}}{M_B}\right)^{-1}
\label{vAd}
\end{equation}
Using the analytic expression for the EQTS given in Eq.(\ref{eqts_g_dopo_ad}) and in \citet{EQTS_ApJL2}, Eq.(\ref{Sist}) takes the form (\citet{beaming}):
\begin{equation}
\left\{
\begin{array}{rcl}
\cos\vartheta_\circ & = & \sqrt{\gamma_\circ^2-1}\left(\gamma_\circ+\frac{M_{ism}}{M_B}\right)^{-1} \\[18pt]
\cos\vartheta_\circ & = & \frac{m_i^\circ}{4M_B\sqrt{\gamma_\circ^2-1}}\left[\left(\frac{r}{r_\circ}\right)^3  - \frac{r_\circ}{r}\right] + \frac{ct_\circ}{r} \\[6pt]
& - & \frac{c(t_a^d)_{jet}}{r\left(1+z\right)} + \frac{r^\star}{r} - \frac{\gamma_\circ-\left(m_i^\circ/M_B\right)}{\sqrt{\gamma_\circ^2-1}}\left[\frac{r_\circ}{r} - 1\right]
\end{array}
\right.
\label{SistAd}
\end{equation}
where all the quantities have the same definition as in Eq.(\ref{SistRad}).

\subsubsection{The comparison between the two solutions}

\begin{figure}
\includegraphics[width=\hsize,clip]{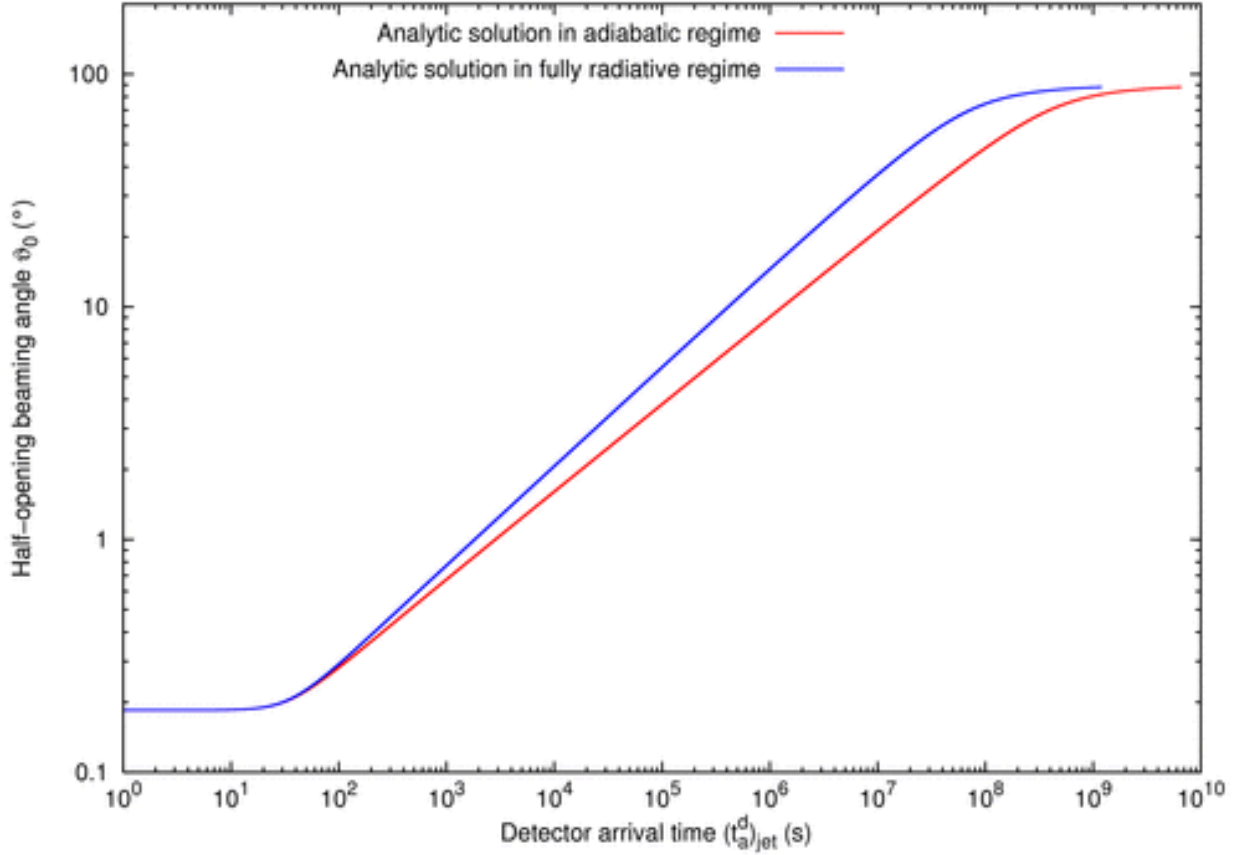}
\caption{Comparison between the numerical solution of Eq.(\ref{SistRad}) assuming fully radiative regime (blue line) and the corresponding one of Eq.(\ref{SistAd}) assuming adiabatic regime (red line). The departure from power-law behavior at small arrival time follows from the constant Lorentz $\gamma$ factor regime, while the one at large angles follows from the approach to the non-relativistic regime (see details in Fig. \ref{Beam_Comp_Num_Fit_Log}, as well as in \citet{PowerLaws,beaming}).}
\label{Beam_Comp_Rad_Ad}
\end{figure}

In Fig. \ref{Beam_Comp_Rad_Ad} we plot the numerical solutions of both Eq.(\ref{SistRad}), corresponding to the fully radiative regime, and Eq.(\ref{SistAd}), corresponding to the adiabatic one. Both curves have been plotted assuming the same initial conditions, namely the ones of GRB 991216 (see \citet{rubr}).

\subsection{Comparison with the existing literature}

\begin{figure}
\begin{minipage}{\hsize}
\includegraphics[width=0.5\hsize,clip]{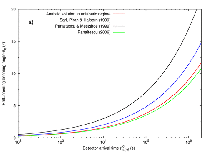}
\includegraphics[width=0.5\hsize,clip]{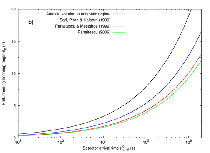}\\
\includegraphics[width=0.5\hsize,clip]{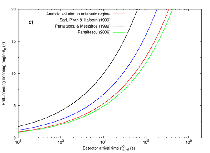}
\includegraphics[width=0.5\hsize,clip]{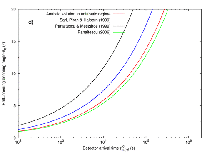}
\end{minipage}
\caption{Comparison between the numerical solution of Eq.(\ref{SistAd}) (red line) and the corresponding approximate formulas given in Eq.(\ref{ThetaSPH99}) (blue line), in Eq.(\ref{ThetaPM99}) (black line), and in Eq.(\ref{ThetaP06}) (green line). All four curves have been plotted for four different GRBs: a) GRB 991216 (\citet{rubr}), b) GRB 980519 (\citet{980519}), c) GRB 031203 (\citet{031203}), d) GRB 980425 (\citet{cospar02,Mosca_Orale}). The ranges of the two axes have been chosen to focus on the sole domains of application of the approximate treatments in the current literature. Details in \citet{beaming}.}
\label{Beam_Comp}
\end{figure}

Three different approximated formulas for the relation between $(t_a^d)_{jet}$ and $\vartheta_\circ$ have been given in the current literature, all assuming the adiabatic regime. \citet{pm99} proposed:
\begin{equation}
\cos\vartheta_\circ \simeq 1 - 5.9\times10^7 \left(\frac{n_{ism}}{E}\right)^{1/4} \left[\frac{(t_a^d)_{jet}}{1+z}\right]^{3/4} \, ,
\label{ThetaPM99}
\end{equation}
\citet{sph99}, instead, advanced:
\begin{equation}
\vartheta_\circ \simeq 7.4\times 10^3 \left(\frac{n_{ism}}{E}\right)^{1/8} \left[\frac{(t_a^d)_{jet}}{1+z}\right]^{3/8} \, .
\label{ThetaSPH99}
\end{equation}
In both Eq.(\ref{ThetaPM99}) and Eq.(\ref{ThetaSPH99}), $(t_a^d)_{jet}$ is measured in seconds, $E$ is the source initial energy measured in ergs and $n_{ism}$ is the ISM number density in particles/cm$^3$. The formula by \citet{sph99} has been applied quite often in the current literature (see e.g. \citet{fa01,ggl04,fa05}).

Both Eq.(\ref{ThetaPM99}) and Eq.(\ref{ThetaSPH99}) compute the arrival time of the photons at the detector assuming that all the radiation is emitted at $\vartheta=0$ (i.e. on the line of sight), neglecting the full shape of the EQTSs. Recently, a new expression has been proposed by \citet{p06}, again neglecting the full shape of the EQTSs but assuming that all the radiation is emitted from $\vartheta = 1/\gamma$, i.e. from the boundary of the visible region. Such an expression is:
\begin{equation}
\vartheta_\circ \simeq 5.4\times 10^3 \left(\frac{n_{ism}}{E}\right)^{1/8} \left[\frac{(t_a^d)_{jet}}{1+z}\right]^{3/8} \, .
\label{ThetaP06}
\end{equation}

In Fig. \ref{Beam_Comp} we plot Eq.(\ref{ThetaPM99}), Eq.(\ref{ThetaSPH99}) and Eq.(\ref{ThetaP06}) together with the numerical solution of Eq.(\ref{SistAd}) relative to the adiabatic regime. All four curves have been plotted assuming the same initial conditions for four different GRBs, encompassing more than $5$ orders of magnitude in energy and more than $2$ orders of magnitude in ISM density: a) GRB 991216 (\citet{rubr}), b) GRB 980519 (\citet{980519}), c) GRB 031203 (\citet{031203}), d) GRB 980425 (\citet{cospar02,Mosca_Orale}). The approximate Eq.(\ref{ThetaSPH99}) by \citet{sph99} and Eq.(\ref{ThetaP06}) by \citet{p06} both imply a power-law relation between $\vartheta_\circ$ and $(t_a^d)_{jet}$ with constant index $3/8$ for any value of $\vartheta_\circ$, while Eq.(\ref{ThetaPM99}) by \citet{pm99} implies a power-law relation with constant index $3/8$ only for $\vartheta_\circ \to 0$ (for greater $\vartheta_\circ$ values the relation is trigonometric).

All the above three approximate treatments are based on the approximate power-law solutions of the GRB afterglow dynamics which have been shown in the previous sections and in \citet{PowerLaws} to be not applicable to GRBs. They also do not take fully into account the structure of the EQTSs, although in different ways. Both Eq.(\ref{ThetaPM99}) and Eq.(\ref{ThetaSPH99}), which assume all the radiation coming from $\vartheta=0$, overestimate the behavior of the exact solution. On the other hand, Eq.(\ref{ThetaP06}), which assumes all the radiation coming from $\vartheta\sim 1/\gamma$, is a better approximation than the previous two, but still slightly underestimates the exact solution.

\subsection{An empirical fit of the numerical solution}

\begin{figure}
\includegraphics[width=\hsize,clip]{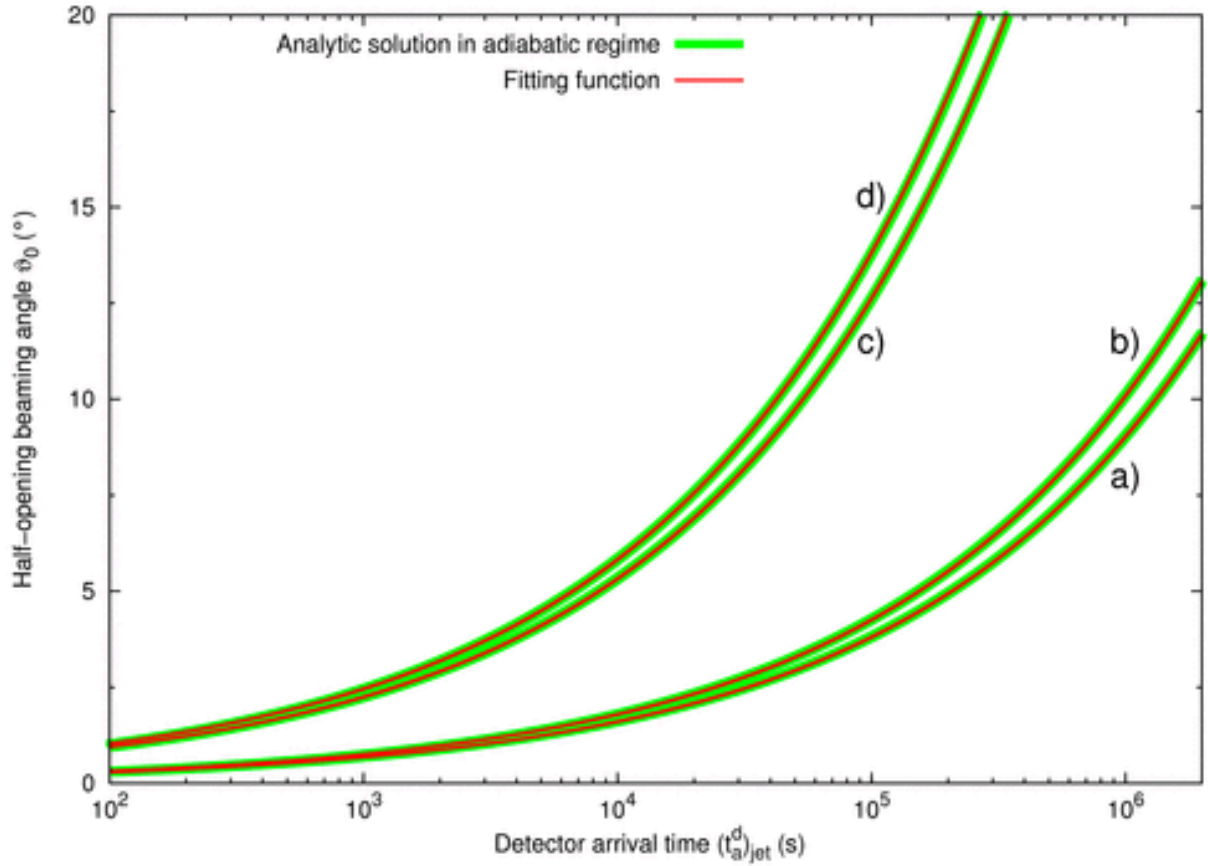}
\caption{The overlapping between the numerical solution of Eq.(\ref{SistAd}) (thick green lines) and the approximate fitting function given in Eq.(\ref{ThetaFIT}) (thin red lines) is shown in the four cases (a--d) represented in Fig. \ref{Beam_Comp}.}
\label{Beam_Comp_Num_Fit}
\end{figure}

\begin{figure}
\includegraphics[width=\hsize,clip]{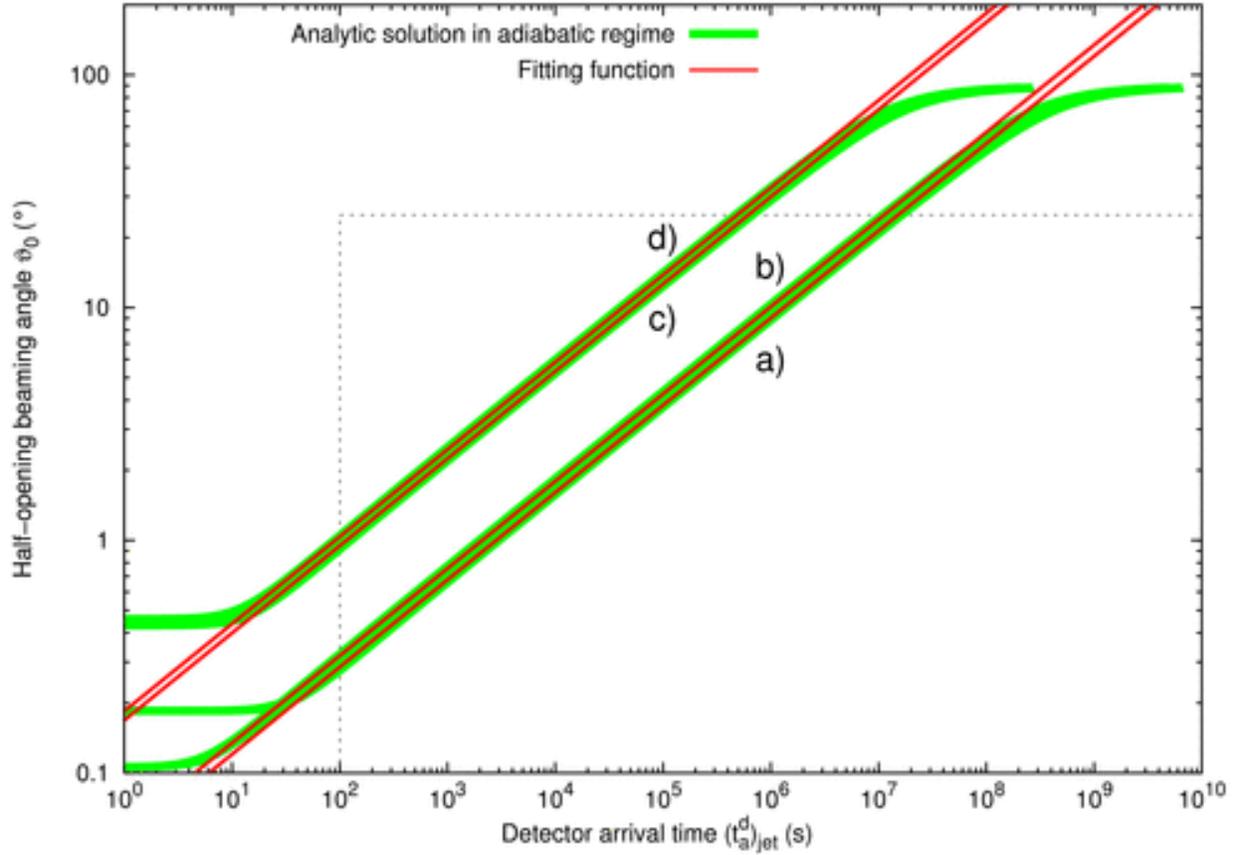}
\caption{Comparison between the numerical solution of Eq.(\ref{SistAd}) (think green lines) and the approximate fitting function given in Eq.(\ref{ThetaFIT}) (thin red lines) in all the four cases (a--d) represented in Fig. \ref{Beam_Comp}. The ranges of the two axes have been chosen to have their full extension (i.e. the one of Fig. \ref{Beam_Comp_Rad_Ad}). The dashed gray lines are the boundaries of the region where the empirical fitting function can be applied. Details in \citet{beaming}.}
\label{Beam_Comp_Num_Fit_Log}
\end{figure}

For completeness, we now fit our exact solution with a suitable explicit functional form in the four cases considered in Fig. \ref{Beam_Comp}. We chose the same functional form of Eq.(\ref{ThetaP06}), which is the closer one to the numerical solution, using the numerical factor in front of it (i.e. $5.4\times 10^3$) as the fitting parameter. We find that the following approximate expression (\citet{beaming}):
\begin{equation}
\vartheta_\circ \simeq 5.84\times 10^3 \left(\frac{n_{ism}}{E}\right)^{1/8} \left[\frac{(t_a^d)_{jet}}{1+z}\right]^{3/8}
\label{ThetaFIT}
\end{equation}
is in agreement with the numerical solution in all the four cases presented in Fig. \ref{Beam_Comp} (see Fig. \ref{Beam_Comp_Num_Fit}). However, if we enlarge the axis ranges to their full extension (i.e. the one of Fig. \ref{Beam_Comp_Rad_Ad}), we see that such approximate empirical fitting formula can only be applied for $\vartheta_\circ < 25^\circ$ \emph{and} $(t_a^d)_{jet} > 10^2$ s (see the gray dashed rectangle in Fig. \ref{Beam_Comp_Num_Fit_Log}).

An equivalent empirical fit in the fully radiative regime is not possible. In this case, indeed, there is a domain in the $((t_a^d)_{jet},\vartheta_\circ)$ plane where the numerical solution shows a power-law dependence on time, with an index $\sim 0.423$ (see Fig. \ref{Beam_Comp_Rad_Ad}). However, the dependence on the energy cannot be factorized out with a simple power-law. Therefore, in the fully radiative regime, which is the relevant one for our GRB model (see e.g. \citet{rubr}), the application of the full Eq.(\ref{SistRad}) does not appear to be avoidable.

\section{The afterglow bolometric luminosity and the ISM discrete structure}

We assume that the internal energy due to kinetic collision is instantly radiated away and that the corresponding emission is isotropic. Let $\Delta \varepsilon$ be the internal energy density developed in the collision. In the comoving frame the energy per unit of volume and per solid angle is simply
\begin{equation} 
\left(\frac{dE}{dV d\Omega}\right)_{\circ}  =  \frac{\Delta \varepsilon}{4 
\pi} 
\label{dEo} 
\end{equation} 
due to the fact that the emission is isotropic in this frame. The total number of photons emitted is an invariant quantity independent of the frame used. Thus we can compute this quantity as seen by an observer in the comoving frame (which we denote with the subscript ``$\circ$'') and by an observer in the laboratory frame (which we denote with no subscripts). Doing this we find:
\begin{equation} 
\frac{dN_\gamma}{dt d \Omega d \Sigma}= \left(\frac{dN_\gamma}{dt d \Omega 
d \Sigma} \right)_{\circ} \Lambda^{-3} 
\cos \vartheta 
\, , 
\end{equation} 
where $\cos\vartheta$ comes from the projection of the elementary surface of the shell on the direction of propagation and $\Lambda = \gamma ( 1 - \beta \cos \vartheta )$ is the Doppler factor introduced in the two following differential transformation
\begin{equation} 
d \Omega_{\circ} = d \Omega \times \Lambda^{-2} 
\end{equation} 
for the solid angle transformation and 
\begin{equation} 
d t_{\circ} = d t \times \Lambda^{-1} 
\end{equation} 
for the time transformation. The integration in $d \Sigma$ is performed over the visible area of the ABM pulse at laboratory time $t$, namely with $0\le\vartheta\le\vartheta_{max}$ and $\vartheta_{max}$ is the boundary of the visible region defined in Eq.(\ref{bound}) (see also Figs. \ref{openang}--\ref{opening}). An extra $\Lambda$ factor comes from the energy transformation:
\begin{equation} 
E_{\circ} = E \times \Lambda\, . 
\end{equation} 
See also \citet{cd99}. Thus finally we obtain:
\begin{equation} 
\frac{dE}{dt d \Omega d \Sigma} = \left(\frac{dE}{dt d \Omega d \Sigma} 
\right)_{\circ} \Lambda^{-4} \cos \vartheta \, . 
\end{equation} 
Doing this we clearly identify  $  \left(\frac{dE}{dt d \Omega d \Sigma} \right)_{\circ} $ 
as the energy density in the comoving frame up to a factor $\frac{v}{4\pi}$ (see Eq.(\ref{dEo})). Then we have: 
\begin{equation} 
\frac{dE}{dt d \Omega } = \int_{shell} \frac{\Delta \varepsilon}{4 \pi} \; 
v \; \cos \vartheta \; \Lambda^{-4} \; d \Sigma\, , 
\label{fluxlab} 
\end{equation} 
where the integration in $d \Sigma$ is performed over the ABM pulse visible area at laboratory time $t$, namely with $0\le\vartheta\le\vartheta_{max}$ and $\vartheta_{max}$ is the boundary of the visible region defined in Eq.(\ref{bound}). Eq.(\ref{fluxlab}) gives us the energy emitted toward the observer per unit solid angle and per unit laboratory time $t$ in the laboratory frame.

What we really need is the energy emitted per unit solid angle and per unit detector arrival time $t_a^d$, so we must use the complete relation between $t_a^d$ and $t$ given in Eq.(\ref{ta_g}). First we have to multiply the integrand in Eq.(\ref{fluxlab}) by the factor $\left(dt/dt_a^d\right)$ to transform the energy density generated per unit of laboratory time $t$ into the energy density generated per unit arrival time $t_a^d$. Then we have to integrate with respect to $d \Sigma$ over the EQTS corresponding to arrival time $t_a^d$ instead of the ABM pulse visible area at laboratory time $t$. The analog of Eq.(\ref{fluxlab}) for the source luminosity in detector arrival time is then:
\begin{equation} 
\frac{dE_\gamma}{dt_a^d d \Omega } = \int_{EQTS} \frac{\Delta 
\varepsilon}{4 \pi} \; v \; \cos \vartheta \; \Lambda^{-4} \; 
\frac{dt}{dt_a^d} d \Sigma\, . 
\label{fluxarr} 
\end{equation} 
It is important to note that, in the present case of GRB 991216, the Doppler factor $\Lambda^{-4}$ in Eq.(\ref{fluxarr}) enhances the apparent luminosity of the burst, as compared to the intrinsic luminosity, by a factor which at the peak of the afterglow is in the range between $10^{10}$ and $10^{12}$!

We are now able to reproduce in Fig. \ref{bolum} the general behavior of the luminosity starting from the P-GRB to the latest phases of the afterglow as a function of the arrival time. It is generally agreed that the GRB afterglow originates from an ultrarelativistic shell of baryons with an initial Lorentz factor $\gamma_\circ\sim 200$--$300$ with respect to the interstellar medium (see e.g. \citet{rubr,EQTS_ApJL} and references therein). Using GRB 991216 as a prototype, in \citet{lett1,lett2} we have shown how from the time varying bolometric intensity of the afterglow it is possible to infer the average density $\left<n_{ism}\right>=1$ particle/cm$^3$ of the InterStellar Medium (ISM) in a region of approximately $10^{17}$ cm surrounding the black hole giving rise to the GRB phenomenon.

It was shown in \citet{r02} that the theoretical interpretation of the intensity variations in the prompt phase in the afterglow implies the presence in the ISM of inhomogeneities of typical scale $10^{15}$ cm. Such inhomogeneities were there represented for simplicity as spherically symmetric over-dense regions with $\left<n_{ism}^{od}\right> \simeq 10^2\left<n_{ism}\right>$ separated by under-dense regions with $\left<n_{ism}^{ud}\right> \simeq 10^{-2}\left<n_{ism}\right>$ also of typical scale $\sim 10^{15}$ cm in order to keep $\left<n_{ism}\right>$ constant.

The summary of these general results are shown in Fig. \ref{grb991216}, where the P-GRB, the emission at the peak of the afterglow in relation to the ``prompt emission'' and the latest part of the afterglow are clearly identified for the source GRB 991216. Details in \citet{rubr}.

\subsection{On the structures in the afterglow peak emission of gamma ray bursts}

\begin{figure}
\includegraphics[width=10cm,clip]{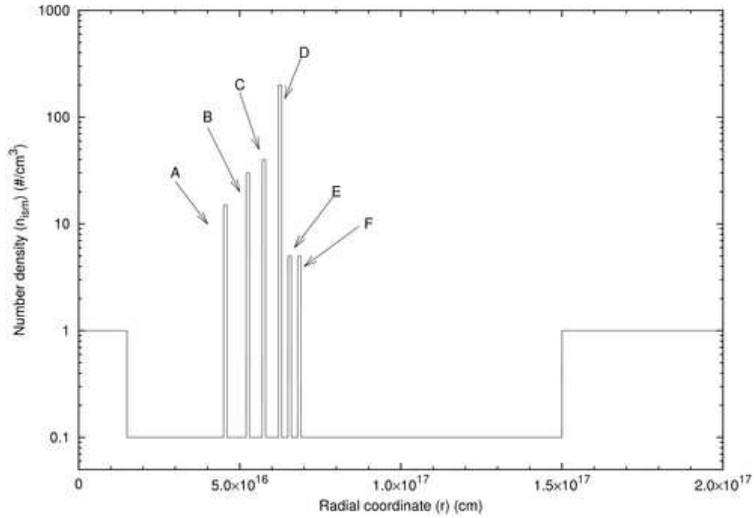}
\caption{The density profile (``mask") of an ISM cloud used to reproduce the GRB~991216 temporal structure. As before, the radial coordinate is measured from the black hole. In this cloud we have six ``spikes" with overdensity separated by low density regions. Each spike has the same spatial extension of $10^{15}\, {\rm cm}$. The cloud average density is $<n_{ism}>=1\, {\rm particle}/{\rm cm}^3$.}
\label{maschera}
\end{figure}

We are now ready to reconsider the problem of the ISM inhomogeneity generating the temporal substructures in the E-APE by integrating on the EQTS surfaces and improving on the considerations based on the purely radial approximation. We have created (see details in \citet{r02}) an ISM inhomogeneity ``mask" (see Fig.~\ref{maschera} and Tab.~\ref{tab3}) with the main criteria that the density inhomogeneities and their spatial distribution still fulfill $<n_{ism}>=1\, {\rm particle}/{\rm cm}^3$.

\begin{table}
\centering
\caption{For each ISM density peak represented in Fig.~\ref{maschera} we give the initial radius $r$, the corresponding comoving time $\tau$, laboratory time $t$, arrival time at the detector $t_a^d$, diameter of the ABM pulse visible area $d_{v}$, Lorentz factor $\gamma$ and observed duration $\Delta t_a^d$ of the afterglow luminosity peaks generated by each density peak. In the last column, the apparent motion in the radial coordinate, evaluated in the arrival time at the detector, leads to an enormous ``superluminal" behavior, up to $9.5\times 10^4\,c$. \label{tab3}}
\begin{tabular}{c|e{8}|e{8}|e{8}|e{1}|e{8}|e{3}|e{2}|e{9}}
Peak & r (cm) & \tau (s) & t (s) & t_a^d (s) & d_v (cm) & \Delta t_a^d (s) & \gamma &  \begin{array}{c} {\rm ``Superluminal"} \\ v\equiv\frac{r}{t_a^d} \\ \\ \end{array}\\
\hline
A & 4.50\times10^{16} & 4.88\times10^3 & 1.50\times10^6 & 15.8 & 2.95\times10^{14} & 0.400 & 303.8 & 9.5\times10^4c\\
B & 5.20\times10^{16} & 5.74\times10^3 & 1.73\times10^6 & 19.0 & 3.89\times10^{14} & 0.622 & 265.4 & 9.1\times10^4c\\
C & 5.70\times10^{16} & 6.54\times10^3 & 1.90\times10^6 & 22.9 & 5.83\times10^{14} & 1.13  & 200.5 & 8.3\times10^4c\\
D & 6.20\times10^{16} & 7.64\times10^3 & 2.07\times10^6 & 30.1 & 9.03\times10^{14} & 5.16  & 139.9 & 6.9\times10^4c\\
E & 6.50\times10^{16} & 9.22\times10^3 & 2.17\times10^6 & 55.9 & 2.27\times10^{15} & 10.2  & 57.23 & 3.9\times10^4c\\
F & 6.80\times10^{16} & 1.10\times10^4 & 2.27\times10^6 & 87.4 & 2.42\times10^{15} & 10.6  & 56.24 & 2.6\times10^4c\\
\end{tabular}
\end{table}

\begin{figure}
\includegraphics[width=8.5cm,clip]{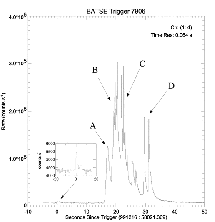}
\includegraphics[width=8.5cm,clip]{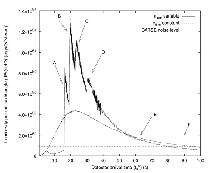}
\caption{{\bf Left)} The BATSE data on the E-APE of GRB~991216 (source: \citet{grblc99}) together with an enlargement of the P-GRB data (source: \citet{brbr99}). For convenience each E-APE peak has been labeled by a different uppercase Latin letter. {\bf Right)} The source luminosity connected to the mask in Fig.~\ref{maschera} is given as a function of the detector arrival time (solid ``spiky" line) with the corresponding curve for the case of constant $n_{ism}=1\, {\rm particle}/{\rm cm}^3$ (dashed smooth line) and the BATSE noise level (dotted horizontal line). The ``noise" observed in the theoretical curves is due to the discretization process adopted, described in \citet{r02}, for the description of the angular spreading of the scattered radiation. For each fixed value of the laboratory time we have summed $500$ different contributions from different angles. The integration of the equation of motion of this system is performed in $22,314,500$ contributions to be considered. An increase in the number of steps and in the precision of the numerical computation would lead to a smoother curve.}
\label{substr_peak}
\end{figure}

The results are given in Fig.~\ref{substr_peak}. We obtain, in perfect agreement with the observations:
\begin{enumerate}
\item the theoretically computed intensity of the A, B, C peaks as a function of the ISM inhomogneities;
\item the fast rise and exponential decay shape for each peak;
\item a continuous and smooth emission between the peaks.
\end{enumerate}

Interestingly, the signals from shells E and F, which have a density inhomogeneity comparable to A, are undetectable. The reason is due to a variety of relativistic effects and partly to the spreading in the arrival time, which for A, corresponding to $\gamma=303.8$ is $0.4 s$ while for E (F) corresponding to $\gamma=57.23$ $(56.24)$ is of $10.2\, {\rm s}$ $(10.6\, {\rm s})$ (see Tab.~\ref{tab3} and \citet{r02}).

In the case of D, the agreement with the arrival time is reached, but we do not obtain the double peaked structure. The ABM pulse visible area diameter at the moment of interaction with the D shell is $\sim 1.0\times 10^{15}\, {\rm cm}$, equal to the extension of the ISM shell (see Tab.~\ref{tab3} and \citet{r02}). Under these conditions, the concentric shell approximation does not hold anymore: the disagreement with the observations simply makes manifest the need for a more detailed description of the three dimensional nature of the ISM cloud.

The physical reasons for these results can be simply summarized: we can distinguish two different regimes corresponding in the afterglow of GRB~991216 respectively to $\gamma > 150$ and to $\gamma < 150$. For different sources this value may be slightly different. In the E-APE region ($\gamma > 150$) the GRB substructure intensities indeed correlate with the ISM inhomogeneities. In this limited region (see peaks A, B, C) the Lorentz gamma factor of the ABM pulse ranges from $\gamma\sim 304$ to $\gamma\sim 200$. The boundary of the visible region is smaller than the thickness $\Delta R$ of the inhomogeneities (see Fig.~\ref{opening} and Tab.~\ref{tab3}). Under this condition the adopted spherical approximation is not only mathematically simpler but also fully justified. The angular spreading is not strong enough to wipe out the signal from the inhomogeneity spike.

As we descend in the afterglow ($\gamma < 150$), the Lorentz gamma factor decreases markedly and in the border line case of peak D $\gamma\sim 140$. For the peaks E and F we have $\gamma\sim 50$ and, under these circumstances, the boundary of the visible region becomes much larger than the thickness $\Delta R$ of the inhomogeneities (see Fig.~\ref{opening} and Tab.~\ref{tab3}). A three dimensional description would be necessary, breaking the spherical symmetry and making the computation more difficult. However we do not need to perform this more complex analysis for peaks E and F: any three dimensional description would {\em a fortiori} augment the smoothing of the observed flux. The spherically symmetric description of the inhomogeneities is already enough to prove the overwhelming effect of the angular spreading (\citet{r02}).

From our analysis we show that the \citet{dm99} conclusions are correct for $\gamma\sim 300$ and do indeed hold for $\gamma > 150$. However, as the gamma factor drops from $\gamma\sim 150$ to $\gamma\sim 1$ (see Fig~\ref{gamma}), the intensity due to the inhomogeneities markedly decreases also due to the angular spreading (events E and F). The initial Lorentz factor of the ABM pulse $\gamma\sim 310$ decreases very rapidly to $\gamma\sim 150$ as soon as a fraction of a typical ISM cloud is engulfed (see Tab.~\ref{tab3}). We conclude that the ``tomography'' is indeed effective, but uniquely in the first ISM region close to the source and for GRBs with $\gamma > 150$.

One of the most striking feature in our analysis is clearly represented by the fact that the inhomogeneities of a mask of radial dimension of the order of $10^{17}\, {\rm cm}$ give rise to arrival time signals of the order of $20\, {\rm s}$. This outstanding result implies an apparent ``superluminal velocity'' of $\sim 10^5c$ (see Tab.~\ref{tab3}). The ``superluminal velocity'' here considered, first introduced in \citet{lett1}, refers to the motion along the line of sight. This effect is proportional to $\gamma^2$. It is much larger than the one usually considered in the literature, within the context of radio sources and microquasars (see e.g. \citet{mr95}), referring to the component of the velocity at right angles to the line of sight (see details in \citet{r02}). This second effect is in fact proportional to $\gamma$ (see \citet{r66}). We recall that this ``superluminal velocty'' was the starting point for the enunciation of the RSTT paradigm (\citet{lett1}), emphasizing the need of the knowledge of the {\em entire} past worldlines of the source. This need has been further clarified here in the determination of the EQTS surfaces (see Fig.~\ref{opening} which indeed depend on an integral of the Lorentz gamma factor extended over the {\em entire} past worldlines of the source. In turn, therefore, the agreement between the observed structures and the theoretical predicted ones (see Figs.~\ref{grb991216}--\ref{substr_peak}) is also an extremely stringent additional test on the values of the Lorentz gamma factor determined as a function of the radial coordinate within the EMBH theory (see Fig.~\ref{gamma}).

\section{The theory of the luminosity in fixed energy bands and spectra of the afterglow}

In our approach we focus uniquely on the X- and $\gamma$-ray radiation, which appears to be conceptually much simpler than the optical and radio emission. It is perfectly predictable by a set of constitutive equations (see next section), which leads to directly verifiable and very stable features in the spectral distribution of the observed GRB afterglows. In line with the observations of GRB 991216 and other GRB sources, we assume in the following that the X- and $\gamma$-ray luminosity represents approximately $90$\% of the energy flux of the afterglow, while the optical and radio emission represents only the remaining $10$\%.

This approach differs significantly from the other ones in the current literature, where attempts are made to explain at once all the multi-wavelength emission in the radio, optical, X and gamma ray as coming from a common origin which is linked to boosted synchrotron emission. Such an approach has been shown to have a variety of difficulties (\citet{gcg02,pa98}) and cannot anyway have the instantaneous variability needed to explain the structure in the ``prompt radiation'' in an external shock scenario, which is indeed confirmed by our model.

\subsection{The equations determining the luminosity in fixed energy bands} 

Here the fundamental new assumption is adopted (see also \citet{Spectr1}) that the X- and gamma ray radiation during the entire afterglow phase has a thermal spectrum in the co-moving frame. The temperature is then given by:
\begin{equation} 
T_s=\left[\Delta E_{\rm int}/\left(4\pi r^2 \Delta \tau \sigma 
\mathcal{R}\right)\right]^{1/4}\, , 
\label{TdiR} 
\end{equation} 
where $\Delta E_{\rm int}$ is the internal energy developed in the collision with the ISM in a time interval $\Delta \tau$ in the co-moving frame, $\sigma$ is the Stefan-Boltzmann constant and
\begin{equation} 
\mathcal{R}=A_{eff}/A_{vis}\, , 
\label{Rdef} 
\end{equation} 
is the ratio between the ``effective emitting area'' of the ABM pulse of radius $r$ and its total visible area. Due to the ISM inhomogeneities, composed of clouds with filamentary structure, the ABM emitting region is in fact far from being homogeneous. In GRB 991216 such a factor is observed to be decreasing during the afterglow between: $3.01\times 10^{-8} \ge \mathcal{R} \ge 5.01 \times 10^{-12}$ (\citet{Spectr1}). 

\begin{figure}
\centering
\includegraphics[width=\hsize,clip]{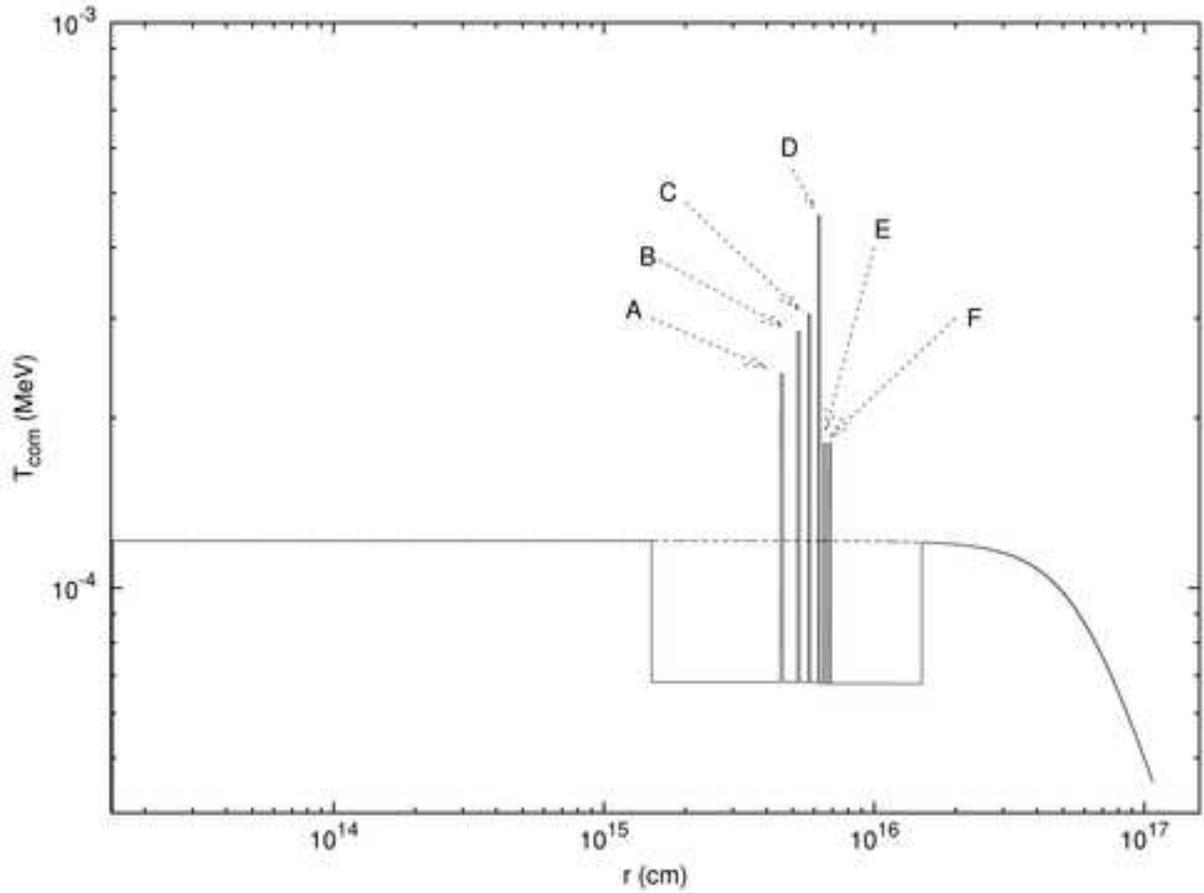}
\caption{The temperature in the comoving frame of the shock front corresponding to the density distribution with the six spikes A,B,C,D,E,F presented in Ruffini et al.$^5$. The dashed line corresponds to an homogeneous distribution with $n_{ism}=1$. Details in \citet{fil}.}
\label{tcom_fig}
\end{figure}

The temperature in the comoving frame corresponding to the density distribution described in \citet{r02} is shown in Fig. \ref{tcom_fig}. 

We are now ready to evaluate the source luminosity in a given energy band. The source luminosity at a detector arrival time $t_a^d$, per unit solid angle $d\Omega$ and in the energy band $\left[\nu_1,\nu_2\right]$ is given by (see \citet{rubr,Spectr1}): 
\begin{equation} 
\frac{dE_\gamma^{\left[\nu_1,\nu_2\right]}}{dt_a^d d \Omega } = 
\int_{EQTS} \frac{\Delta \varepsilon}{4 \pi} \; v \; \cos \vartheta \; 
\Lambda^{-4} \; \frac{dt}{dt_a^d} W\left(\nu_1,\nu_2,T_{arr}\right) d 
\Sigma\, , 
\label{fluxarrnu} 
\end{equation} 
where $\Delta \varepsilon=\Delta E_{int}/V$ is the energy density released in the interaction of the ABM pulse with the ISM inhomogeneities measured in the comoving frame, $\Lambda=\gamma(1-(v/c)\cos\vartheta)$ is the Doppler factor, $W\left(\nu_1,\nu_2,T_{arr}\right)$ is an ``effective weight'' required to evaluate only the contributions in the energy band $\left[\nu_1,\nu_2\right]$, $d\Sigma$ is the surface element of the EQTS at detector arrival time $t_a^d$ on which the integration is performed (see also \citet{r02}) and $T_{arr}$ is the observed temperature of the radiation emitted from $d\Sigma$: 
\begin{equation} 
T_{arr}=T_s/\left[\gamma 
\left(1-(v/c)\cos\vartheta\right)\left(1+z\right)\right]\, . 
\label{Tarr} 
\end{equation} 

The ``effective weight'' $W\left(\nu_1,\nu_2,T_{arr}\right)$ is given by the ratio of the integral over the given energy band of a Planckian distribution at a temperature $T_{arr}$ to the total integral $aT_{arr}^4$: 
\begin{equation} 
W\left(\nu_1,\nu_2,T_{arr}\right)=\frac{1}{aT_{arr}^4}\int_{\nu_1}^{\nu_2}\rho\left(T_{arr},\nu\right)d\left(\frac{h\nu}{c}\right)^3\, , 
\label{effweig} 
\end{equation} 
where $\rho\left(T_{arr},\nu\right)$ is the Planckian distribution at temperature $T_{arr}$: 
\begin{equation} 
\rho\left(T_{arr},\nu\right)=\left(2/h^3\right)h\nu/\left(e^{h\nu/\left(kT_{arr}\right)}-1\right) 
\label{rhodef} 
\end{equation} 

\subsection{On the time integrated spectra and the hard-to-soft spectral transition} 

\begin{figure}
\centering
\includegraphics[width=0.7\hsize,clip]{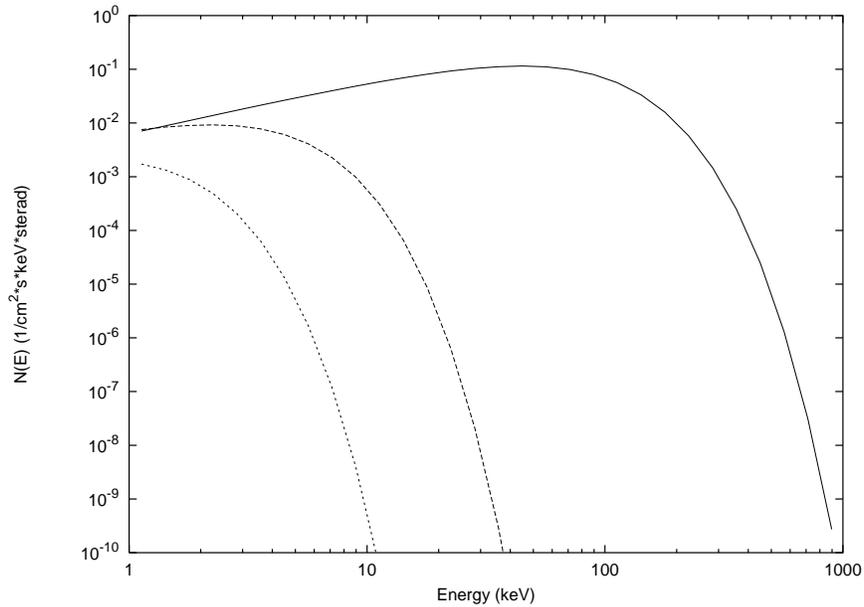} 
\caption{The instantaneous spectra of the radiation observed in GRB~991216 at three different EQTS respectively, from top to bottom, for $t_a^d=10$ s, $t_a^d=10^4$ s and $t_a^d=1.45\times10^5$ s. These diagrams have been computed assuming a constant $\left<n_{ism}\right>\simeq 1$ particle/cm$^3$ and clearly explains the often quoted hard-to-soft spectral evolution in GRBs. Details in \citet{Spectr1}.}
\label{spectrum} 
\end{figure}

We turn now to the much debated issue of the origin of the observed hard-to-soft spectral transition during the GRB observations (see e.g. \citet{fa00,gcg02,p99,p99b}). We consider the instantaneous spectral distribution of the observed radiation for three different EQTSs:
\begin{itemize} 
\item $t_a^d=10$ s, in the early radiation phase near the peak of the luminosity, 
\item $t_a^d=1.45\times 10^5$ s, in the last observation of the afterglow by the Chandra satellite, and 
\item $t_a^d=10^4$ s, chosen in between the other two (see Fig. \ref{spectrum}). 
\end{itemize} 
The observed hard-to-soft spectral transition is then explained and traced back to: 
\begin{enumerate} 
\item a time decreasing temperature of the thermal spectrum measured in the comoving frame, 
\item the GRB equations of motion, 
\item the corresponding infinite set of relativistic transformations. 
\end{enumerate} 
A clear signature of our model is the existence of a common low-energy behavior of the instantaneous spectrum represented by a power-law with index $\alpha = +0.9$. This prediction will be possibly verified in future observations.

\begin{figure}
\centering
\includegraphics[width=0.7\hsize,clip]{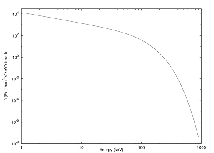} 
\caption{The time-integrated spectrum of the radiation observed in GRB~991216. The low energy part of the curve below $10$ keV is fit by a power-law with index $\alpha = -1.05$ and the high energy part above $500$ keV is fit by a power-law with an index $\beta < -16$. Details in \citet{Spectr1}.}
\label{spectband} 
\end{figure}

Starting from these instantaneous values, we integrate the spectra in arrival time obtaining what is usually fit in the literature by the ``Band relation'' (\citet{b93}). Indeed we find for our integrated spectra a low energy spectral index $\alpha=-1.05$ and an high energy spectral index $\beta < -16$ when interpreted within the framework of a Band relation (see Fig. \ref{spectband}). This theoretical result can be submitted to a direct confrontation with the observations of GRB 991216 and, most importantly, the entire theoretical framework which we have developed can now be applied to any GRB source. The theoretical predictions on the luminosity in fixed energy bands so obtained can be then straightforwardly confronted with the observational data.

\subsection{Evidence for isotropic emission in GRB991216}

\begin{figure}
\begin{minipage}{\hsize}
\includegraphics[width=0.8\hsize,clip]{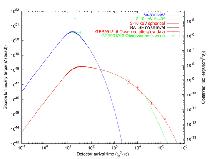}\\
\includegraphics[width=0.8\hsize,clip]{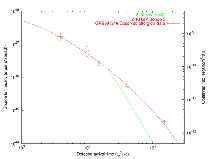}
\end{minipage}
\caption{\textbf{Above:} Best fit of the afterglow data of GRB991216. The blue line is the luminosity in the $50$--$300$ keV energy band. The red line is the luminosity in the $2$--$10$ keV band computed assuming spherical symmetry. The observational data from R-XTE and Chandra (see \citet{ha00}) are perfectly consistent with such an assumption. The presence of a $\vartheta_\circ=3^\circ$ half-opening beaming angle (green line) is ruled out. \textbf{Below:} Enlargement of the plot in the region of the afterglow observational data from R-XTE and Chandra.}
\label{991216}
\end{figure}

We give in Fig. \ref{991216} the results of the fit of the GRB 991216 light curves in the two energy bands $50$--$300$ keV (observed by BATSE) and $2$--$10$ keV (observed by R-XTE and Chandra). We already pointed out in the previous section the agreement with the data of the ``prompt'' radiation obtained by BATSE in the energy range $50$--$300$ keV (see dotted line in Fig. \ref{991216}). We here show the fit of the data obtained by the R-XTE and Chandra satellites (\citet{ha00}) in the energy range $2$--$10$ keV (see dashed line in Fig. \ref{991216}). These data refer to the decaying part of the afterglow and cover a time span of $\sim 10^6$ s.

We have also computed, within our global self-consistent approach which fits both the ``prompt'' radiation and the decaying part of the afterglow, the flux in the $2$--$10$ keV range which would be expected for a beamed emission with half opening angle $\vartheta_\circ = 3^\circ$, which is the value claimed in the current literature for GRB 991216 (see \citet{ha00} and Fig. \ref{991216}). The presence of beaming manifest itself, as expected, in the decaying part of the afterglow and is incompatible with the data. In fact, the actual afterglow luminosity in fixed energy bands, in spherical symmetry, does not have a simple power law dependence on arrival time (see Fig. \ref{991216}). This circumstance has been erroneously interpreted, in the usual presentation in the literature, as a broken power-law supporting the existence of jet-like structures in GRBs. Moreover, the slope of the beamed emission and the arrival time at which the break occurs have been there computed using the approximate equations (see previous sections and \citet{EQTS_ApJL,EQTS_ApJL2,PowerLaws,beaming}). If one assumes the presence of jets in a consistent afterglow theory, one finds that the break corresponding to the purported beaming appears at an arrival time incompatible with the observations (see Fig. \ref{991216} and \citet{cospar04}).

\section{Analysis of GRB 031203}

GRB 031203 was observed by IBIS, on board of the INTEGRAL satellite (\citet{mg}), as well as by XMM (\citet{Wa04}) and Chandra (\citet{sod}) in the $2-10$ keV band, and by VLT (\citet{sod}) in the radio band. It appears as a typical long burst (\citet{saz}), with a simple profile and a duration of $\approx 40$ s. The burst fluence in the $20-200$ keV band is $(2.0\pm 0.4)\times 10^{-6}$ erg/cm$^2$ (\citet{saz}), and the measured redshift is $z=0.106$ (\citet{proch}). We analyze in the following the gamma-ray signal received by INTEGRAL. The observations in other wavelengths, in analogy with the case of GRB 980425 (\citet{pian00,cospar02,Mosca_Orale}), could be related to the supernova event, as also suggested by \citet{sod}, and they will be examined elsewhere.

The INTEGRAL observations find a direct explanation in our theoretical model. We determine the values of the two free parameters which characterize our model: the total energy stored in the Dyadosphere $E_{dya}$ and the mass of the baryons left by the collapse $M_Bc^2 \equiv B E_{dya}$. We follow the expansion of the pulse, composed by the electron-positron plasma initially created by the vacuum polarization process in the Dyadosphere. The plasma self-propels itself outward and engulfs the baryonic remnant left over by the collapse of the progenitor star. As such pulse reaches transparency, the P-GRB is emitted (\citet{rswx99,rswx00,lett2}). The remaining accelerated baryons, interacting with the interstellar medium (ISM), produce the afterglow emission. The ISM is described by the two additional parameters of the theory: the average particle number density $<n_{ISM}>$ and the ratio $<\mathcal{R}>$ between the effective emitting area and the total area of the pulse (\citet{Spectr1}), which take into account the ISM filamentary structure (\citet{fil}).

We reproduce correctly in several GRBs and in this specific case (see e.g. Fig. \ref{fig1a}) the observed time variability of the prompt emission (see e.g.\citet{r02,rubr,rubr2} and references therein). The radiation produced by the interaction of the accelerated baryons with the ISM agrees with observations both for intensity and time structure.

The progress in reproducing the X and $\gamma-$ray emission as originating from a thermal spectrum in the co-moving frame of the burst (\citet{Spectr1}) leads to the characterization of the instantaneous spectral properties which are shown to drift from hard to soft during the evolution of the system. The convolution of these instantaneous spectra over the observational time scale is in very good agreement with the observed power-law spectral shape.

As shown in previous cases (see \citet{rubr,cospar04}), also for GRB 031203, using the correct equations of motion, there is no need to introduce a collimated emission to fit the afterglow observations (see also \citet{sod} who find this same conclusion starting from different considerations).

\subsection{The initial conditions}

The best fit of the observational data leads to a total energy of the Dyadosphere $E_{e^\pm}^{tot}=E_{dya}=1.85\times10^{50}$ erg. Assuming a black hole mass $M=10M_{\odot}$, we then have a black hole charge to mass ratio $\xi=6.8\times 10^{-3}$; the plasma is created between the radii $r_1=2.95\times10^6$ cm and $r_2=2.81\times10^7$ cm with an initial temperature of $1.52$ MeV and a total number of pairs $N_{e^+e^-}=2.98\times10^{55}$. The amount of baryonic matter in the remnant is $B = 7.4\times10^{-3}$.

After the transparency point and the P-GRB emission, the initial Lorentz gamma factor of the accelerated baryons is $\gamma=132.8$ at an arrival time at the detector $t^d_a=8.14\times 10^{-3}$ s and a distance from the Black Hole $r=6.02\times 10^{12}$ cm. This corresponds to an apparent superluminal velocity along the line of sight of $2.5 \times 10^4c$. The ISM parameters are: $<n_{ism}>=0.3$ particle/$cm^3$ and $<\mathcal{R}>=7.81\times 10^{-9}$.

\subsection{The GRB luminosity in fixed energy bands}\label{par3}

The aim of our model is to derive from first principles both the luminosity in selected energy bands and the time resolved/integrated spectra. The luminosity in selected energy bands is evaluated integrating over the EQTSs (see \citet{EQTS_ApJL,EQTS_ApJL2}) the energy density released in the interaction of the accelerated baryons with the ISM measured in the co-moving frame, duly boosted in the observer frame. The radiation viewed in the co-moving frame of the accelerated baryonic matter is assumed to have a thermal spectrum and to be produced by the interaction of the ISM with the front of the expanding baryonic shell.

In order to evaluate the contributions in the band $[\nu_1,\nu_2]$ we have to multiply the bolometric luminosity with an ``effective weight'' $W(\nu_1,\nu_2,T_{arr})$, where $T_{arr}$ is the observed temperature. $W(\nu_1,\nu_2,T_{arr})$ is given by the ratio of the integral over the given energy band of a Planckian distribution at temperature $T_{arr}$ to the total integral $aT_{arr}^4$ (\citet{Spectr1}). The resulting expression for the emitted luminosity is Eq.(\ref{fluxarrnu}).

\subsection{The GRB 031203 ``prompt emission''}

\begin{figure}
\includegraphics[width=\hsize,clip]{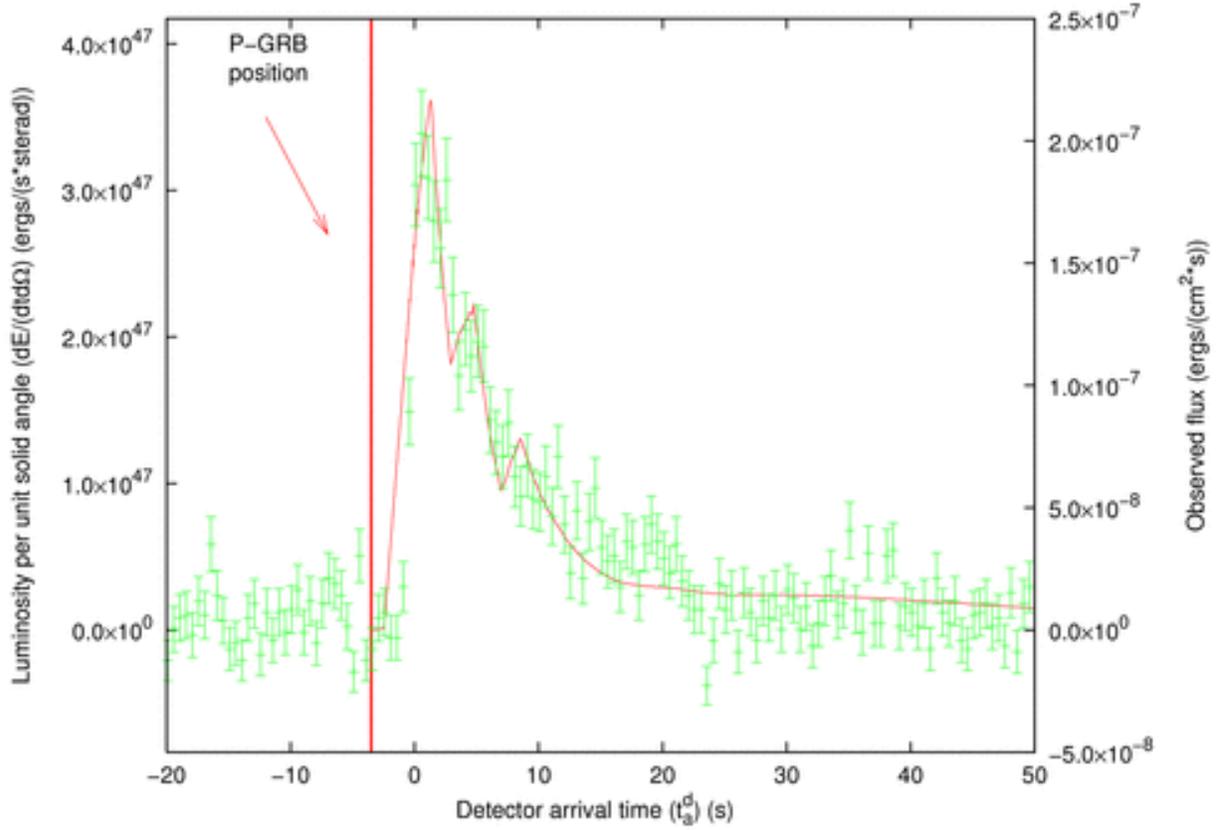}
\caption{Theoretically simulated light curve of the GRB 031203 prompt emission in the $20-200$ keV energy band (solid red line) is compared with the observed data (green points) from \citet{saz}. The vertical bold red line indicates the time position of P-GRB.}
\label{fig1a}
\end{figure}

In order to compare our theoretical prediction with the observations, it is important to notice that there is a shift between the initial time of the GRB event and the moment in which the satellite instrument has been triggered. In fact, in our model the GRB emission starts at the transparency point when the P-GRB is emitted. If the P-GRB is under the threshold of the instrument, the trigger starts a few seconds later with respect to the real beginning of the event. Therefore it is crucial, in the theoretical analysis, to estimate and take into due account this time delay. In the present case it results in $\Delta t^d_a=3.5$ s (see the bold solid line in Fig. \ref{fig1a}). In what follows, the detector arrival time is referred to the onset of the instrument.

The structure of the prompt emission of GRB 031203, which is a single peak with a slow decay, is reproduced assuming an ISM which has not a constant density but presents several density spikes with $<n_{ISM}>=0.16$ particle/cm$^3$. Such density spikes corresponding to the main peak are modeled as three spherical shells with width $\Delta$ and density contrast $\Delta n/n$: we adopted for the first peak $\Delta=3.0\times10^{15}$ cm and $\Delta n/n=8$, for the second peak $\Delta=1.0\times10^{15}$ cm and $\Delta n/n=1.5$ and for the third one $\Delta=7.0\times10^{14}$ cm and $\Delta n/n=1$. To describe the details of the ISM filamentary structure we would require an intensity vs. time information with an arbitrarily high resolving power. With the finite resolution of the INTEGRAL instrument, we can only describe the average density distribution compatible with the given accuracy. Only structures at scales of $10^{15}$ cm can be identified. Smaller structures would need a stronger signal and/or a smaller time resolution of the detector. The three clouds here considered are necessary and sufficient to reproduce the observed light curve: a smaller number would not fit the data, while a larger number is unnecessary and would be indeterminable.

The result (see Fig. \ref{fig1a}) shows a good agreement with the light curve reported by \citet{saz}, and it provides a further evidence for the possibility of reproducing light curves with a complex time variability through ISM inhomogeneities (\citet{r02,rubr,rubr2}, see also the analysis of the prompt emission of GRB 991216 in \citet{r02}).

\subsection{The GRB 031203 instantaneous spectrum}\label{inst}

\begin{figure}
\includegraphics[width=\hsize,clip]{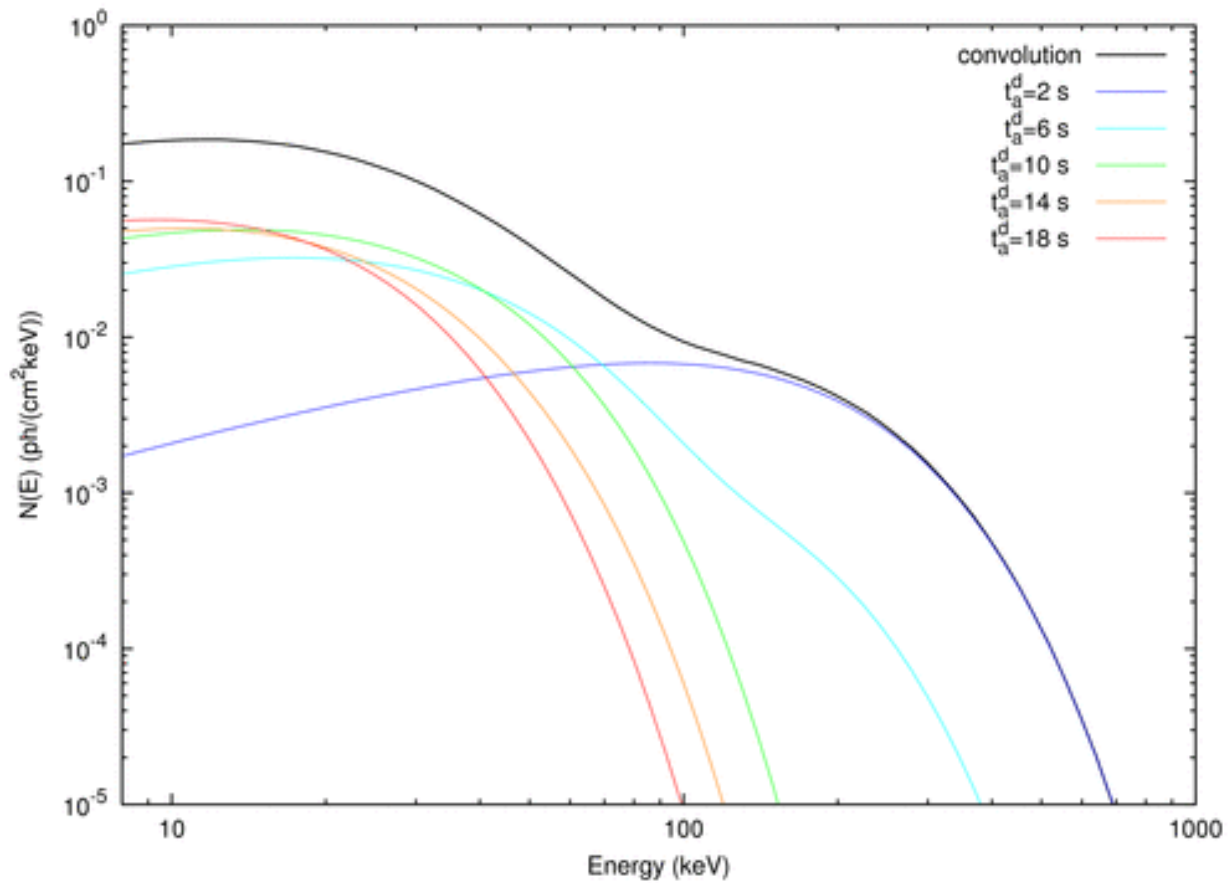}
\caption{Five different theoretically predicted instantaneous photon number spectrum $N(E)$ for $t_a^d=2$, $6$, $10$, $14$, $18$ s are here represented (colored curves) together with their own temporal convolution (black bold curve). The shapes of the instantaneous spectra are not blackbodies due to the spatial convolution over the EQTS (see text).}
\label{fig2a}
\end{figure}

As outlined in section \ref{par3}, in addition to the the luminosity in fixed energy bands we can derive also the instantaneous photon number spectrum $N(E)$. In Fig. \ref{fig2a} are shown samples of time-resolved spectra for five different values of the arrival time which cover the whole duration of the event.

It is manifest from this picture that, although the spectrum in the co-moving frame of the expanding pulse is thermal, the shape of the final spectrum in the laboratory frame is clearly non thermal. In fact, as explained in \citet{Spectr1}, each single instantaneous spectrum is the result of an integration of hundreds of thermal spectra over the corresponding EQTS. This calculation produces a non thermal instantaneous spectrum in the observer frame (see Fig. \ref{fig2a}).

\begin{figure}
\includegraphics[width=\hsize,clip]{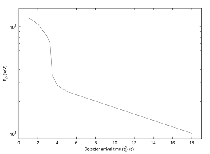}
\caption{The energy of the peak of the instantaneous photon number spectrum $N(E)$ is here represented as a function of the arrival time during the ``prompt emission'' phase. The clear hard to soft behavior is shown.}
\label{fig3a}
\end{figure}

Another distinguishing feature of the GRBs spectra which is also present in these instantaneous spectra, as shown in Fig. \ref{fig2a}, is the hard to soft transition during the evolution of the event (\citet{cri97,p99,fa00,gcg02}). In fact the peak of the energy distributions $E_p$ drift monotonically to softer frequencies with time (see Fig. \ref{fig3a}). This feature explains the change in the power-law low energy spectral index $\alpha$ (\citet{b93}) which at the beginning of the prompt emission of the burst ($t_a^d=2$ s) is $\alpha=0.75$, and progressively decreases for later times (see Fig. \ref{fig2a}). In this way the link between $E_p$ and $\alpha$ identified by \citet{cri97} is explicitly shown. This theoretically predicted evolution of the spectral index during the event unfortunately cannot be detected in this particular burst by INTEGRAL because of the not sufficient quality of the data (poor photon statistics, see \citet{saz}).

\subsection{The GRB 031203 time-integrated spectrum and the comparison with the observed data}

The time-integrated observed GRB spectra show a clear power-law behavior. Within a different framework Shakura, Sunyaev and Zel'dovich (see e.g. \citet{poz83} and references therein) argued that it is possible to obtain such power-law spectra from a convolution of many non power-law instantaneous spectra evolving in time. This result was recalled and applied to GRBs by \citet{bk99} assuming for the instantaneous spectra a thermal shape with a temperature changing with time. They showed that the integration of such energy distributions over the observation time gives a typical power-law shape possibly consistent with GRB spectra.

Our specific quantitative model is more complicated than the one considered by \citet{bk99}: as pointed out in section \ref{inst}, the instantaneous spectrum here is not a black body. Each instantaneous spectrum is obtained by an integration over the corresponding EQTS (\citet{EQTS_ApJL,EQTS_ApJL2}): it is itself a convolution, weighted by appropriate Lorentz and Doppler factors, of $\sim 10^6$ thermal spectra with variable temperature. Therefore, the time-integrated spectra are not plain convolutions of thermal spectra: they are convolutions of convolutions of thermal spectra (see Fig. \ref{fig2a}).

\begin{figure}
\includegraphics[width=\hsize,clip]{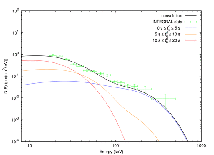}
\caption{Three theoretically predicted time-integrated photon number spectra $N(E)$ are here represented for $0 \le t_a^d \le 5$ s, $5 \le t_a^d \le 10$ s and $10 \le t_a^d \le 20$ s (colored curves). The hard to soft behavior presented in Fig. \ref{fig3a} is confirmed. Moreover, the theoretically predicted time-integrated photon number spectrum $N(E)$ corresponding to the first $20$ s of the ``prompt emission'' (black bold curve) is compared with the data observed by INTEGRAL (green points, see \citet{saz,saz2}). This curve is obtained as a convolution of 108 instantaneous spectra, which are enough to get a good agreement with the observed data.}
\label{fig4a}
\end{figure}

The simple power-law shape of the integrated spectrum is more evident if we sum tens of instantaneous spectra, as in Fig. \ref{fig4a}. In this case we divided the prompt emission in three different time interval, and for each one we integrated on time the energy distribution. The resulting three time-integrated spectra have a clear non-thermal behavior, and still present the characteristic hard to soft transition.

Finally, we integrated the photon number spectrum $N(E)$ over the whole duration of the prompt event (see again Fig. \ref{fig4a}): in this way we obtain a typical non-thermal power-law spectrum which results to be in good agreement with the INTEGRAL data (see \citet{saz,saz2}) and gives a clear evidence of the possibility that the observed GRBs spectra are originated from a thermal emission.

The precise knowledge we have here acquired on GRB 031203 will help in clarifying the overall astrophysical system GRB 031203 - SN 2003lw - the $2-10$ keV XMM and Chandra data (see next sections, where the late $2-10$ keV XMM and Chandra data are also discussed).

\section{Analysis of GRB 050315}

GRB 050315 (\citet{va05}) has been triggered and located by the BAT instrument (\citet{b04,ba05}) on board of the {\em Swift} satellite (\citet{ga04}) at 2005-March-15 20:59:42 UT (\citet{pa05}). The narrow field instrument XRT (\citet{bua04,bua05}) began observations $\sim 80$ s after the BAT trigger, one of the earliest XRT observations yet made, and continued to detect the source for $\sim 10$ days (\citet{va05}). The spectroscopic redshift has been found to be $z = 1.949$ (\citet{kb05}).

We present here the results of the fit of the \emph{Swift} data of this source in $5$ energy bands in the framework of our theoretical model, pointing out a new step toward the uniqueness of the explanation of the overall GRB structure. We first recall the essential features of our theoretical model; then we fit the GRB 050315 observations by both the BAT and XRT instruments; we also present the instantaneous spectra for selected values of the detector arrival time ranging from $60$ s (i.e. during the so called ``prompt emission'') all the way to $3.0\times 10^4$ s (i.e. the latest afterglow phases).

\subsection{Our theoretical model}\label{model}

A major difference between our theoretical model and the ones in the current literature (see e.g. \citet{p04} and references therein) is that what is usually called ``prompt emission'' in our case coincides with the peak of the afterglow emission and is not due to the prolonged activity of an ``inner engine'' which, clearly, would introduce an additional and independent physical process to explain the GRB phenomenon (\citet{lett2}). A basic feature of our model consists, in fact, in a sharp distinction between two different components in the GRB structure: {\bf 1)} the P-GRB, emitted at the moment of transparency of the self-accelerating $e^\pm$-baryons plasma (see e.g. \citet{g86,p86,sp90,psn93,mlr93,gw98,rswx99,rswx00,lett1,lett2,Monaco_RateEq}); {\bf 2)} an afterglow described by external shocks and composed of three different regimes (see \citet{rswx99,rswx00,lett2,rubr} and references therein). The first afterglow regime corresponds to a bolometric luminosity monotonically increasing with the photon detector arrival time, corresponding to a substantially constant Lorentz gamma factor of the accelerated baryons. The second regime consists of the bolometric luminosity peak, corresponding to the ``knee'' in the decreasing phase of the baryonic Lorentz gamma factor. The third regime corresponds to a bolometric luminosity decreasing with arrival time, corresponding to the late deceleration of the Lorentz gamma factor. In some sources the P-GRB is under the observability threshold. In \citet{lett2} we have chosen as a prototype the source GRB 991216 which clearly shows the existence of the P-GRB and the three regimes of the afterglow. Unfortunately, data from BATSE existed only up to $ 36 $ s, and data from R-XTE and Chandra only after $ 3500 $ s, leaving our theoretical predictions in the whole range between $ 36 $ s and $ 3500 $ s without the support of the comparison with observational data. Nevertheless, both the relative intensity of the P-GRB to the peak of the afterglow in such source, as well as their corresponding temporal lag, were theoretically predicted within a few percent (see Fig. 11 in \citet{rubr}).

The verification of the validity of our model has been tested in a variety of other sources, beside GRB 991216 (\citet{rubr}), like GRB 980425 (\citet{cospar02,Mosca_Orale}), GRB 030329 (\citet{030329}), GRB 031203 (\citet{031203}). In all such sources, again, the observational data were available only during the prompt emission and the latest afterglow phases, leaving our theoretical predictions of the in-between evolution untested. Now, thanks to the data provided by the \emph{Swift} satellite, we are finally able to confirm, by direct confrontation with the observational data, our theoretical predictions on the GRB structure with a detailed fit of the complete afterglow light curve of GRB 050315, from the peak, including the ``prompt emission'', all the way to the latest phases without any gap in the observational data.

\subsection{The fit of the observations}

The best fit of the observational data leads to a total energy of the black hole dyadosphere, generating the $e^\pm$ plasma, $E_{e^\pm}^{tot}=E_{dya} = 1.46\times 10^{53}$ erg (the observational \emph{Swift} $E_{iso}$ is $> 2.62\times 10^{52}$ erg, see \citet{va05}), so that the plasma is created between the radii $r_1 = 5.88\times 10^6$ cm and $r_2 = 1.74 \times 10^8$ cm with an initial temperature $T = 2.05 MeV$ and a total number of pairs $N_{e^+e^-} = 7.93\times 10^{57}$. The second parameter of the theory, the amount $M_B$ of baryonic matter in the plasma, is found to be such that $B \equiv M_Bc^2/E_{dya} = 4.55 \times 10^{-3}$. The transparency point and the P-GRB emission occurs then with an initial Lorentz gamma factor of the accelerated baryons $\gamma_\circ = 217.81$ at a distance $r = 1.32 \times 10^{14}$ cm from the black hole.

\subsubsection{The BAT data}

\begin{figure}
\begin{minipage}{\hsize}
\includegraphics[width=0.5\hsize,clip]{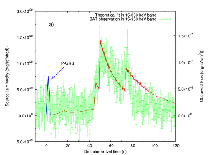}
\includegraphics[width=0.5\hsize,clip]{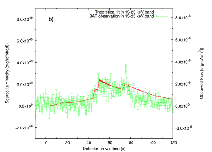}\\
\includegraphics[width=0.5\hsize,clip]{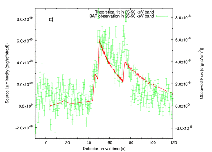}
\includegraphics[width=0.5\hsize,clip]{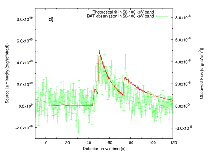}
\end{minipage}
\caption{Our theoretical fit (red line) of the BAT observations (green points) of GRB 050315 in the $15$--$350$ keV (a), $15$--$25$ keV (b), $25$--$50$ keV (c), $50$--$100$ keV (d) energy bands (\citet{va05}). The blue line in panel (a) represents our theoretical prediction for the intensity and temporal position of the P-GRB.}
\label{tot}
\end{figure}

In Fig. \ref{tot} we represent our theoretical fit of the BAT observations in the three energy channels $15$--$25$ keV, $25$--$50$ keV and $50$--$100$ keV and in the whole $15$--$350$ keV energy band.

In our model the GRB emission starts at the transparency point when the P-GRB is emitted; this instant of time is often different from the moment in which the satellite instrument triggers, due to the fact that sometimes the P-GRB is under the instrumental noise threshold or comparable with it. In order to compare our theoretical predictions with the observations, it is important to estimate and take into account this time shift. In the present case of GRB 050315 it has been observed (see \citet{va05}) a possible precursor before the trigger. Such a precursor is indeed in agreement with our theoretically predicted P-GRB, both in its isotropic energy emitted (which we theoretically predict to be $E_{P-GRB} = 1.98 \times 10^{51}$ erg) and its temporal separation from the peak of the afterglow (which we theoretically predicted to be $\Delta t^d_a = 51$ s). In Fig. \ref{tot}a the blue line shows our theoretical prediction for the P-GRB in agreement with the observations.

After the P-GRB emission, all the observed radiation is produced by the interaction of the expanding baryonic shell with the interstellar medium. In order to reproduce the complex time variability of the light curve of the prompt emission as well as of the afterglow, we describe the ISM filamentary structure, for simplicity, as a sequence of overdense spherical regions separated by much less dense regions. Such overdense regions are nonhomogeneously filled, leading to an effective emitting area $A_{eff}$ determined by the dimensionless parameter ${\cal R}$ (see previous sections and \citet{Spectr1,fil} for details). Clearly, in order to describe any detailed structure of the time variability an authentic three dimensional representation of the ISM structure would be needed. However, this finer description would not change the substantial agreement of the model with the observational data. Anyway, in the ``prompt emission'' phase, the small angular size of the source visible area due to the relativistic beaming makes such a spherical approximation an excellent one (see also for details \citet{r02}).

The structure of the ``prompt emission'' has been reproduced assuming three overdense spherical ISM regions with width $\Delta$ and density contrast $\Delta n/\langle n\rangle$: we chose for the first region, at $r = 4.15\times 10^{16}$ cm, $\Delta = 1.5\times 10^{15}$ cm and $\Delta n/\langle n\rangle = 5.17$, for the second region, at $r = 4.53\times 10^{16}$ cm, $\Delta = 7.0\times 10^{14}$ cm and $\Delta n/\langle n\rangle = 36.0$ and for the third region, at $r = 5.62\times 10^{16}$ cm, $\Delta = 5.0\times 10^{14}$ cm and $\Delta n/\langle n\rangle = 85.4$. The ISM mean density during this phase is $\left\langle n_{ISM} \right\rangle=0.81$ particles/cm$^3$ and $\left\langle {\cal R} \right\rangle = 1.4 \times 10^{-7}$. With this choice of the density mask we obtain agreement with the observed light curve, as shown in Fig. \ref{tot}. A small discrepancy occurs in coincidence with the last peak: this is due to the fact that at this stage the source visible area due to the relativistic beaming is comparable with the size of the clouds, therefore the spherical shell approximation should be duly modified by a detailed analysis of a full three-dimensional treatment of the ISM filamentary structure. Such a topic is currently under investigation (see also for details \citet{r02}). Fig. \ref{tot} shows also the theoretical fit of the light curves in the three BAT energy channels in which the GRB has been detected ($15$--$25$ keV in Fig. \ref{tot}b, $25$--$50$ keV in Fig. \ref{tot}c, $50$--$100$ keV in Fig. \ref{tot}d).

\subsubsection{The XRT data}

\begin{figure}
\includegraphics[width=\hsize,clip]{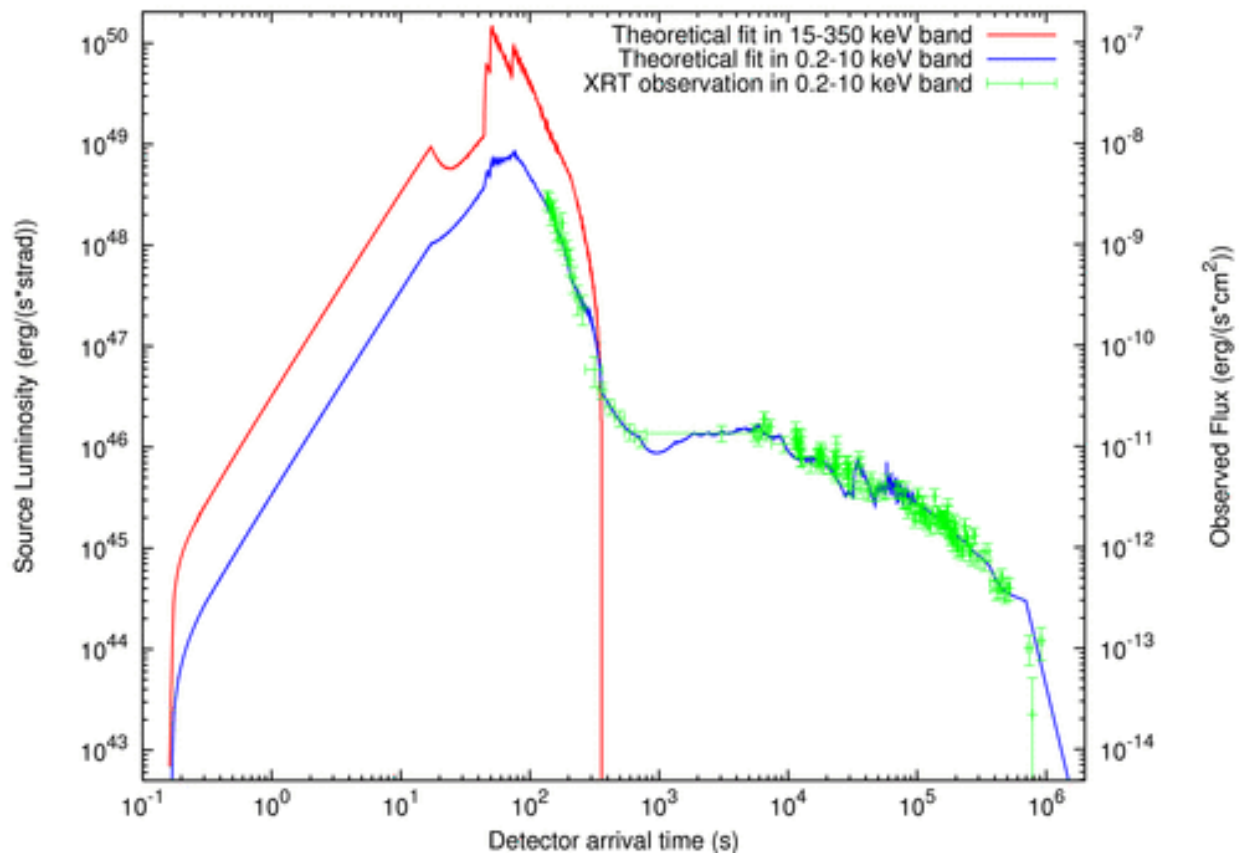}
\caption{Our theoretical fit (blue line) of the XRT observations (green points) of GRB 050315 in the $0.2$--$10$ keV energy band (\citet{va05}). The theoretical fit of the BAT observations (see Fig. \ref{tot}a) in the $15$--$350$ keV energy band is also represented (red line).}
\label{global}
\end{figure}

The same analysis can be applied to explain the features of the XRT light curve in the afterglow phase. It has been recently pointed out (\citet{nousek}) that almost all the GRBs observed by {\em Swift} show a ``canonical behavior'': an initial very steep decay followed by a shallow decay and finally a steeper decay. In order to explain these features many different approaches have been proposed (\citet{meszaros,nousek,panaitescu,zhang}). In our treatment these behaviors are automatically described by the same mechanism responsible for the prompt emission described above: the baryonic shell expands in an ISM region, between $r = 9.00\times 10^{16}$ cm and $r = 5.50\times 10^{18}$ cm, which is significantly at lower density ($\left\langle n_{ISM} \right\rangle=4.76 \times 10^{-4}$ particles/cm$^3$, $\left\langle {\cal R} \right\rangle = 7.0 \times 10^{-6}$) then the one corresponding to the prompt emission, and this produces a slower decrease of the velocity of the baryons with a consequent longer duration of the afterglow emission. The initial steep decay of the observed flux is due to the smaller number of collisions with the ISM. In Fig. \ref{global} is represented our theoretical fit of the XRT data, together with the theoretically computed $15$--$350$ keV light curve of Fig. \ref{tot}a (without the BAT observational data to not overwhelm the picture too much).

What is impressive is that no different scenarios need to be advocated in order to explain the features of the light curves: both the prompt and the afterglow emission are just due to the thermal radiation in the comoving frame produced by inelastic collisions with the ISM duly boosted by the relativistic transformations over the EQTSs.

\subsection{The instantaneous spectrum}\label{spectra}

\begin{figure}
\includegraphics[width=\hsize,clip]{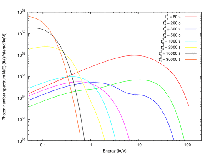}
\caption{Eight theoretically predicted instantaneous photon number spectra $N(E)$ are here represented for different values of the arrival time (colored curves). The hard to soft behavior is confirmed.}
\label{spettro}
\end{figure}

In addition to the the luminosity in fixed energy bands we can derive also the instantaneous photon number spectrum $N(E)$ starting from the same assumptions. In Fig. \ref{spettro} are shown samples of time-resolved spectra for eight different values of the arrival time which cover the whole duration of the event. It is manifest from this picture that, although the spectrum in the co-moving frame of the expanding pulse is thermal, the shape of the final spectrum in the laboratory frame is clearly non thermal. In fact, as explained in \citet{Spectr1}, each single instantaneous spectrum is the result of an integration of thousands of thermal spectra over the corresponding EQTS. This calculation produces a non thermal instantaneous spectrum in the observer frame (see Fig. \ref{spettro}).

A distinguishing feature of the GRBs spectra which is also present in these instantaneous spectra is the hard to soft transition during the evolution of the event (\citet{cri97,fa00,gcg02}). In fact the peak of the energy distribution $E_p$ drifts monotonically to softer frequencies with time. This feature is linked to the change in the power-law low energy spectral index $\alpha$ (\citet{b93}), so the correlation between $\alpha$ and $E_p$ (\citet{cri97}) is explicitly shown.

It is important to stress that there is no difference in the nature of the spectrum during the prompt and the afterglow phases: the observed energy distribution changes from hard to soft, with continuity, from the ``prompt emission'' all the way to the latest phases of the afterglow.

\subsection{Problems with the definition of ``long'' GRBs}

\begin{figure}
\includegraphics[width=\hsize,clip]{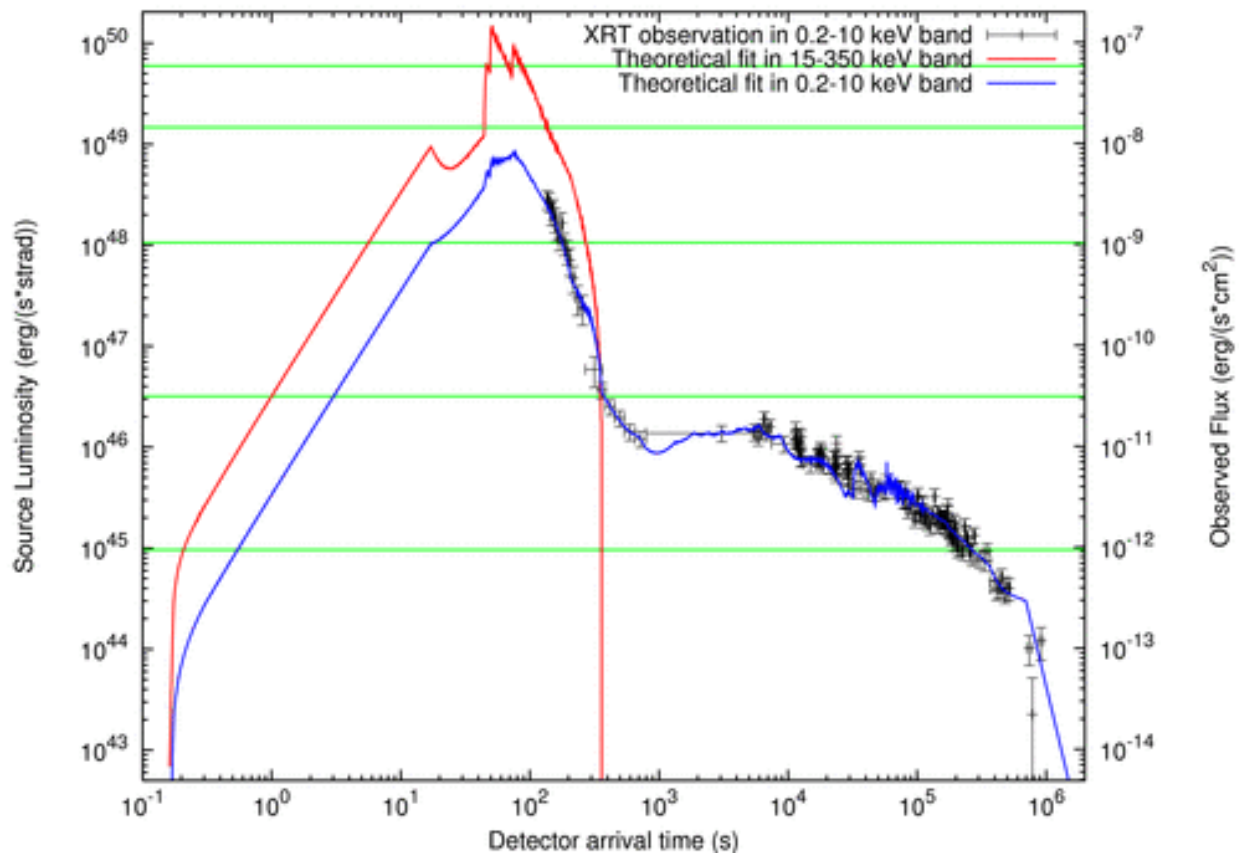}
\caption{Same as Fig. \ref{global}. The horizontal dashed lines corresponds to different possible instrumental thresholds. It is clear that long GRB durations are just functions of the observational threshold.}
\label{global_th}
\end{figure}

The confirmation by \emph{Swift} of our prediction of the overall afterglow structure, and especially the coincidence of the ``prompt emission'' with the peak of the afterglow, opens a new problematic in the definition of the long GRBs. It is clear, in fact, that the identification of the ``prompt emission'' in the current GRB literature is not at all intrinsic to the phenomenon but is merely due to the threshold of the instruments used in the observations (e.g. BATSE in the $50$--$300$ keV energy range, or BeppoSAX GRBM in $40$--$700$ keV, or \emph{Swift} BAT in $15$--$350$ keV). As it is clear from Fig. \ref{global_th}, there is no natural way to identify in the source a special extension of the peak of the afterglow that is not the one purely defined by the experimental threshold. It is clear, therefore, that long GRBs, as defined till today, are just the peak of the afterglow and there is no way, as explained above, to define their ``prompt emission'' duration as a characteristic signature of the source. As the \emph{Swift} observations show, the duration of the long GRBs has to coincide with the duration of the entire afterglow. A Kouveliotou - Tavani plot of the long GRBs, done following our interpretation which is clearly supported by the recent \emph{Swift} data (see Fig. \ref{global_th}), will present enormous dispersion on the temporal axis.

We recall that in our theory both ``short'' and ``long'' GRBs originate from the same process of black hole formation. The major difference between the two is the value of the baryon loading parameter $B$ (see Fig. \ref{bcross}). In the limit of small baryon loading, all the plasma energy is emitted at the transparency in the P-GRB, with negligible afterglow observed flux. For higher values of the baryon loading, the relative energy content of the P-GRB with respect to the afterglow diminishes (see e.g. \citet{rubr2} and references therein).

\section{On the GRB-SN association}

Models of GRBs based on a single source (the ``collapsar'') generating both the SN and the GRB abounds in the literature (see e.g. \citet{wb06}). In our approach we have emphasized the concept of induced gravitational collapse, which occurs strictly in a binary system. The SN originates from a star evolved out of the main sequence and the GRB from the collapse to a black hole. The two phenomena are qualitatively very different. There is still much to be discovered about SNe due to their complexity, while the GRB is much better known since its collapse to a black hole is now understood. The concept of induced collapse implies at least two alternative scenarios. In the first, the GRB triggers a SN explosion in the very last phase of the thermonuclear evolution of a companion star (\citet{lett3}). In the second, the early phases of the SN induce gravitational collapse of a companion neutron star to a black hole (\citet{mg11}). Of course, in absence of SN, there is also the possibility that the collapse to a black hole, generating the GRB, occurs in a single star system or in the final collapse of a binary neutron star system. Still, in such a case there is also the possibility that the black hole progenitor is represented by a binary system composed by a white dwarf and/or a neutron star and/or a black hole in various combinations. What is most remarkable is that, following the ``uniqueness of the black hole'' (see \citet{RuKerr}), all these collapses lead to a common GRB independently of the nature of their progenitors.

We have already outlined in the previous sections that we have successfully used as a prototype the source GRB 991216 which clearly shows the existence of the P-GRB and the three regimes of the afterglow (see Fig. \ref{grb991216}). Unfortunately, data from BATSE existed only up to $ 36 $ s, and data from R-XTE and Chandra only after $ 3500 $ s, leaving our theoretical predictions in the whole range between $ 36 $ s and $ 3500 $ s without the support of the comparison with observational data. Nevertheless, both the relative intensity of the P-GRB to the peak of the afterglow in such source, as well as their corresponding temporal lag, were theoretically predicted within a few percent (see Fig. 11 in \citet{rubr}).

Having obtained success in the fit of GRB 991216, GRB 031203 and GRB 050315, we turn to the application of our theoretical analysis to the other closest GRB sources associated with SNe. We start with GRB 980425 / SN 1998bw. We have however to caution about the validity of this fit. From the available data of BeppoSAX, BATSE, XMM and Chandra, only the data of the prompt emission ($t_a^d < 10^2$ s) and of the latest afterglow phases ($t_a^d > 10^5$ s all the way to more than $10^8$ s!) were available. Our fit refers only to the prompt emission, as usually interpreted as the peak of the afterglow. The fit, therefore, represents an underestimate of the GRB 980425 total energy and in this sense it is not surprising that it does not fit the \citet{aa02} relation. The latest afterglow emission, the URCA-1 emission, presents a different problematic which we will shortly address (see below).

\subsection{GRB 980425 / SN 1998bw / URCA-1}\label{980425}

\begin{figure}
\includegraphics[width=\hsize,clip]{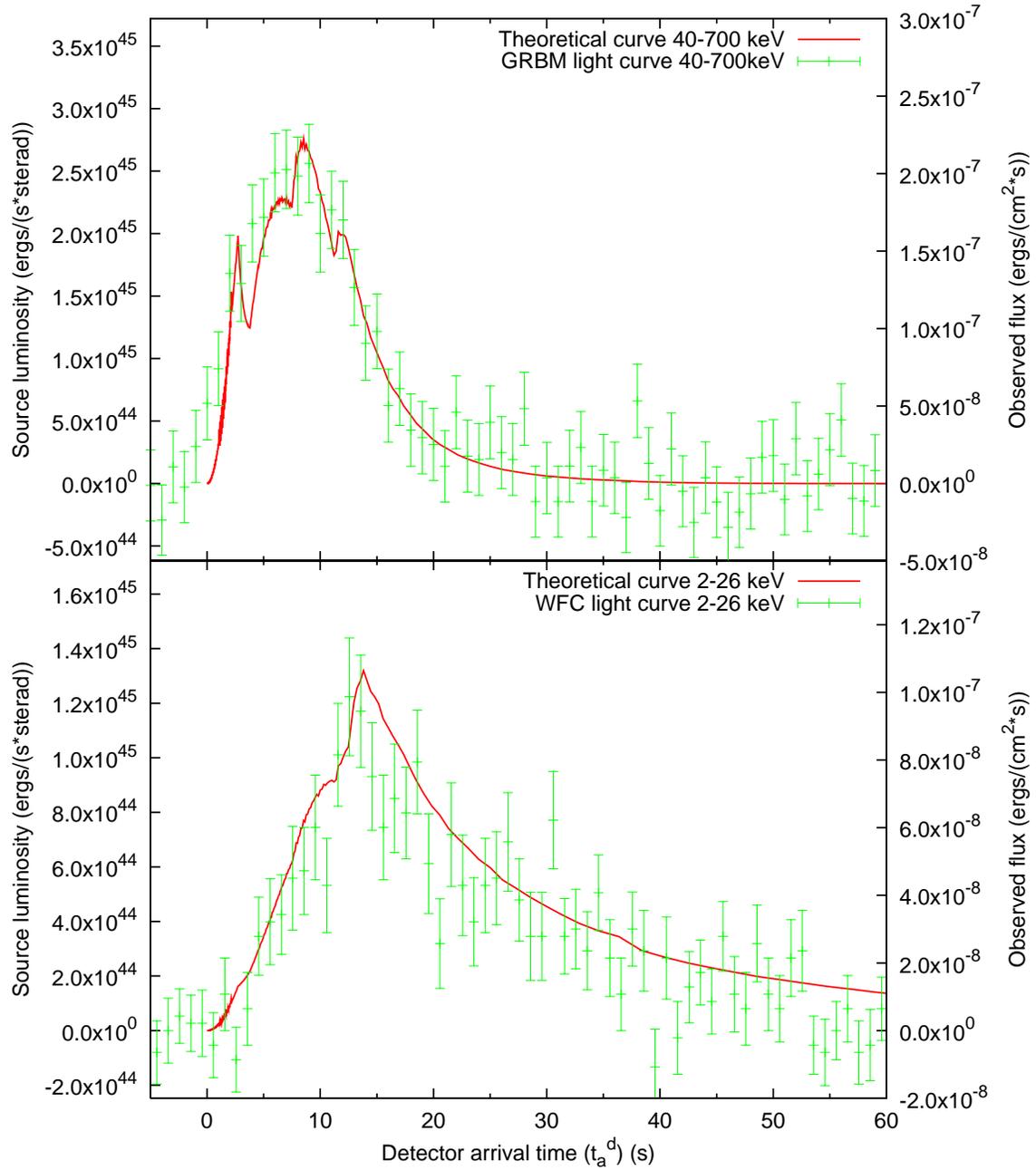}
\caption{Theoretical light curves of GRB 980425 prompt emission in the $40$--$700$ keV and $2$--$26$ keV energy bands (solid line), compared with the observed data respectively from Beppo-SAX GRBM and WFC (see \citet{pian00,fa00}).}
\label{980425_picco}
\end{figure}

\begin{figure}
\includegraphics[width=\hsize,clip]{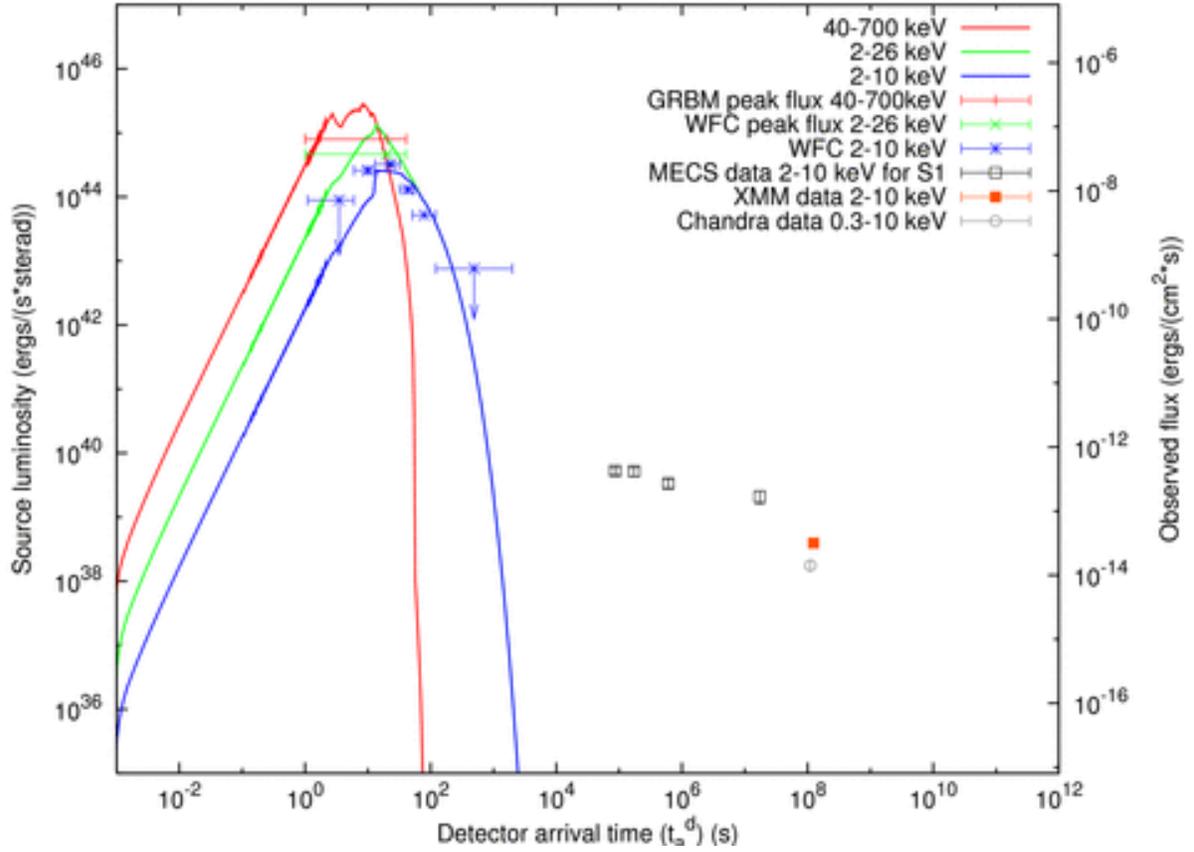}
\caption{Theoretical light curves of GRB 980425 in the $40$--$700$ keV (solid line), $2$--$26$ keV (dashed line), $2$--$10$ keV (dotted line) energy bands, represented together with URCA-1 observational data. All observations are by BeppoSAX (\citet{pian00}), with the exception of the last two URCA-1 points, which is observed by XMM and \textit{Chandra} (\citet{pian04,k04}).}
\label{980425_global}
\end{figure}

The best fit of the observational data of GRB 980425 (\citet{pian00,fa00}) leads to $E_{e^\pm}^{tot}=E_{dya}=1.2\times10^{48}$ erg and $B = 7.7\times10^{-3}$. This implies an initial $e^\pm$ plasma with $N_{e^+e^-} = 3.6\times10^{53}$ and with an initial temperature $T = 1.2$ MeV. After the transparency point, the initial Lorentz gamma factor of the accelerated baryons is $\gamma = 124$. The variability of the luminosity, due to the inhomogeneities of the ISM (\citet{r02}), is characterized by a density contrast $\delta n / n \sim 10^{-1}$ on a length scale of $\Delta \sim 10^{14}$ cm. We determine the effective ISM parameters to be: $\langle n_{ism} \rangle = 2.5\times 10^{-2}$ particle/$cm^3$ and $\langle \mathcal{R} \rangle = 1.2\times 10^{-8}$.

In Fig. \ref{980425_picco} we address the first $60$ s of data, which in the current literature are generally called ``prompt emission'' as due to an unidentified ``inner engine'' and neglected in the theoretical modeling of the source. In our approach, we test our specific theoretical assumptions comparing and contrasting our theoretically computed light curves in the $40$--$700$ and $2$--$26$ keV energy bands with the observations by the BeppoSAX GRBM and WFC during such time interval (see \citet{pian00,fa00}). As in the previous works, we have used our exact analytic solution for the equations of motion of the baryons (\citet{PowerLaws}). The agreements in Fig. \ref{980425_picco} shows the very satisfactory predictive power of our theory.

In Fig. \ref{980425_global} we summarize some of the problematic implicit in the old pre-\emph{Swift} era: data are missing in the crucial time interval between $60$ s and $10^5$ s, when the BeppoSAX NFI starts to point the GRB 980425 location. In this region we have assumed, for the effective ISM parameters, constant values inferred by the last observational data. Currently we are relaxing this condition, also in view of the interesting paper by \citet{ga06}. In this respect, we are currently examining GRB 060218 / SN 2006aj (see \citet{caa06,060218}). We then represent the URCA-1 observations performed by BeppoSAX-NFI in the energy band $2$--$10$ keV (\citet{pian00}), by XMM-EPIC in the band $0.2$--$10$ keV (\citet{pian04}) and by \textit{Chandra} in the band $0.3$--$10$ keV (\citet{k04}). The separation between the light curves of GRB 980425 in the $2$--$700$ keV energy band, of SN 1998bw in the optical band (\citet{nomoto,pian06}), and of the above mentioned URCA-1 observations is given in Fig. \ref{urca123+GRB_full}A.

\subsection{GRB 030329 / SN 2003dh / URCA-2}\label{030329}

For GRB 030329 we have obtained (see \citet{030329,030329_la,Mosca_Orale}) a total energy $E_{e^\pm}^{tot}=E_{dya}=2.12\times10^{52}$ erg and a baryon loading $B = 4.8\times10^{-3}$. This implies an initial $e^\pm$ plasma with $N_{e^+e^-}=1.1\times10^{57}$ and with an initial temperature $T=2.1$ MeV. After the transparency point, the initial Lorentz gamma factor of the accelerated baryons is $\gamma = 206$. The effective ISM parameters are $\langle n_{ism} \rangle = 2.0$ particle/$cm^3$ and $\langle \mathcal{R} \rangle = 2.8\times 10^{-9}$, with a density contrast $\delta n / n \sim 10$ on a length scale of $\Delta \sim 10^{14}$ cm. The resulting fit of the observations, both of the prompt phase and of the afterglow have been presented in (\citet{030329,030329_la}). We compare in Fig. \ref{urca123+GRB_full}B the light curves of GRB 030329 in the $2$--$400$ keV energy band, of SN 2003dh in the optical band (\citet{nomoto,pian06}) and of URCA-2 observed by XMM-EPIC in $2$--$10$ keV energy band (\citet{tiengo03,tiengo04}).

\subsection{GRB 031203 / SN 2003lw / URCA-3}\label{031203}

\begin{figure}
\includegraphics[width=0.5\hsize,clip]{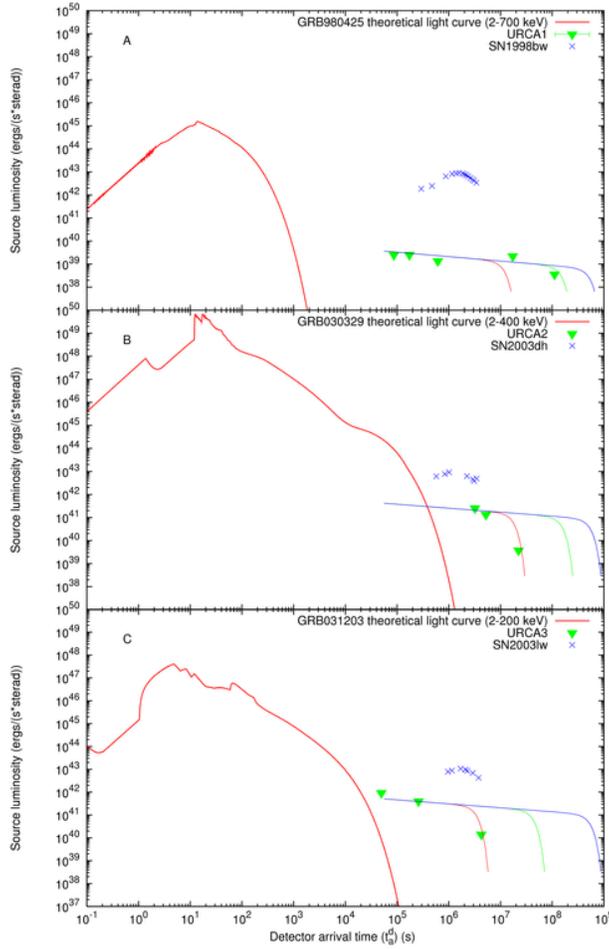}
\caption{Theoretically computed light curves of GRB 980425 in the $2$--$700$ keV band (A), of GRB 030329 in the $2$--$400$ keV band (B) and of GRB 031203 in the $2$--$200$ keV band (C) are represented, together with the URCA observational data and qualitative representative curves for their emission, fitted with a power law followed by an exponentially decaying part. The luminosity of the SNe in the $3000-24000$ {\AA} are also represented (\citet{nomoto,pian06}).}
\label{urca123+GRB_full}
\end{figure}

We recall that, in the previous sections, we have shown how the analysis of GRB 031203 leads to a total energy $E_{e^\pm}^{tot}=E_{dya}=1.85\times10^{50}$ erg and to a baryon loading $B = 7.4\times10^{-3}$. This implies an initial $e^\pm$ plasma with $N_{e^+e^-}=3.0\times 10^{55}$ and with an initial temperature $T=1.5$ MeV. After the transparency point, the initial Lorentz gamma factor of the accelerated baryons is $\gamma = 132$. The effective ISM parameters are $\langle n_{ism} \rangle = 1.6\times 10^{-1}$ particle/$cm^3$ and $\langle \mathcal{R} \rangle = 3.7\times 10^{-9}$, with a density contrast $\delta n / n \sim 10$ on a length scale of $\Delta \sim 10^{15}$ cm. In Fig. \ref{urca123+GRB_full}C we compare the light curves of GRB 031203 in the $2$--$200$ keV energy band, of SN 2003lw in the optical band (\citet{nomoto,pian06}) and of URCA-3 observed by XMM-EPIC in the $0.2$--$10$ keV energy band (\citet{Wa04}) and by \textit{Chandra} in the $2$--$10$ keV energy band (\citet{sod}).

\subsection{The GRB / SN / URCA connection}

\begin{table}
\centering
\caption{a) see \citet{yk}; b) Mazzali, P., private communication at MG11 meeting in Berlin, July 2006; c) evaluated fitting the URCAs with a power law followed by an exponentially decaying part; d) evaluated assuming a mass of the neutron star $M=1.5 M_\odot$ and $T \sim 5$--$7$ keV in the source rest frame; e) see \citet{galama98,greiner,proch,mh}.}
{\footnotesize
\begin{tabular}{ccccccccccc}
\hline
GRB & $\begin{array}{c}E_{e^\pm}^{tot}\\ \mathrm{(erg)}\end{array}$ & $B$ & $\gamma_0$ & $\begin{array}{c}E_{SN}^{bolom}\\ \mathrm{(erg)^a}\end{array}$ & $\begin{array}{c}E_{SN}^{kin}\\ \mathrm{(erg)^b}\end{array}$ & $\begin{array}{c}E_{URCA}\\ \mathrm{(erg)^c}\end{array}$ & $\displaystyle\frac{E_{e^\pm}^{tot}}{E_{URCA}}$ & $\displaystyle\frac{E_{SN}^{kin}}{E_{URCA}}$ & $\begin{array}{c}R_{NS}\\ \mathrm{(km)^d}\end{array}$ & $z^e$ \\
\hline
980425 & $1.2\times 10^{48}$ & $7.7\times10^{-3}$ & $124$ & $2.3\times 10^{49}$ & $1.0\times 10^{52}$ & $3\times 10^{48}$ & $0.4$ & $1.7\times10^{4}$ & $ 8$ & $0.0085$\\
030329 & $2.1\times 10^{52}$ & $4.8\times10^{-3}$ & $206$ & $1.8\times 10^{49}$ & $8.0\times10^{51}$ & $3\times10^{49}$ & $6\times 10^{2}$ & $1.2\times10^{3}$ & $14$ & $0.1685$\\
031203 & $1.8\times 10^{50}$ & $7.4\times10^{-3}$ & $133$ & $3.1\times 10^{49}$ & $1.5\times10^{52}$ & $2\times10^{49}$ & $8.2$ & $3.0\times10^{3}$ & $20$ & $0.105$\\
060218 & $1.8\times 10^{50}$ & $1.0\times10^{-2}$ & $99$ & $9.2\times 10^{48}$ & $2.0\times10^{51}$ & $?$ & $?$ & $?$ & $?$ & $0.033$\\
\end{tabular}}
\label{tabella}
\end{table}

In Tab. \ref{tabella} we summarize the representative parameters of the above four GRB-SN systems, including the very large kinetic energy observed in all SNe (\citet{mazzaliVen}). Some general conclusions on these weak GRBs at low redshift, associated to SN Ib/c, can be established on the ground of our analysis:\\
{\bf 1)} From the detailed fit of their light curves, as well as their accurate spectral analysis, it follows that all the above GRB sources originate consistently from the formation of a black hole. This result extends to this low-energy GRB class at small cosmological redshift the applicability of our model, which now spans over a range of energy of six orders of magnitude from $10^{48}$ to $10^{54}$ ergs (\citet{rubr,cospar02,Mosca_Orale,030329,030329_la,031203,050315}). Distinctive of this class is the very high value of the baryon loading which in one case (GRB 060218) is very close to the maximum limit compatible with the dynamical stability of the adiabatic optically thick acceleration phase of the GRBs (\citet{rswx00}). Correspondingly, the maximum Lorentz gamma factors are systematically smaller than the ones of the more energetic GRBs at large cosmological distances. This in turn implies the smoothness of the observed light curves in the so-called ``prompt phase''. The only exception to this is the case of GRB 030329.\\
{\bf 2)} The accurate fits of the GRBs allow us to infer also some general properties of the ISM. While the size of the clumps of the inhomogeneities is $\Delta \approx 10^{14}$ cm, the effective ISM average density is consistently smaller than in the case of more energetic GRBs: we have in fact $\langle n_{ism} \rangle$ in the range between $\sim 10^{-6}$ particle/$cm^3$ (GRB 060218) and $\sim 10^{-1}$ particle/$cm^3$ (GRB 031203), while only in the case of GRB 030329 it is $\sim 2$ particle/$cm^3$. We are also currently studying a characteristic trend in the variability of $\mathcal{R}$  during some specific bursts as well as the physical origin of the consistently smaller effective ISM density $\langle n_{ism} \rangle$ values observed in these sources (see \citet{060218}).\\
{\bf 3)} Still within their weakness these four GRB sources present a large variability in their total energy: a factor $10^4$ between GRB 980425 and GRB 030329. Remarkably, the SNe emission both in their very high kinetic energy and in their bolometric energy appear to be almost constant respectively $10^{52}$ erg and $10^{49}$ erg. The URCAs present also a remarkably steady behavior around a ``standard luminosity'' and a typical temporal evolution. The weakness in the energetics of GRB 980425 and GRB 031203, and the sizes of their dyadospheres, suggest that they originate from the formation of the smallest possible black hole, just over the critical mass of the neutron star (see Fig. \ref{IndColl06} and \citet{mg11}).

\subsection{URCA-1, URCA-2 and URCA-3}

Before closing, we turn to the search for the nature of URCA-1, URCA-2 and URCA-3. These systems are not yet understood and may have an important role in the comprehension of the astrophysical scenario of GRB sources. It is important to perform additional observations in order to verify if the URCA sources are related to the black hole originating the GRB phenomenon or to the SN. Even a single observation of an URCA source with a GRB in absence of a SN would prove their relation with the black hole formation. Such a result is today theoretically unexpected and would open new problematics in relativistic astrophysics and in the physics of black holes. Alternatively, even a single observation of an URCA source during the early expansion phase of a type Ib/c SN in absence of a GRB would prove the early expansion phases of the SN remnants. In the case that none of such two conditions are fulfilled, then the URCA sources must be related to the GRBs occurring in presence of a SN. In such a case, one of the possibilities would be that for the first time we are observing a newly born neutron star out of the supernova phenomenon unveiled by the GRB. This last possibility would offer new fundamental information about the outcome of the gravitational collapse, and especially about the equations of state at supranuclear densities and about a variety of fundamental issues of relativistic astrophysics of neutron stars.

The names of ``URCA-1'' and ``URCA-2'' for the peculiar late X-ray emission of GRB 980425 and GRB 030329 were given in the occasion of the Tenth Marcel Grossmann meeting held in Rio de Janeiro (Brazil) in the Village of Urca (see \citet{r03mg10}). Their identification was made at that time and presented at that meeting. However, there are additional reasons for the choice of these names. Another important physical phenomenon was indeed introduced in 1941 in the same Village of Urca by George Gamow and Mario Schoenberg (see \citet{gs41}). The need for a rapid cooling process due to neutrino anti-neutrino emission in the process of gravitational collapse leading to the formation of a neutron star was there considered for the first time. It was Gamow who named this cooling as ``Urca process'' (see \citet{GamowBook-MyWorldlines}). Since then, a systematic analysis of the theory of neutron star cooling was advanced by \citet{t64,t79,tc66,t02,c78}. The coming of age of X-ray observatories such as Einstein (1978-1981), EXOSAT (1983-1986), ROSAT (1990-1998), and the contemporary missions of Chandra and XMM-Newton since 1999 dramatically presented an observational situation establishing very embarrassing and stringent upper limits to the surface temperature of neutron stars in well known historical supernova remnants (see e.g. \citet{r87}). It was so that, for some remnants, notably SN 1006 and the Tycho supernova, the upper limits to the surface temperatures were significantly lower than the temperatures given by standard cooling times (see e.g. \citet{r87}). Much of the theoretical works has been mainly directed, therefore, to find theoretical arguments in order to explain such low surface temperature $T_s \sim 0.5$--$1.0\times 10^6$ K --- embarrassingly low, when compared to the initial hot ($\sim 10^{11}$ K) birth of a neutron star in a supernova explosion (see e.g. \citet{r87}). Some important contributions in this researches have been presented by \citet{vr88,vr91,bl86,lvrpp94,yp04}. The youngest neutron star to be searched for thermal emission has been the pulsar PSR J0205+6449 in 3C 58 (see e.g. \citet{yp04}), which is $820$ years old! \citet{t05} reported evidence for the detection of thermal emission from the crab nebula pulsar which is, again, $951$ years old.

URCA-1, URCA-2 and URCA-3 may explore a totally different regime: the X-ray emission possibly from a recently born neutron star in the first days -- months of its existence. The thermal emission from the young neutron star surface would in principle give information on the equations of state in the core at supranuclear densities and on the detailed mechanism of the formation of the neutron star itself with the related neutrino emission. It is also possible that the neutron star is initially fast rotating and its early emission could be dominated by the magnetospheric emission or by accretion processes from the remnant which would overshadow the thermal emission. A periodic signal related to the neutron star rotational period should in principle be observable in a close enough GRB-SN system. In order to attract attention to this problematic, we have given in Tab. \ref{tabella} an estimate of the corresponding neutron star radius for URCA-1, URCA-2 and URCA-3. It has been pointed out (see e.g. \citet{pian00}) the different spectral properties between the GRBs and the URCAs. It would be also interesting to compare and contrast the spectra of all URCAs in order to evidence any analogy among them. Observations of a powerful URCA source on time scales of $0.1$--$10$ seconds would be highly desirable.

\section{Pair production in Coulomb potential of nuclei and heavy-ion collisions}

\subsection{$Z=137$ catastrophe and critical value $Z_{cr}$}

We discuss the pair production in the Coulomb potential of a bare nucleus with super-critical charge 
$Z>Z_{cr}$. 
In a Coulomb potential, there are discrete energy-levels indicating bound states of electrons in the energy-spectrum,
which differs from the continuous energy-spectrum  of electron states in an external 
constant electric field. This makes differences not only in the rate of pair production, but also in
the produced final states. Although atoms
with such large $Z$ are hardly synthesized for a time larger than $\hbar/(m_ec^2)$ needed for pair production
even in recent experiments, it is of great interest their theoretical study within the QED framework.

\begin{figure}
\centering 
\includegraphics[width=\hsize,clip]{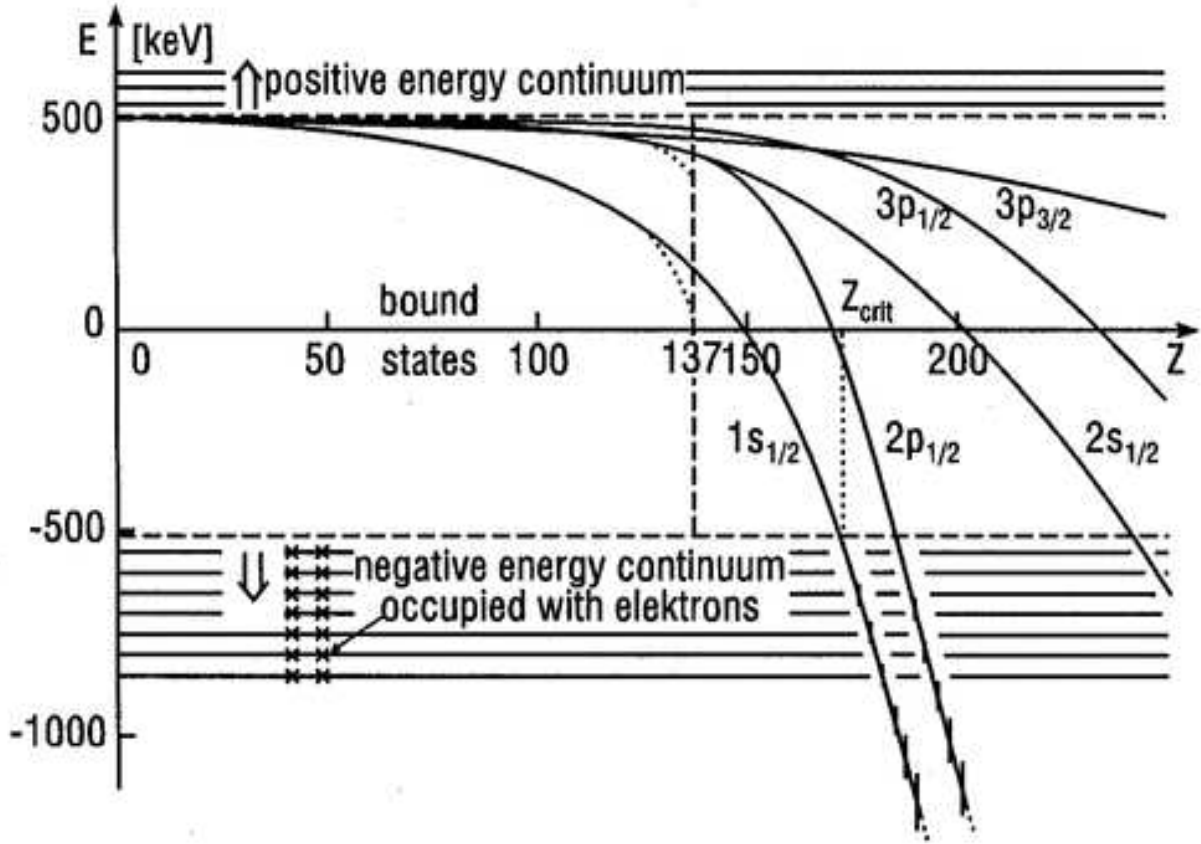}
\caption{Atomic binding energies as function of nuclear charge Z. This figure is reproduced from
Fig. 1 in \citet{grc98}.}%
\label{zspectrum}%
\end{figure}

Very soon after the Dirac equation for a relativistic electron was discovered (\citet{z4a,z4b}), see also \citet{z5}, \citet{z7} (for all $Z< 137$) 
and \citet{z6} (for $Z=1$) found its solution in the point-like Coulomb potential $V(r)=-Z\alpha/r, \quad 0<r<\infty$. 
Solving the differential equations for the Dirac wave function, they obtained the well-known
Sommerfeld's formula \citep{Sommerfeld} for energy-spectrum, 
\begin{equation}
{\mathcal E}(n,j)=m_ec^2\left[1+\left(\frac{Z\alpha}{ n-|K|+(K^2-Z^2\alpha^2)^{1/2}}\right)^2\right]^{-1/2}.
\label{dirac}
\end{equation}
Here the principle quantum number $n=1,2,3,\cdot\cdot\cdot$ and 
\begin{equation}
K=\left\{\begin{array}{ll} -(j+1/2)= -(l+1), & {\rm if}\quad j=l+\frac{1}{2}, \quad l\ge 0 \\
 (j+1/2)= l, & {\rm if}\quad j=l-\frac{1}{2}, \quad l\ge 1
 \end{array}\right.
\label{dirac-k}
\end{equation}
where $l=0,1,2,\cdot\cdot\cdot$ is the orbital angular momentum corresponding to the upper component of Dirac bi-spinor, 
$j$ is the total angular momentum, and the states with 
$K=\mp 1,\mp 2,\mp 3,\cdot\cdot\cdot, \mp (n-1)$ are doubly degenerate, while the state $K=-n$ is a singlet (\citet{z7,z6}). 
The integer values $n$ and $K$ label bound states whose energies are ${\mathcal E}(n,j)\in (0,m_ec^2)$. For the example, 
in the case of the lowest energy states, one has     
\begin{eqnarray}
{\mathcal E}(1S_{\frac{1}{2}})&=& \sqrt {1-(Z\alpha)^2},\label{dirac-k1}\\
{\mathcal E}(2S_{\frac{1}{2}})&=&{\mathcal E}(2P_{\frac{1}{2}})
= \sqrt{\frac{1+\sqrt{1-(Z\alpha)^2}}{2}},\label{dirac-k2}\\
{\mathcal E}(2P_{\frac{3}{2}})&=& \sqrt {1-\frac{1}{4}(Z\alpha)^2}.
\label{dirac-k3}
\end{eqnarray}
For all states of the discrete spectrum, the binding energy
$mc^2-{\mathcal E}(n,j)$ increases as the nuclear charge $Z$ increases, as shown in
Fig.~\ref{zspectrum}.  When $Z=137$, ${\mathcal E}(1S_{1/2})=0$, 
${\mathcal E}(2S_{1/2})={\mathcal E}(2P_{1/2})=1/\sqrt{2}$ 
and ${\mathcal E}(2S_{3/2})=\sqrt{3}/2$. 
No regular solutions with $n=1,j=1/2,l=0,$ and $K=-1$ (the $1S_{1/2}$ ground state)
are found beyond $Z=137$ \footnote{Gordon noticed this in his pioneer paper \citep{z7}}. 
This phenomenon is the so-called ``$Z=137$ catastrophe'' and it is associated with the
assumption that the nucleus is point-like in calculating the electronic energy-spectrum.
In fact, it was shown by \citet{g4} that in nature there cannot
be a point-like charged object with effective coupling constant $Z\alpha >1$, because any surplus charge is screened by the over critical vacuum-polarization.

Some aspects of this problem were solved by considering the fact that the
nucleus is not point-like and
has an extended charge distribution (\citet{g1a,g1b,g1c,g1d,z10,z11a,z11b,z12,z}). When doing so, the $Z=137$ catastrophe disappears
and the energy-levels ${\mathcal E}(n,j)$ of the bound states $1S$, $2P$ and $2S$, $\cdot\cdot\cdot$ smoothly continue
to drop toward the negative energy continuum, as $Z$ increases to values larger than $137$, as shown in Fig.~\ref{zspectrum}.
The reason is that the finite size $R$
of nucleus charge distribution provides a cutoff $\Lambda$ for the boundary condition
at the origin $r\rightarrow 0$
and the energy-levels ${\mathcal E}(n,j)$ of the Dirac equation are shifted due to
the cutoff. In order to determine the critical value $Z_{cr}$
when the negative energy continuum (${\mathcal E}<- m_ec^2$) is encountered
(see Fig.~\ref{zspectrum}), Zel'dovich and Popov (see \citet{z10,z11a,z11b,z12,z}) solved the
Dirac equation corresponding to a nucleus of finite
extended charge distribution, i.e. the Coulomb potential is modified as
\begin{equation}
V(r)=\left\{\begin{array}{ll} -\frac{Ze^2}{ r}, &  r>R, \\
 -\frac{Ze^2}{ R}f\left(\frac{r}{ R}\right), &  r<R,
 \end{array}\right.
\label{extpotential}
\end{equation}
where $R\sim 10^{-12}$cm is the size of the nucleus. The form of the cutoff function $f(x)$ depends on the
distribution of the electric charge over the volume of the nucleus $(x=r/R, 0<x<1$, with $f(1)=1)$.
Thus, $f(x)=(3-x^2)/2$ corresponds to a constant volume density of charge. 
Solving the Dirac equation with the modified Coulomb potential (\ref{extpotential}) and calculating
the corresponding perturbative shift $\Delta {\mathcal E}_R$ of the lowest energy level (\ref{dirac-k1})
one obtains (\citet{z10,z}),
\begin{equation}
\Delta {\mathcal E}_R= m_ec^2\frac{(\xi)^2(2\xi e^{-\Lambda})^{2\gamma_z}}{\gamma_z (1+2\gamma_z)}
\left[1-2\gamma_z\int_0^1f(x)x^{2\gamma_z} dx\right],
\label{zshift}
\end{equation}
where $\xi=Z\alpha$, $\gamma_z=\sqrt{1-\xi^2}$ and the $\Lambda=\ln(\hbar / m_ecR)\gg 1$ is logarithmic parameter 
in the problem considered, However, note that Eq.~(\ref{zshift}) is inapplicable 
at $\xi\rightarrow 1 (Z\rightarrow 137)$.
The asymptotic expressions of the $1S_{1/2}$ energy were obtained (\citet{z12,z}), 
\begin{equation}
{\mathcal E}(1S_{1/2})=m_ec^2\left\{\begin{array}{ll} \sqrt {1-\xi^2}\coth(\Lambda\sqrt {1-\xi^2}), &  0<\xi<1, \\
 \Lambda^{-1}, &  \xi=1,\\
 \sqrt {\xi^2-1}\cot(\Lambda\sqrt {\xi^2-1}), &  \xi>1.
 \end{array}\right.
\label{easymptotic}
\end{equation}
This asymptotic expressions for ${\mathcal E}(1S_{1/2})$ practically coincides with Eq.~(\ref{dirac-k1}) for a point-like charge, 
if $(1-\xi)\gg 1/(8\Lambda^2)$, or $Z\leq 137-17/\Lambda^2$, however, it shows that Eq.~(\ref{dirac-k1}) 
is incorrect for larger values of nuclear charge $Z$. As a result, the ``$Z=137$ catastrophe'' in Eq.~(\ref{dirac}) disappears, 
${\mathcal E}(1S_{1/2})=0$ gives
\begin{equation}
\xi_0=1+\frac{\pi^2}{8\Lambda}+{\mathcal O}(\Lambda^{-4});
\label{0xi}
\end{equation}
the state $1S_{1/2}$ energy continuously
goes down to negative energy continuum as $Z\alpha >1$, 
and ${\mathcal E}(1S_{1/2})=-1$ gives
\begin{equation}
\xi_{cr}=1+\frac{\pi^2}{2\Lambda(\Lambda +2)}+{\mathcal O}(\Lambda^{-4}),
\label{0xicr}
\end{equation}
as shown in Fig.~\ref{zspectrum}. 
In \citet{z10,z} it is found that the critical values $\xi_c^{(n)}= Z_c\alpha$
for the energy-levels $nS_{1/2}$ and $nP_{1/2}$ reaching negative energy continuum
are equal to
\begin{equation}
\xi_c^{(n)}= 1+ \frac{n^2\pi^2}{ 2\Lambda^2}+{\mathcal O}(\Lambda^{-3}).
\label{criticalxi}
\end{equation}
The critical value increases rapidly with
increasing $n$. As a result, it is found that
$Z_{cr}\simeq 173$ is a critical value at which the lowest energy-level of the bound state
$1S_{1/2}$ encounters the negative energy continuum, while other bound states encounter the negative energy continuum at $Z_{cr}>173$ (see also \citet{g1c} for a numerical estimation of the same spectrum). 
It is shown that $Z_{cr}\simeq 170$ for a bare nucleus and $Z_{cr}\simeq 173$ (\citet{popov1972a,popov1972b}) for a nucleus with only 
$K$-shell unoccupied. Note that two nuclei with charges $Z_1$ and $Z_2$ respectively, if $Z_1>Z_2$ and $K$-shell of 
the $Z_1$-nucleus is empty, then $Z_2$ may be neutral atom. In this case two nuclei make a quasi molecular state for
which the ground term $(1s\sigma)$ is unoccupied by electrons:  
so spontaneous production of positrons is also possible (\citet{popov1973a,popov1973b}).    
We refer the readers to \citet{z10,z11a,z11b,z12,z,popov2001} for mathematical and numerical details.

When $Z>Z_{cr}=173$, the lowest energy-level of the bound state $1S_{1/2}$ enters the negative energy continuum, 
its energy-level can be estimated as follow,
\begin{equation}
{\mathcal E}(1S_{1/2})=m_ec^2 - \frac{Z \alpha}{\bar r}<-m_ec^2,
\label{1S}
\end{equation}
where $\bar r $ is the average radius of the $1S_{1/2}$ state's orbit, and the binding energy of this
state $Z\alpha/\bar r > 2 m_ec^2$. If this bound state is unoccupied, 
the bare nucleus gains a binding energy $Z\alpha/\bar r$ larger than 
$2m_ec^2$, and becomes unstable against the production of an electron-positron pair. Assuming this 
pair-production occur around the radius $\bar r$, we have energies of electron ($\epsilon_-$) and positron ($\epsilon_+$):
\begin{equation}
\epsilon_-=\sqrt{|c{\bf p}_-|^2+m_e^2c^4}-\frac{Z \alpha}{\bar r};\ \quad \epsilon_+=\sqrt{|c{\bf p}_+|^2+m_e^2c^4}+\frac{Z \alpha}{\bar r},
\label{eofep}
\end{equation}
where ${\bf p}_\pm$ are electron and positron momenta, and ${\bf p}_-=-{\bf p}_+$. 
The total energy required for a pair production is,
\begin{equation}
\epsilon_{-+}=\epsilon_-+\epsilon_+=2\sqrt{|c{\bf p}_-|^2+m_e^2c^4},
\label{totaleofep}
\end{equation}
which is independent of the potential $V(\bar r)$. The potential energies $\pm eV(\bar r)$ of electron 
and positron cancel 
each other and do not contribute to the total energy (\ref{totaleofep}) required for pair production. 
This energy (\ref{totaleofep}) is acquired from the binding energy ($Z\alpha/\bar r > 2 m_ec^2$)
by the electron filling into the bound state $1S_{1/2}$. A part of the binding energy becomes 
the kinetic energy of positron that goes out.  
This is analogous to the familiar case that a proton ($Z=1$)
catches an electron into the ground state $1S_{1/2}$, and a photon is emitted with the energy not less than
13.6 eV.   
In the same way, 
more electron-positron pairs are produced, when $Z\gg Z_{cr}=173$ the energy-levels of 
the next bound states $2P_{1/2},2S_{3/2},\cdot\cdot\cdot$ 
enter the negative energy continuum, provided these bound states of bare nucleus are unoccupied.  

\subsection{Positron production}

\citet{GZ69,GZ70} proposed that when $Z>Z_{cr}$ the bare nucleus produces spontaneously pairs of electrons and positrons: the two positrons\footnote{Hyperfine structure of $1S_{1/2}$ state: single and triplet.} 
go off to infinity and the effective charge of the bare nucleus decreases
by two electrons, which corresponds exactly to filling the K-shell\footnote{The supposition was made 
by \citet{GZ69,GZ70} that the electron density of $1S_{1/2}$ state, as well as the vacuum polarization density, 
is delocalized at $Z\rightarrow Z_{cr}$. Further it was proved to be incorrect (\citet{z11a,z11b,z}).}
A more detailed investigation was made for the solution of
the Dirac equation at $Z\sim Z_{cr}$, when the lowest electron level $1S_{1/2}$ merges with the
negative energy continuum, by \citet{z10,z11a,z11b,z12,z65}. It was there further clarified the situation, showing that at
$Z\gtrsim Z_{cr}$, an imaginary resonance energy of Dirac equation appears,
\begin{equation}
\epsilon = \epsilon_0 - i\frac{\Gamma}{2},
\label{zimaginary}
\end{equation}
where
\begin{eqnarray}
\epsilon_0 &=&-m_ec^2- a(Z-Z_{cr}),
\label{epsilon0}\\
\Gamma &\sim& \theta(Z-Z_{cr})\exp \left(-b\sqrt\frac{Z_{cr}}{ Z-Z_{cr}}\right),
\label{zprobability}
\end{eqnarray}
and $a,b$ are constants, depending on the cutoff $\Lambda$ (for example, $b=1.73$ for $Z=Z_{cr}=173$, see \citet{z11a,z11b,z}). The
energy and momentum of emitted positrons are $|\epsilon_0|$ and $|{\bf p}|=\sqrt{|\epsilon_0|-m_ec^2}$.

The kinetic energy of the two positrons at infinity is given by
\begin{equation}
\varepsilon_p = |\epsilon_0| - m_ec^2 = a(Z-Z_{cr})+\cdot\cdot\cdot,
\label{zkinetic}
\end{equation}
which is proportional to $Z-Z_{cr}$ (so long as $(Z-Z_{cr})\ll Z_{cr}$) and tends to zero as $Z\rightarrow Z_{cr}$.
The pair-production resonance at the energy (\ref{zimaginary}) is extremely narrow
and practically all positrons are emitted with almost same kinetic energy for $Z\sim Z_{cr}$, i.e. nearly
monoenergetic spectra (sharp line structure).
Apart from a pre-exponential factor, $\Gamma$ in Eq.~(\ref{zprobability}) coincides with the probability of positron production, i.e., the penetrability of the Coulomb barrier.
The related problems of vacuum charge density due to electrons filling into the K-shell and charge renormalization due to the change of wave function of electron states are discussed by \citet{z20,z21,z22,z23,z24}.
An extensive and detailed review on this theoretical issue can be found in \citet{grc98,z,popov2001,Greinerbook}.

On the other hand, some theoretical work has been done studying the possibility that
pair production due to bound states encountering the negative energy continuum is
prevented from occurring by higher order processes of quantum
field theory, such as charge renormalization, electron self-energy and nonlinearities in electrodynamics and even Dirac field itself (\citet{g3a,g3b,g3c,g3d,grc98-5a,grc98-5b,grc98-6}).
However, these studies show that various effects modify $Z_{cr}$ by a few percent, but have no way to prevent the binding energy from increasing to $2m_ec^2$ as $Z$ increases, without simultaneously contradicting the existing precise experimental data on stable
atoms (\citet{g}). Contrary claim (\citet{grc98-7}) according to which bound states are repelled by the lower continuum through some kind of self screening appear to be unfounded
(\citet{grc98}).

It is worth noting that an over critical nucleus ($Z\ge Z_{cr}$) can be formed for example in the collision
of two heavy nuclei (\citet{z65,GZ69,GZ70,greiner1972a,greiner1972b}).
To observe the emission of positrons originated from pair production occurring near to an overcritical nucleus
temporally formed by two nuclei, the following necessary conditions have to be full filled:
(i) the atomic number of an over critical nucleus is larger than $Z_{cr}=173$;
(ii) the lifetime of the over critical
nucleus must be much longer than the characteristic time $(\hbar/m_ec^2)$ of pair production;
(iii) inner shells (K-shell) of the over critical nucleus should be unoccupied.

\subsection{A transient super heavy ``quasimolecules''}

Experiments on heavy-ion collisions are expected to observe spontaneously emitted positrons
from pair production associated with the overcritical field of the two colliding nuclei.
An essential idea is to use heavy-ion collisions to form transient super heavy
``quasimolecules'' that is a metastable nuclear complex (\citet{z65,g,GZ69,GZ70,greiner1972a,greiner1972b}).
Due to the heavy masses ($M_n$) of the nuclei, the relative velocity required to bring two nuclei into contact is non-relativistic and much smaller than the velocity of the fast-moving inner-shell electrons. At small internuclear distances, well within the electron's orbiting
radii, the electrons cannot distinguish between the two nuclear centers and they evolve as if they were bounded by all $Z_1+Z_2$ protons of the two nuclear charges. Thus, one would expect electrons to evolve quasi 
statically through a series
of well defined quasimolecule states in the two-center field of the nuclei as the internuclear separation decreases and then increases again. 

Moreover, the time-varying overcritical electric field of quasimolecules is supposed to be
created by the collision of two nuclei, whose total atomic number $Z_1+Z_2$ is larger than the
critical value $Z_{cr}=173$, for instance the Uranium-Uranium collision. The
time scale of such time-varying electric field, which is related to the time of two nuclei collision and reaction, is much longer than the evolution time of electrons in
quasimolecule states. As consequence, the adiabatic approximation can be applied, that is
the Coulomb potential of the two colliding nuclei varies sufficiently slowly for electrons in inner shells to adjust adiabatically. Note that our discussion focus only on the lowest lying energy-level $1s\sigma$ of quasimolecule.

A critical parameter is the internuclear separation $R=R_{cr}$
at which the electron binding energy of Coulomb potential $Z_1+Z_2$ exceeds the energy-gap $2m_ec^2$ and the level $1s\sigma$ of quasimolecule begins to penetrate into the
negative-energy continuum. It is then assumed that at the minimum $R_{min}$
of internuclear separation, where $R_{min} < R_{cr}$, such a metastable quasimolecule
forms for the sticking time $\Delta t_s$ that is associated with the nuclear delay time
due to nuclear reactions (\citet{grc98}). If this sticking time $\Delta t_s$ is much longer than
$\hbar/m_ec^2=1.288\cdot 10^{-21}$ sec, pair-production processes of electron and positron
have enough time to occurs. Moreover, due to the Pauli principle the quasimolecular
state $1s\sigma$ must be vacant for pair-production to occur. Two electrons then fill in the vacancy of the energy-level $1s\sigma$ of the quasimolecule. As discussed in the previous section, at the same time two positrons
are kicked out to infinity with the nearly monoenergetic spectrum (a sharp line structure) (see Eq.~(\ref{zkinetic})),
\begin{equation}
\varepsilon_{\rm peak}=|\varepsilon_{1s\sigma}(Z)|-m_ec^2,
\label{pspectrum}
\end{equation}
where $\varepsilon_{\rm 1s\sigma}$ is the energy-level of the $1s\sigma$ state
and the combined nuclear charge $Z=Z_1+Z_2 \gtrsim Z_{cr}$. While, the vacancy of the energy-level $1s\sigma$
is provided by both a radial variation and a rotation of the
internuclear axis that induce the ionization and excitation of electrons
in the quasimolecular state $1s\sigma$, during the heavy-nuclei collision
(\citet{g,grc98-31,grc98-32,grc98-33a,grc98-33b,grc98-34,g14,g15,g16,g17}).
 One would expect the nearly monoenergetic spectrum (a sharp line structure)
from pair production positron emission alone in a quasi static
situation. The detection of positrons from pair production becomes challenging (\citet{g7a,g7b,g7c,g8,g9}).

\subsection{Numerical simulations}

The collision of two Uranium nuclei: $Z=92$ was considered (\citet{z}).
The conservation of energy in the collision reads:
\begin{equation}
M_nv_0^2=(Ze)^2/R_{min},
\label{zmotion}
\end{equation}
where $v_0$ is the relative velocity of the nuclei at infinity,
$R_{min}$ is the smallest distance, and $M_n$ is the Uranium atomic mass. In order to have $R_{min}\simeq 30$fm it needs $v_0\simeq 0.034c$, the
characteristic collision time is then $\Delta t_c= R_{min}/v_0\simeq 10^{-20}$s. On the other hand, the typical velocity of an electron in the inner shell ($r \sim 115.8$fm) is $v \sim c$ and therefore its characteristic time $\Delta\tau_0\sim r/v\sim 4 \cdot 10^{-22}$s. This means that the characteristic collision time
$\Delta t_c$ in which the two colliding nuclei are brought into contact and separated again is much larger than the time scale $\Delta\tau_0$ of electron evolution. In general, it is
required that bombarding energies ($M_nv_0^2/2$) of nuclei are not much above the nuclear
Coulomb barrier. This gives justification for an adiabatic description of the collision
in terms of {\it quasimolecules}. The formation of ``quasimolecules'' can also be
verified by the characteristic molecular-orbital X-rays radiation due to the electron
transitions between ``quasimolecules'' orbits (\citet{g,g11,g12a,g12b,g12c,g12d,g13a,g13b}).

In addition, \citet{grc98-18a,grc98-18b} showed that the critical binding of $2m_ec^2$
should be reached and the level $1s\sigma$ begins to penetrate into the negative-energy
continuum in two Uranium nuclei collisions at a distance $R_{cr}\simeq 30$fm by solving the Dirac equation with two Coulomb centers.
However, it is necessary to quantitatively solve the {\it time-dependent}
two-center Dirac equation,
since the nuclei move on their Rutherford trajectories which causes the wave functions and
binding energies to vary rapidly with time and also leads to strong dynamically induced transitions.

\begin{figure}
\centering 
\includegraphics[width=\hsize,clip]{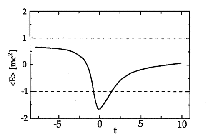}
\caption{Energy expectation values of the $1s\sigma$ state in a U+U collision at 10 GeV/nucleon.
The unit of time is $\hbar/m_ec^2$. This figure is reproduced from Fig.~4 in \citet{grc98}.}%
\label{example}%
\end{figure}

In \citet{grc98}, it is briefly discussed the dynamics of the electron field
in ``slow'' collisions of heavy nuclei where the total charge is sufficiently large to let the quaimolecular $1s\sigma$-state enter the negative energy
continuum at critical distance $R_{cr}$. The time-dependent electronic wave function
$\Psi_i({\bf r},t)$, which satisfy the usual boundary conditions as $t\rightarrow -\infty$,
can be expanded as follows
\begin{eqnarray}
\Psi_i({\bf r},t)&=&\sum _j a_{ij}(t)\psi_i({\bf r},{\bf R}(t)) e^{-i\chi_j(t)},\label{wave0}\\
\chi_j(t) &=& \int_{t_0}^t dt\langle\psi_j|\hat H_{TC}|\psi_j\rangle
\label{wave}
\end{eqnarray}
where $\{\psi_j\}$ is a basis (containing bound states and two sets of continuum states)
of adiabatic, quasimolecular eigenstates of the two-center Dirac Hamiltonian $\hat H_{TC}$,
\begin{equation}
\left(\hat H_{TC}({\bf r},{\bf R}(t))-E_j({\bf R})\right)\psi_i({\bf r},{\bf R}(t))= 0.
\label{eigenequation}
\end{equation}
The time dependent expansion coefficients $a_{ij}(t)$ are determined by solving a
set of coupled ordinary differential equations, the coupled channel equations.
A considerable number of approaches has been developed to attack this problem,
we refer the readers to the review article by \citet{grc98} and references
\citet{grc98-19,grc98-20a,grc98-20b,grc98-21a,grc98-21b,grc98-21c,grc98-22,grc98-23a,grc98-23b,grc98-24,grc98-25} for details.
As an example Fig.~\ref{example} shows the result of a calculation in which the
time-dependent two-center Dirac equation was solved with a finite-difference method
(\citet{grc98-24}). The calculation demonstrates that the mean binding energy
increases very strongly and well exceeds the critical value $2m_ec^2$.

It is worth noting that several other dynamical processes contribute to
the production of positrons in under critical as well as in overcritical collision systems (\citet{g3a,g3b,g3c,g3d}).
Due to the time-energy uncertainty relation (collision broadening),
the energy-spectrum of such positrons has a rather broad and
oscillating structure, considerably different from a sharp line structure
that we would expect from pair-production positron emission alone.

When in the course of a heavy-ion collision the two nuclei come into contact, i.e.,
deep-inelastic reactions, a nuclear reaction can occur that lasts a certain time $\Delta t_s$. 
The length of this contact (sticking) or delay time
depends on the nuclei involved in the reaction and on beam energy. For very heavy nuclei, the Coulomb
interaction is the dominant force between the nuclei,
so that the sticking times $\Delta t_s$ are typically much shorter and in the mean
probably do not exceed $1\sim 2\cdot 10^{-21}$ sec (\citet{grc98}). Accordingly the calculations (see
Fig.~\ref{example})
also show that the time when the binding energy is over critical is very short,
about $1.2\cdot 10^{-21}$ sec. Up to now no conclusive theoretical or experimental
evidence exists for long nuclear delay times in very heavy collision systems.

While the pair production rate should increase
dramatically when the sticking time exceeds about $3\hbar/m_ec^2\sim 3\cdot 10^{-21}$
sec. It would be worthwhile task to study in details the nuclear aspects
of heavy-ion collisions at energies very close to the Coulomb barrier and search
for conditions, which would serve as a trigger for prolonged nuclear reaction times,
(the sticking time $\Delta t_s$) to enhance the amplitude of pair production (\citet{grc98,g,g5,g6,grc98-30}).

\subsection{Experiments}

As already remarked, if the sticking time $\Delta t_s$ is prolonged, the probability of pair production in vacuum around the super heavy nucleus is enhanced. As a consequence, the spectrum of emitted positrons develops a sharp line structure, indicating the
spontaneous vacuum decay caused by the overcritical electric field of a forming super heavy nuclear system with $Z\ge Z_{cr}$. If the striking time $\Delta t_s$ is not long enough
and the sharp line of pair production positrons has not yet well-developed, in observed
positron spectrum it is difficult to distinguish the pair production
positrons from positrons created through other different mechanisms.
Prolonging the sticking time and identifying pair production positrons among all other
particles (\citet{g10,g11}) created in the collision process are important experimental tasks (\citet{g7a,g7b,g7c,g8,g9,g18-22a,g18-22b,g18-22c}).

\begin{figure}
\centering 
\includegraphics[width=\hsize,clip]{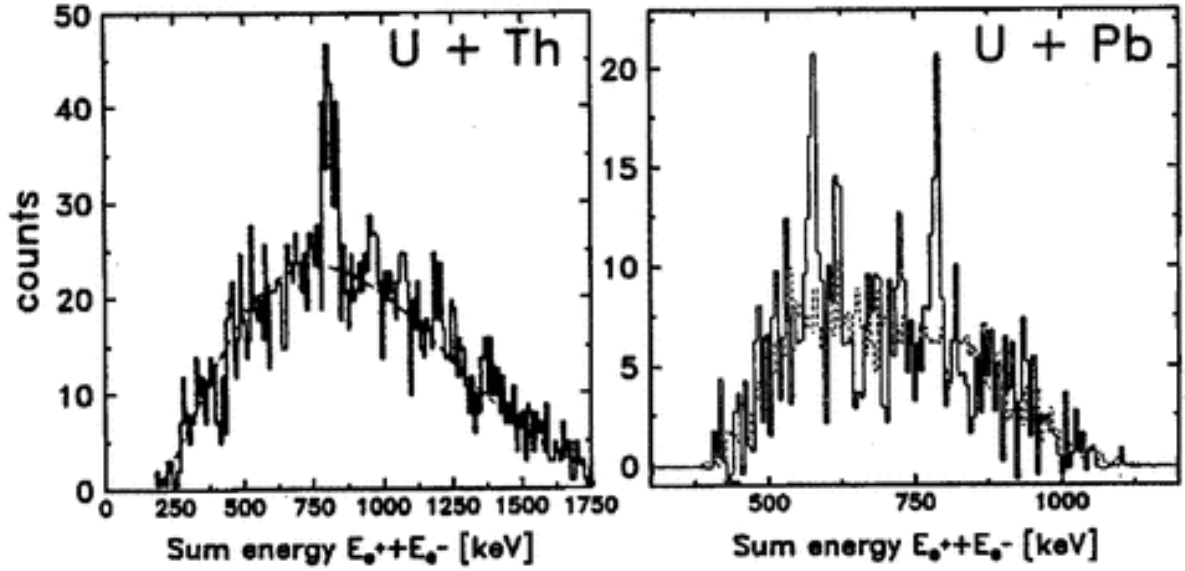}
\caption{Two typical example of coincident electron-positron spectra measured by the
Epose group in the system U+Th (left) and by the Orange group in U+Pb collisions (right).
When plotted as a function of the sum energy of electron and positron very narrow line structures were observed.
This figure is reproduced from Fig.~7 in \citet{grc98}.}%
\label{linestructure}%
\end{figure}

For nearly 20 years the study of atomic excitation processes and in particular of positron creation in heavy-ion collisions has been a major research topic at GSI (Darmstadt) (\citet{zexp1,G...96,L...97,zexp4,H...98}).
The Orange and Epos groups at GSI (Darmstadt) discovered narrow line structures
(see Fig.~\ref{linestructure})
of unexplained origin, first in the single positron energy spectra and later in coincident
electron-positron pair emission. Studying more collision systems with a wider range of the
combined nuclear charge $Z=Z_1+Z_2$ shows that narrow line structures is essentially
independent of $Z$. This rules out the explanation of pair-production positron,
since the line would be expected at the position of the $1s\sigma$ resonance, i.e. at a
kinetic energy given by Eqs.~(\ref{zkinetic}) and (\ref{pspectrum}), which is strongly $Z$ dependent.
Attempts to link this positron line to spontaneous pair production have failed.
Other attempts to explain this positron line in term of atomic
physics and new particle scenario were not successful as well (\citet{grc98}).

The anomalous positron line problem has perplexed experimentalists and theorists alike for more than a decade.
Moreover, later results
obtained by the Apex collaboration at Argonne National Laboratory showed no statistically significant positron line structures (\citet{A...95,grc98-52b}). This is in strong contradiction with
the former results obtained by the Orange and Epos groups. However, the analysis of
Apex data was challenged in the comment by \citet{grc98-53a,grc98-53b} for the Apex measurement
would have been less sensitive to extremely narrow positron lines. A new generation of
experiments (Apex at Argonne and the new Epos and Orange setups at GSI) with much improved
counting statistics has failed to reproduce the earlier results (\citet{grc98}).

To overcome the problem posed by the short time scale of
pair production ($10^{-21}$ sec), hopes rest on the idea to select collision systems
in which a nuclear reaction with sufficient sticking time occurs. Whether such situation
can be realized still is an open question (\citet{grc98}). In addition, the anomalous positron line
problem and its experimental contradiction overshadow on the field of studying
the pair production in heavy ion collisions.
In summary, clear experimental signals for electron-positron pair production in heavy ion 
collisions are still missing (\citet{grc98}) at the present time.

\section{Vacuum polarization in uniform electric field and in Kerr-Newman geometries}

\subsection{Early works on pair production}

\subsubsection{Klein and Sauter works}

\begin{figure}
\begin{picture}(350,350)
\put(60,50){\includegraphics[height=7.8cm,width=10.8cm]{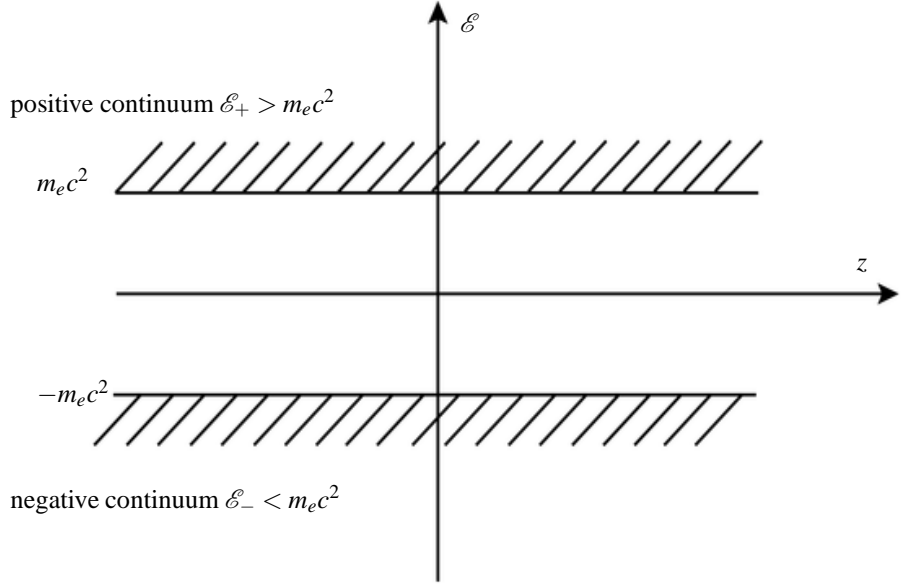}}
\put(350,170){$z$}
\put(200,260){${\mathcal E}$}
\put(30,230){positive continuum ${\mathcal E}_+>m_ec^2$}
\put(30,80){negative continuum ${\mathcal E}_-<m_ec^2$}
\put(40,200){$m_ec^2$}
\put(40,120){$-m_ec^2$}
\end{picture}
\caption{The mass-gap $2m_ec^2$  that separates the positive continuum spectrum ${\mathcal E}_+$
from the negative continuum spectrum ${\mathcal E}_-$.}%
\label{gap}%
\end{figure}

It is well known that every relativistic wave equation of a free relativistic 
particle of mass $m_e$, momentum ${\bf p}$ and energy ${\mathcal E}$,
admits symmetrically ``positive energy'' and ``negative energy'' solutions. Namely the wave-function 
\begin{equation}
\psi^{\pm}({\bf x},t)\sim e^{\frac{i}{\hbar}({\bf k}\cdot{\bf x}-{\mathcal E}_{\pm}t)}
\label{freeparticle}
\end{equation}
describes a relativistic particle, whose energy, mass and momentum must satisfy,
\begin{equation}
{\mathcal E}_{\pm}^2=m_e^2c^4 +c^2|{\bf p}|^2;\quad {\mathcal E}_\pm=\pm\sqrt{m_e^2c^4 +c^2|{\bf p}|^2},
\label{klein0} 
\end{equation}
this gives rise to the familiar positive and negative energy spectrum (${\mathcal E}_\pm$) of positive and negative energy states ($\psi^{\pm}({\bf x},t)$) of the relativistic particle, as represented in Fig.~\ref{gap}. 
In such free particle situation (flat space, no external field), 
all the quantum states are stable; that is, there is no possibility of ``positive'' (``negative'') energy states decaying into a ``negative'' (``positive'') energy states, since all negative energy states are fully filled and
there is an energy gap $2m_ec^2$ separating the negative energy spectrum from the positive energy spectrum. This is the 
view of Dirac theory on the spectrum of a relativistic particle (\citet{Dir30,Dir33}). 

Klein studied a relativistic particle moving in an external {\it constant} potential $V$ and in this case 
Eq.~(\ref{klein0}) is modified as 
\begin{equation}
[{\mathcal E}-V]^2=m_e^2c^4 +c^2|{\bf p}|^2, ;\quad {\mathcal E}_\pm=V\pm\sqrt{m_e^2c^4 +c^2|{\bf p}|^2}.
\label{klein0bis} 
\end{equation}
He solved this relativistic wave equation by considering an incident free relativistic wave of positive energy states
scattered by the constant potential $V$, leading to reflected and transmitted waves. He found a paradox that in 
the case $V\ge {\mathcal E}+m_ec^2$, the reflected flux is larger than the incident flux $j_{\rm ref} >j_{\rm inc}$, 
although the total flux is conserved, i.e. $j_{\rm inc}=j_{\rm ref}+j_{\rm tran}$. This was known as the 
Klein paradox (see \citet{klein}). This implies that negative energy states have contributions to both the 
transmitted flux $j_{\rm tran}$ and reflected flux $j_{\rm ref}$.     

\begin{figure}
\includegraphics[width=\hsize,clip]{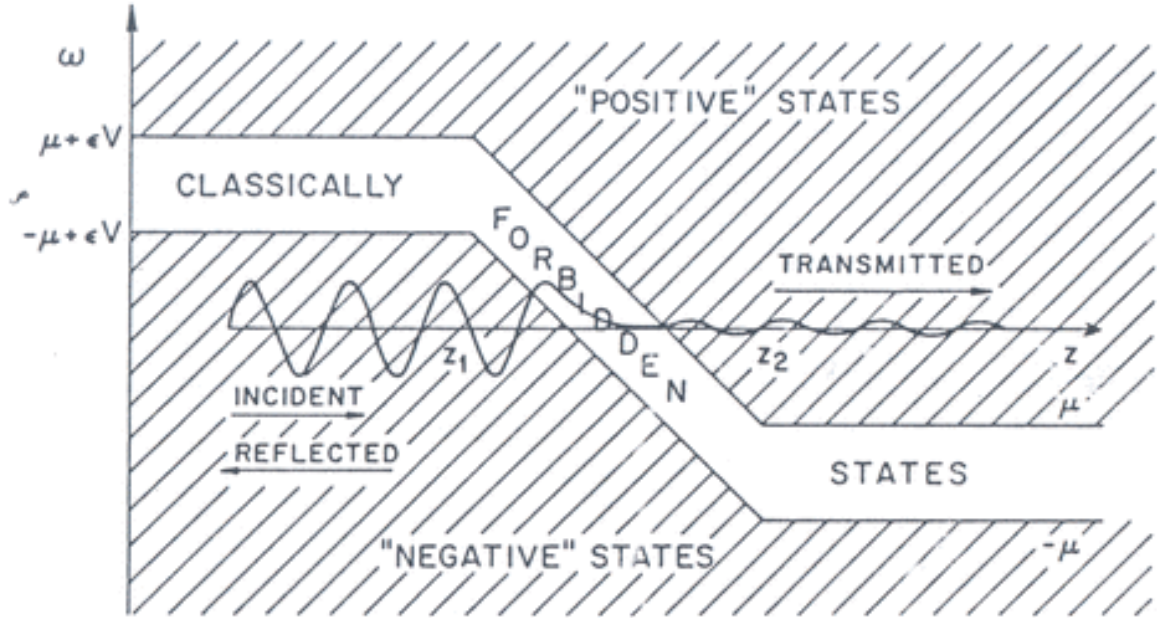}
\caption{In presence of a strong enough electric field the boundaries of the classically allowed states 
(``positive'' or ``negative'') can also be so tilted that a ``negative'' is at the same level as a ``positive'' 
(level crossing). Therefore a ``negative'' wave-packet from the left will be partially transmitted,
after an exponential damping due to the tunneling through the classically forbidden states, as s ``positive''
wave-packet outgoing to the right. This figure
is reproduced from Fig.~II in \citet{demourwkb}, and $\mu=m_ec^2, \epsilon V=V(z), \omega={\mathcal E}$.}%
\label{demourgape}%
\end{figure}

Sauter studied this problem by considering an electric potential of an external constant 
electric field $E$ in the $\hat {\bf z}$ direction (\citet{s31}). In this case the energy 
${\mathcal E}$ is shifted by the amount $V(z)=-e Ez$, where $e$ is the electron charge. 
In the case of the electric field $E$ uniform between $z_1$ and $z_2$ and null outside,
Fig.~\ref{demourgape} represents the corresponding sketch of allowed states.
The key point now, which is the essence of the Klein paradox (\citet{klein}), is that the above mentioned stability of the ``positive energy'' states is lost for sufficiently strong electric fields. The same is true for ``negative energy'' states. Some ``positive energy'' and ``negative energy'' states have the same energy-levels, i.e. the crossing of
energy-levels occurs. Thus, these ``negative energy'' waves incident from the left will be both reflected back by the electric field and partly transmitted to the right as a ```positive energy'' wave, as shown in Fig.~\ref{demourgape} (\citet{demourwkb}). This transmission is nothing else but a quantum tunneling of the wave function through the 
electric potential barrier, where classical states are forbidden. This is the same as the so-called 
the Gamow tunneling of the wave function through nuclear potential barrier (Gamow-wall, see \citet{gamow-book}). 

Sauter first solved the relativistic Dirac equation (\citet{Dir30,Dir33})
in the presence of the constant electric field by the ansatz,
\begin{equation}
\psi_s({\bf x},t)= e^{\frac{i}{\hbar}(k_xx+k_yy-{\mathcal E}_\pm t)}\chi_{s_3}(z)
\label{Eparticle}
\end{equation}
where spinor function $\chi_{s_3}(z)$ obeys the following equation ($\gamma_0, \gamma_i $ are Dirac matrices)
\begin{equation}
\left[\hbar c\gamma_3\frac{d}{dz}+\gamma_0(V(z)-{\mathcal E}_\pm)+(m_ec^2+ic\gamma_2p_y
+ic\gamma_1 p_x)\right]\chi_{s_3}(z)=0,
\label{sauterfunction}
\end{equation}
and the solution $\chi_{s_3}(z)$ can be expressed in terms of hyper geometric functions (\citet{s31}). 
Using this wave-function $\psi_s({\bf x},t)$ (\ref{Eparticle}) and the flux 
$ic\psi_s^\dagger\gamma_3\psi_s$, Sauter computed the transmitted flux of 
positive energy states, the incident and reflected
fluxes of negative energy states, as well as exponential decaying flux of classically forbidden states, 
as indicated in Fig.~\ref{demourgape}. Using continuous conditions of wave functions and fluxes 
at boundaries of the potential, Sauter found that         
the transmission coefficient $|T|^2$ of the wave through the electric potential 
barrier from the negative energy state to positive energy states:
\begin{equation}
|T|^2=\frac{|{\rm transmission\hskip0.2cm flux}|}{|{\rm incident\hskip0.2cm flux}|}
\sim e^{-\pi\frac{m_e^2c^3}{\hbar e E}}.
\label{transmission}       
\end{equation}
This is the probability of negative energy states decaying to positive energy states, caused by an external
electric field. The method that Sauter adopted to calculate the transmission coefficient $|T|^2$ is the same as
the one Gamow used at that time to calculate quantum tunneling of the wave function 
through nuclear potential barrier (Gamow-wall), leading to the $\alpha$-particle emission (\citet{gamow-book}).

\subsubsection{Heisenberg-Euler-Weisskopf effective theory}

To be able to explain elastic light-light scattering (\citet{ek1935}),
Heisenberg, Euler and Weisskopf proposed a  theory that
attributes to the vacuum certain
non-linear electromagnetic properties, as if it were
a dielectric and
permeable medium (\citet{he35,weisskopf0}).

Let ${\mathcal L}$ to be the Lagrangian density of electromagnetic fields ${\bf E},{\bf B}$,
a Legendre transformation
produces the Hamiltonian density:
\begin{equation}
{\mathcal H}=E_i{\frac{\delta {\mathcal L}}{ \delta E_i}} - {\mathcal L}.
\label{hlrelation}
\end{equation}
In  Maxwell's theory, the two densities are given by
\begin{equation}
{\mathcal L}_M={\frac{1}{8\pi}}({\bf  E}^2-{\bf B}^2),\hskip0.5cm {\mathcal H}_M
={\frac{1}{8\pi}}({\bf E}^2+{\bf B}^2).
\label{maxwell}
\end{equation}
To quantitatively describe non-linear electromagnetic properties of the vacuum
based on the Dirac theory, the above authors
introduced the concept of an
 effective Lagrangian ${\mathcal L}_{\rm eff}$
of the vacuum state in the presence of electromagnetic fields, and an associated
Hamiltonian density
\begin{equation}
{\mathcal L}_{\rm eff}={\mathcal L}_M+\Delta {\mathcal L},\hskip0.5cm
{\mathcal H}_{\rm eff}={\mathcal H}_M + \Delta {\mathcal H}.
\label{effectivelh}
\end{equation}
>From these one derives
induced fields $\bf  D,\bf  H$ as the derivatives
\begin{equation}
D_i=4\pi{\frac{\delta {\mathcal L}_{\rm eff}}{\delta E_i}},\hskip0.5cm H_i=-4\pi{\frac{\delta {\mathcal L}_{\rm eff}}{\delta B_i}}.
\label{dh}
\end{equation}
In Maxwell's theory, $\Delta {\mathcal L}\equiv 0$ in the vacuum, so that
$\bf  D={\bf E}$ and $\bf  H={\bf B}$.
In Dirac's theory, however, $\Delta {\mathcal L}$ is
a complex function of ${\bf E}$ and ${\bf B}$.
Correspondingly, the vacuum behaves as a dielectric and permeable medium (\citet{he35,weisskopf0}) in which,
\begin{equation}
D_i=\sum_k\epsilon_{ik}E_k,\hskip0.5cm H_i=\sum_k\mu_{ik}B_k,
\label{dh1}
\end{equation}
where complex $\epsilon_{ik}$ and $\mu_{ik}$ are the field-dependent dielectric and permeability tensors of the vacuum. 

The discussions on complex dielectric and permeability tensors ($\epsilon_{ik}$ and $\mu_{ik}$) 
can be found for example in \citet{landaumedium}.
The effective Lagrangian and Hamiltonian densities in such a medium is given by,
\begin{equation}
{\mathcal L}_{\rm eff}={\frac{1}{8\pi}}({\bf E}\cdot {\bf  D}-{\bf B}\cdot \bf  H),\hskip0.5cm {\mathcal H}_{\rm eff}
={\frac{1}{8\pi}}({\bf E}\cdot {\bf  D}+{\bf B}\cdot {\bf  H}).
\label{effmaxwell}
\end{equation}
In this medium, the conservation of electromagnetic energy
has the form
\begin{equation}
-{\rm div}{\bf  S}={\frac{1}{4\pi}}\left({\bf E}\cdot{\frac{\partial {\bf  D}}{\partial t}}+{\bf B}\cdot {\frac{\partial {\bf  H}}{\partial t}}\right),\hskip0.5cm \bf  S={\frac{c}{4\pi}}{\bf E}\times {\bf B},
\label{effcons}
\end{equation}
where $\bf  S$ is the Poynting vector  describing
the density of electromagnetic energy flux.
Let us consider that electromagnetic fields are complex and monochromatic
\begin{equation}
{\bf E}={\bf E}(\omega) \exp-i(\omega t);\quad 
{\bf B}={\bf B}(\omega) \exp-i(\omega t),
\label{monochromaticfields}
\end{equation} 
of frequency $\omega$, and dielectric and permeability tensors are frequency-dependent, i.e., $\epsilon_{ik}(\omega)$ 
and $\mu_{ik}(\omega)$. 
Substituting these fields and tensors into the r.h.s.~of Eq.~(\ref{effcons}),
one obtains the dissipation of electromagnetic energy per time into the medium,
\begin{equation}
Q={\frac{\omega}{8\pi}}\left\{{\rm Im}\left[\epsilon_{ik}(\omega)\right]E_i E^*_k
+{\rm Im}\left[\mu_{ik}(\omega)\right]B_iB^*_k\right\}.
\label{dissipation}
\end{equation}
This is nonzero if $\epsilon_{ik}(\omega)$ and $\mu_{ik}(\omega)$ contain an imaginary part. The dissipation of electromagnetic 
energy is accompanied by heat production. In the light of the third thermodynamical law of entropy increase,
The energy lost $Q$ of electromagnetic fields in the medium is always positive, i.e., $Q>0$. As a consequence,
${\rm Im}[\epsilon_{ik}(\omega)]>0$ and ${\rm Im}[\mu_{ik}(\omega)]>0$.  
The real parts of $\epsilon_{ik}(\omega)$ and $\mu_{ik}(\omega)$
represent an electric and magnetic polarizability of the vacuum and
leads, for example, to the refraction of light in an electromagnetic field, 
or to the elastic scattering of light from light (\citet{ek1935}). 
The $n_{ij}(\omega)=\sqrt{\epsilon_{ik}(\omega)\mu_{kj}(\omega)}$ is the reflection index of the medium.
The field-dependence of $\epsilon_{ik}$ and $\mu_{ik}$ implies
non-linear electromagnetic properties of the vacuum as a dielectric and permeable medium.

The effective Lagrangian density (\ref{effectivelh}) is a relativistically
invariant function of the
field strengths ${\bf E}$ and ${\bf B}$.
Since $({\bf E}^2-{\bf B}^2)$ and $({\bf E}\cdot {\bf B})^2$ are relativistic invariants, one can formally expand
$\Delta {\mathcal L}$ in powers of weak field strengths:
\begin{equation}
\Delta {\mathcal L} = \kappa_{20}
 ({\bf E}^2-{\bf B}^2)^2+\kappa_{02}
 ({\bf E}\cdot {\bf B})^2
+ \kappa _{30} ({\bf E}^2 -{\bf B}^2)^3 + \kappa _{12} ({\bf E}^2- {\bf B}^2)({\bf E}\cdot {\bf B})^2+\dots~ ,
\label{leffective}
\end{equation}
where $\kappa_{ij}$ are field-independent constants
whose subscripts indicate the powers of
$({\bf E}^2-{\bf B}^2)$ and
${\bf E}\cdot{\bf B}$, respectively.
Note that the invariant
${\bf E}\cdot{\bf B}$ appears only in even powers since it is odd under parity
and electromagnetism is
parity invariant.
The Lagrangian density (\ref{leffective}) corresponds, via
relation (\ref{hlrelation}), to
\begin{eqnarray}
\Delta {\mathcal H}&=&\kappa_{2,0}
 ({\bf E}^2-{\bf B}^2)(3{\bf E}^2+{\bf B}^2)+ \kappa _{0,2}
 ({\bf E}\cdot {\bf B})^2 \nonumber\\
&&+ \kappa _{3,0} ({\bf E}^2 -{\bf B}^2)^2(5{\bf E}^2+{\bf B}^2)+
 \kappa _{1,2}(3{\bf E}^2- {\bf B}^2)({\bf E}\cdot {\bf B})^2+\dots~ .
\label{heffective}
\end{eqnarray}
To obtain ${\mathcal H}_{\rm eff}$
in Dirac's theory,
one has to calculate
\begin{equation} \Delta {\mathcal H}=\sum_k\left\{\psi^*_k,\Big[
\mbox{\boldmath$\alpha$}\cdot(-i h c{\bf \nabla} + e{\bf A}\
)) +\beta m_ec^2\Big]\psi_k\right\},
\label{wcal}
\end{equation}
where $\alpha_i,\beta$ are Dirac matrices, ${\bf  A}$
is the vector potential, and
$\{\psi_k(x)\}$
are the wave functions
of the occupied negative-energy states.
When performing the sum,
one encounters infinities
which were removed
by \citet{weisskopf0,dirac1934,heisenberg1934,weisskopf1934}
by a suitable subtraction.

\citet{heisenberg1934}
expressed the
Hamiltonian density in terms of the
density matrix
$ \rho (x,x')=\sum_k\psi^*_k(x)\psi_k(x')$ (\citet{dirac1934}).
\citet{ek1935},
\citet{he35}
calculated the
coefficients
$ \kappa _{ij}$.
They did so by solving the Dirac equation
in the presence of parallel electric and magnetic fields
${\bf E}$ and ${\bf B}$ in
a specific direction,
\begin{equation}
\psi_k(x)\rightarrow
\psi_{p_z,n,s_3}\equiv
e^{{\frac{i}{\hbar}}(zp_z-{\mathcal E}t)}u_{n}(y)\chi_{s_3}(x),
\hskip0.5cm  n=0,1,2,\dots~
\label{diracsolution}
\end{equation}
where $\{u_n(y)\}$ are the Landau states (\citet{LanLif75a,LanLif75b}) depending on the magnetic field and
$\chi_{s_3}(x)$
are the spinor functions
calculated
by \citet{s31}.
Heisenberg and Euler used the Euler-Maclaurin formula
to perform the sum over $ n$, and
obtained for the additional Lagrangian  in (\ref{effectivelh})
the integral representation,     
\begin{eqnarray}
 \Delta {\mathcal L}_{\rm eff}&=&\frac{e^2}{16\pi^2\hbar c}\int^\infty_0
e^{-s}\frac{ds}{s^3}\Big[is^2\,\bar E \bar B
\frac{\cos(s[\bar E^2-\bar B^2+2i(\bar E\bar B)]^{1/2})
+{\rm c.c.}}{\cos(s[\bar E^2-\bar B^2+2i(\bar E\bar B)]^{1/2})-{\rm c.c.}}\nonumber\\
&&+ \left(\frac{m_e^2c^3}{e\hbar}\right)^2
+\frac{s^2}{3}(|\bar B|^2-|\bar E|^2)\Big],
\label{effectiveint}
\end{eqnarray}
where
$\bar E,\bar B$ are the dimensionless reduced fields in the unit of the critical field $E_c$,
\begin{equation}
\bar E=\frac{|{\bf E}|}{E_c},~~~~\bar B=\frac{|{\bf B}|}{E_c};~~~~E_c\equiv \frac{m_e^2c^3}{e\hbar}.
\label{dimenlessEB}
\end{equation}
Expanding this in powers of $ \alpha $ up to $ \alpha ^3$
yields
the following values for the four constants:
\begin{equation}
\kappa_{2,0}=\frac{\alpha}{90\pi^2}E_c^{-2},\hskip0.3cm \kappa _{0,2} = 7\kappa_{2,0} ,\hskip0.3cm
\kappa _{3,0}= \frac{32\pi \alpha}{315}E_c^{-4},
\hskip0.3cm \kappa_{1,2} = {\frac{13}{2}} \kappa _{3,0} .
\label{lconstant}
\end{equation}
\citet{weisskopf0} adopted a simpler method.
He considered first
the special case in which ${\bf E}=0, {\bf B}\not=0$
and used the Landau states
to find
$\Delta {\mathcal H}$ of Eq.~(\ref{heffective}),
extracting from this
$\kappa_{2,0}$ and $ \kappa _{3,0}$.
Then he  added a
weak electric field ${\bf E}\not=0$
to calculate perturbatively
its contributions to $\Delta {\mathcal H}$
in the Born approximation (see for example
\citet{LanLif75a,LanLif75b}). This led again to
the coefficients (\ref{lconstant}).

The above results will receive higher corrections in QED
and are correct only up to order $ \alpha ^2$.
Up to this order,
the field-dependent dielectric and permeability tensors $\epsilon_{ik}$
and $\mu_{ik}$ (\ref{dh1}) have the following real parts
for weak fields
\begin{eqnarray}
{\rm Re}(\epsilon_{ik})&=&\delta_{ik} +
\frac{4\alpha}{45}\big[2(\bar E^2-\bar B^2)\delta_{ik}+7\bar B_i\bar B_k\big]
+{\mathcal O}( \alpha ^2),
\nonumber\\
{\rm Re}(\mu_{ik})&=&\delta_{ik} +
\frac{4\alpha}{45}\big[2(\bar E^2-\bar B^2)\delta_{ik}+7\bar E_i\bar B_k\big]
+{\mathcal O}( \alpha ^2).
\label{dielectricity}
\end{eqnarray}

\subsubsection{Imaginary part of the effective Lagrangian}

\citet{he35}  were the first
to  realize that for ${\bf E}\not=0$ the powers
series expansion (\ref{leffective}) is not convergent,
due to singularities of the integrand in (\ref{effectiveint}) at
$s=\pi/\bar E, 2\pi/\bar E,\dots~$. They concluded
that the powers series
expansion (\ref{leffective}) does not yield
all corrections to the Maxwell
Lagrangian, calling for a more careful evaluation
of the integral representation (\ref{effectiveint}).
Selecting an integration path that avoids
these singularities, they found an imaginary term.
Motivated by Sauter's work \citep{s31}
on Klein paradox \citep{klein}, Heisenberg and Euler
estimated the size of the imaginary term  in the effective Lagrangian as
\begin{equation}
-{\frac{8i}{\pi}}{\bar E}^2 m_ec^2\left(\frac{m_ec}{h}\right)^3 e^{-\pi/\bar E},
\label{herate}
\end{equation}
and pointed out that it is associated with pair production by the electric field. This imaginary term
in the effective Lagrangian is related to the imaginary parts of field-dependent dielectric $\epsilon$ 
and permeability $\mu$ of the vacuum. 

In 1950's, \citet{s51,s54a,s54b}  derived the
same formula (\ref{effectiveint}) once more
within the quantum field theory of 
Quantum Electromagnetics (QED),
\begin{equation} \!\!\!\!\!\!\!\!
\frac{ \tilde\Gamma }{V}
=\frac{\alpha E^2}{ \pi^2\hbar}\sum_{n=1}^\infty \frac{1}{ n^2}\exp
\left(-\frac{n\pi E_c}{ E}\right).
\label{probability1}
\end{equation}
and its Lorentz-invariant expression in terms of electromagnetic fields  ${\bf E}$ and ${\bf B}$,
\begin{equation}
\frac{ \tilde\Gamma }{V}=\frac{  \alpha   \varepsilon^2}{ \pi^2 }
\sum_{n=1}  \frac{1}{n^2}
\frac{ n\pi\beta / \varepsilon }
{\tanh {n\pi \beta/ \varepsilon}}\exp\left(-\frac{n\pi E_c}{ \varepsilon}\right),
\label{probabilityeh}
\end{equation}
where
\begin{eqnarray} \!\!\!\!\!\!\!\!\!\!\!\!
\left\{
{ \frac{\varepsilon}{\beta} }\right\}
 & \equiv\!&
\frac{1}{ \sqrt{2} }
 \sqrt{
 \sqrt{
 ({\bf E}^2-{\bf B}^2)^2+4( {\bf E}\cdot {\bf B})^2
}
\pm ({{\bf E}^2-{\bf B}^2})
}.
\label{fieldinvariant}
\end{eqnarray}

The exponential factor $e^{\pi\frac{m_e^2c^3}{\hbar e E}}$ in Eqs.~(\ref{transmission}) and (\ref{herate})
characterizes the transmission coefficient of quantum tunneling, 
\citet{he35} introduced the critical field strength $E_c=\frac{m_e^2c^3}{\hbar e}$ (\ref{dimenlessEB}). 
They compared
it with the field strength $E_e$ of an electron at its classical radius, $E_e=e/r_e^2$ where 
$r_e=\alpha \hbar/(m_ec)$ and $\alpha=1/137$. They found the field strength $E_e$ is 137 time larger than 
the critical field strength $E_c$, i.e. $E_e=\alpha ^{-1}E_c$. At a critical radius $r_c=\alpha^{1/2}\hbar/(m_ec)< r_e$, the
field strength of the electron would be equal to the critical field strength $E_c$. 

As shown in Fig.~\ref{gap}, the negative-energy
spectrum
of solutions of the Dirac equation
has energies ${\mathcal E}_-<-m_ec^2$, and  is separated from the
positive energy-spectrum ${\mathcal E}_+> m_ec^2$
by a gap $2m_ec^2\approx1.02$MeV.
The negative-energy states are all filled.
The energy gap
 is by a factor $4/ \alpha^2
\approx 10^5 $ larger
than the typical binding energy  of atoms
($\sim 13.6$eV).
In order to create an
electron-positron
pair, one must spend this large amount of energy.
The source of this energy can  be an external field.

If an electric field attempts to tear an electron out of the filled state
the gap energy must be gained over the distance of
two electron radii.
The virtual particles
give an electron a radius of the order of the
Compton wavelength $ \lambda \equiv \hbar/m_ec$.
Thus we expect  a significant
creation
of electron-positron pairs
if the work done by the electric field $E$
over twice the Compton wave length $\hbar/m_ec$ is larger than $2m_ec^2$
\begin{equation}
eE\left({\frac{2\hbar}{ m_ec}}\right)>2m_ec^2 .
\nonumber
\end{equation}
This condition defines a critical electric field
\begin{equation}
E_c\equiv{\frac{m_e^2c^3}{ e\hbar}}\simeq 1.3\cdot 10^{16}\, {\rm V/cm},
\label{critical1}
\end{equation}
above which
pair creation becomes abundant.
To have an idea how large this critical electric field is, we
compare it with the value of the electric field required to ionize a hydrogen atom.
There the above inequality
holds for twice of the Bohr radius and the Rydberg energy
\begin{equation}
eE^{\rm ion}\left({\frac{2\hbar}{  {\alpha  m_ec}}}\right)> \alpha ^2{m_ec^2},
\nonumber
\end{equation}
so that $E_c\approx E^{\rm ion}_c/ \alpha^3 $ is about $10^6$
times as large,
a value that has so far
not been reached in a laboratory on Earth.

\subsection{Vacuum polarization around a black hole with electromagnetic structure}\label{ruffini}

We already discussed the phenomenon of
electron-positron pair production in a strong electric field over a flat
space-time. We study the same phenomenon
occurring around a black hole with electromagnetic structure (EMBH).
In order not to involve the complex dynamics of gravitational collapse at this
stage, for simplicity, we postulate that the collapse has already occurred and
has leaded to the formation of an EMBH. The spacetime around the EMBH is
described by the Kerr-Newman geometry which we rewrite here, for convenience,
in Boyer-Lindquist coordinates $(t,r,\theta,\phi)$
\begin{equation}
ds^{2}={\frac{\Sigma}{\Delta}}dr^{2}+\Sigma d\theta^{2}+{\frac{\Delta}{\Sigma
}}(dt-a\sin^{2}\theta d\phi)^{2}+{\frac{\sin^{2}\theta}{\Sigma}}\left[
(r^{2}+a^{2})d\phi-adt\right]  ^{2},\label{kerrnewmannBL}%
\end{equation}
where $\Delta=r^{2}-2Mr+a^{2}+Q^{2}$ and $\Sigma=r^{2}+a^{2}\cos^{2}\theta$,
as before and as usual $M$ is the mass, $Q$ the charge and $a$ the angular
momentum per unit mass of the EMBH. We recall that the Reissner-Nordstr{\o}m
geometry is a particular case for a non-rotating black holes: $a=0$, and the natural unit $G=\hbar=c=1$ is adopted this 
section. The
electromagnetic vector potential around the Kerr-Newman black hole is given,
in Boyer-Lindquist coordinates, by
\begin{equation}
\mathbf{A}=-Q\Sigma^{-1}r(dt-a\sin^{2}\theta d\phi).
\label{potentialcurve}
\end{equation}
The electromagnetic field tensor is then
\begin{align}
\mathbf{F}= &\  d\mathbf{A}=   2Q\Sigma^{-2}[(r^{2}-a^{2}\cos^{2}\theta)dr\wedge
dt-2a^{2}r\cos\theta\sin\theta d\theta\wedge dt\nonumber\\
& -a\sin^{2}\theta(r^{2}-a^{2}\cos^{2}\theta)dr\wedge d\phi+2ar(r^{2}%
+a^{2})\cos\theta\sin\theta d\theta\wedge d\phi].
\end{align}
In a Kerr-Newman geometry, the occurrence of pair production has been
predicted with the classical example of superradiance by \citet{dr12z} and has been confirmed for massless fields by \citet{dr12s}. The formulation of this problem in the framework of
second-quantized massless field (spin 0 and spin 1/2) has been given in \citet{dr12unruh}. Finally the detailed study of this process for massive
particles has been addressed in \citet{dr12mg1} where the rate of pair
production was also computed.

In a Reissner-Nordstr\"om geometry, the QED pair production has been studied
by \citet{dr13zaumen,dr13gibbons}. \citet{dr75} studied QED pair production in the Kerr-Newman geometry and they
obtained the rate of pair production with particular emphasis on:

\begin{itemize}
\item the limitations imposed by pair production on the strength of the
electromagnetic field of a black hole (\citet{r70});
\item the efficiency of extracting rotational and Coulomb energy from a black
hole by pair production;
\item the possibility of having observational consequences of astrophysical interest.
\end{itemize}

The third point was in fact a far-reaching prevision of possible energy
sources of gamma ray bursts that are most important phenomena under current
theoretical and observational studies. In the following, we discuss the
fundamental work of Damour and Ruffini in some details.

In order to study the pair production in the Kerr-Newman geometry, we introduce
at each event $(t,r,\theta,\phi)$ a local Lorentz frame, associated with a
stationary observer ${\mathcal{O}}$ at the event $(t,r,\theta,\phi)$. A
convenient frame is defined by the following orthogonal tetrad (\citet{dr15})
\begin{align}
\boldsymbol{\omega}^{(0)} &  =(\Delta/\Sigma)^{1/2}(dt-a\sin^{2}\theta
d\phi),\label{tetrad1}\\
\boldsymbol{\omega}^{(1)} &  =(\Sigma/\Delta)^{1/2}dr,\label{tetrad2}\\
\boldsymbol{\omega}^{(2)} &  =\Sigma^{1/2}d\theta,\label{tetrad3}\\
\boldsymbol{\omega}^{(3)} &  =\sin\theta\Sigma^{-1/2}((r^{2}+a^{2}%
)d\phi-adt).\label{tetrad4}%
\end{align}
In the so fixed Lorentz frame, the electric potential $A_{0}$, the electric
field ${\bf E}$ and the magnetic field ${\bf B}$ are given by the following
formulas (c.e.g. \citet{mtw73}),
\begin{align*}
A_{0} &  =\boldsymbol{\omega}_{a}^{(0)}A^{a}\\
{\bf E}^{\alpha} &  =\boldsymbol{\omega}_{\beta}^{(0)}F^{\alpha\beta}\\
{\bf B}^{\beta} &  ={\frac{1}{2}}\boldsymbol{\omega}_{\gamma}^{(0)}%
\epsilon^{\alpha\gamma\delta\beta}F_{\gamma\delta}.
\end{align*}
We then obtain
\begin{equation}
A_{0}=-Qr(\Sigma\Delta)^{-1/2},\label{gaugepotential}%
\end{equation}
while the electromagnetic fields ${\bf E}$ and ${\bf B}$ are parallel to the
direction of $\boldsymbol{\omega}^{(1)}$ and have got strengths given by
\begin{align}
E_{(1)} &  =Q\Sigma^{-2}(r^{2}-a^{2}\cos^{2}\theta),\label{e1}\\
B_{(1)} &  =Q\Sigma^{-2}2ar\cos\theta,\label{b1}%
\end{align}
respectively. The maximal strength $E_{\mathrm{\max}}$ of the electric field
is obtained in the case $a=0$ at the horizon of the EMBH: $r=r_{+}$. We have
\begin{equation}
E_{\max}=Q^{2}/r_{+}^{2}\label{emax2}%
\end{equation}
Equating the maximal strength of electric field (\ref{emax2}) to the critical
value (\ref{critical1}), one obtains the maximal black hole mass
$M_{\mathrm{max}}\simeq7.2\cdot10^{6}M_{\odot}$ for pair production to occur.
For any black hole with mass smaller than $M_{\mathrm{max}}$, the pair
production process can drastically modify its electromagnetic structure.

Both the gravitational and the electromagnetic background fields of the
Kerr-Newman black hole are stationary. We consider the quantum field of the
electron, which has mass $m_e$ and charge $e$. If $m_eM\gg1$, i.e. the spatial
variation scale $GM/c^{2}$ of the background fields is much larger than the
typical wavelength $\hbar/m_ec$ of the quantum field, then, for what concern
purely QED phenomena, such as pair production, it is possible to consider the
electric and magnetic fields defined by Eqs.~(\ref{e1},\ref{b1}) as constants
in a neighborhood of a few wavelengths around any events $(r,\theta,\phi,t)$.
Thus, our analysis and discussion on the Sauter-Euler-Heisenberg-Schwinger
process over a flat space-time can be locally applied to the case of the
curved Kerr-Newman geometry, based on the equivalence principle.

The rate of pair production around a Kerr-Newman black hole can be obtained from the Schwinger 
formula (\ref{probabilityeh}) for parallel electromagnetic fields 
$\varepsilon =E_{(1)}$ and $\beta= B_{(1)}$ as:
\begin{equation}
\frac{\tilde\Gamma}{V}={\frac{e^2E_{(1)}B_{(1)}}{4\pi^{2}}}%
\sum_{n=1}^{\infty}{\frac{1}{n}}%
\coth\left(  {\frac{n\pi B_{(1)}}{E_{(1)}}}\right)  \exp\left(
-{\frac{n\pi E_{\mathrm{c}}}{E_{(1)}}}\right)  .\label{drw}%
\end{equation}
The total number of pair
produced in a region $D$ of the space-time is
\begin{equation}
N=\int_{D}d^{4}x\sqrt{-g}\frac{\tilde\Gamma}{V},\label{drn}%
\end{equation}
where $\sqrt{-g}=\Sigma\sin\theta$. In \citet{dr75}, it was assumed that for
each created pair the particle (or antiparticle) with the same sign of charge
as the black hole is expelled at infinity with charge $e$, energy
$\omega$ and angular momentum $l_{\phi}$
while the antiparticle is absorbed by the black hole. This implies the
decrease of charge, mass and angular momentum of the black hole and a
corresponding extraction of all three quantities. The rates of the three
quantities are then determined by the rate of pair production (\ref{drw}) and
by the conservation laws of total charge, energy and angular momentum,
\begin{align}
\dot{Q} &  =-Re\nonumber\\
\dot{M} &  =-R\langle\omega\rangle\label{damour1}\\
\dot{L} &  =-R\langle l_{\phi}\rangle,\nonumber
\end{align}
where $R=\dot{N}$ is the rate of pair production, $\langle\omega\rangle$ and
$\langle l_{\phi}\rangle$ represent some suitable mean values for the energy
and angular momentum carried by the pairs.

Supposing the maximal variation of black hole charge to be $\Delta Q=-Q$, one
can estimate the maximal number of pairs created and the maximal mass-energy
variation. It was concluded in \citet{dr75} that
the maximal mass-energy variation in the pair production process is larger than
$10^{41}$erg and up to $10^{58}$erg, as a function of the black hole mass.
This was immediately viewed as a most simple model for the explanation of
gamma-ray bursts discovered at that time.

\section{Description of the electron-positron plasma oscillations}

In a treatment based on electro-fluidodynamics approach we conclude that, for $\mathcal{E}>\mathcal{E}_{\mathrm{c}}$ (with $\mathcal{E}_{\mathrm{c}}\equiv m_{e}^{2}c^{3}/(e\hbar )$, where as usual $m_{e}$ and $e$ are the electron mass and charge), the vacuum polarization process transform the electromagnetic energy of
the field mainly in the creation of pairs, with moderate contribution to
their kinetic energy. For $\mathcal{E}<\mathcal{E}_{\mathrm{c}}$ the kinetic
energy contribution is maximized. Due to the existence of plasma
oscillations, for every initial value of the electric field, there exists an
upper limit to the maximum value of the Lorentz gamma factor of the
electrons and positrons. Explicit relation between the distance of the
creation of electron and positron in the pair and the length of oscillations
in terms of electric field are given. The asymptotic behavior for large time
is explored using the phase portrait technique of the dynamical systems
theory. The existence of an infinite series of plasma oscillations is
discovered. It can be concluded that electron-positron pairs created by
vacuum polarization in any uniform unbounded electric field never reach
larger Lorentz gamma factor than the one constrained by the oscillations.
The collective effects are always predominant. The timescale of the
relaxation is estimated to be $\sim 10^{3}-10^{4}\ \hbar /m_{e}c^{2}$.

\subsection{On the observability of electron-positron pairs created in
vacuum polarization in Earth bound experiment and in astrophysics}

Three different earth-bound experiments and one astrophysical observation
have been proposed for identifying the polarization of the electronic vacuum
due to a supercritical electric field postulated by
Sauter-Heisenberg-Euler-Schwinger (see \citet{s31,he35,s51,n70}):

\begin{enumerate}
\item In central collisions of heavy ions near the Coulomb barrier, as first
proposed in \citet{GZ69,GZ70} (see also \citet{PR71,z65,z}). Despite some
apparently encouraging results (see \citet{S...83}), such efforts have failed so
far due to the small contact time of the colliding ions (\citet{A...95,G...96,L...97,B...95,H...98}). Typically the electromagnetic energy
involved in the collisions of heavy ions with impact parameter $%
l_{1}\sim10^{-12}$cm is $E_{1}\sim10^{-6}$erg and the lifetime of the
diatomic system is $t_{1}\sim10^{-22}$s.
\item In collisions of an electron beam with optical laser pulses: a signal
of positrons above background has been observed in collisions of a 46.6 GeV
electron beam with terawatt pulses of optical laser in an experiment at the
Final Focus Test Beam at SLAC (\citet{B...97}); it is not clear if this
experimental result is an evidence for the vacuum polarization phenomenon.
The energy of the laser pulses was $E_{2}\sim10^{7}$erg, concentrated in a
space-time region of spacial linear extension (focal length) $l_{2}\sim
10^{-3}$cm and temporal extension (pulse duration) $t_{2}\sim10^{-12}$s (\citet{B...97}).
\item At the focus of an X-ray free electron laser (XFEL) (see \citet{R01,AHRSV01,RSV02} and references therein). Proposals for this experiment
exist at the TESLA collider at DESY and at the LCLS facility at SLAC (\citet{R01}). Typically the electromagnetic energy at the focus of an XFEL can be $%
E_{3}\sim10^{6}$erg, concentrated in a space-time region of spacial linear
extension (spot radius) $l_{3}\sim10^{-8}$cm and temporal extension
(coherent spike length) $t_{3}\sim10^{-13}$s (\citet{R01}).
\end{enumerate}

and from astrophysics

\begin{enumerate}
\item around an electromagnetic black hole (black hole) (\citet{dr75,prx98,prx02}), giving rise to the observed phenomenon of GRBs (\citet{lett1,lett2,lett3,r02}). The electromagnetic
energy of an black hole of mass $M\sim10M_{\odot}$ and charge $Q\sim0.1M/%
\sqrt{G}$ is $E_{4}\sim10^{54}$erg and it is deposited in a space-time
region of spacial linear extension $l_{4}\sim10^{8}$cm (\citet{prx98,rv02a})
and temporal extension (collapse time) $t_{4}\sim10^{-2}$s (\citet{rvx03}).
\end{enumerate}

\subsection{On the role of transparency condition in the electron-positron
plasma}

In addition to their marked quantitative difference in testing the same
basic physical phenomenon, there is a very important conceptual difference
among these processes: the first three occur in a transparency condition in
which the created electron-positron pairs and, possibly, photons freely
propagate to infinity, while the one in the black hole occurs in an opacity
condition (\citet{rswx00}). Under the opacity condition a relaxation effect
occurs and a final equipartition between the $e^{+}e^{-}$ and $\gamma $ is
reached. Far from being just an academic issue, this process and its
characteristic timescale is of the greatest importance in physics and
astrophysics.

The evolution of a system of particle-antiparticle pairs created by the
Schwinger process has been often described by a transport Vlasov equation
(see, for example, \citet{KM85,GKM87}). More recently it has been showed that
such an equation can be derived from quantum field theory (\citet{SRS...97,KME98,SBR...98}). In the homogeneous case, the equations have been
numerically integrated taking into account the back reaction on the external
electric field (\citet{KESCM91,KESCM92,CEKMS93,BMP...99}). In many papers (see 
\citet{V...01} and references therein) a phenomenological term describing
equilibrating collisions is introduced in the transport equation which is
parameterized by an effective relaxation time $\tau $. In \citet{V...01} one
further step is taken by allowing time variability of $\tau $; the ignorance
on the collision term is then parameterized by a free dimensionless
constant. The introduction of a relaxation time corresponds to the
assumption that the system rapidly evolves towards thermal equilibrium. In
this paper we focus on the evolution of a system of $e^{+}e^{-}$ pairs,
explicitly taking into account the scattering processes $e^{+}e^{-}%
\rightleftarrows \gamma \gamma $. Since we are mainly interested in a system
in which the electric field varies on macroscopic length scale ($l\sim
10^{8} $cm, above), we can limit ourselves to a homogeneous electric field.
Also, we will use transport equations for electrons, positrons and photons,
with collision terms, coupled to Maxwell equations. There is no free
parameter here: the collision terms can be exactly computed, since the QED
cross sections are known. Starting from a regime which is far from thermal
equilibrium, we find that collisions do not prevent plasma oscillations in
the initial phase of the evolution and analyse the issue of the timescale of
the approach to a $e^{+}e^{-}\gamma $ plasma equilibrium configuration,
which is the most relevant quantity in the process of gravitational collapse 
(\citet{rvx03}).

\subsection{The continuity, energy-momentum and Maxwell equations}

Consider electrons and positrons created at rest in pairs, due to vacuum
polarization in homogeneous electric field with strength $\mathcal{E}$ (\citet{s31,he35,s51,books,books2}), with the average rate per unit volume and per unit time 
\begin{equation}
S\equiv \frac{dN}{dVdt}=\frac{m^{4}}{4\pi ^{3}}\left( \frac{\mathcal{E}}{%
\mathcal{E}_{c}}\right) ^{2}\exp \left( -\pi \frac{\mathcal{E}_{c}}{\mathcal{%
E}}\right) .  \label{rate2}
\end{equation}%
For the moment let us neglect interactions between electrons and positrons.

As the result of abundant pair creation homogeneous isotropic neutral plasma
consisting of electrons and positrons with comoving number density of
electrons $\bar{n}$ appears. Electrons and positrons move along the electric
field line with average velocity $v$ in opposite directions and their
dynamics has to be followed as well as their backreaction on the initial
uniform electric field. For this reason we study the continuity,
energy-momentum conservation and Maxwell equations written for electrons,
positrons and electromagnetic field: 
\begin{align}
\frac{\partial \left( \bar{n}U^{\mu }\right) }{\partial x^{\mu }}& =S,
\label{cont} \\
\frac{\partial T^{\mu \nu }}{\partial x^{\nu }}& =-F^{\mu \nu }J_{\nu },
\label{em2} \\
\frac{\partial F^{\mu \nu }}{\partial x^{\nu }}& =-4\pi J^{\mu },  \label{me}
\end{align}%
where $T^{\mu \nu }$ is energy-momentum tensor of electrons and positrons 
\begin{equation}
T^{\mu \nu }=m\bar{n}\left( U_{(+)}^{\mu }U_{(+)}^{\nu }+U_{(-)}^{\mu
}U_{(-)}^{\nu }\right) ,  \label{emten}
\end{equation}%
where $F^{\mu \nu }$ is electromagnetic field tensor, $J^{\mu }$ is the
total four-current density, $\gamma $ is relativistic Lorentz factor $\gamma
=\left( 1-v^{2}\right) ^{-1/2}$, $U^{\mu }$ is four velocity respectively of
positrons and electrons%
\begin{equation}
U_{(+)}^{\mu }=U^{\mu }=\gamma \left( 1,v,0,0\right) ,\qquad U_{(-)}^{\mu
}=\gamma \left( 1,-v,0,0\right) .
\end{equation}

We choose a coordinate frame where pairs are created at rest. Electric field
in this frame is directed along $x$-axis and introduce coordinate number
density $n=\bar{n}\gamma $. In spatially homogeneous case from (\ref{cont})
we have%
\begin{equation*}
\dot{n}=S.
\end{equation*}%
With our definitions (\ref{emten}) from (\ref{em2}) and equation of motion
for positrons and electrons%
\begin{equation*}
m\frac{\partial U_{(\pm )}^{\mu }}{\partial x^{\nu }}=\mp eF_{\nu }^{\mu },
\end{equation*}%
we find 
\begin{equation*}
\frac{\partial T^{\mu \nu }}{\partial x^{\nu }}=-e\bar{n}\left( U_{(+)}^{\nu
}-U_{(-)}^{\nu }\right) F_{\nu }^{\mu }+mS\left( U_{(+)}^{\mu }+U_{(-)}^{\mu
}\right) =-F_{\nu }^{\mu }J^{\nu },
\end{equation*}%
where the total current density is the sum of conducting $J_{cond}^{\mu }$
and polarization $J_{pol}^{\mu }$ currents densities%
\begin{align}
\qquad J^{\mu }& =J_{cond}^{\mu }+J_{pol}^{\mu }, \\
J_{cond}^{\mu }& =e\bar{n}\left( U_{(+)}^{\mu }-U_{(-)}^{\mu }\right) , \\
J_{pol}^{\mu }& =\frac{2mS}{\mathcal{E}}\gamma \left( 0,1,0,0\right) .
\end{align}

Energy-momentum tensor in (\ref{em2}) and electromagnetic field tensor in (%
\ref{me}) change for two reasons: 1)\ electrons and positrons acceleration
in the electric field, given by the term $J_{cond}^{\mu}$, 2)\ particle
creation, described by the term $J_{pol}^{\mu}$. Equation (\ref{cont}) is
satisfied separately for electrons and positrons.

Defining energy density of positrons%
\begin{equation*}
\rho =\frac{1}{2}T^{00}=mn\gamma ,
\end{equation*}%
we find from (\ref{em2}) 
\begin{equation*}
\dot{\rho}=envE+\frac{m\gamma S}{\mathcal{E}}.
\end{equation*}%
Due to homogeneity of the electric field and plasma, electrons and positrons
have the same energy density but opposite momentum $p$. Our definitions also
imply for velocity and momentum densities of electrons and positrons%
\begin{equation}
v=\frac{p}{\rho },  \label{veleq}
\end{equation}%
and%
\begin{equation}
\rho ^{2}=p^{2}+m^{2}n^{2},  \label{rhopn}
\end{equation}%
which is just relativistic relation between the energy, momentum and mass
densities of particles.

Gathering together the above equations we then have the following equations%
\begin{align}
\frac{dn}{dt}& =S,  \label{ndot} \\
\frac{d\rho }{dt}& =\mathcal{E}\left( env+\frac{m\gamma S}{\mathcal{E}}%
\right) ,  \label{rhodot} \\
\frac{dp}{dt}& =en\mathcal{E}+mv\gamma S,  \label{pdot} \\
\frac{d\mathcal{E}}{dt}& =-8\pi \left( env+\frac{m\gamma S}{\mathcal{E}}%
\right) .  \label{Edot}
\end{align}

\subsection{The Vlasov-Boltzmann-Maxwell equations}

Consider now the more general problem when electron-positron pairs interact
through annihilation into photons $e^{+}e^{-}\rightarrow \gamma \gamma $,
and its inverse process: pair production $\gamma \gamma \rightarrow
e^{+}e^{-}$.

The motion of positrons (electrons) is the resultant of three contributions:
the pair creation, the electric acceleration and the annihilation damping.
The homogeneous system consisting of electric field, electrons, positrons
and photons can be described by the equations 
\begin{align}
\partial _{t}f_{e}+e\mathbf{E}\partial _{\mathbf{p}}f_{e}& =\mathcal{S}%
\left( \mathbf{E},\mathbf{p}\right) -\tfrac{1}{\left( 2\pi \right) ^{5}}%
\epsilon _{\mathbf{p}}^{-1}\mathcal{C}_{e}\left( t,\mathbf{p}\right) ,
\label{pairs} \\
\partial _{t}f_{\gamma }& =\tfrac{2}{\left( 2\pi \right) ^{5}}\epsilon _{%
\mathbf{k}}^{-1}\mathcal{C}_{\gamma }\left( t,\mathbf{k}\right) ,
\label{photons} \\
\partial _{t}\mathbf{E}& =-\mathbf{j}_{p}\left( \mathbf{E}\right) -\mathbf{j}%
_{c}\left( t\right) ,  \label{Maxwell}
\end{align}%
where $f_{e}=f_{e}\left( t,\mathbf{p}\right) $ is the distribution function
in the phase-space of positrons (electrons), $f_{\gamma }=f_{\gamma }\left(
t,\mathbf{k}\right) $ is the distribution function in the phase-space of
photons, $\mathbf{E}$ is the electric field, $\epsilon _{\mathbf{p}}=\left( 
\mathbf{p}\cdot \mathbf{p}+m_{e}^{2}\right) ^{1/2}$ is the energy of an
electron of 3-momentum $\mathbf{p}$ ($m_{e}$ is the mass of the electron)
and $\epsilon _{\mathbf{k}}=\left( \mathbf{k}\cdot \mathbf{k}\right) ^{1/2}$
is the energy of a photon of 3-momentum $\mathbf{k}$. $f_{e}$ and $f_{\gamma
}$ are normalized so that $\int \tfrac{d^{3}\mathbf{p}}{\left( 2\pi \right)
^{3}}\ f_{e}\left( t,\mathbf{p}\right) =n_{e}\left( t\right) $, $\int \tfrac{%
d^{3}\mathbf{k}}{\left( 2\pi \right) ^{3}}\ f_{\gamma }\left( t,\mathbf{k}%
\right) =n_{\gamma }\left( t\right) $ , where $n_{e}$ and $n_{\gamma }$ are
number densities of positrons (electrons) and photons, respectively. The
term 
\begin{equation}
\mathcal{S}\left( \mathbf{E},\mathbf{p}\right) =\left( 2\pi \right) ^{3}%
\tfrac{dN}{dtd^{3}\mathbf{x}d^{3}\mathbf{p}}=-\left\vert e\mathbf{E}%
\right\vert \log \left[ 1-\exp \left( -\tfrac{\pi (m_{e}^{2}+\mathbf{p}%
_{\perp }^{2})}{\left\vert e\mathbf{E}\right\vert }\right) \right] \delta
(p_{\parallel })  \label{S}
\end{equation}%
is the Schwinger source for pair creation (see \citet{KESCM91,KESCM92}): $%
p_{\parallel }$ and $\mathbf{p}_{\perp }$ are the components of the
3-momentum $\mathbf{p}$ parallel and orthogonal to $\mathbf{E}$. We assume
that the pairs are produced at rest in the direction parallel to the
electric field (\citet{KESCM91,KESCM92}). We also have, in Eqs. (\ref{pairs}), (%
\ref{photons}) and (\ref{Maxwell}), 
\begin{align}
\mathcal{C}_{e}\left( t,\mathbf{p}\right) & \simeq \int \tfrac{d^{3}\mathbf{p%
}_{1}}{\epsilon _{\mathbf{p}_{1}}}\tfrac{d^{3}\mathbf{k}_{1}}{\epsilon _{%
\mathbf{k}_{1}}}\tfrac{d^{3}\mathbf{k}_{2}}{\epsilon _{\mathbf{k}_{2}}}%
\delta ^{\left( 4\right) }\left( p+p_{1}-k_{1}-k_{2}\right)  \notag \\
& \times \left\vert \mathcal{M}\right\vert ^{2}\left[ f_{e}\left( \mathbf{p}%
\right) f_{e}\left( \mathbf{p}_{1}\right) -f_{\gamma }\left( \mathbf{k}%
_{1}\right) f_{\gamma }\left( \mathbf{k}_{2}\right) \right] ,  \label{Ce} \\
\mathcal{C}_{\gamma }\left( t,\mathbf{k}\right) & \simeq \int \tfrac{d^{3}%
\mathbf{p}_{1}}{\epsilon _{\mathbf{p}_{1}}}\tfrac{d^{3}\mathbf{p}_{2}}{%
\epsilon _{\mathbf{p}_{2}}}\tfrac{d^{3}\mathbf{k}_{1}}{\epsilon _{\mathbf{k}%
_{1}}}\delta ^{\left( 4\right) }\left( p_{1}+p_{2}-k-k_{1}\right)  \notag \\
\times & \left\vert \mathcal{M}\right\vert ^{2}\left[ f_{e}\left( \mathbf{p}%
_{1}\right) f_{e}\left( \mathbf{p}_{2}\right) -f_{\gamma }\left( \mathbf{k}%
\right) f_{\gamma }\left( \mathbf{k}_{1}\right) \right] ,  \label{Cf}
\end{align}%
which describe probability rates for pair creation by photons and pair
annihilation into photons, $\mathcal{M}=\mathcal{M}_{e^{+}\left( \mathbf{p}%
_{1}\right) e^{-}\left( \mathbf{p}_{2}\right) \rightleftarrows \gamma \left( 
\mathbf{k}\right) \gamma \left( \mathbf{k}_{1}\right) }$ being the matrix
element for the process $e^{+}\left( \mathbf{p}_{1}\right) e^{-}\left( 
\mathbf{p}_{2}\right) \rightarrow \gamma \left( \mathbf{k}\right) \gamma
\left( \mathbf{k}_{1}\right) $. Note that the collisional terms (\ref{Ce})
and (\ref{Cf}) are either inapplicable or negligible in the case of the
above three earth-bound experiments where the created pairs do not originate
a dense plasma. They have been correctly neglected in previous works (see e.
g. \citet{RSV02}). Collisional terms have also been considered in the
different physical context of vacuum polarization by strong chromoelectric
fields. Unlike the present QED case, where expressions for the cross
sections are known exactly, in the QCD case the cross sections are yet
unknown and such collisional terms are of a phenomenological type and useful
uniquely near the equilibrium regime (\citet{V...01}). Finally $\mathbf{j}%
_{p}\left( \mathbf{E}\right) =2\tfrac{\mathbf{E}}{\mathbf{E}^{2}}\int \tfrac{%
d^{3}\mathbf{p}}{\left( 2\pi \right) ^{3}}\epsilon _{\mathbf{p}}\mathcal{S}%
\left( \mathbf{E},\mathbf{p}\right) $ and $\mathbf{j}_{c}\left( t\right)
=2en_{e}\int \tfrac{d^{3}\mathbf{p}}{\left( 2\pi \right) ^{3}}\tfrac{\mathbf{%
p}}{\epsilon _{\mathbf{p}}}f_{e}\left( \mathbf{p}\right) $ are polarization
and conduction current respectively (see \citet{GKM87}). In Eqs. (\ref{Ce})
and (\ref{Cf}) we neglect, as a first approximation, Pauli blocking and Bose
enhancement (see e.g. \citet{KESCM92}). By suitably integrating (\ref{pairs})
and (\ref{photons}) over the phase spaces of positrons (electrons) and
photons, we find the following exact equations for mean values: 
\begin{align}
\tfrac{d}{dt}n_{e}& =S\left( \mathbf{E}\right) -n_{e}^{2}\left\langle \sigma
_{1}v^{\prime }\right\rangle _{e}+n_{\gamma }^{2}\left\langle \sigma
_{2}v^{\prime \prime }\right\rangle _{\gamma },  \notag \\
\tfrac{d}{dt}n_{\gamma }& =2n_{e}^{2}\left\langle \sigma _{1}v^{\prime
}\right\rangle _{e}-2n_{\gamma }^{2}\left\langle \sigma _{2}v^{\prime \prime
}\right\rangle _{\gamma },  \notag \\
\tfrac{d}{dt}n_{e}\left\langle \epsilon _{\mathbf{p}}\right\rangle _{e}&
=en_{e}\mathbf{E}\cdot \left\langle \mathbf{v}\right\rangle _{e}+\tfrac{1}{2}%
\mathbf{E\cdot j}_{p}-n_{e}^{2}\left\langle \epsilon _{\mathbf{p}}\sigma
_{1}v^{\prime \prime }\right\rangle _{e}+n_{\gamma }^{2}\left\langle
\epsilon _{\mathbf{k}}\sigma _{2}v^{\prime \prime }\right\rangle _{\gamma },
\notag \\
\tfrac{d}{dt}n_{\gamma }\left\langle \epsilon _{\mathbf{k}}\right\rangle
_{\gamma }& =2n_{e}^{2}\left\langle \epsilon _{\mathbf{p}}\sigma
_{1}v^{\prime }\right\rangle _{e}-2n_{\gamma }^{2}\left\langle \epsilon _{%
\mathbf{k}}\sigma _{2}v^{\prime \prime }\right\rangle _{\gamma },  \notag \\
\tfrac{d}{dt}n_{e}\left\langle \mathbf{p}\right\rangle _{e}& =en_{e}\mathbf{E%
}+\tfrac{1}{2}\left( \mathbf{E\cdot j}_{p}\right) \left\langle \mathbf{p}%
\right\rangle _{e}\left\langle \epsilon _{\mathbf{p}}\right\rangle
_{e}^{-1}-n_{e}^{2}\left\langle \mathbf{p}\sigma _{1}v^{\prime
}\right\rangle _{e},  \notag \\
\tfrac{d}{dt}\mathbf{E}& =-2en_{e}\left\langle \mathbf{v}\right\rangle _{e}-%
\mathbf{j}_{p}\left( \mathbf{E}\right) ,  \label{System1}
\end{align}%
where, for any function of the momenta 
\begin{align}
\left\langle F\left( \mathbf{p}_{1},...,\mathbf{p}_{n}\right) \right\rangle
_{e}& \equiv n_{e}^{-n}\int \tfrac{d^{3}\mathbf{p}_{1}}{\left( 2\pi \right)
^{3}}...\tfrac{d^{3}\mathbf{p}_{n}}{\left( 2\pi \right) ^{3}}\ F\left( 
\mathbf{p}_{1},...,\mathbf{p}_{n}\right) \cdot f_{e}\left( \mathbf{p}%
_{1}\right) \cdot ...\cdot f_{e}\left( \mathbf{p}_{n}\right) , \\
\left\langle G\left( \mathbf{k}_{1},...,\mathbf{k}_{l}\right) \right\rangle
_{\gamma }& \equiv n_{\gamma }^{-l}\int \tfrac{d^{3}\mathbf{k}_{1}}{\left(
2\pi \right) ^{3}}...\tfrac{d^{3}\mathbf{k}_{l}}{\left( 2\pi \right) ^{3}}\
G\left( \mathbf{k}_{1},...,\mathbf{k}_{l}\right) \cdot f_{\gamma }\left( 
\mathbf{k}_{1}\right) \cdot ...\cdot f_{\gamma }\left( \mathbf{k}_{l}\right)
.
\end{align}%
Furthermore $v^{\prime }$ is the relative velocity between electrons and
positrons, $v^{\prime \prime }$ is the relative velocity between photons, $%
\sigma _{1}=\sigma _{1}\left( \epsilon _{\mathbf{p}}^{\mathrm{CoM}}\right) $
is the total cross section for the process $e^{+}e^{-}\rightarrow \gamma
\gamma $ and $\sigma _{2}=\sigma _{2}\left( \epsilon _{\mathbf{k}}^{\mathrm{%
CoM}}\right) $ is the total cross section for the process $\gamma \gamma
\rightarrow e^{+}e^{-}$ (here $\epsilon ^{\mathrm{CoM}}$ is the energy of a
particle in the reference frame of the center of mass).

In order to evaluate the mean values in system (\ref{System1}) we need some
further hypotheses on the distribution functions. Let us define $\bar{p}%
_{\parallel }$, $\bar{\epsilon}_{\mathbf{p}}$ and $\mathbf{\bar{p}}_{\perp
}^{2}$ such that $\left\langle p_{\parallel }\right\rangle _{e}\equiv \bar{p}%
_{\parallel },~\left\langle \epsilon _{\mathbf{p}}\right\rangle _{e}\equiv 
\bar{\epsilon}_{\mathbf{p}}\equiv (\bar{p}_{\parallel }^{2}+\mathbf{\bar{p}}%
_{\perp }^{2}+\ m_{e}^{2})^{1/2}$. We assume 
\begin{equation}
f_{e}\left( t,\mathbf{p}\right) \propto n_{e}\left( t\right) \delta \left(
p_{\parallel }-\bar{p}_{\parallel }\right) \delta \left( \mathbf{p}_{\perp
}^{2}-\mathbf{\bar{p}}_{\perp }^{2}\right) .  \label{fe}
\end{equation}%
Since in the scattering $e^{+}e^{-}\rightarrow \gamma \gamma $ the
coincidence of the scattering direction with the incidence direction is
statistically favored, we also assume 
\begin{equation}
f_{\gamma }\left( t,\mathbf{k}\right) \propto n_{\gamma }\left( t\right)
\delta \left( \mathbf{k}_{\perp }^{2}-\mathbf{\bar{k}}_{\perp }^{2}\right) %
\left[ \delta \left( k_{\parallel }-\bar{k}_{\parallel }\right) +\delta
\left( k_{\parallel }+\bar{k}_{\parallel }\right) \right] ,  \label{fgamma}
\end{equation}%
where $k_{\parallel }$ and $\mathbf{k}_{\perp }$ have analogous meaning as $%
p_{\parallel }$ and $\mathbf{p}_{\perp }$ and the terms $\delta \left(
k_{\parallel }-\bar{k}_{\parallel }\right) $ and $\delta \left( k_{\parallel
}+\bar{k}_{\parallel }\right) $ account for the probability of producing,
respectively, forwardly scattered and backwardly scattered photons. Since
the Schwinger source term (\ref{S}) implies that the positrons (electrons)
have initially fixed $p_{\parallel }$, $p_{\parallel }=0$, assumption (\ref%
{fe}) ((\ref{fgamma})) means that the distribution of $p_{\parallel }$ ($%
k_{\parallel }$) does not spread too much with time and, analogously, that
the distribution of energies is sufficiently peaked to be describable by a $%
\delta -$function. The dependence on the momentum of the distribution
functions has been discussed in \citet{KESCM92,KME98}. Approximations (\ref%
{fe}), (\ref{fgamma}) reduce Eqs. (\ref{System1}) to a system of ordinary
differential equations. In average, since the inertial reference frame we
fix coincides with the center of mass frame for the processes $%
e^{+}e^{-}\rightleftarrows \gamma \gamma $, $\epsilon ^{\mathrm{CoM}}\simeq 
\bar{\epsilon}$ for each species. Substituting (\ref{fe}) and (\ref{fgamma})
into (\ref{System1}) we find 
\begin{align}
\tfrac{d}{dt}n_{e}& =S\left( \mathcal{E}\right) -2n_{e}^{2}\sigma _{1}\rho
_{e}^{-1}\left\vert {\pi }_{e\parallel }\right\vert +2n_{\gamma }^{2}\sigma
_{2},  \notag \\
\tfrac{d}{dt}n_{\gamma }& =4n_{e}^{2}\sigma _{1}\rho _{e}^{-1}\left\vert {%
\pi }_{e\parallel }\right\vert -4n_{\gamma }^{2}\sigma _{2},  \notag \\
\tfrac{d}{dt}\rho _{e}& =en_{e}\mathcal{E}\rho _{e}^{-1}\left\vert {\pi }%
_{e\parallel }\right\vert +\tfrac{1}{2}\mathcal{E}j_{p}-2n_{e}\rho
_{e}\sigma _{1}\rho _{e}^{-1}\left\vert {\pi }_{e\parallel }\right\vert
+2n_{\gamma }\rho _{\gamma }\sigma _{2},  \notag \\
\tfrac{d}{dt}\rho _{\gamma }& =4n_{e}\rho _{e}\sigma _{1}\rho
_{e}^{-1}\left\vert {\pi }_{e\parallel }\right\vert -4n_{\gamma }\rho
_{\gamma }\sigma _{2},  \notag \\
\tfrac{d}{dt}{\pi }_{e\parallel }& =en_{e}\mathcal{E}+\tfrac{1}{2}\mathcal{E}%
j_{p}\left\vert {\pi }_{e\parallel }\right\vert \rho _{e}^{-1}-2n_{e}{\pi }%
_{e\parallel }\sigma _{1}\rho _{e}^{-1}\left\vert {\pi }_{e\parallel
}\right\vert ,  \notag \\
\tfrac{d}{dt}\mathcal{E}& =-2en_{e}\rho _{e}^{-1}\left\vert {\pi }%
_{e\parallel }\right\vert -j_{p}\left( \mathcal{E}\right) ,  \label{System2}
\end{align}%
where $\rho _{e}=n_{e}\bar{\epsilon}_{\mathbf{p}}$, $\rho _{\gamma
}=n_{\gamma }\bar{\epsilon}_{\mathbf{k}}$, ${\pi }_{e\parallel }=n_{e}\bar{p}%
_{\parallel }$ are the energy density of positrons (electrons), the energy
density of photons and the density of \textquotedblleft parallel
momentum\textquotedblright\ of positrons (electrons), $\mathcal{E}$ is the
electric field strength and $j_{p}$ the unique component of $\mathbf{j}_{p}$
parallel to $\mathbf{E}$. $\sigma _{1}$ and $\sigma _{2}$ are evaluated at $%
\epsilon ^{\mathrm{CoM}}=\bar{\epsilon}$ for each species. Note that Eqs.(%
\ref{System2}) are \textquotedblleft classical\textquotedblright\ in the
sense that the only quantum information is encoded in the terms describing
pair creation and scattering probabilities. Eqs.(\ref{System2}) are
consistent with energy density conservation: $\tfrac{d}{dt}\left( \rho
_{e}+\rho _{\gamma }+\tfrac{1}{2}\mathcal{E}^{2}\right) =0.$

When the interaction between electrons and positrons is neglected in (\ref%
{System2}) these equations are in full agreement with (\ref{ndot})-(\ref%
{Edot}).

\subsection{Plasma oscillations}

Consider again only electrons and positrons neglecting their interaction.
From (\ref{rhodot}) and (\ref{Edot}) we obtain the energy conservation
equation%
\begin{equation}
\frac{\mathcal{E}_{0}^{2}-\mathcal{E}^{2}}{8\pi }+2\rho =0,  \label{energy}
\end{equation}%
so the particle energy density vanishes for electric field $E_{0}$.

These equations give also the maximum number of the pair density
asymptotically attainable consistently with the above rate equation and
energy conservation%
\begin{equation*}
n_{0}=\frac{\mathcal{E}_{0}^{2}}{8\pi m}.
\end{equation*}

For simplicity we introduce dimensionless variables $n=m^{3}\tilde{n}$, $%
\rho =m^{4}\tilde{\rho}$, $p=m^{4}\tilde{p}$, $\mathcal{E}=\mathcal{E}_{c}%
\mathcal{\tilde{E}}$, and $t=m^{-1}\tilde{t}$. With these variables our
system of equations (\ref{ndot})-(\ref{Edot}) takes the form 
\begin{align}
\frac{d\tilde{n}}{d\tilde{t}}& =\tilde{S},  \notag \\
\frac{d\tilde{\rho}}{d\tilde{t}}& =\tilde{n}\mathcal{\tilde{E}}\tilde{v}+%
\tilde{\gamma}\tilde{S},  \label{numsys} \\
\frac{d\tilde{p}}{d\tilde{t}}& =\tilde{n}\mathcal{\tilde{E}}+\tilde{\gamma}%
\tilde{v}\tilde{S},  \notag \\
\frac{d\mathcal{\tilde{E}}}{d\tilde{t}}& =-8\pi \alpha \left( \tilde{n}%
\tilde{v}+\frac{\tilde{\gamma}\tilde{S}}{\mathcal{\tilde{E}}}\right) , 
\notag
\end{align}%
where $\tilde{S}=\frac{1}{4\pi ^{3}}\mathcal{\tilde{E}}^{2}\exp \left( -%
\frac{\pi }{\mathcal{\tilde{E}}}\right) $, $\tilde{v}=\frac{\tilde{p}}{%
\tilde{\rho}}$ and $\tilde{\gamma}=\left( 1-\tilde{v}^{2}\right) ^{-1/2}$.

We solve numerically the system of equations (\ref{numsys}) with the initial
conditions $n(0)=\rho (0)=v(0)=0$, and the electric field $E(0)=E_{0}$.

\begin{figure}
\centering 
\includegraphics[width=4in]{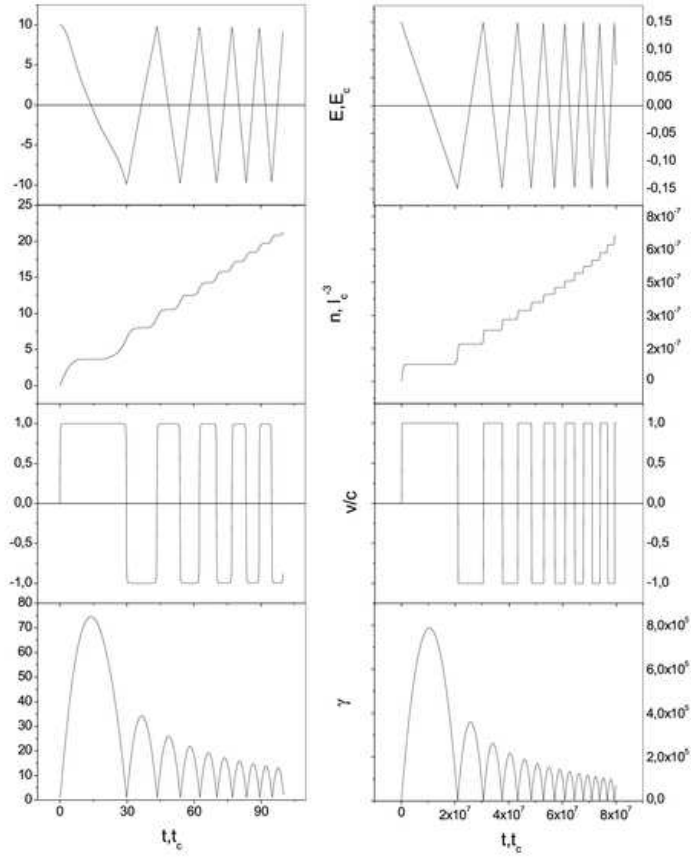}
\caption{Electric field strength, number density of electrons, their
velocity and Lorentz gamma factor depending on time with $\mathcal{E}_{0}=10%
\mathcal{E}_{c}$ (left column) and $\mathcal{E}_{0}=0.15\mathcal{E}_{c}$
(right column). Electric field, number density and velocity of positron are
measured respectively in terms of the critical field $\mathcal{E}_{c}$,
Compton volume $l_{c}^{3}=\left( \frac{\hbar }{mc}\right)^{3}$, and the
speed of light $c$.}
\label{fig1b}
\end{figure}

In fig. \ref{fig1b} we provide diagrams for electric field strength, number
density, velocity and Lorentz gamma factor of electrons as functions of
time, for initial values of the electric field $E=10E_{c}$ (left column) and 
$E=0.15E_{c}$ (right column). 

\begin{figure}
\centering 
\includegraphics[width=4in]{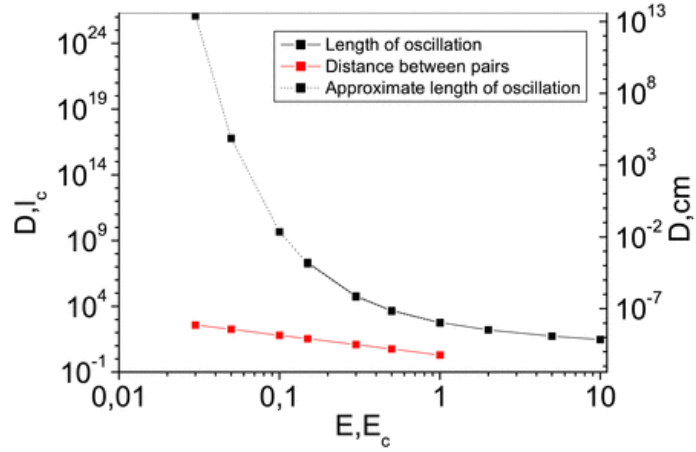}
\caption{Maximum length of oscillations (solid and dotted above lines) together with the
distance between electron and positron in a pair (below line) computed from (%
\protect\ref{D}), depending on initial value of electric field strength. The
solid above line is obtained from solutions of exact equations (\protect\ref%
{numsys}), while the dotted above line corresponds to solutions of
approximate equation (\protect\ref{eeq}).}
\label{fig2b}
\end{figure}

At fig. \ref{fig2b} characteristic length of oscillations is shown together
with the distance between pairs (\citet{n69,k00})
\begin{equation}
D^{\ast }=\frac{2}{m}\left( \frac{\mathcal{E}_{c}}{\mathcal{E}}\right)
^{3/2}.  \label{D}
\end{equation}

Thus, given initial electric field strength we define two characteristic
distances: $D^{\ast }$ above which pair creation is possible, and $D$ above
which plasma oscillations occur in a uniform electric field. Clearly $D\gg
D^{\ast }$. 
\begin{figure}
\centering 
\includegraphics[width=4in]{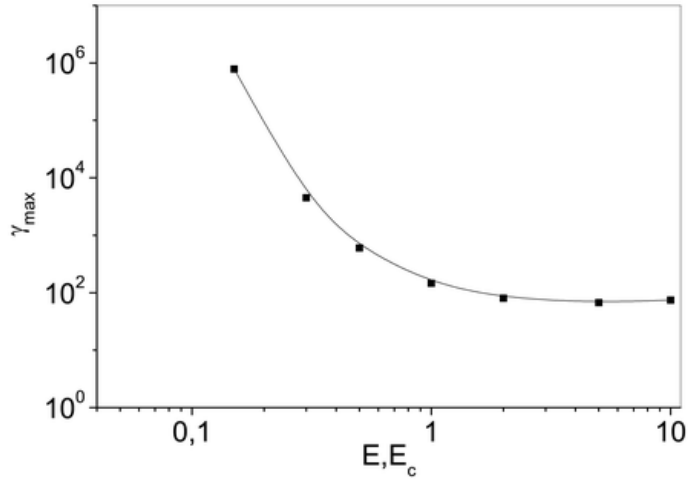}
\caption{Maximum Lorentz gamma factor reached at the first oscillation
depending on initial value of the electric field strength.}
\label{fig3b}
\end{figure}

At fig. \ref{fig3b} maximum gamma factor in the first oscillation is
presented depending on initial value of the electric field. Since in the
successive oscillations the maximal value of the Lorentz gamma factor is
monotonically decreasing (see fig. \ref{fig1b}) we conclude that for every
initial value of the electric field there exist a maximum Lorentz gamma
factor attainable by the electrons and positrons in the plasma. This clearly
shows that never in this process the test particle approximation for the
electrons and positrons motion can be applied and the collective effects are
always predominant both in the case of $\mathcal{E}>\mathcal{E}_{c}$ and $%
\mathcal{E}<\mathcal{E}_{c}$.

We estimated the half-life of oscillations to be $966t_{c}$ for $\mathcal{E}%
_{0}=10\mathcal{E}_{c}$ and $1.43\times 10^{5}t_{c}$ for $\mathcal{E}_{0}=0.8%
\mathcal{E}_{c}$ respectively.

\begin{figure}
\centering
\includegraphics[width=4in]{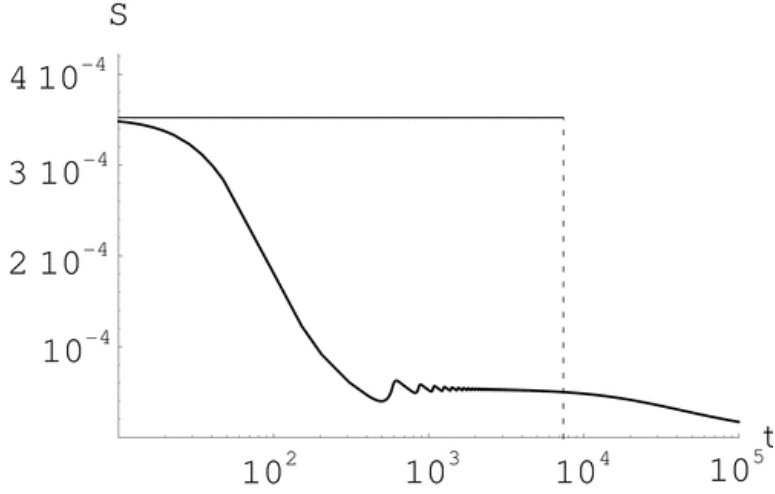}
\caption{The average rate of pair production $n/t$ is shown as function of
time (thick curve), comparing to its initial value $S(\mathcal{E}_{0})$
(thin line) for $\mathcal{E}_{0}=\mathcal{E}_{c}$. The dashed line marks the
time when the energy of electric field would have exhaused if the rate kept
constant.}
\label{fig4b}
\end{figure}

We also compare the average rate of pair creation for two cases:\ when the
electric field value is constant in time (an external energy source keeps
the field unchanged) and when it is self-regulated by equations (\ref{numsys}%
). The result is represented at fig. \ref{fig4b}. 
It is clear from the structure of the above equations that for $\mathcal{E}<%
\mathcal{E}_{c}$ the number of pairs is small, electrons and positrons are
accelerated in electric field and the conducting current is dominating.
Consequently, the polarization current can be neglected in (\ref{Edot}).

Assuming electric field to be weak we neglect polarization current in energy
conservation (\ref{rhodot}) and in Maxwell equation (\ref{Edot}). This means
energy density change due to acceleration is much larger than the one due to
pair creation,%
\begin{equation}
\mathcal{E}env\gg m\gamma S.  \label{weak}
\end{equation}%
In this case oscillations equations (\ref{ndot})-(\ref{Edot}) simplify. From
(\ref{rhodot}) and (\ref{pdot}) we have $\dot{\rho}=v\dot{p}$, and using (%
\ref{veleq}) obtain $v=\pm 1$. This is the limit when rest mass energy is
much smaller than the kinetic energy, $\gamma \gg 1$.

One may therefore use only the first and the last equations from the above
set. Taking time derivative of the Maxwell equation we arrive to a single
second order differential equation 
\begin{equation}
\frac{d^{2}\mathcal{E}}{dt^{2}}+\frac{2em^{4}}{\pi ^{2}}\left( \frac{%
\mathcal{E}}{\mathcal{E}_{c}}\right) \left\vert \frac{\mathcal{E}}{\mathcal{E%
}_{c}}\right\vert \exp \left( -\pi \left\vert \frac{\mathcal{E}_{c}}{%
\mathcal{E}}\right\vert \right) =0.  \label{eeq}
\end{equation}%
This is an equation for nonlinear oscillator. Solution of (\ref{eeq}) is
some periodic function which cannot be found analytically. Notice that
condition (\ref{weak}) means ultrarelativistic approximation for electrons
and positrons, so that although according to (\ref{ndot}) there is creation
of pairs with rest mass $2m$ for each pair, the corresponding increase of
plasma energy is neglected, as can be seen from (\ref{weak}).

Now we turn to qualitative properties of the system (\ref{ndot})-(\ref{Edot}%
). These nonlinear ordinary differential equations describe certain
dynamical system which can be studied by using methods of qualitative
analysis of dynamical systems. The presence of the two integrals (\ref%
{energy}) and (\ref{rhopn}) allows reduction of the system to two
dimensions. It is useful to work with the variables $v$ and $E$. In these
variables we have%
\begin{align*}
\frac{d\tilde{v}}{d\tilde{t}}& =\left( 1-\tilde{v}^{2}\right) ^{3/2}\mathcal{%
\tilde{E}}, \\
\frac{d\mathcal{\tilde{E}}}{d\tilde{t}}& =-\frac{1}{2}\tilde{v}\left( 1-%
\tilde{v}^{2}\right) ^{1/2}\left( \mathcal{\tilde{E}}_{0}^{2}-\mathcal{%
\tilde{E}}^{2}\right) -8\pi \alpha \frac{\tilde{S}}{\mathcal{\tilde{E}}%
\left( 1-\tilde{v}^{2}\right) ^{1/2}}.
\end{align*}%
Introducing the new time variable $\tau $%
\begin{equation*}
\frac{d\tau }{d\tilde{t}}=\left( 1-\tilde{v}^{2}\right) ^{-1/2}
\end{equation*}%
we arrive at%
\begin{align}
\frac{d\tilde{v}}{d\tau }& =\left( 1-\tilde{v}^{2}\right) ^{2}\mathcal{%
\tilde{E}},  \label{vdeq} \\
\frac{d\mathcal{\tilde{E}}}{d\tau }& =-\frac{1}{2}\tilde{v}\left( 1-\tilde{v}%
^{2}\right) \left( \mathcal{\tilde{E}}_{0}^{2}-\mathcal{\tilde{E}}%
^{2}\right) -8\pi \alpha \frac{\tilde{S}}{\mathcal{\tilde{E}}},  \label{edeq}
\end{align}%
where $\alpha $ is the parameter. Clearly the phase space is bounded by the
two curves $\tilde{v}=\pm 1$. Moreover, physical requirement $\rho \geq 0$
leads to existence of two other bounds $\mathcal{\tilde{E}}=\pm \mathcal{%
\tilde{E}}_{0}$. This system has only one singular point in the physical
region, of the type focus at $\mathcal{\tilde{E}}=0$ and $\tilde{v}=0$.

\begin{figure}
\centering
\includegraphics[width=4in]{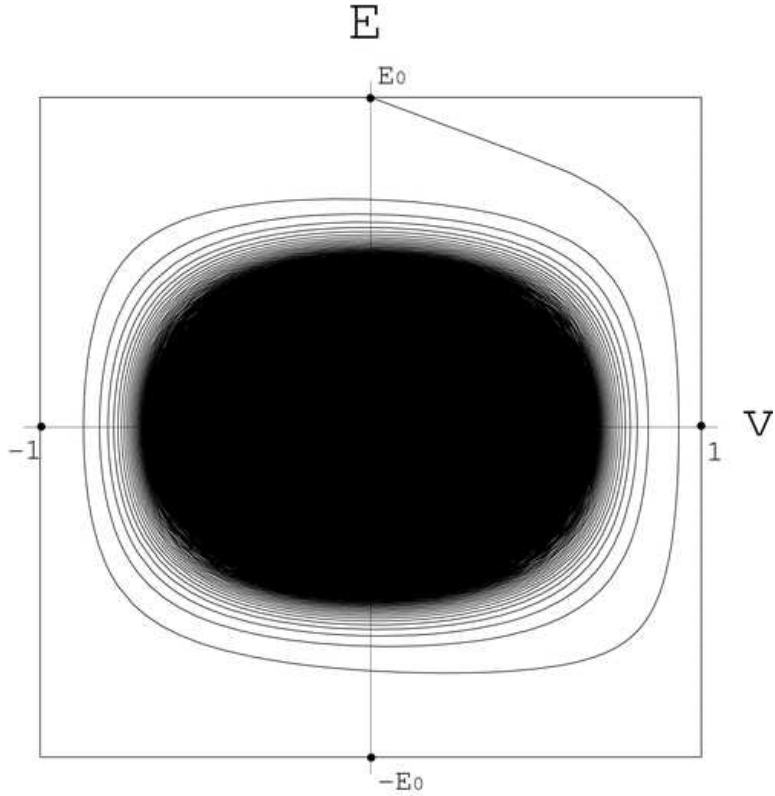}
\caption{Phase portrait of the two-dimensional dynamical system (\protect\ref%
{vdeq}),(\protect\ref{edeq}). Tildes are ommitted. Notice that phase
trajectories are not closed curves and with each cycle they approach the
point with $\mathcal{\tilde{E}}=0$ and $\tilde{v}=0$.}
\label{fig5}
\end{figure}

The phase portrait of the dynamical system (\ref{vdeq}),(\ref{edeq}) is
represented at fig. {\ref{fig5}}. 
Thus, every phase trajectory tends asymptotically to the only singular point
at $\mathcal{\tilde{E}}=0$ and $\tilde{v}=0$. This means oscillations stop
only when electric field vanishes. At that point clearly 
\begin{equation}
\rho =mn.  \label{rest}
\end{equation}%
is valid. i.e. all the energy in the system transform just to the rest mass
of the pairs.

\begin{figure}
\centering
\includegraphics[width=4in]{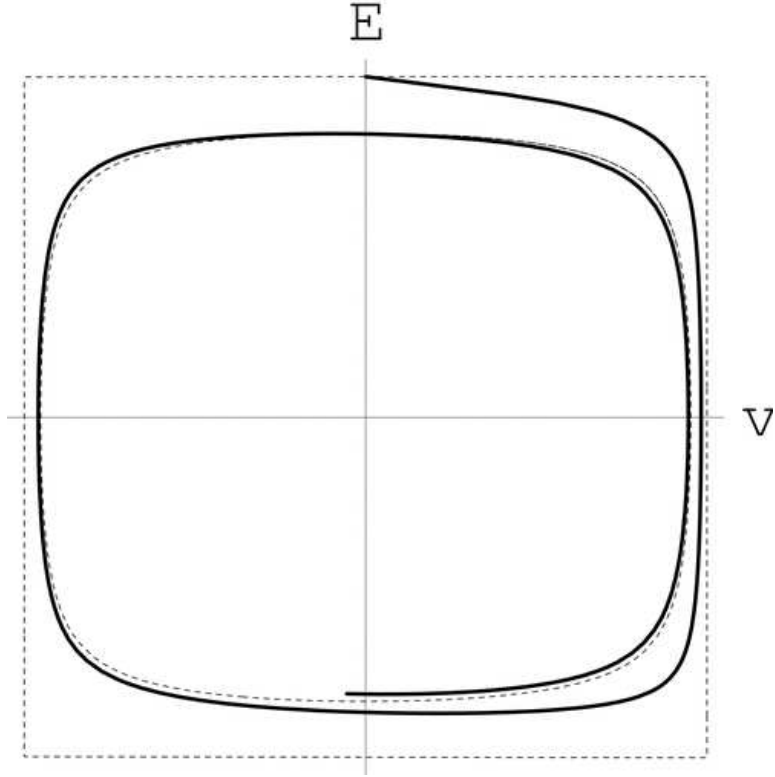}
\caption{Phase trajectory for 1.5 cycles (thick curve) compared with
solutions where the Schwinger pair production is switched off (dashed
curves).}
\label{fig5a}
\end{figure}

In order to illustrate details of the phase trajectories shown at fig. \ref%
{fig5} we plot only 1.5 cycles at fig. \ref{fig5a}. One can see that the
deviation from closed curves shown by dashed curves is maximal when the
field peaks, namely when the pair production rate is maximal. 
The Schwinger formula (\ref{rate2}) used in this analysis is derived for
uniform constant in time electric field. However, it still can be used for a
time-variable electric field providing the inverse adiabaticity parameter 
(\citet{b70,p71,po73,books,books2}) is much larger than one,%
\begin{equation*}
\eta =\frac{m}{\omega }\frac{\mathcal{E}_{peak}}{\mathcal{E}_{c}}=\tilde{T}%
\mathcal{\tilde{E}}_{peak}\gg 1,
\end{equation*}%
where $\omega $ is the frequency of oscillations, $\tilde{T}=m/\omega $ is
dimensionless period of oscillations. In the two cases we considered in the
paper, $\mathcal{E}=10\mathcal{E}_{c}$ and $\mathcal{E}=0.15\mathcal{E}_{c}$
we find for the first oscillation $\eta =334$ and $\eta =3.1\times 10^{6}$
respectively. As can be seen for the fig. \ref{fig1b} the period of
oscillations decrease with time which means the parameter $\eta $ becoming
smaller. Eventually it may reach unity so the Schwinger formula (\ref{rate2})
becomes inapplicable.

\subsection{Bremsstrahlung}

All the above treatment has been done by considering uniquely
electron-positron pairs neglecting the bremsstrahlung radiation and the
electron-positron annihilation into photons.

In order to estimate the effect of bremsstrahlung we recall the classical
formula for the radiation loss in electric field%
\begin{equation*}
I=\frac{2}{3}\frac{e^{4}}{m^{2}}\mathcal{E}^{2}=\frac{2}{3}\alpha
m^{2}\left( \frac{\mathcal{E}}{\mathcal{E}_{c}}\right) ^{2}.
\end{equation*}%
Thus to take into account radiative loss we need to correct our equations as
follows%
\begin{align}
\frac{d\rho }{dt}& =\mathcal{E}\left( env+\frac{m\gamma S}{\mathcal{E}}%
\right) -\frac{2}{3}\frac{e^{4}}{m^{2}}\mathcal{E}^{2},  \label{energy_br} \\
\frac{dp}{dt}& =en\mathcal{E}+mv\gamma S-\frac{2}{3}\frac{e^{4}}{m^{2}}%
\mathcal{E}^{2}v.  \label{momentum_br}
\end{align}%
while the rest equations remain unchanged.

\begin{figure}
\centering
\includegraphics[width=4in]{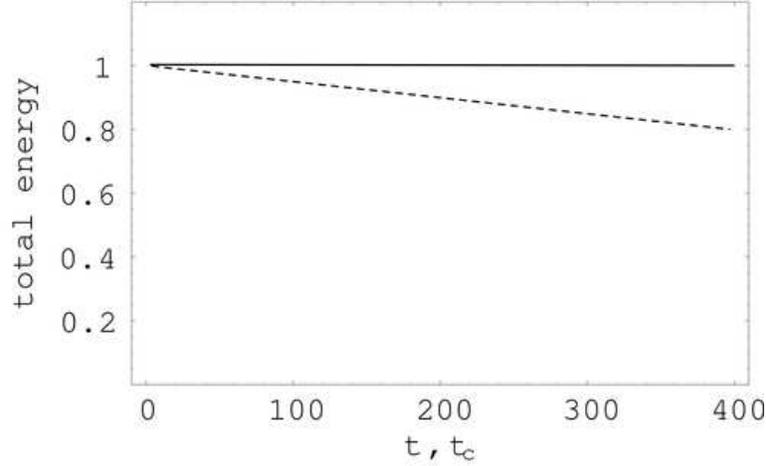}
\caption{Losses of the energy due to classical bremsstrahlung radiation. The
energy density of the system of electrons, positrons and the electric field
normalized to the initial energy density is shown without (solid line) and
with (dashed line) the effect of bremsstrahlung.}
\label{fig6}
\end{figure}
The result is shown at fig. \ref{fig6} where we plot the energy of electric
field, electrons and positrons normalized to the initial energy as a
function of time when bremsstrahlung is accounted for (dashed line) and
switched off (solid line). We find that for initial electric field strength $%
\mathcal{E}=10\mathcal{E}_{c}$ the radiation loss is very significant: $20$
per cent of the initial energy goes to soft bremsstrahlung radiation at $400$
Compton times. Thus one of the main source of damping of the plasma
oscillations discussed above is the radiation loss due to bremsstrahlung
radiation.

\subsection{Pair annihilation and pair production}

Now consider the effect of pair production and annihilation solving
numerically (\ref{System2}).

\begin{figure}
\centering
\includegraphics[width=8.5cm]{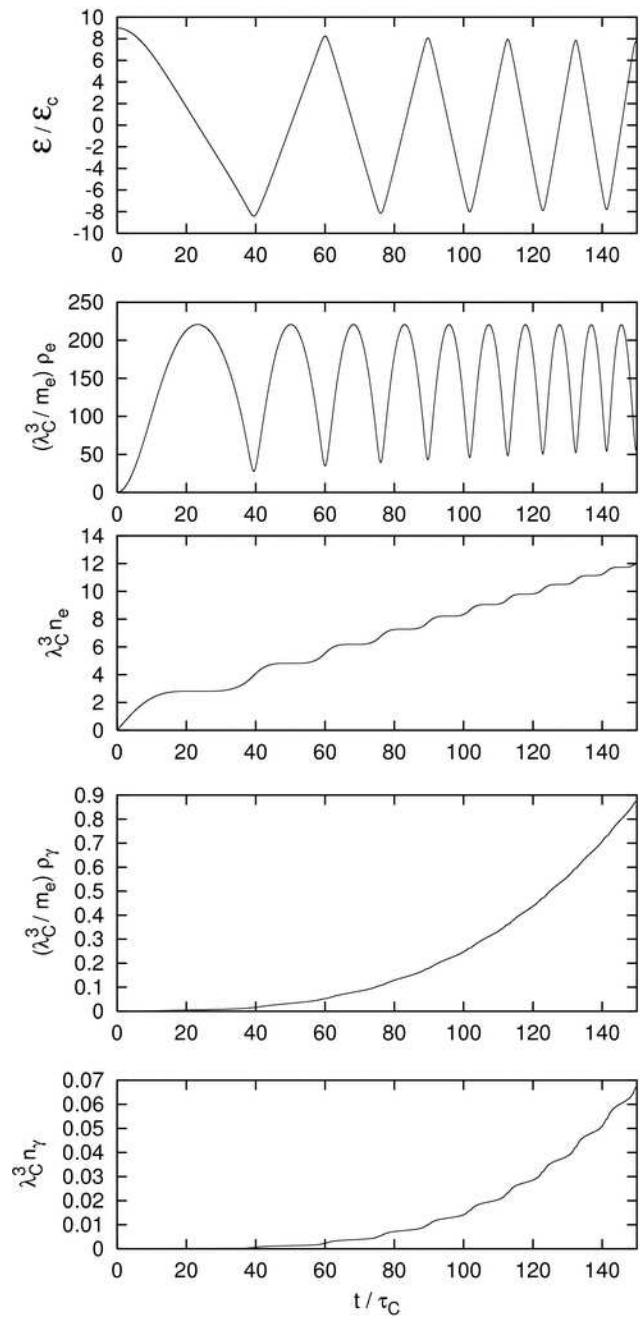}
\caption{Plasma oscillations. We set $\mathcal{E}_{0}=9\mathcal{E}_{\mathrm{c%
}}$, $t<150\protect\tau_{\mathrm{C}}$ and plot: a) electromagnetic field
strength; b) electrons energy density; c) electrons number density; d)
photons energy density; e) photons number density as functions of time.}
\label{fig1luca}
\end{figure}

The initial conditions for Eqs.(\ref{System2}) are $n_{e}=n_{\gamma}=\rho
_{e}=\rho_{\gamma}=\pi_{e\parallel}=0,~\mathcal{E}=\mathcal{E}_{0}$. In Fig. %
\ref{fig1luca} the results of the numerical integration for $\mathcal{E}%
_{0}=9\mathcal{E}_{\mathrm{c}}$ is showed. The integration stops at $t=150\
\tau_{\mathrm{C}}$ (where $\tau_{\mathrm{C}}=\hbar/m_{e}c^{2}$). Each
variable is represented in units of $m_{e}$ and $\lambda_{\mathrm{C}%
}=\hbar/m_{e}c$. The numerical integration confirms (\citet{KESCM91,KESCM92})
that the system undergoes plasma oscillations: a) the electric field
oscillates with decreasing amplitude rather than abruptly reaching the
equilibrium value; b) electrons and positrons oscillates in the electric
field direction, reaching ultrarelativistic velocities; c) the role of the $%
e^{+}e^{-}\rightleftarrows$ $\gamma\gamma$ scatterings is marginal in the
early time of the evolution, the electrons are too extremely relativistic
and consequently the density of photons builds up very slowly (see. details
in Fig. \ref{fig1luca}).

\begin{figure}
\centering
\includegraphics[width=8.5cm]{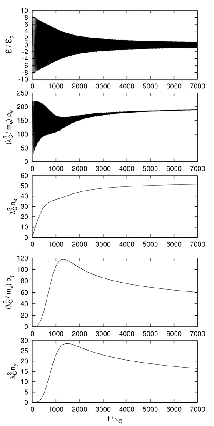}
\caption{Plasma oscillations. We set $\mathcal{E}_{0}=9\mathcal{E}_{\mathrm{c%
}}$, $t<7000\protect\tau _{\mathrm{C}}$ and plot: a) electromagnetic field
strength; b) electrons energy density; c) electrons number density; d)
photons energy density; e) photons number density as functions of time - the
oscillation period is not resolved in these plots. The model used should
have a breakdown at a time much earlier than $7000\protect\tau _{\mathrm{C}}$
and therefore this plot contains no more than qualitative informations.}
\end{figure}
At late times the system is expected to relax and assumptions (\ref{fe}) and
(\ref{fgamma}) have to be generalized to take into account quantum spreading
of the distribution functions. It is nevertheless interesting to look at the
solutions of Eqs.(\ref{System2}) in this regime. In Fig. \ref{fig2b} we plot
the numerical solution of Eqs.(\ref{System2}) but the integration extends
here all the way up to $t=7000\ \tau _{\mathrm{C}}$ (the time scale of
oscillations is not resolved in these plots). It is interesting that the
leading term recovers the expected asymptotic behavior: a) the electric
field is screened to about the critical value: $\mathcal{E}\simeq \mathcal{E}%
_{\mathrm{c}}$ for $t\sim 10^{3}-10^{4}\tau _{\mathrm{C}}\gg \tau _{\mathrm{C%
}}$; b) the initial electromagnetic energy density is distributed over
electron-positron pairs and photons, indicating energy equipartition; c)
photons and electron-positron pairs number densities are asymptotically
comparable, indicating number equipartition. At such late times a regime of
thermalized electrons-positrons-photons plasma begins and the system is
describable by hydrodynamic equations (\citet{rvx03,rswx00}).

We studied oscillations in neutral plasma created by the vacuum polarization
process in uniform electric field. Our analysis shows that in strong
electric field with $\mathcal{E}>\mathcal{E}_{c}$ the original electric
field discharge on timescale much larger than the Compton time:\ for $%
\mathcal{E}=10\mathcal{E}_{c}$ the half-life of oscillations is $966$
Compton times. In weak field with $\mathcal{E}<\mathcal{E}_{c}$ strong
currents develop, where charge carriers aquire large but finite Lorentz
gamma factors and electric field survives on even longer time:\ for $%
\mathcal{E}=0.8\mathcal{E}_{c}$ the half-life of oscillations is $1.43\times
10^{5}$ Compton times and $\gamma _{max}=205$. For given initial value of
electric field strength we give the maximal Lorentz gamma factor reached by
electrons and positrons as well as the maximal length of oscillations. We
conclude that in a uniform unbounded electric field the collective effects
are always predominant.

We provided a simple formalism apt to describe simultaneously the creation
of electron-positron pairs by a strong electric field $\mathcal{E}\gtrsim 
\mathcal{E}_{c}$, their annihilation into photons and accounting also for
bremsstrahlung radiation. We find that the collisions do not prevent plasma
oscillations. This is because the momentum of electrons (positrons) is very
high, therefore the cross section for the process $e^{+}e^{-}\rightarrow
\gamma \gamma $ is small and the annihilation into photons is negligible in
the very first phase of the evolution. As a result, the system takes some
time ($t\sim 10^{3}-10^{4}\tau _{\mathrm{C}}$) to relax to a $%
e^{+}e^{-}\gamma $ plasma configuration. We remark that, at least in the
case of electromagnetic Schwinger mechanism, the picture could be quite
different from the one previously depicted in literature, where the system
is assumed to thermalize in a very short time (see \citet{V...01} and
references therein).

It is conceivable that in the race to first identify the vacuum polarization
process \emph{\`{a} la} Sauter-Euler-Heisenberg-Schwinger, the astrophysical
observations will reach a positive result before earth-bound experiments,
much like in the case of the discovery of lines in the Sun chromosphere by
J. N. Lockyer in 1869, later identified with the Helium spectral lines by W.
Ramsay in 1895 (\citet{G03}).

\quad M. Schwarzschild, \textquotedblleft Structure and evolution of the
stars\textquotedblright , Dover Publications (New York, 1965).

\section{On the irreducible mass of the black hole and the role of subcritical and overcritical electric fields}

While the formation in time of the dyadosphere is the fundamental phenomena we are interested in, we can get an insight on the issue of gravitational collapse of an electrically charged star core studying in details a simplified model, namely a thin shell of charged dust.

In \citet{i66,idlc67} it is shown that the problem of a collapsing charged shell in general relativity can be reduced to a set of ordinary differential equations. We reconsider here the following relativistic system: a spherical shell of electrically charged dust which is moving radially in the Reissner-Nordstr{\o}m background of an already formed nonrotating black hole of mass $M_{1}$ and charge $Q_{1}$, with $Q_{1}\leq M_{1}$.

The world surface spanned by the shell divides the space-time into two regions: an internal one $\mathcal{M}_{-}$ and an external one $\mathcal{M}_{+}$. The line element in Schwarzschild like coordinate is (\citet{crv02})
\begin{equation}
ds^{2}=\left\{
\begin{array}[c]{l}
-f_{+}dt_{+}^{2}+f_{+}^{-1}dr^{2}+r^{2}d\Omega^{2}\qquad\text{in } \mathcal{M}_{+}\\
-f_{-}dt_{-}^{2}+f_{-}^{-1}dr^{2}+r^{2}d\Omega^{2}\qquad\text{in } \mathcal{M}_{-}
\end{array}
\right.  , \label{E0}
\end{equation}
where $f_{+}=1-\tfrac{2M}{r}+\tfrac{Q^{2}}{r^{2}}$, $f_{-}=1-\tfrac{2M_{1}} {r}+\tfrac{Q_{1}^{2}}{r^{2}}$ and $t_{-}$ and $t_{+}$ are the Schwarzschild-like time coordinates in $\mathcal{M}_{-}$ and $\mathcal{M}_{+}$ respectively. $M$ is the total mass-energy of the system formed by the shell and the black hole, measured by an observer at rest at infinity and $Q=Q_{0}+Q_{1}$ is the total charge: sum of the charge $Q_{0}$ of the shell and the charge $Q_{1}$ of the internal black hole.

Indicating by $R$ the radius of the shell and by $T_{\pm}$ its time coordinate, the equations of motion of the shell become (\citet{rv02a})
\begin{align}
\left(  \tfrac{dR}{d\tau}\right)  ^{2}  &  =\tfrac{1}{M_{0}^{2}}\left(M-M_{1}+\tfrac{M_{0}^{2}}{2R}-\tfrac{Q_{0}^{2}}{2R}-\tfrac{Q_{1}Q_{0}}{R}\right)  ^{2}-f_{-}\left(  R\right) \nonumber\\
&  =\tfrac{1}{M_{0}^{2}}\left(  M-M_{1}-\tfrac{M_{0}^{2}}{2R}-\tfrac{Q_{0}^{2}}{2R}-\tfrac{Q_{1}Q_{0}}{R}\right)  ^{2}-f_{+}\left(  R\right),\label{EQUYa}\\
\tfrac{dT_{\pm}}{d\tau}  &  =\tfrac{1}{M_{0}f_{\pm}\left(  R\right)  }\left(M-M_{1}\mp\tfrac{M_{0}^{2}}{2R}-\tfrac{Q_{0}^{2}}{2R}-\tfrac{Q_{1}Q_{0}}{R}\right)  , \label{EQUYb}
\end{align}
where $M_{0}$ is the rest mass of the shell and $\tau$ is its proper time. Eqs.(\ref{EQUYa},\ref{EQUYb}) (together with Eq.(\ref{E0})) completely describe a 5-parameter ($M$, $Q$, $M_{1}$, $Q_{1}$, $M_{0}$) family of solutions of the Einstein-Maxwell equations. Note that Eqs.(\ref{EQUYa},\ref{EQUYb}) imply that
\begin{equation}
M-M_{1}-\tfrac{Q_{0}^{2}}{2R}-\tfrac{Q_{1}Q_{0}}{R}>0
\label{Constraint}
\end{equation}
holds for $R>M+\sqrt{M^{2}-Q^{2}}$ if $Q<M$ and for $R>M_{1}+\sqrt{M_{1}^{2}-Q_{1}^{2}}$ if $Q>M$.

For astrophysical applications (\citet{rvx03b}) the trajectory of the shell $R=R\left(  T_{+}\right)  $ is obtained as a function of the time coordinate $T_{+}$ relative to the space-time region $\mathcal{M}_{+}$. In the following we drop the $+$ index from $T_{+}$. From Eqs.(\ref{EQUYa},\ref{EQUYb}) we have
\begin{equation}
\tfrac{dR}{dT}=\tfrac{dR}{d\tau}\tfrac{d\tau}{dT}=\pm\tfrac{F}{\Omega} \sqrt{\Omega^{2}-F},
\label{EQUAISRDLC}
\end{equation}
where
\begin{align}
F  & \equiv f_{+}\left(  R\right)  =1-\tfrac{2M}{R}+\tfrac{Q^{2}}{R^{2}},\\
\Omega & \equiv\bar\Gamma-\tfrac{M_{0}^{2}+Q^{2}-Q_{1}^{2}}{2M_{0}R},\\
\bar\Gamma & \equiv\tfrac{M-M_{1}}{M_{0}}.
\end{align}
Since we are interested in an imploding shell, only the minus sign case in (\ref{EQUAISRDLC}) will be studied. We can give the following physical interpretation of $\bar\Gamma$. If $M-M_{1}\geq M_{0}$, $\bar\Gamma$ coincides with the Lorentz $\gamma$ factor of the imploding shell at infinity; from Eq.(\ref{EQUAISRDLC}) it satisfies
\begin{equation}
\bar\Gamma=\tfrac{1}{\sqrt{1-\left(  \frac{dR}{dT}\right)  _{R=\infty}^{2}}}\geq1.
\end{equation}
When $M-M_{1}<M_{0}$ then there is a \emph{turning point} $R^{\ast}$, defined by $\left.  \tfrac{dR}{dT}\right|  _{R=R^{\ast}}=0$. In this case $\bar\Gamma$ coincides with the ``effective potential'' at $R^{\ast}$ :
\begin{equation}
\bar\Gamma=\sqrt{f_{-}\left(  R^{\ast}\right)  }+M_{0}^{-1}\left(  -\tfrac{M_{0}^{2}}{2R^{\ast}}+\tfrac{Q_{0}^{2}}{2R^{\ast}}+\tfrac{Q_{1}Q_{0}}{R^{\ast}}\right)
\leq1.
\end{equation}
The solution of the differential equation (\ref{EQUAISRDLC}) is given by:
\begin{equation}
\int dT=-\int\tfrac{\Omega}{F\sqrt{\Omega^{2}-F}}dR.
\label{GRYD}
\end{equation}
The functional form of the integral (\ref{GRYD}) crucially depends on the degree of the polynomial $P\left(  R\right)  =R^{2}\left(  \Omega^{2}-F\right)  $, which is generically two, but in special cases has lower values. We therefore distinguish the following cases:

\begin{enumerate}
\item {\boldmath$M=M_{0}+M_{1}$}; {\boldmath$Q_{1}=M_{1}$}; {\boldmath $Q=M$}: $P\left(  R\right)  $ is equal to $0$, we simply have
\begin{equation}
R(T)=\mathrm{{const}.}
\end{equation}
\item {\boldmath$M=M_{0}+M_{1}$}; {\boldmath$M^{2}-Q^{2}=M_{1}^{2}-Q_{1}^{2}$}; {\boldmath$Q\neq M$}: $P\left(  R\right)  $ is a constant, we have
\begin{align}
T  & =\mathrm{const}+\tfrac{1}{2\sqrt{M^{2}-Q^{2}}}\left[  \left(
R+2M\right)  R\right.  \nonumber\\
& \left.  +r_{+}^{2}\log\left(  \tfrac{R-r_{+}}{M}\right)  +r_{-}^{2}%
\log\left(  \tfrac{R-r_{-}}{M}\right)  \right]  .\label{CASO1}%
\end{align}
\item {\boldmath$M=M_{0}+M_{1}$}; {\boldmath$M^{2}-Q^{2}\neq M_{1}^{2}-Q_{1}^{2}$}: $P\left(  R\right)  $ is a first order polynomial and
\begin{align}
T &  =\mathrm{const}+2R\sqrt{\Omega^{2}-F}\left[  \tfrac{M_{0}R}{3\left(
M^{2}-Q^{2}-M_{1}^{2}+Q_{1}^{2}\right)  }\right.  \nonumber\\
&  \left.  +\tfrac{\left(  M_{0}^{2}+Q^{2}-Q_{1}^{2}\right)  ^{2}%
-9MM_{0}\left(  M_{0}^{2}+Q^{2}-Q_{1}^{2}\right)  +12M^{2}M_{0}^{2}%
+2Q^{2}M_{0}^{2}}{3\left(  M^{2}-Q^{2}-M_{1}^{2}+Q_{1}^{2}\right)  ^{2}%
}\right]  \nonumber\\
&  -\tfrac{1}{\sqrt{M^{2}-Q^{2}}}\left[  r_{+}^{2}\mathrm{arctanh}\left(
\tfrac{R}{r_{+}}\tfrac{\sqrt{\Omega^{2}-F}}{\Omega_{+}}\right)  \right.
\nonumber\\
&  \left.  -r_{-}^{2}\mathrm{arctanh}\left(  \tfrac{R}{r_{-}}\tfrac
{\sqrt{\Omega^{2}-F}}{\Omega_{-}}\right)  \right]  ,\label{CASO2}%
\end{align}
where $\Omega_{\pm}\equiv\Omega\left(  r_{\pm}\right)  $.
\item {\boldmath$M\neq M_{0}+M_{1}$}: $P\left(  R\right)  $ is a second order polynomial and
\begin{align}
T &  =\mathrm{const}-\tfrac{1}{2\sqrt{M^{2}-Q^{2}}}\left\{  \tfrac
{2\bar\Gamma\sqrt{M^{2}-Q^{2}}}{\bar\Gamma^{2}-1}R\sqrt{\Omega^{2}-F}\right.
\nonumber\\
&  +r_{+}^{2}\log\left[  \tfrac{R\sqrt{\Omega^{2}-F}}{R-r_{+}}+\tfrac
{R^{2}\left(  \Omega^{2}-F\right)  +r_{+}^{2}\Omega_{+}^{2}-\left(  \bar\Gamma
^{2}-1\right)  \left(  R-r_{+}\right)  ^{2}}{2\left(  R-r_{+}\right)
R\sqrt{\Omega^{2}-F}}\right]  \nonumber\\
&  -r_{-}^{2}\log\left[  \tfrac{R\sqrt{\Omega^{2}-F}}{R-r_{-}}+\tfrac
{R^{2}\left(  \Omega^{2}-F\right)  +r_{-}^{2}\Omega_{-}^{2}-\left(  \bar\Gamma
^{2}-1\right)  \left(  R-r_{-}\right)  ^{2}}{2\left(  R-r_{-}\right)
R\sqrt{\Omega^{2}-F}}\right]  \nonumber\\
&  -\tfrac{\left[  2MM_{0}\left(  2\bar\Gamma^{3}-3\bar\Gamma\right)  +M_{0}^{2}%
+Q^{2}-Q_{1}^{2}\right]  \sqrt{M^{2}-Q^{2}}}{M_{0}\left(  \bar\Gamma^{2}-1\right)
^{3/2}}\log\left[  \tfrac{R\sqrt{\Omega^{2}-F}}{M}\right.  \nonumber\\
&  \left.  \left.  +\tfrac{2M_{0}\left(  \bar\Gamma^{2}-1\right)  R-\left(
M_{0}^{2}+Q^{2}-Q_{1}^{2}\right)  \bar\Gamma+2M_{0}M}{2M_{0}M\sqrt{\bar\Gamma^{2}-1}%
}\right]  \right\}  .\label{CASO3}%
\end{align}
\end{enumerate}

\begin{figure}
\centering
\includegraphics[width=8.5cm,clip]{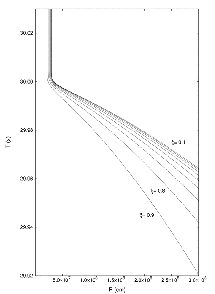}
\includegraphics[width=8.5cm,clip]{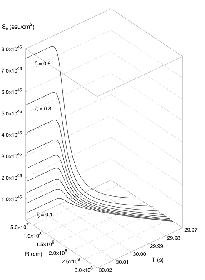}
\caption{{\bf Left)} Collapse curves in the plane $(T,R)$ for $M=20M_{\odot}$ and for different values of the parameter $\xi$. The asymptotic behavior is the clear manifestation of general relativistic effects as the horizon of the black hole is approached. {\bf Right)} Electric field behavior at the surface of the shell for $M=20M_{\odot}$ and for different values of the parameter $\xi$. The asymptotic behavior is the clear manifestation of general relativistic effects as the horizon of the black hole is approached.}
\label{elec3d}
\end{figure}

Of particular interest is the time varying electric field $\mathcal{E}%
_{R}=\tfrac{Q}{R^{2}}$ on the external surface of the shell. In order to study
the variability of $\mathcal{E}_{R}$ with time it is useful to consider in the
tridimensional space of parameters $(R,T,\mathcal{E}_{R})$ the parametric
curve $\mathcal{C}:\left(  R=\lambda,\quad T=T(\lambda),\quad\mathcal{E}%
_{R}=\tfrac{Q}{\lambda^{2}}\right)  $. In astrophysical applications
(\citet{rvx03b}) we are specially interested in the family of solutions such that
$\frac{dR}{dT}$ is 0 when $R=\infty$ which implies that $\bar\Gamma=1$. In Fig. 
\ref{elec3d} we plot the collapse curves in the plane $(T,R)$ for different
values of the parameter $\xi\equiv\frac{Q}{M}$, $0<\xi<1$. The initial data
$\left(  T_{0},R_{0}\right)  $ are chosen so that the integration constant in
equation (\ref{CASO2}) is equal to 0. In all the cases we can follow the
details of the approach to the horizon which is reached in an infinite
Schwarzschild time coordinate. In Fig. \ref{elec3d} we plot the parametric
curves $\mathcal{C}$ in the space $(R,T,\mathcal{E}_{R})$ for different values
of $\xi$. Again we can follow the exact asymptotic behavior of the curves
$\mathcal{C}$, $\mathcal{E}_{R}$ reaching the asymptotic value $\frac{Q}%
{r_{+}^{2}}$. The detailed knowledge of this asymptotic behavior is of great
relevance for the observational properties of the black hole formation (see e.g. \citet{rv02a}).

In the case of a shell falling in a flat background ($M_{1}=Q_{1}=0$) Eq.(\ref{EQUYa}) reduces to
\begin{equation}
\left(  \tfrac{dR}{d\tau}\right)  ^{2}=\tfrac{1}{M_{0}^{2}}\left(
M+\tfrac{M_{0}^{2}}{2R}-\tfrac{Q^{2}}{2R}\right)  ^{2}-1. \label{EQUY2}%
\end{equation}
Introducing the total radial momentum $P\equiv M_{0}u^{r}=M_{0}\tfrac
{dR}{d\tau}$ of the shell, we can express the kinetic energy of the shell as
measured by static observers in $\mathcal{M}_{-}$ as $T\equiv-M_{0}u_{\mu}%
\xi_{-}^{\mu}-M_{0}=\sqrt{P^{2}+M_{0}^{2}}-M_{0}$. Then from equation
(\ref{EQUY2}) we have
\begin{equation}
M=-\tfrac{M_{0}^{2}}{2R}+\tfrac{Q^{2}}{2R}+\sqrt{P^{2}+M_{0}^{2}}%
=M_{0}+T-\tfrac{M_{0}^{2}}{2R}+\tfrac{Q^{2}}{2R}. \label{EQC}%
\end{equation}
where we choose the positive root solution due to the constraint
(\ref{Constraint}). Eq.(\ref{EQC}) is the \emph{mass formula} of the shell,
which depends on the time-dependent radial coordinate $R$ and kinetic energy
$T$. If $M\geq Q$, a black hole is formed and we have
\begin{equation}
M=M_{0}+T_{+}-\tfrac{M_{0}^{2}}{2r_{+}}+\tfrac{Q^{2}}{2r_{+}}\,, \label{EQL}%
\end{equation}
where $T_{+}\equiv T\left(  r_{+}\right)  $ and $r_{+}=M+\sqrt{M^{2}-Q^{2}}$
is the radius of external horizon of the black hole. 

\subsection{On the physical origin of the terms in mass formula of the black hole}

We know from the
Christodoulou-Ruffini black hole mass formula that
\begin{equation}
M=M_{\mathrm{irr}}+\tfrac{Q^{2}}{2r_{+}}, \label{irrmass}%
\end{equation}
so it follows that {}
\begin{equation}
M_{\mathrm{irr}}=M_{0}-\tfrac{M_{0}^{2}}{2r_{+}}+T_{+}, \label{EQM}
\end{equation}
namely that $M_{\mathrm{irr}}$ is the sum of only three contributions: the rest
mass $M_{0}$, the gravitational potential energy and the kinetic energy of the
rest mass evaluated at the horizon. $M_{\mathrm{irr}}$ is independent of the
electromagnetic energy, a fact noticed by \citet{b71}. We have taken
one further step here by identifying the independent physical contributions to
$M_{\mathrm{irr}}$.

Next we consider the physical interpretation of the electromagnetic term
$\tfrac{Q^{2}}{2R}$, which can be obtained by evaluating the conserved Killing
integral
\begin{align}
\int_{\Sigma_{t}^{+}}\xi_{+}^{\mu}T_{\mu\nu}^{\mathrm{(em)}}d\Sigma^{\nu}  &
=\int_{R}^{\infty}r^{2}dr\int_{0}^{1}d\cos\theta\int_{0}^{2\pi}d\phi
\ T^{\mathrm{(em)}}{}{}_{0}{}^{0}\nonumber\\
& =\tfrac{Q^{2}}{2R}\,,\label{EQR}%
\end{align}
where $\Sigma_{t}^{+}$ is the space-like hypersurface in $\mathcal{M}_{+}$
described by the equation $t_{+}=t=\mathrm{const}$, with $d\Sigma^{\nu}$ as
its surface element vector and where $T_{\mu\nu}^{\mathrm{(em)}}=-\tfrac
{1}{4\pi}\left(  F_{\mu}{}^{\rho}F_{\rho\nu}+\tfrac{1}{4}g_{\mu\nu}%
F^{\rho\sigma}F_{\rho\sigma}\right)  $ is the energy-momentum tensor of the
electromagnetic field. The quantity in Eq.(\ref{EQR}) differs from the purely
electromagnetic energy
\[
\int_{\Sigma_{t}^{+}}n_{+}^{\mu}T_{\mu\nu}^{\mathrm{(em)}}d\Sigma^{\nu}%
=\tfrac{1}{2}\int_{R}^{\infty}dr\sqrt{g_{rr}}\tfrac{Q^{2}}{r^{2}},
\]
where $n_{+}^{\mu}=f_{+}^{-1/2}\xi_{+}^{\mu}$ is the unit normal to the
integration hypersurface and $g_{rr}=f_{+}$. This is similar to the analogous
situation for the total energy of a static spherical star of energy density
$\epsilon$ within a radius $R$, $m\left(  R\right)  =4\pi\int_{0}^{R}%
dr\ r^{2}\epsilon$, which differs from the pure matter energy $m_{\mathrm{p}%
}\left(  R\right)  $ $=4\pi\int_{0}^{R}dr\sqrt{g_{rr}}r^{2}\epsilon$ by the
gravitational energy (see \citet{mtw73}). Therefore the term $\tfrac{Q^{2}}%
{2R}$ in the mass formula (\ref{EQC}) is the \emph{total} energy of the
electromagnetic field and includes its own gravitational binding energy. This
energy is stored throughout the region $\Sigma_{t}^{+}$, extending from $R$ to infinity.

\subsection{On the energy extraction process of blackholic energy}

We now turn to the problem of extracting the blackholic energy from a black hole (see \citet{cr71}). We can distinguish between two conceptually physically
different processes, depending on whether the electric field strength
$\mathcal{E}=\frac{Q}{r^{2}}$ is smaller or greater than the critical value
$\mathcal{E}_{\mathrm{c}}=\tfrac{m_{e}^{2}c^{3}}{e\hbar}$. Here $m_{e}$ and
$e$ are the mass and the charge of the electron. As already mentioned in this
paper an electric field $\mathcal{E}>\mathcal{E}_{\mathrm{c}}$ polarizes the
vacuum creating electron-positron pairs (see \citet{he35}). The maximum value
$\mathcal{E}_{+}=\tfrac{Q}{r_{+}^{2}}$ of the electric field around a black hole is
reached at the horizon. We then have the following:

\begin{enumerate}
\item For $\mathcal{E}_{+}<\mathcal{E}_{\mathrm{c}}$ the leading energy extraction mechanism consists of a sequence of discrete elementary decay processes of a neutral system, e.g. an atom, into two oppositely charged systems, e.g. a nucleus and electrons. We do not address here the creation of pairs due to vacuum polarization process, which is especially relevant for astrophysical applications only in the case $\mathcal{E}_{+} > \mathcal{E}_{\mathrm{c}}$ (see below). We are instead focusing on the above mentioned phenomenon since it may be particularly relevant to the creation of UHECRs. The condition $\mathcal{E}_{+}<\mathcal{E}_{\mathrm{c}}$ implies
\begin{align}
\xi & \equiv\tfrac{Q}{\sqrt{G}M}\nonumber\\
& \lesssim\left\{
\begin{array}
[c]{r}%
\tfrac{GM/c^{2}}{\lambda_{\mathrm{C}}}\tfrac{\sqrt{G}m_{e}}{e}\sim
10^{-6}\tfrac{M}{M_{\odot}}\quad\text{if }\tfrac{M}{M_{\odot}}\leq10^{6}\\
1\quad\quad\quad\quad\quad\quad\quad\quad\text{if }\tfrac{M}{M_{\odot}}>10^{6}%
\end{array}
\right.  ,\label{critical3}%
\end{align}
where $\lambda_{\mathrm{C}}$ is the Compton wavelength of the electron. \citet{dr73} and \citet{dhr74} have defined as the \emph{effective ergosphere} the region around a black hole where the energy extraction processes occur. This region extends from the horizon $r_{+}$ up to a radius
\begin{align}
r_{\mathrm{Eerg}}  & =\tfrac{GM}{c^{2}}\left[  1+\sqrt{1-\xi^{2}\left(
1-\tfrac{e^{2}}{G{m_{e}^{2}}}\right)  }\right]  \nonumber\\
& \simeq\tfrac{e}{m_{e}}\tfrac{Q}{c^{2}}\,.\label{EffErg}%
\end{align}
The energy extraction occurs in a finite number $N_{\mathrm{PD}}$ of such
discrete elementary processes, each one corresponding to a decrease of the black hole
charge. We have
\begin{equation}
N_{\mathrm{PD}}\simeq\tfrac{Q}{e}\,.
\end{equation}
Since the total extracted energy is (see Eq.(\ref{irrmass})) $E^{\mathrm{tot}%
}=\tfrac{Q^{2}}{2r_{+}}$, we obtain for the mean energy per accelerated
particle $\left\langle E\right\rangle _{\mathrm{PD}}=\tfrac{E^{\mathrm{tot}}%
}{N_{\mathrm{PD}}}$
\begin{equation}
\left\langle E\right\rangle _{\mathrm{PD}}=\tfrac{Qe}{2r_{+}}=\tfrac{1}%
{2}\tfrac{\xi}{1+\sqrt{1-\xi^{2}}}\tfrac{e}{\sqrt{G}m_{e}}\ m_{e}c^{2}%
\simeq\tfrac{1}{2}\xi\tfrac{e}{\sqrt{G}m_{e}}\ m_{e}c^{2},
\end{equation}
which gives
\begin{equation}
\left\langle E\right\rangle _{\mathrm{PD}}\lesssim\left\{
\begin{array}
[c]{r}%
\left(  \tfrac{M}{M_{\odot}}\right)  \times10^{21}eV\quad\text{if }\tfrac
{M}{M_{\odot}}\leq10^{6}\\
10^{27}eV\quad\quad\text{if }\tfrac{M}{M_{\odot}}>10^{6}%
\end{array}
\right.  . \label{UHECR}%
\end{equation}
One of the crucial aspects of the energy extraction process from a black hole is
its back reaction on the irreducible mass expressed in \citet{cr71}. Although
the energy extraction processes can occur in the entire effective ergosphere
defined by Eq. (\ref{EffErg}), only the limiting processes occurring on the
horizon with zero kinetic energy can reach the maximum efficiency while
approaching the condition of total reversibility (see Fig. 2 in \citet{cr71} for details). 
The farther from the horizon that a decay occurs, the more it
increases the irreducible mass and loses efficiency. Only in the complete
reversibility limit (\citet{cr71}) can the energy extraction process from an
extreme black hole reach the upper value of $50\%$ of the total black hole energy.
\item  For $\mathcal{E}_{+}\geq\mathcal{E}_{\mathrm{c}}$ the leading
extraction process is a \emph{collective} process based on an
electron-positron plasma generated by the vacuum polarization,
as discussed in section III in \citet{rubr}. The condition $\mathcal{E}_{+}%
\geq\mathcal{E}_{\mathrm{c}}$ implies
\begin{equation}
\tfrac{GM/c^{2}}{\lambda_{\mathrm{C}}}\left(  \tfrac{e}{\sqrt{G}m_{e}}\right)
^{-1}\simeq2\cdot10^{-6}\tfrac{M}{M_{\odot}}\leq\xi\leq1\,.
\end{equation}
This vacuum polarization process can occur only for a black hole with mass smaller
than $2\cdot10^{6}M_{\odot}$. The electron-positron pairs are now produced in
the dyadosphere of the black hole, (note that the dyadosphere is a subregion of the
effective ergosphere) whose radius $r_{ds}$ is given in Eq.(\ref{rc}).
We have $r_{ds}\ll r_{\mathrm{Eerg}}$. The number of particles
created and the total energy stored in dyadosphere are given in Eqs.(17,18) of \citet{rv02a} respectively and we have approximately
\begin{align}
N^\circ_{e^+e^-}  & \simeq\left(  \tfrac{r_{ds}}%
{\lambda_{\mathrm{C}}}\right)  \tfrac{Q}{e}\,,\label{numdya}\\
E_{dya}  & \simeq\tfrac{Q^{2}}{2r_{+}}\,
\end{align}
The mean energy per particle produced in the dyadosphere $\left\langle
E\right\rangle _{\mathrm{ds}}=\tfrac{E_{dya}}{N^\circ_{e^+e^-}}$ is
then
\begin{equation}
\left\langle E\right\rangle _{\mathrm{ds}}\simeq\tfrac{3}{8}\left(
\tfrac{\lambda_{\mathrm{C}}}{r_{ds}}\right)  \tfrac{Qe}{r_{+}%
}\,,\label{meanenedya}%
\end{equation}
which can be also rewritten as
\begin{equation}
\left\langle E\right\rangle _{\mathrm{ds}}\simeq\tfrac{1}{2}\left(
\tfrac{r_{\mathrm{ds}}}{r_{+}}\right)  \ m_{e}c^{2}\sim\sqrt{\tfrac{\xi
}{M/M_{\odot}}}10^{5}keV\,.\label{GRB}%
\end{equation}
Such a process of vacuum polarization, occurring not at the horizon but in the
extended dyadosphere region ($r_{+}\leq r\leq r_{\mathrm{ds}}$) around a
black hole, has been observed to reach the maximum efficiency limit of $50\%$ of the
total mass-energy of an extreme black hole (see e.g. \citet{prx98}). The conceptual
justification of this result follows from the present work: the $e^{+}e^{-}$
creation process occurs at the expense of the Coulomb energy given by Eq.
(\ref{EQR}) and does not affect the irreducible mass given by Eq. (\ref{EQM}),
which indeed, as we have proved, does not depend of the electromagnetic
energy. In this sense, $\delta M_{\mathrm{irr}}=0$ and the transformation is
fully reversible. This result will be further validated by the study of the
dynamical formation of the dyadosphere, which we have obtained using the
present work and \citet{crv02} (see \citet{rvx03a,rvx03b}).
\end{enumerate}

Let us now compare and contrast these two processes. We have
\begin{align}
r_{\mathrm{Eerg}}  & \simeq\left(  \tfrac{r_{ds}}{\lambda
_{\mathrm{C}}}\right)  r\\
N_{\mathrm{dya}}  & \simeq\left(  \tfrac{r_{ds}}{\lambda
_{\mathrm{C}}}\right)  N_{\mathrm{PD}},\\
\left\langle E\right\rangle _{\mathrm{dya}}  & \simeq\left(  \tfrac
{\lambda_{\mathrm{C}}}{r_{ds}}\right)  \left\langle E\right\rangle
_{\mathrm{PD}}.
\end{align}
Moreover we see (Eqs. (\ref{UHECR}), (\ref{GRB})) that $\left\langle
E\right\rangle _{\mathrm{PD}}$ is in the range of energies of UHECR, while for
$\xi\sim0.1$ and $M\sim10M_{\odot}$, $\left\langle E\right\rangle
_{\mathrm{ds}}$ is in the gamma ray range. In other words, the discrete
particle decay process involves a small number of particles with ultra high
energies ($\sim10^{21}eV$), while vacuum polarization involves a much larger
number of particles with lower mean energies ($\sim10MeV$).

\begin{figure}
\centering
\includegraphics[width=10cm,clip]{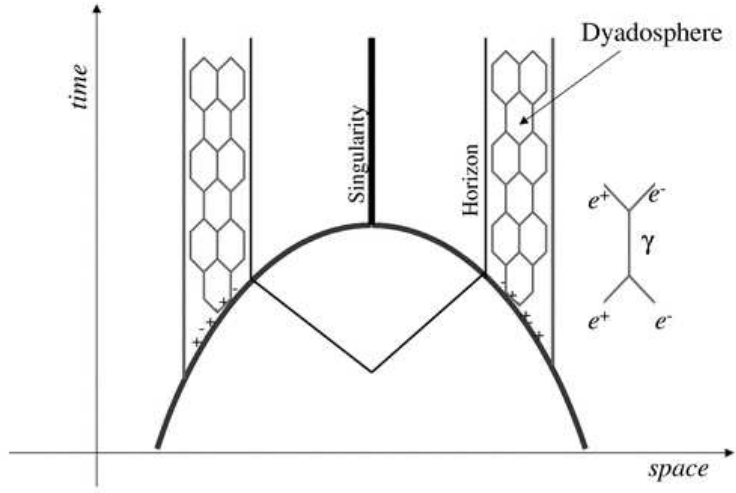}
\caption{Space-time diagram of the collapse process leading to the formation of the dyadosphere. As the collapsing core crosses the dyadosphere radius the pair creation process starts, and the pairs thermalize in a neutral plasma configuration. Then also the horizon is crossed and the singularity is formed.}
\label{dyaform}
\end{figure}

Having so established and clarified the basic conceptual processes of the energetic of the black hole, we are now ready to approach, using the new analytic solution obtained, the dynamical process of vacuum polarization occurring during the formation of a black hole as qualitatively represented in Fig. \ref{dyaform}. The study of the dyadosphere dynamical formation as well as of the electron-positron plasma dynamical evolution will lead to the first possibility of directly observing the general relativistic effects approaching the black hole horizon.

\section{Contributions of GRBs to the black hole theory}

\subsection{On the gravitational binding energy of white dwarf and neutron stars}

The aim of this section is to point out how the knowledge obtained from the black hole model is of relevance also for the basic theory of black holes and further how very high precision verification of general relativistic effects in the very strong field near the formation of the horizon should be expected in the near future.

We shall first see how Eq.(\ref{EQM}) for $M_{\mathrm{irr}}$,
\begin{equation}
M_{\mathrm{irr}}=M_{0}-\tfrac{M_{0}^{2}}{2r_{+}}+T_{+}\, ,
\end{equation}
leads to a deeper physical understanding of the role of the gravitational interaction in the maximum energy extraction process of a black hole. This formula can also be of assistance in clarifying some long lasting epistemological issue on the role of general relativity, quantum theory and thermodynamics.

It is well known that if a spherically symmetric mass distribution without any electromagnetic structure undergoes free gravitational collapse, its total mass-energy $M$ is conserved according to the Birkhoff theorem: the increase in the kinetic energy of implosion is balanced by the increase in the gravitational energy of the system. If one considers the possibility that part of the kinetic energy of implosion is extracted then the situation is very different: configurations of smaller mass-energy and greater density can be attained without violating Birkhoff theorem.

We illustrate our considerations with two examples: one has found confirmation from astrophysical observations, the other promises to be of relevance for gamma ray bursts (GRBs) (see \citet{rv02a}). Concerning the first example, it is well known from the work of \citet{l32} that at the endpoint of thermonuclear evolution, the gravitational collapse of a spherically symmetric star can be stopped by the Fermi pressure of the degenerate electron gas (white dwarf). A configuration of equilibrium can be found all the way up to the critical number of particles
\begin{equation}
N_{\mathrm{crit}}=0.775\tfrac{m_{Pl}^{3}}{m_{0}^{3}},
\end{equation}
where the factor $0.775$ comes from the coefficient $\tfrac{3.098}{\mu^{2}}$ of the solution of the Lane-Emden equation with polytropic index $n=3$, and $m_{Pl}=\sqrt{\tfrac{\hbar c}{G}}$ is the Planck mass, $m_{0}$ is the nucleon mass and $\mu$ the average number of electrons per nucleon. As the kinetic energy of implosion is carried away by radiation the star settles down to a configuration of mass
\begin{equation}
M=N_{\mathrm{crit}}m_{0}-U, \label{BE}%
\end{equation}
where the gravitational binding energy $U$ can be as high as $5.72\times 10^{-4}N_{\mathrm{crit}}m_{0}$.

Similarly \citet{g51} has shown that a gravitational collapse process to still higher densities can be stopped by the Fermi pressure of the neutrons (neutron star) and \citet{ov39} has shown that, if the effects of strong interactions are neglected, a configuration of equilibrium exists also in this case all the way up to a critical number of particles
\begin{equation}
N_{\mathrm{crit}}=0.398\tfrac{m_{Pl}^{3}}{m_{0}^{3}},
\end{equation}
where the factor $0.398$ comes now from the integration of the
Tolman-Oppenheimer-Volkoff equation (see e.g. \citet{htww65}). If the kinetic energy of implosion is again carried away by radiation of photons or neutrinos and antineutrinos the final configuration is characterized by the formula (\ref{BE}) with $U\lesssim2.48\times10^{-2}N_{\mathrm{crit}}m_{0}$. These considerations and the existence of such large values of the gravitational binding energy have been at the heart of the explanation of astrophysical phenomena such as red-giant stars and supernovae: the corresponding measurements of the masses of neutron stars and white dwarfs have been carried out with unprecedented accuracy in binary systems (\citet{gr75}).

\subsection{On the minimum value of the reducible mass of a black hole formed in a spherically symmetric gravitational collapse}

From a theoretical physics point of view it is still an open question how far such a sequence can go: using causality nonviolating interactions, can one find a sequence of braking and energy extraction processes by which the density and the gravitational binding energy can increase indefinitely and the mass-energy of the collapsed object be reduced at will? This question can also be formulated in the mass-formula language of a black hole given in \citet{cr71} (see also \citet{rv02a}): given a collapsing core of nucleons with a given rest mass-energy $M_{0}$, what is the minimum irreducible mass of the black hole which is formed?

Following \citet{crv02} and \citet{rv02a}, consider a spherical shell of rest mass $M_{0}$ collapsing in a flat space-time. In the neutral case the irreducible mass of the final black hole satisfies the equation (see \citet{rv02a})
\begin{equation}
M_{\mathrm{irr}}=M=M_{0}-\tfrac{M_{0}^{2}}{2r_{+}}+T_{+}, \label{Mirr2}%
\end{equation}
where $M$ is the total energy of the collapsing shell and $T_{+}$ the kinetic energy at the horizon $r_{+}$. Recall that the area $S$ of the horizon is \citet{cr71}
\begin{equation}
S=4\pi r_{+}^{2}=16\pi M_{\mathrm{irr}}^{2} \label{Sbis}%
\end{equation}
where $r_{+}=2M_{\mathrm{irr}}$ is the horizon radius. The minimum irreducible mass $M_{\mathrm{irr}}^{\left(  {\mathrm{min}}\right)  }$ is obtained when the kinetic energy at the horizon $T_{+}$ is $0$, that is when the entire kinetic energy $T_{+}$ has been extracted. We then obtain the simple result
\begin{equation}
M_{\mathrm{irr}}^{\left(  \mathrm{min}\right)  }=\tfrac{M_{0}}{2}.
\label{Mirrmin}%
\end{equation}
We conclude that in the gravitational collapse of a spherical shell of rest mass $M_{0}$ at rest at infinity (initial energy $M_{\mathrm{i}}=M_{0}$), an energy up to $50\%$ of $M_{0}c^{2}$ can in principle be extracted, by braking processes of the kinetic energy. In this limiting case the shell crosses the horizon with $T_{+}=0$. The limit $\tfrac{M_{0}}{2}$ in the extractable kinetic energy can further increase if the collapsing shell is endowed with kinetic energy at infinity, since all that kinetic energy is in principle extractable.

\begin{figure}
\centering
\includegraphics[width=10cm,clip]{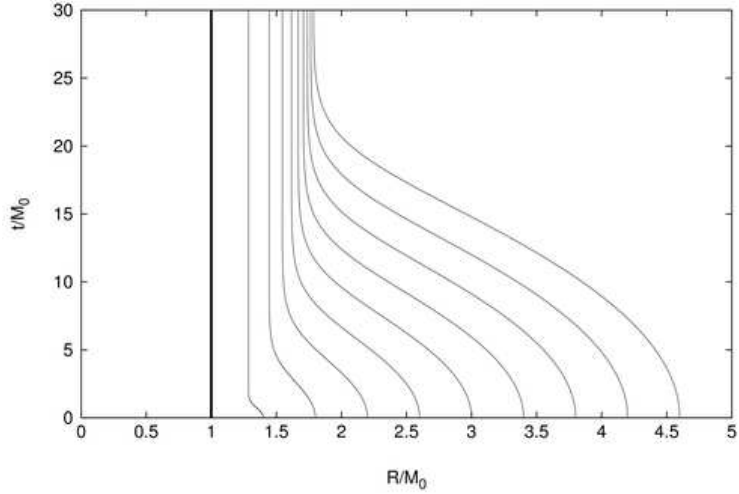}
\caption{Collapse curves for neutral shells with rest mass $M_{0}$ starting at rest at selected radii $R^{\ast}$ computed by using the exact solutions given in \citet{crv02}. A different value of $M_{\mathrm{irr}}$ (and therefore of $r_{+}$) corresponds to each curve. The time parameter is the Schwarzschild time coordinate $t$ and the asymptotic behavior at the respective horizons is evident. The limiting configuration $M_{\mathrm{irr}}=\tfrac{M_{0}}{2}$ (solid line) corresponds to the case in which the shell is trapped, at the very beginning of its motion, by the formation of the horizon.}
\label{fig1l2}
\end{figure}

\begin{figure}
\centering
\includegraphics[width=10cm,clip]{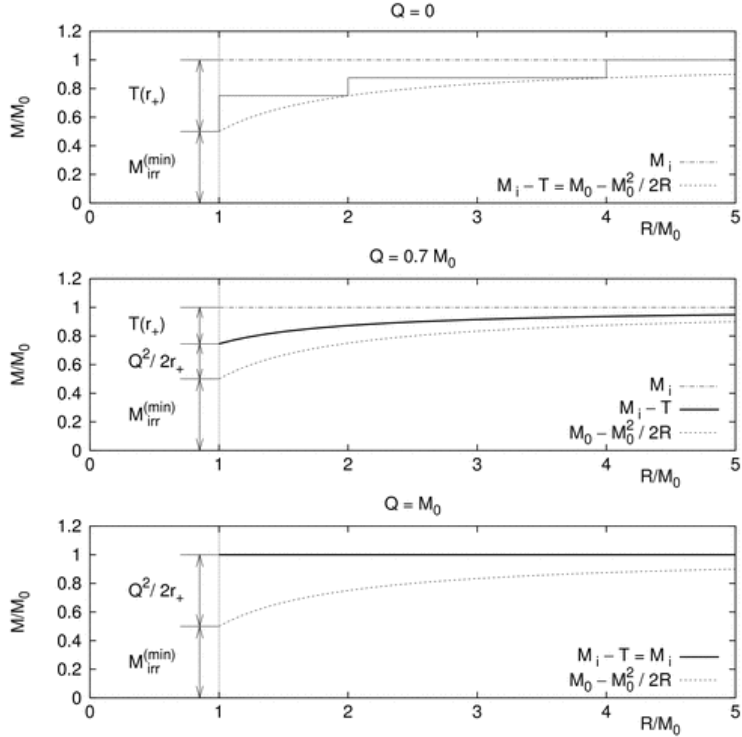}
\caption{Energetics of a shell such that $M_{\mathrm{i}}=M_{0}$, for selected values of the charge. In the first diagram $Q=0$; the dashed line represents the total energy for a gravitational collapse without any braking process as a function of the radius $R$ of the shell; the solid, stepwise line represents a collapse with suitable braking of the kinetic energy of implosion at selected radii; the dotted line represents the rest mass energy plus the gravitational binding energy. In the second and third diagram $Q/M_{0}=0.7$, $Q/M_{0}=1$ respectively; the dashed and the dotted lines have the same meaning as above; the solid lines represent the total energy minus the kinetic energy. The region between the solid line and the dotted line corresponds to the stored electromagnetic energy. The region between the dashed line and the solid line corresponds to the kinetic energy of collapse. In all the cases the sum of the kinetic energy and the electromagnetic energy at the horizon is 50\% of $M_{0}$. Both the electromagnetic and the kinetic energy are extractable. It is most remarkable that the same underlying process occurs in the three cases: the role of the electromagnetic interaction is twofold: a) to reduce the kinetic energy of implosion by the Coulomb repulsion of the shell; b) to store such an energy in the region around the black hole. The stored electromagnetic energy is extractable as shown in \citet{rv02a}.}
\label{fig3l2}
\end{figure}

In order to illustrate the physical reasons for this result, using the formulas of \citet{crv02}, we have represented in Fig. \ref{fig1l2} the world lines of spherical shells of the same rest mass $M_{0}$, starting their gravitational collapse at rest at selected radii $R^{\ast}$. These initial conditions can be implemented by performing suitable braking of the collapsing shell and concurrent kinetic energy extraction processes at progressively smaller radii (see also Fig. \ref{fig3l2}). The reason for the existence of the minimum (\ref{Mirrmin}) in the black hole mass is the ``self closure'' occurring by the formation of a horizon in the initial configuration (thick line in Fig. \ref{fig1l2}).

Is the limit $M_{\mathrm{irr}}\rightarrow\tfrac{M_{0}}{2}$ actually attainable without violating causality? Let us consider a collapsing shell with charge $Q$. If $M\geq Q$ a black hole is formed. As pointed out in \citet{rv02a} the irreducible mass of the final black hole does not depend on the charge $Q$. Therefore Eqs.(\ref{Mirr2}) and (\ref{Mirrmin}) still hold in the charged case with $r_{+}=M+\sqrt{M^{2}-Q^{2}}$. In Fig. \ref{fig3l2} we consider the special case in which the shell is initially at rest at infinity, i.e. has initial energy $M_{\mathrm{i}}=M_{0}$, for three different values of the charge $Q$. We plot the initial energy $M_{i}$, the energy of the system when all the kinetic energy of implosion has been extracted as well as the sum of the rest mass energy and the gravitational binding energy $-\tfrac{M_{0}^{2}}{2R}$ of the system (here $R$ is the radius of the shell). In the extreme case $Q=M_{0}$, the shell is in equilibrium at all radii (see \citet{crv02}) and the kinetic energy is identically zero. In all three cases, the sum of the extractable kinetic energy $T$ and the electromagnetic energy $\tfrac{Q^{2}}{2R}$ reaches $50\%$ of the rest mass energy at the horizon, according to Eq.(\ref{Mirrmin}).

What is the role of the electromagnetic field here? If we consider the case of a charged shell with $Q\simeq M_{0}$, the electromagnetic repulsion implements the braking process and the extractable energy is entirely stored in the electromagnetic field surrounding the black hole (see \citet{rv02a}). In \citet{rv02a} we have outlined two different processes of electromagnetic energy extraction. We emphasize here that the extraction of $50\%$ of the mass-energy of a black hole is not specifically linked to the electromagnetic field but depends on three factors: a) the increase of the gravitational energy during the collapse, b) the formation of a horizon, c) the reduction of the kinetic energy of implosion. Such conditions are naturally met during the formation of an extreme black hole but are more general and can indeed occur in a variety of different situations, e.g. during the formation of a Schwarzschild black hole by a suitable extraction of the kinetic energy of implosion (see Fig. \ref{fig1l2} and Fig. \ref{fig3l2}).

\subsection{On the Bekenstein-Hawking consideration of incompatibility between general relativity and thermodynamics}

Now consider a test particle of mass $m$ in the gravitational field of an already formed Schwarzschild black hole of mass $M$ and go through such a sequence of braking and energy extraction processes. \citet{k49} found for the energy $E$ of the particle as a function of the radius $r$
\begin{equation}
E=m\sqrt{1-\tfrac{2M}{r}}.\label{pointtest}
\end{equation}
It would appear from this formula that the entire energy of a particle could be extracted in the limit $r\rightarrow2M$. Such $100\%$ efficiency of energy extraction has often been quoted as evidence for incompatibility between General Relativity and the second principle of Thermodynamics (see \citet{b73} and references therein). J. Bekenstein and S. Hawking have gone as far as to consider General Relativity not to be a complete theory and to conclude that in order to avoid inconsistencies with thermodynamics, the theory should be implemented through a quantum description (\citet{b73,h74}). Einstein himself often expressed the opposite point of view (see e.g. \citet{d02}).

The analytic treatment presented in \citet{crv02} can clarify this fundamental issue. It allows to express the energy increase $E$ of a black hole of mass $M_{1}$ through the accretion of a shell of mass $M_{0}$ starting its motion at rest at a radius $R$ in the following formula which generalizes Eq.(\ref{pointtest}):
\begin{equation}
E\equiv M-M_{1}=-\tfrac{M_{0}^{2}}{2R}+M_{0}\sqrt{1-\tfrac{2M_{1}}{R}},
\end{equation}
where $M=M_{1}+E$ is clearly the mass-energy of the final black hole. This formula differs from the Kaplan formula (\ref{pointtest}) in three respects: a) it takes into account the increase of the horizon area due to the accretion of the shell; b) it shows the role of the gravitational self energy of the imploding shell; c) it expresses the combined effects of a) and b) in an exact closed formula.

The minimum value $E_{\mathrm{\min}}$ of $E$ is attained for the minimum value of the radius $R=2M$: the horizon of the final black hole. This corresponds to the maximum efficiency of the energy extraction. We have
\begin{equation}
E_{\min}=-\tfrac{M_{0}^{2}}{4M}+M_{0}\sqrt{1-\tfrac{M_{1}}{M}}=-\tfrac
{M_{0}^{2}}{4(M_{1}+E_{\min})}+M_{0}\sqrt{1-\tfrac{M_{1}}{M_{1}+E_{\min}}},
\end{equation}
or solving the quadratic equation and choosing the positive solution for physical reasons
\begin{equation}
E_{\min}=\tfrac{1}{2}\left(  \sqrt{M_{1}^{2}+M_{0}^{2}}-M_{1}\right)  .
\end{equation}
The corresponding efficiency of energy extraction is
\begin{equation}
\eta_{\max}=\tfrac{M_{0}-E_{\min}}{M_{0}}=1-\tfrac{1}{2}\tfrac{M_{1}}{M_{0}%
}\left(  \sqrt{1+\tfrac{M_{0}^{2}}{M_{1}^{2}}}-1\right)  , \label{efficiency}%
\end{equation}
which is strictly \emph{smaller than} 100\% for \emph{any} given $M_{0}\neq0$. It is interesting that this analytic formula, in the limit $M_{1}\ll M_{0}$, properly reproduces the result of equation (\ref{Mirrmin}), corresponding to an efficiency of $50\%$. In the opposite limit $M_{1}\gg M_{0}$ we have
\begin{equation}
\eta_{\max}\simeq1-\tfrac{1}{4}\tfrac{M_{0}}{M_{1}}.
\end{equation}
Only for $M_{0}\rightarrow0$, Eq.(\ref{efficiency}) corresponds to an efficiency of 100\% and correctly represents the limiting reversible transformations introduced in \citet{cr71}. It seems that the difficulties of reconciling General Relativity and Thermodynamics are ascribable not to an incompleteness of General Relativity but to the use of the Kaplan formula in a regime in which it is not valid. The generalization of the above results to stationary black holes is being considered.

\section{On a separatrix in an overcritical collapse}

We are now ready to analyze the dynamical properties of an electron--positron--photon plasma created by the vacuum polarization process occurring around a charged gravitationally collapsing core of an initially neutral star are examined within the framework of General Relativity and Quantum Field Theory. The Reissner-Nordstr{\o}m geometry is assumed to apply between the collapsing core and the oppositely charged remnant of the star. The appearance of a separatrix at radius $\bar{R}$, well outside the asymptotic approach to the horizon, is evidenced. The neutral electron--positron--photon plasma created at radii $r>\bar{R}$ self-propels outwards to infinity, following the classical PEM--pulse analysis (\citet{rswx99,rswx00}). The plasma created at $r<\bar{R}$ remains trapped and follows the gravitational collapse of the core only contributing to the reduction of the electromagnetic energy of the black hole and to the increase of its irreducible mass. This phenomenon has consequences for the observational properties of gamma--ray bursts and is especially relevant for the theoretical prediction of the temporal and spectral structure of the short bursts.

The formulation of the physics of the \emph{dyadosphere} of an electromagnetic
black hole (black hole) has been until now approached by assuming the vacuum
polarization process \emph{\`{a} l\`{a}} Sauter--Heisenberg--Euler--Schwinger
\citep{s31,he35,s51} in the field of an already formed Kerr--Newman
(\citet{dr75}) or Reissner-Nordstr{\o}m black hole (\citet{prx98,rv02a}). This
acausal approach is certainly valid in order to describe the overall
energetics and the time development of the gamma--ray bursts (GRBs) reaching a
remarkable agreement between the observations and the theoretical prediction,
in particular with respect to: a) the existence of a P-GRB (\citet{lett1}),
b) the afterglow detailed luminosity function and
spectral properties (\citet{rbcfx03a,rubr,r02}) and c) the relative
intensity of the P-GRB to the afterglow (\citet{lett2,rbcfx03a,rubr}).

This acausal approach has to be improved by taking into account the causal
dynamical process of the formation of the dyadosphere as soon as the detailed
description on timescales of $10^{-4}-10^{-3}$s of the P--GRB are considered.
Such a description leads to theoretical predictions on the time variability of
the P--GRB spectra which may become soon testable by a new class of specially
conceived space missions.

We report progress in this theoretically
challenging process which is marked by distinctive and precise quantum and
general relativistic effects. These new results have been made possible by the
recent progress in \citet{crv02,rv02a} and especially
\citet{rvx03}. There it was demonstrated the intrinsic stability of the
gravitational amplification of the electromagnetic field at the surface of a
charged star core collapsing to a black hole. The $e^{+}e^{-}$ plasma generated by
the vacuum polarization process around the core is entangled in the
electromagnetic field (\citet{rvx03a}). The $e^{+}e^{-}$ pairs do thermalize in
an electron--positron--photon plasma on a time scale $10^{2}-10^{4}$ times
larger than $\hbar/m_{e}c$ (\citet{rvx03}), where $c$ is the speed of light and
$m_{e}$ the electron mass. As soon as the thermalization has occurred, a
dynamical phase of this electrically neutral plasma starts following the
considerations already discussed in \citet{rswx99,rswx00}. While the temporal
evolution of the $e^{+}e^{-}\gamma$ plasma takes place, the gravitationally
collapsing core moves inwards, giving rise to a further amplified
supercritical field, which in turn generates a larger amount of $e^{+}e^{-}$
pairs leading to a yet higher temperature in the newly formed $e^{+}%
e^{-}\gamma$ plasma. We report, in the following, progress in the
understanding of this crucial dynamical process: the main difference from the
previous treatments is the fact that we do not consider an already formed black hole
but we follow the dynamical phase of the formation of dyadosphere and of the
asymptotic approach to the horizon by examining the time varying process at
the surface of the gravitationally collapsing core.

The space--time external to the surface of the spherically symmetric
collapsing core is described by the Reissner-Nordstr{\o}m geometry (\citet{P74})
with line element
\begin{equation}
ds^{2}=-\alpha^{2}dt^{2}+\alpha^{-2}dr^{2}+r^{2}d\Omega^{2},\label{ds}%
\end{equation}
with $d\Omega^{2}=d\theta^{2}+\sin^{2}\theta d\phi^{2}$, $\alpha^{2}%
=\alpha^{2}\left(  r\right)  =1-2M/r+Q^{2}/r^{2}$, where $M$ and $Q$ are the
total energy and charge of the core as measured at infinity. On the core
surface, which at the time $t_{0}$ has radial coordinate $r_{0}$, the
electromagnetic field strength is $\mathcal{E}=\mathcal{E}\left(
r_{0}\right)  =Q/r_{0}^{2}$. The equation of core's collapse is (see
\citet{crv02}):
\begin{equation}
\tfrac{dr_{0}}{dt_{0}}=-\tfrac{\alpha^{2}\left(  r_{0}\right)  }{H\left(
r_{0}\right)  }\sqrt{H^{2}\left(  r_{0}\right)  -\alpha^{2}\left(
r_{0}\right)  }\label{Motion}%
\end{equation}
where $H\left(  r_{0}\right)  =\tfrac{M}{M_{0}}-\tfrac{M_{0}^{2}+Q^{2}}%
{2M_{0}r_{0}}$ and $M_{0}$ is the core rest mass. Analytic expressions for the
solution of Eq.(\ref{Motion}) were given in \citet{crv02}. We here recall that
the dyadosphere radius is defined by $\mathcal{E}\left(  r_{\mathrm{ds}%
}\right)  =\mathcal{E}_{\mathrm{c}}=$ $m_{e}^{2}c^{3}/e\hbar$ (\citet{prx98}) as
$r_{\mathrm{ds}}=\sqrt{eQ\hbar/m_{e}^{2}c^{3}}$, where $e$ is the electron
charge. In the following we assume that the dyadosphere starts to be formed at
the instant $t_{\mathrm{ds}}=t_{0}\left(  r_{\mathrm{ds}}\right)  =0$.

Having formulated the core collapse in General Relativity in Eq.(\ref{Motion}%
), in order to describe the quantum phenomena, we consider, at each value of
$r_{0}$ and $t_{0}$, a slab of constant coordinate thickness $\Delta r$ small
in comparison with $r_{\mathrm{ds}}$ and larger than $\hbar/m_{e}c^{2}$. All
the results will be shown to be independent on the choice of the value of
$\Delta r$. In each slab the process of vacuum polarization leading to
$e^{+}e^{-}$ pair creation is considered. As shown in \citet{rvx03,rvx03a} the
pairs created oscillate (\citet{KESCM91,KESCM92,CEKMS93,BMP...99}) with
ultrarelativistic velocities and partially annihilate into photons; the
electric field oscillates around zero and the amplitude of such oscillations
decreases with a characteristic time of the order of $10^{2}-10^{4}$
$\hbar/m_{e}c^{2}$. The electric field is effectively screened to the critical
value $\mathcal{E}_{\mathrm{c}}$ and the pairs thermalize to an $e^{+}%
e^{-}\gamma$ plasma. While the average of the electric field $\mathcal{E}$
over one oscillation is $0$, the average of $\mathcal{E}^{2}$ is of the order
of $\mathcal{E}_{c}^{2}$, therefore the energy density in the pairs and
photons, as a function of $r_{0}$, is given by \citet{rv02a}
\begin{equation}
\epsilon_{0}\left(  r_{0}\right)  =\tfrac{1}{8\pi}\left[  \mathcal{E}%
^{2}\left(  r_{0}\right)  -\mathcal{E}_{c}^{2}\right]  =\tfrac{\mathcal{E}%
_{c}^{2}}{8\pi}\left[  \left(  \tfrac{r_{\mathrm{ds}}}{r_{0}}\right)
^{4}-1\right]  .\label{eps0}%
\end{equation}
For the number densities of $e^{+}e^{-}$ pairs and photons at thermal
equilibrium we have $n_{e^{+}e^{-}}\simeq n_{\gamma}$; correspondingly the
equilibrium temperature $T_{0}$, which is clearly a function of $r_{0}$ and is
different for each slab, is such that
\begin{equation}
\epsilon\left(  T_{0}\right)  \equiv\epsilon_{\gamma}\left(  T_{0}\right)
+\epsilon_{e^{+}}\left(  T_{0}\right)  +\epsilon_{e^{-}}\left(  T_{0}\right)
=\epsilon_{0},\label{eq0}%
\end{equation}
with $\epsilon$ and $n$ given by Fermi (Bose) integrals (with zero chemical
potential) already given in Eqs.\eqref{enpho}--\eqref{enel1}:
\begin{align}
\epsilon_{e^{+}e^{-}}\left(  T_{0}\right)   &  =\tfrac{2}{\pi^{2}\hbar^{3}%
}\int_{m_{e}}^{\infty}\tfrac{\left(  E^{2}-m_{e}^{2}\right)  ^{1/2}}%
{\exp\left(  E/kT_{0}\right)  +1}E^{2}dE,\quad\epsilon_{\gamma}\left(
T_{0}\right)  =\tfrac{\pi^{2}}{15\hbar^{3}}\left(  kT_{0}\right)
^{4},\label{Integrals1}\\
n_{e^{+}e^{-}}\left(  T_{0}\right)   &  =\tfrac{1}{\pi^{2}\hbar^{3}}%
\int_{m_{e}}^{\infty}\tfrac{\left(  E^{2}-m_{e}^{2}\right)  ^{1/2}}%
{\exp\left(  E/kT_{0}\right)  +1}EdE,\quad n_{\gamma}\left(  T_{0}\right)
=\tfrac{2\zeta\left(  3\right)  }{\hbar^{3}}\left(  kT_{0}\right)
^{3},\label{Integrals2}%
\end{align}
where $k$ is the Boltzmann constant. From the conditions set by Eqs.(\ref{eq0}%
), (\ref{Integrals1}), (\ref{Integrals2}), we can now turn to the dynamical
evolution of the $e^{+}e^{-}\gamma$ plasma in each slab. We use the covariant
conservation of energy momentum and the rate equation for the number of pairs
in the Reissner-Nordstr{\o}m geometry external to the star core:
\begin{equation}
\nabla_{a}T^{ab}=0,\quad\nabla_{a}\left(  n_{e^{+}e^{-}}u^{a}\right)
=\overline{\sigma v}\left[  n_{e^{+}e^{-}}^{2}\left(  T\right)  -n_{e^{+}%
e^{-}}^{2}\right]  ,\label{na}%
\end{equation}
We follow closely the treatment which we developed for the consideration of a
plasma generated in the dyadosphere of an already formed black hole
(\citet{rswx99,rswx00}). It was shown in \citet{rswx99,rswx00} that the plasma
expands as a pair--electromagnetic pulse (PEM pulse) of constant thickness in
the laboratory frame. Since the expansion, hydrodynamical timescale is much
larger than the pair creation ($\hbar/m_{e}c^{2}$) and the thermalization
($10^{2}-10^{4}\hbar/m_{e}c^{2}$) time-scales, in each slab the plasma remains
at thermal equilibrium in the initial phase of the expansion and the right
hand side of the rate Eq.(\ref{na}) is effectively $0$, see Fig. 24 (second
panel) of \citet{rubr} for details.

If we denote by $\xi^{a}$ the static Killing vector field normalized at unity
at spacial infinity and by $\left\{  \Sigma_{t}\right\}  _{t}$ the family of
space-like hypersurfaces orthogonal to $\xi^{a}$ ($t$ being the Killing time)
in the Reissner-Nordstr{\o}m geometry, from Eqs.(\ref{na}), the following
integral conservation laws can be derived (see for instance \citet{D79,S60})
\begin{equation}
\int_{\Sigma_{t}}\xi_{a}T^{ab}d\Sigma_{b}=E,\quad\int_{\Sigma_{t}}%
n_{e^{+}e^{-}}u^{b}d\Sigma_{b}=N_{e^{+}e^{-}},\label{Ne}%
\end{equation}
where $d\Sigma_{b}=\alpha^{-2}\xi_{b}r^{2}\sin\theta drd\theta d\phi$ is the
vector surface element, $E$ the total energy and $N_{e^{+}e^{-}}$ the total
number of pairs which remain constant in each slab. We then have
\begin{equation}
\left[  \left(  \epsilon+p\right)  \gamma^{2}-p\right]  r^{2}=\mathfrak
{E},\quad n_{e^{+}e^{-}}\gamma\alpha^{-1}r^{2}=\mathfrak{N}_{e^{+}e^{-}%
},\label{ne}%
\end{equation}
where $\mathfrak{E}$ and $\mathfrak{N}_{e^{+}e^{-}}$ are constants and
\begin{equation}
\gamma\equiv\alpha^{-1}u^{a}\xi_{a}=\left[  1-\alpha^{-4}\left(  \tfrac
{dr}{dt}\right)  ^{2}\right]  ^{-1/2}%
\end{equation}
is the Lorentz $\gamma$ factor of the slab as measured by static observers. We
can rewrite Eqs.(\ref{Ne}) for each slab as
\begin{align}
\left(  \tfrac{dr}{dt}\right)  ^{2} &  =\alpha^{4}f_{r_{0}},\label{eq17}\\
\left(  \tfrac{r}{r_{0}}\right)  ^{2} &  =\left(  \tfrac{\epsilon+p}%
{\epsilon_{0}}\right)  \left(  \tfrac{n_{e^{+}e^{-}0}}{n_{e^{+}e^{-}}}\right)
^{2}\left(  \tfrac{\alpha}{\alpha_{0}}\right)  ^{2}-\tfrac{p}{\epsilon_{0}%
}\left(  \tfrac{r}{r_{0}}\right)  ^{4},\label{eq18}\\
f_{r_{0}} &  =1-\left(  \tfrac{n_{e^{+}e^{-}}}{n_{e^{+}e^{-}0}}\right)
^{2}\left(  \tfrac{\alpha_{0}}{\alpha}\right)  ^{2}\left(  \tfrac{r}{r_{0}%
}\right)  ^{4}\label{eq19}%
\end{align}
where pedex $_{0}$ refers to quantities evaluated at selected initial times
$t_{0}>0$, having assumed $r\left(  t_{0}\right)  =r_{0}$, $\left.
dr/dt\right|  _{t=t_{0}}=0$, $T\left(  t_{0}\right)  =T_{0}$.

Eq.(\ref{eq17}) is only meaningful when $f_{r_{0}}\left(  r\right)  \geq0$.
From the structural analysis of such equation it is clearly identifiable a
critical radius $\bar{R}$ such that:

\begin{itemize}
\item for any slab initially located at $r_{0}>\bar{R}$ we have $f_{r_{0}%
}\left(  r\right)  \geq0$ for any value of $r\geq r_{0}$ and $f_{r_{0}}\left(
r\right)  <0$ for $r\lesssim r_{0}$; therefore a slab initially located at a
radial coordinate $r_{0}>\bar{R}$ moves outwards,
\item  for any slab initially located at $r_{0}<\bar{R}$ we have $f_{r_{0}%
}\left(  r\right)  \geq0$ for any value of $r_{+}<r\leq r_{0}$ and $f_{r_{0}%
}\left(  r\right)  <0$ for $r\gtrsim r_{0}$; therefore a slab initially
located at a radial coordinate $r_{0}<\bar{R}$ moves inwards and is trapped by
the gravitational field of the collapsing core.
\end{itemize}

We define the surface $r=\bar{R}$, the \emph{dyadosphere trapping surface
}(DTS). The radius $\bar{R}$ of DTS is generally evaluated by the condition
$\left.  \tfrac{df_{\bar{R}}}{dr}\right|  _{r=\bar{R}}=0$.
$\bar{R}$ is so close to the horizon value $r_{+}$ that the initial
temperature $T_{0}$ satisfies $kT_{0}\gg m_{e}c^{2}$ and we can obtain for
$\bar{R}$ an analytical expression. Namely the ultrarelativistic approximation
of all Fermi integrals, Eqs.(\ref{Integrals1}) and (\ref{Integrals2}), is
justified and we have $n_{e^{+}e^{-}}\left(  T\right)  \propto T^{3}$ and
therefore $f_{r_{0}}\simeq1-\left(  T/T_{0}\right)  ^{6}\left(  \alpha
_{0}/\alpha\right)  ^{2}\left(  r/r_{0}\right)  ^{4}$ ($r\leq\bar{R}$).
The defining equation of $\bar{R}$, together with (\ref{eq19}), then gives
\begin{equation}
\bar{R}=2M\left[  1+\left(  1-3Q^{2}/4M^{2}\right)  ^{1/2}\right]  >r_{+}.
\end{equation}

\begin{figure}
\centering
\includegraphics[width=10cm]{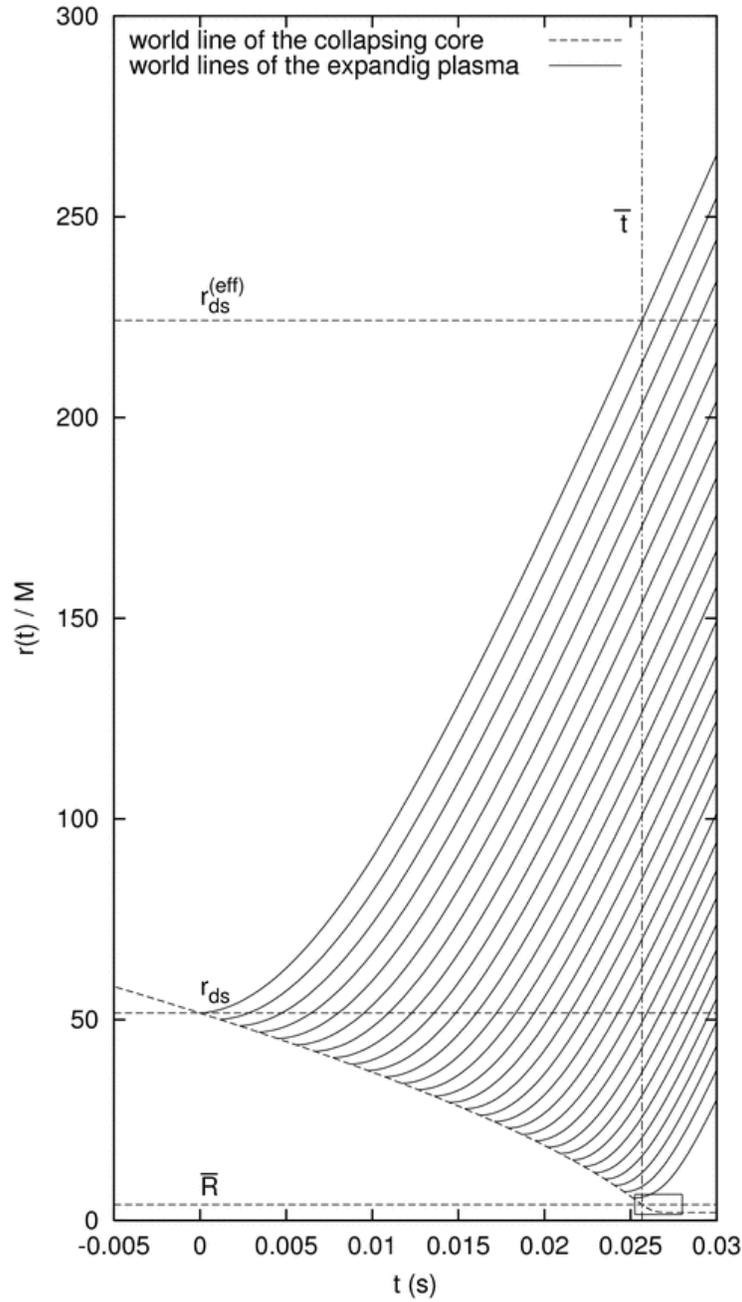}
\caption{World line of the collapsing charged core (dashed line) as derived
from Eq.(\ref{Motion}) for a black hole with $M=20M_{\odot}$, $Q=0.1M$; world
lines of slabs of plasma for selected radii $r_{0}$ in the interval $\left(
\bar{R},r_{\mathrm{ds}}\right)  $. At time $\bar{t}$ the expanding plasma
extends over a region which is almost one order of magnitude larger than the
dyadosphere. The small rectangle in the right bottom is enlarged in
Fig. \ref{f2}.}
\label{f1}
\end{figure}

\begin{figure}
\centering
\includegraphics[width=13cm]{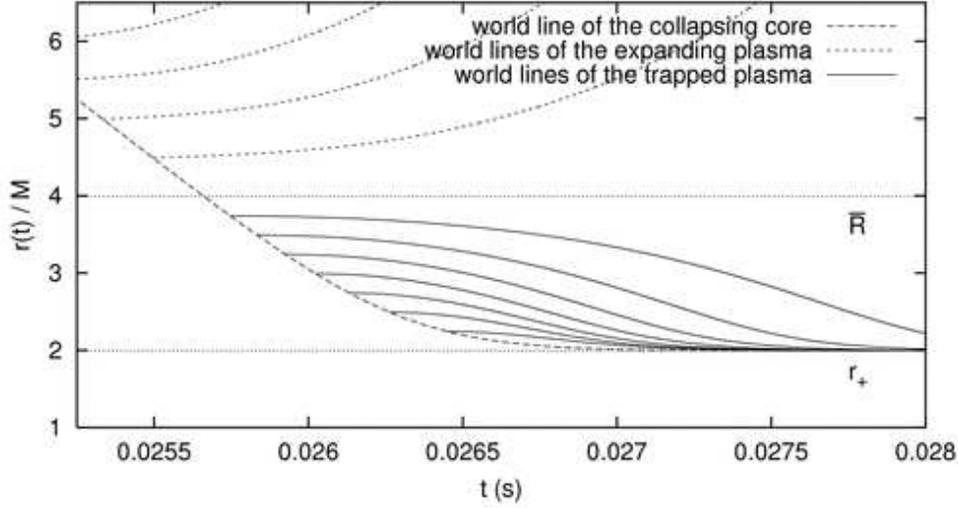}
\caption{Enlargement of the small rectangle in the right bottom of
Fig. \ref{f1}. World--lines of slabs of plasma for selected radii $r_{0}$ in
the interval $\left(  r_{+,}\bar{R}\right)  $.}
\label{f2}
\end{figure}

In the case of a black hole with $M=20M_{\odot}$, $Q=0.1M$, we compute:
\begin{itemize}
\item  the fraction of energy trapped in DTS:
\begin{equation}
\bar{E}=\int_{r_{+}<r<\bar{R}}\alpha\epsilon_{0}d\Sigma\simeq0.53\int
_{r_{+}<r<r_{\mathrm{ds}}}\alpha\epsilon_{0}d\Sigma;
\end{equation}
\item  the world--lines of slabs of plasma for selected $r_{0}$ in the
interval $\left(  \bar{R},r_{\mathrm{ds}}\right)  $ (see Fig. \ref{f1});
\item  the world--lines of slabs of plasma for selected $r_{0}$ in the
interval $\left(  r_{+},\bar{R}\right)  $ (see Fig. \ref{f2}).
\end{itemize}

\begin{figure}
\centering
\includegraphics[width=13cm]{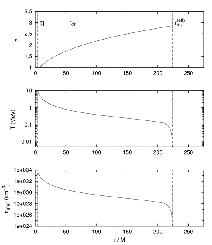}
\caption{Physical parameters in the effective dyadosphere: Lorentz $\gamma$
factor, proper temperature and proper $e^{+}e^{-}$ number density as functions
at time $\bar{t}$ for a black hole with $M=20M_{\odot}$ and $Q=0.1M$.}
\label{f3}
\end{figure}

At time $\bar{t}\equiv t_{0}\left(  \bar{R}\right)  $ when the DTS is formed,
the plasma extends over a region of space which is almost one order of
magnitude larger than the dyadosphere and which we define as the
\emph{effective dyadosphere}. The values of the Lorentz $\gamma$ factor, the
temperature and $e^{+}e^{-}$ number density in the effective dyadosphere are
given in Fig. \ref{f3}.

In conclusion we see how the causal description of the dyadosphere formation
can carry important messages on the time variability and spectral distribution
of the P-GRB due to quantum effects as well as precise signature of General Relativity.

\section{Observational signatures and spectral evolution of short GRBs}

The dynamics of the collapse of an electrically-charged stellar core, separating
itself from an oppositely charged remnant in an initially neutral star, was
first modeled by an exact solution of the Einstein-Maxwell equations
corresponding to a shell of charged matter in \citet{crv02}. The
fundamental dynamical equations and their analytic solutions were obtained, 
revealing the amplification of the electromagnetic field strength
during the process of collapse and the asymptotic approach to the final static
configuration. The results, which properly account for general
relativistic effects, are summarized in Fig. 1 and Fig. 2 of \citet{crv02}.

A first step toward the understanding of the process of extracting energy
from a black hole was obtained in \citet{rv02a}, where it
was shown how the extractable electromagnetic energy is not stored behind the
horizon but is actually distributed all around the black hole. Such a stored
energy is in principle extractable, very efficiently, on time-scales
$\sim\hbar/m_{e}c^{2}$, by a vacuum polarization process \emph{\`{a} l\`{a}}
Sauter-Heisenberg-Euler-Schwinger \citep{s31,he35,s51}. Such a process occurs
if the electromagnetic field becomes larger than the critical field strength
$\mathcal{E}_{\mathrm{c}}$ for $e^+e^-$ pair creation. In
\citet{rv02a} we followed the approach of \citet{dr75} in
order to evaluate the energy density and the temperature of the created
$e^+e^-$-photon plasma. As a byproduct, a formula for the
irreducible mass of a black hole was also derived solely in terms of the
gravitational, kinetic and rest mass energies of the collapsing core. This
surprising result allowed us in \citet{rv02b} to obtain a deeper
understanding of the maximum limit for the extractable energy during the process of
gravitational collapse, namely 50\% of the initial energy of
the star: the well known result of a 50\% maximum efficiency for energy extraction in
the case of a Reissner-Nordstr{\o}m black hole (\citet{cr71}) then becomes a particular
case of a process of much more general validity.

The crucial issue
of the survival of the electric charge of the collapsing core in the presence of a
copious process of $e^+e^-$ pair creation was addressed in
\citet{rvx03a,rvx03b}. By using theoretical techniques borrowed from
plasma physics and statistical mechanics
(\citet{GKM87,KESCM91,KESCM92,CEKMS93,KME98,SBR...98,BMP...99}) based on a
generalized Vlasov equation, it was possible to show that while the core keeps
collapsing, the created $e^+e^-$ pairs are entangled in the
overcritical electric field. The electric field itself, due to the back
reaction of the created $e^+e^-$ pairs, undergoes damped
oscillations in sign finally settling down to the critical value
$\mathcal{E}_{\mathrm{c}}$. The pairs fully thermalize to an
$e^+e^-$-photon plasma on time-scales typically of the order of
$10^{2}$--$10^{4}\hbar/m_{e}c^{2}$. During this characteristic damping time,
which we recall is much larger than the pair creation time-scale
$\hbar/m_{e}c^{2}$, the core moves
inwards, collapsing with a speed $0.2$--$0.8c$,
further amplifying the electric field strength at its surface and
enhancing the pair creation process.

Turning now to the dynamical evolution of such an $e^+e^-$ plasma we recall that, after some original attempt to consider a steady state emission (\citet{p86,p90}), the crucial progress was represented by the understanding that during the optically thick phase such a plasma expands as a thin shell. There exists a fundamental relation between the width of the expanding shell and the Lorentz gamma factor. The shell expands, but the Lorentz contraction is such that its width in laboratory frame appears to be constant. Such a result was found in \citet{psn93} on the basis of a numerical approach, further analyzed in \citet{bm95} on the basis of an analytic approach. Attention to the role of the rate equations governing the $e^+e^-$ annihilation were given in \citet{gw98}, where approximations to the full equation were introduced. These results were improved in two important respects in 1999 and 2000 (\citet{rswx99,rswx00}): the initial conditions were made more accurate by the considerations of the dyadosphere as well as the dynamics of the shell was improved by the self-consistent solution of the hydrodynamical equation and the rate equation for the $e^+e^-$ plasma following both an analytic and numerical approach. 

We are now ready to report the result of using the approach in \citet{rswx99,rswx00} in this general framework describing the dynamical formation of the dyadosphere.

The first attempt to analyze the
expansion of the newly generated and thermalized $e^+e^-$-photon
plasma was made in \citet{rvx03a}. The initial dynamical phases of the
expansion were analyzed, using the general relativistic equations of
\citet{crv02} for the gravitational collapse of the core. A {\itshape separatrix} was found in the motion of the plasma
at a critical radius $\bar{R}$: the plasma created at radii larger than
$\bar{R}$ expands to infinity, while the one created at radii smaller than
$\bar{R}$ is trapped by the gravitational field of the collapsing core and
implodes towards the black hole. The value of $\bar{R}$ was found in
\citet{rvx03a} to be 
$\bar{R}=2GM/c^{2}[1+\left(  1-3Q^{2}/4GM^{2}\right)^{1/2}]$, 
where $M$ and $Q$ are the mass and the charge of the core, respectively.

We now pursue further the evolution of such a
system, describing the dynamical phase of the expansion of the pulse of the
optically thick plasma all the way to the point where the transparency
condition is reached. Some pioneering work in this respect were presented by \citet{g86}. In this process the pulse reaches ultrarelativistic
regimes with Lorentz factor $\gamma\sim10^{2}$--$10^{4}$. The spectra, the
luminosities and the time-sequences of the electromagnetic signals captured
by a far-away observer are analyzed here in detail for the first time. 
The relevance of these theoretical results for short-bursts is then discussed.

\subsection{The expansion of the $e^{+}e^{-}\gamma$ plasma as a discrete set of elementary slabs}

We discretize the gravitational collapse of a spherically symmetric
core of mass $M$ and charge $Q$
by considering a set of events along the world line of a point of fixed
angular position on the collapsing core surface. Between each of
these events we consider a spherical shell slab of plasma of constant coordinate
thickness $\Delta r$ so that:

\begin{enumerate}
\item $\Delta r$ is assumed to be a constant which is 
small with respect to the core radius;
\item $\Delta r$ is assumed to be large with respect to the mean free path of
the particles so that the statistical description of the $e^{+}e^{-}\gamma$
plasma can be used;
\item  There is no overlap among the slabs and their union describes the
entirety of the process.
\end{enumerate}

We check that the final results are independent of the special value of the
chosen $\Delta r$.

In order to describe the dynamics of the
expanding plasma pulse the energy-momentum conservation law and the rate equation
for the number of pairs in the Reissner-Nordstr{\o}m geometry external to
the collapsing core have to be integrated. We use
Eqs.\eqref{na} to study the expansion of each slab, following
closely the treatment developed in \citet{rswx99,rswx00} where it was
shown how a homogeneous slab of plasma expands as a pair-electromagnetic
pulse (PEM pulse) of constant thickness in the laboratory frame. Two regimes
can be identified in the expansion of the slabs:

\begin{enumerate}
\item  In the initial phase of expansion the plasma experiences the strong
gravitational field of the core and a fully general relativistic description of
its motion is needed. The plasma is sufficiently hot in this first phase that
the $e^{+}e^{-}$ pairs and the photons remain at thermal equilibrium in it. As
shown in \citet{rvx03a}, under these circumstances, Eqs.\eqref{na} are equivalent to:
\begin{equation}
\left.
\begin{array}
[c]{c}%
\left(  \tfrac{dr}{cdt}\right)  ^{2}=\alpha^{4}\left[  1-\left(
\tfrac{n_{e^{+}e^{-}}}{n_{e^{+}e^{-}0}}\right)  ^{2}\left(  \tfrac{\alpha_{0}%
}{\alpha}\right)  ^{2}\left(  \tfrac{r}{r_{0}}\right)  ^{4}\right]  ,\\
\left(  \tfrac{r}{r_{0}}\right)  ^{2}=\left(  \tfrac{\epsilon+p}{\epsilon_{0}%
}\right)  \left(  \tfrac{n_{e^{+}e^{-}0}}{n_{e^{+}e^{-}}}\right)  ^{2}\left(
\tfrac{\alpha}{\alpha_{0}}\right)  ^{2}-\tfrac{p}{\epsilon_{0}}\left(
\tfrac{r}{r_{0}}\right)  ^{4},
\end{array}
\right.  \label{Eq2}%
\end{equation}
where $r$ is the radial coordinate of a slab of plasma, $\alpha=\left(
1-2MG/c^{2}r\right.$ $\left.+Q^{2}G/c^{4}r^{2}\right)^{1/2}$ is
the gravitational redshift factor and the subscript 
``$\scriptstyle{0}$" refers to quantities evaluated at the initial time.
\item  At asymptotically late times the temperature of the plasma drops below
an equivalent energy of $0.5$ MeV 
and the $e^{+}e^{-}$ pairs and the photons can no longer be
considered to be
in equilibrium: the full rate equation for pair
annihilation needs to be used. However, the plasma is so far from the central
core that gravitational effects can be neglected. In this new regime, as
shown in \citet{rswx99}, Eqs.\eqref{na} reduce to Eqs\eqref{be'}--\eqref{paira'_2} which we here rewrite for simplicity:
\begin{align}
\tfrac{\epsilon_{0}}{\epsilon}  &  =\left(  \tfrac{\gamma\mathcal{V}}%
{\gamma_{0}\mathcal{V}_{0}}\right)  ^{\Gamma},\nonumber\\
\tfrac{\gamma}{\gamma_{0}}  &  =\sqrt{\tfrac{\epsilon_{0}\mathcal{V}_{0}%
}{\epsilon\mathcal{V}}},\label{Eq3}\\
\tfrac{\partial}{\partial t}N_{e^{+}e^{-}}  &  =-N_{e^{+}e^{-}}\tfrac
{1}{\mathcal{V}}\tfrac{\partial\mathcal{V}}{\partial t}+\overline{\sigma
v}\tfrac{1}{\gamma^{2}}\left[  N_{e^{+}e^{-}}^{2}\left(  T\right)
-N_{e^{+}e^{-}}^{2}\right]  ,\nonumber
\end{align}
where $\Gamma=1+p/\epsilon$, $\mathcal{V}$ is the volume of a single
slab as measured in the laboratory frame by an observer at rest with the black hole, $N_{e^{+}e^{-}}=\gamma n_{e^{+}e^{-}}$ is the pair number density as
measured in the laboratory frame by an observer at rest with the black hole,
and $N_{e^{+}e^{-}}\left(  T\right)  $ is the equilibrium laboratory pair
number density.
\end{enumerate}

\subsection{The reaching of transparency and the signature of the outgoing
gamma ray signal}

\begin{figure}
\centering
\includegraphics[width=10cm]{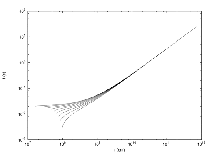}
\includegraphics[width=10cm]{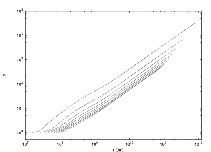}
\caption{Expansion of the plasma created around an overcritical collapsing
stellar core with $M=10M_{\odot}$ and $Q=0.1\sqrt{G}M$. Upper diagram: world
lines of the plasma. Lower diagram: Lorentz $\gamma$ factor as a function of
the radial coordinate $r$.}%
\label{ff1}%
\end{figure}

Eqs.(\ref{Eq2}) and (\ref{Eq3}) must be separately integrated 
and the solutions matched at
the transition between the two regimes. The integration stops when each slab
of plasma reaches the optical transparency condition given by
\begin{equation}
\int_0^{\Delta r}\sigma_{T}n_{e^{+}e^{-}}dr\sim1 \,,
\end{equation}
where $\sigma_{T}$ is the Thomson cross-section and the integral extends over
the radial thickness $\Delta r$ of the slab. The evolution of each slab occurs
without any collision or interaction with the other slabs; see the upper diagram
in Fig. \ref{ff1}. The outer layers are colder than the inner ones and
therefore reach transparency earlier; see the lower diagram in Fig. \ref{ff1}.

In Fig. \ref{ff1}, Eqs.(\ref{Eq2}) and (\ref{Eq3}) have been integrated for a
core with
\begin{equation}
M=10M_{\odot},\quad Q=0.1\sqrt{G}M;
\end{equation}
the upper diagram represents the world lines of the plasma as functions of the radius,
while the lower diagram shows the corresponding Lorentz $\gamma$ factors. The overall independence of the result of the dynamics
on the number $N$ of the slabs adopted in the discretization process or analogously on the value of $\Delta r$ has also
been checked. We have repeated the integration for $N=10$, $N=100$ reaching 
the same result to extremely good accuracy. The results in Fig. \ref{ff1}
correspond to the case $N=10$.

\begin{figure}
\centering
\includegraphics[width=10cm]{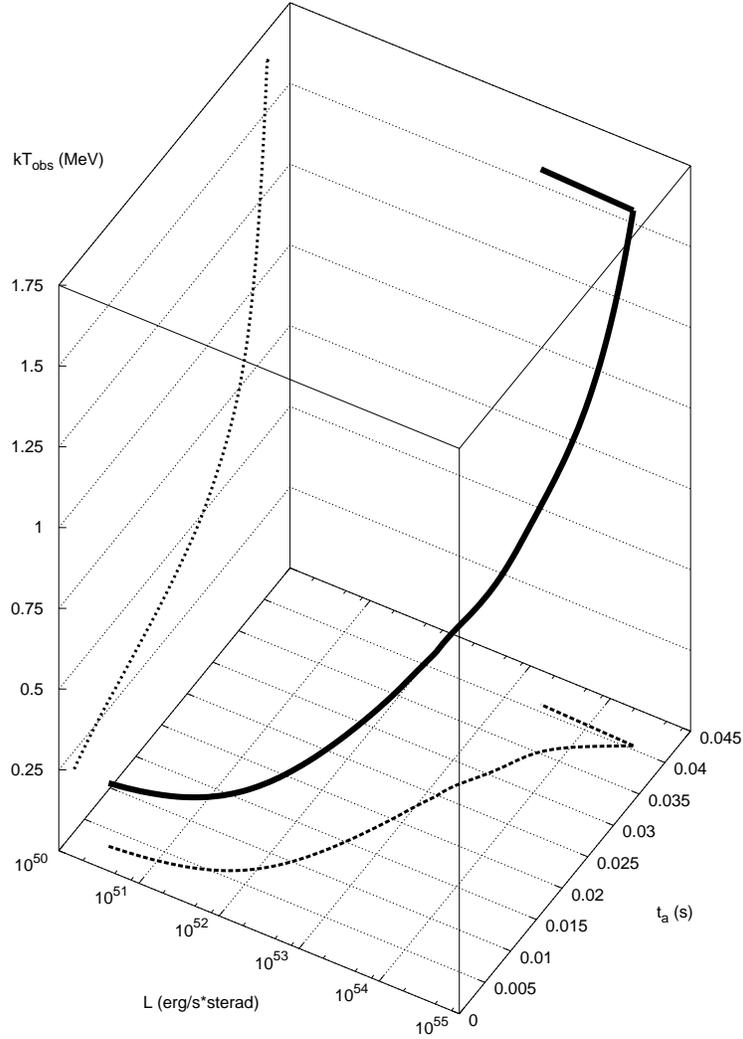}
\caption{Predicted observed luminosity and observed spectral hardness of the
electromagnetic signal from the gravitational collapse of a collapsing core
with $M=10M_{\odot}$, $Q=0.1\sqrt{G}M$ at $z=1$ as functions of the arrival
time $t_{a}$.}%
\label{ff4}%
\end{figure}

 We now turn to the results in
Fig. \ref{ff4}, where we plot both the theoretically predicted luminosity $L$
and the spectral hardness of the signal reaching a far-away observer as
functions of the arrival time $t_{a}$. Since all three of these quantities
depend in an essential way on the cosmological redshift factor $z$, see
\citet{brx01,rbcfx03a}, we have adopted a cosmological
redshift $z=1$ for this figure.

\begin{figure}
\centering
\includegraphics[width=10cm]{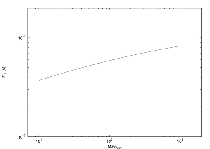}
\caption{Arrival time duration of the electromagnetic signal from the
gravitational collapse of a stellar core with charge $Q=0.1\sqrt{G}M$ as a
function of the mass $M$ of the core.}%
\label{ff5}%
\end{figure}

 As the plasma becomes transparent, gamma ray
photons are emitted. The energy $\hbar\omega$ of the observed photon is
$\hbar\omega=k\gamma T/\left(  1+z\right)  $, where $k$ is the Boltzmann
constant, $T$ is the temperature in the comoving frame of the pulse and
$\gamma$ is the Lorentz factor of the plasma at the transparency time. We also
recall that if the initial zero of time is chosen as the time when the first photon
is observed, then the arrival time $t_{a}$ of a photon at the detector in
spherical coordinates centered on the black hole is given by
(\citet{brx01,rbcfx03a}):
\begin{equation}
t_{a}=\left(  1+z\right)  \left[  t+\tfrac{r_{0}}{c}-\tfrac{r\left(  t\right)
}{c}\cos\theta\right]
\end{equation}
where $\left(  t,r\left(  t\right)  ,\theta,\phi\right)  $ labels the
laboratory emission event along the world line of the emitting slab and
$r_{0}$ is the initial position of the slab. The projection of the plot in
Fig. \ref{ff4} onto the $t_{a}$-$L$ plane gives the total luminosity as the
sum of the partial luminosities of the single slabs. The sudden decrease of
the intensity at the time $t=0.040466$ s corresponds to the creation of the
{\itshape separatrix} introduced in \citet{rvx03b}. We find that the duration of the
electromagnetic signal emitted by the relativistically expanding pulse is
given in arrival time by
\begin{equation}
\Delta t_{a}\sim5\times10^{-2}\mathrm{s}\label{ta1} \,.
\end{equation}
The projection of the plot in Fig. \ref{ff4} onto the $k T_{\mathrm{obs}%
}$, $t_{a}$ plane describes the temporal evolution of the spectral hardness.
We observe a precise soft-to-hard evolution of the spectrum of the gamma ray signal from $\sim10^{2}$ KeV monotonically increasing to $\sim1$ MeV. We
recall that $kT_{\mathrm{obs}}=k\gamma T/\left(  1+z\right)  $. 

 The
above quantities are clearly functions of the cosmological redshift $z$, of
the charge $Q$ and the mass $M$ of the collapsing core. We present in Fig. 3
the arrival time interval for $M$ ranging from $M\sim10M_{\odot}$ to
$10^{3}M_{\odot}$, keeping $Q=0.1\sqrt{G}M$. The arrival time interval is very
sensitive to the mass of the black hole:
\begin{equation}
\Delta t_{a}\sim10^{-2}-10^{-1}\mathrm{s} \,.
\label{ta2}
\end{equation}
Similarly the spectral hardness of the signal is sensitive to the ratio
$Q/\sqrt{G}M$ (\citet{rfvx05}). Moreover the duration, the spectral hardness and
luminosity are all sensitive to the cosmological redshift $z$ (see \citet{rfvx05}).
All the above quantities can also be sensitive to a possible baryonic
contamination of the plasma due to the remnant of the progenitor star which
has undergone the process of gravitational collapse.

\section{Electrodynamics for nuclear matter in bulk}

\subsection{Introduction}

It is well know that the Thomas-Fermi equation is the exact theory for atoms, molecules and solids as $Z\rightarrow\infty$ (\citet{liebsimon}). We show in this chapter that the relativistic Thomas-Fermi theory developed for the study of atoms for heavy nuclei with $Z \simeq 10^6$ (\citet{g1a,ruffinistella81}) gives important basic new information on the study of nuclear matter in bulk in the limit of $N \simeq (m_{\rm Planck}/m_n)^3$ nucleons of mass $m_n$ and on its electrodynamic properties. 
The analysis of nuclear matter bulk in neutron stars composed of degenerate gas of neutrons, protons and electrons, 
has traditionally been approached by implementing microscopically the charge neutrality condition  by requiring the electron density $n_e(x)$ to coincide with the proton density $n_p(x)$,
\begin{eqnarray}
n_e(x)=n_p(x).
\label{localnp}
\end{eqnarray}
It is clear however that especially when conditions close to the gravitational collapse occur, there is an ultra-relativistic component of degenerate electrons whose confinement requires the existence of very strong 
electromagnetic fields, in order to guarantee the overall charge neutrality of the neutron star. Under these conditions equation (\ref{localnp}) will be necessarily violated. We are going to show in this chapter that they will develop electric fields close to the critical value $E_c$ introduced by \citet{s31,he35,s51,s54a,s54b}:
\begin{eqnarray}
E_c=\frac{m^2c^3}{e\hbar}.
\label{ec2}
\end{eqnarray}
Special attention for the existence of critical electric fields and the possible condition for electron-positron ($e^+e^-$)
pair creation out of the vacuum  in the case of heavy bare nuclei, with the atomic number $Z\geq 173$, has been 
given by \citet{g1a,GZ70,z11b,z,g,muller72}.
They analyzed the specific pair creation process of an electron-positron pair around both a point-like 
and extended bare nucleus by direct integration of Dirac equation.
These considerations have been extrapolated to much heavier nuclei $Z\gg 1600$, implying the creation of a large number of  $e^+e^-$ pairs, by using a statistical approach based on the relativistic Thomas-Fermi equation by \citet{muller75,migdal76}. 
Using substantially the same statistical approach based on the relativistic Thomas-Fermi equation, \citet{ruffinistella80,ruffinistella81} have analyzed the electron densities around an extended nucleus in a neutral 
atom all the way  up to $Z\simeq 6000$. They have shown the effect of penetration of the electron orbitals well 
inside the nucleus, leading to a screening of the nuclei positive charge and to the concept of an ``effective'' nuclear charge distribution.
All the above works assumed for the radius of the extended nucleus the semi-empirical formulae (\citet{segrebook}),
\begin{eqnarray}
R_c\approx r_0 A^{1/3},\quad r_0=1.2\cdot 10^{-13}{\rm cm},
\label{dn}
\end{eqnarray}
where the mass number $A=N_n+N_p$, $N_n$ and $N_p$ are the neutron and proton numbers.  
The approximate relation between $A$ and the atomic number $Z=N_p$, 
\begin{eqnarray}
Z \simeq \frac{A}{2},
\label{z2a}
\end{eqnarray}
was adopted in \citet{muller75,migdal76}, or the empirical formulae
\begin{eqnarray}
Z &\simeq & [\frac{2}{A}+\frac{3}{200}\frac{1}{A^{1/3}}]^{-1},
\label{zae}
\end{eqnarray}
was adopted in \citet{ruffinistella80,ruffinistella81}.

\subsection{Electroweak equilibrium in Nuclear Matter in Bulk}
 
The aim of this chapter is to outline an alternative approach of the description of nuclear matter in bulk: it generalizes, to the case of $N \simeq (m_{\rm Planck}/m_n)^3$ nucleons, the above treatments, already developed and tested for the study of heavy nuclei. 
This more general approach differs in many aspects from the ones in the current literature and recovers, in the limiting case of $A$ 
smaller than $10^6$, the above treatments. We shall look for a solution implementing the condition of overall charge neutrality of the star as given by 
\begin{eqnarray}
N_e=N_p,
\label{golbalnp}
\end{eqnarray}
which significantly modifies Eq.~(\ref{localnp}), since now $N_e (N_p)$ is the total number of electrons (protons) of the equilibrium configuration.
Here we present only a
simplified prototype of this approach. We outline the essential relative role
of the four fundamental interactions present in the neutron star physics: the gravitational, weak, strong and
electromagnetic interactions. In addition, we also implement the fundamental role of Fermi-Dirac statistics   
and the phase space blocking due to the Pauli principle in the degenerate configuration. The new results essentially depend from the coordinated action of the five above theoretical components and cannot be obtained if any one of them is neglected.
Let us first recall the role of gravity.
In the case of neutron stars, unlike in the case of nuclei where its effects can be neglected, gravitation has the fundamental role of defining the  basic parameters of the equilibrium configuration. As pointed out by \citet{gamow-book}, at a Newtonian level and by \citet{ov39} in general relativity, configurations of equilibrium exist at approximately one solar mass and at an average density around the nuclear density. This result is obtainable considering only the gravitational interaction of a system of Fermi degenerate self-gravitating neutrons, neglecting all other particles and interactions. It can be formulated within a Thomas-Fermi self-gravitating model (see e.g. \citet{ruffiniphd}). 
In the present case of our simplified prototype model directed at evidencing new electrodynamic properties, the role of gravity is simply taken into account by considering, in line with the generalization of the above results, a mass-radius relation for the baryonic core
\begin{eqnarray}
R^{NS}=R_c\approx \frac{\hbar}{m_\pi c}\frac{m_{\rm Planck}}{m_n} .
\label{dnns}
\end{eqnarray}
This formula generalizes the one given by Eq.~(\ref{dn}) extending its validity  to  $N\approx (m_{\rm Planck}/m_n)^3$, 
leading to a baryonic core radius $R_c\approx  10$km.
We also recall that a more detailed analysis of nuclear matter in bulk in neutron stars (see e.g. \citet{sato1970,cameron1970}) shows that at mass densities larger than the "melting" density of 
\begin{eqnarray}
\rho_c=4.34 \cdot 10^{13} g/cm^3,
\label{melting}
\end{eqnarray}
all nuclei disappear. In the description of nuclear matter in bulk we have to consider then the three Fermi degenerate gas of neutrons, protons and electrons. In turn this naturally leads to consider the role of strong and weak interactions among the nucleons. In the nucleus, the role of the strong and weak interaction, with a short range of one Fermi, is to bind the nucleons, with a binding energy of 8 MeV, in order to balance the Coulomb repulsion of the protons. In the neutron star case we have seen that the neutrons confinement is due to gravity. We still assume that an essential role of the strong interactions is to balance the effective Coulomb repulsion due to the protons, partly screened by the electrons distribution inside the neutron star core. We shall verify, for self-consistency, the validity of this assumption on the final equilibrium solution we are going to obtain.
We now turn to the essential weak interaction role in establishing the relative balance between neutrons, protons and electrons via the direct and inverse 
$\beta$-decay
\begin{eqnarray}
p+ e  &\longrightarrow & n + \nu_e ,
\label{beta}\\
n  &\longrightarrow & p +e + \bar\nu_e.
\label{ibeta}
\end{eqnarray} 
Since neutrinos escape from the star and the Fermi energy of the electrons is null, as we will show below, the only non-vanishing terms in the equilibrium condition given by the weak interactions are: 
\begin{eqnarray}
[(P_n^Fc)^2+M^2_nc^4]^{1/2}-M_nc^2=  [(P_p^Fc)^2+M^2_pc^4]^{1/2}-M_pc^2  + |e|V^p_{\rm coul},  
\label{neq}
\end{eqnarray}
where $P_n^F$ and $P_p^F$  are respectively, the neutron and proton Fermi momenta, and $V^p_{\rm coul}$ is the Coulomb potential of protons. At this point, having fixed all these physical constraints, the main task is to find the electrons distributions fulfilling in addition to the Dirac-Fermi statistics also the Maxwell equations for the electrostatic. The condition of equilibrium of the  Fermi degenerate electrons implies the null value of the Fermi energy:
\begin{eqnarray}
[(P_e^Fc)^2+m^2c^4]^{1/2}-mc^2  + eV_{\rm coul}(r)=0,
\label{eeq2}
\end{eqnarray}
where $P_e^F$ is the electron Fermi momentum and $V_{\rm coul}(r)$ the Coulomb potential.

\subsection{Relativistic Thomas-Fermi Equation for Nuclear Matter in Bulk}

\begin{figure}
\centering 
\includegraphics[width=\hsize,clip]{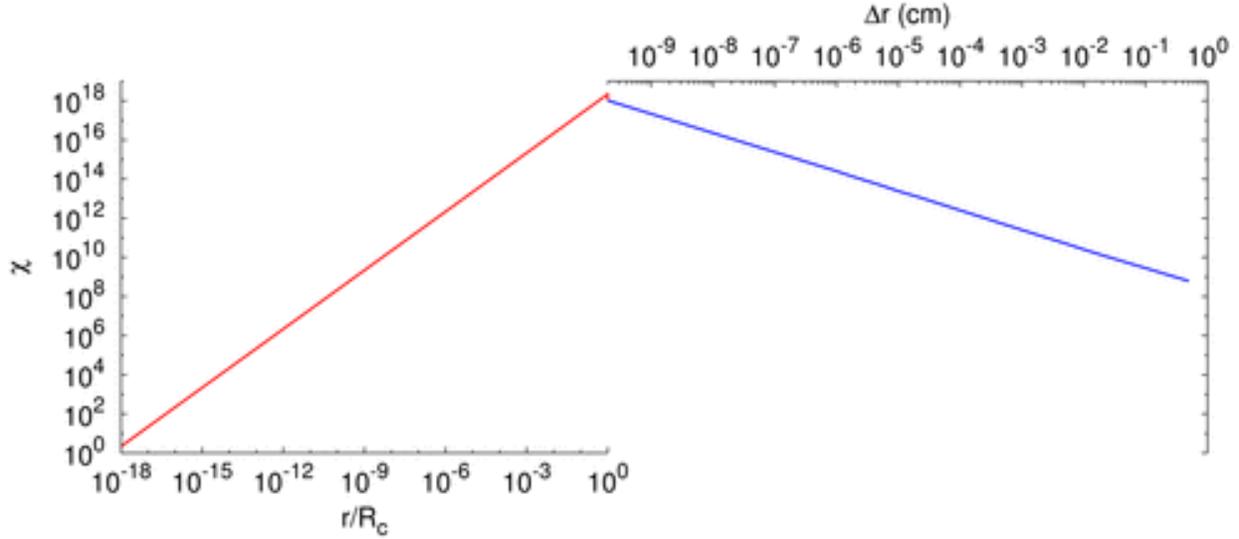}
\caption{The solution $\chi$ of the relativistic Thomas-Fermi Equation for $A=10^{57}$ and core radius $R_c=10$km, is plotted 
as a function of radial coordinate. The left red line corresponds to the internal solution and it is plotted as a function of radial coordinate in unit of $R_c$ in logarithmic scale. The right blue line corresponds to the solution external to the core and it is plotted as function of the distance $\Delta r$ from the surface in the logarithmic scale in centimeter.}%
\label{chif}%
\end{figure}

\begin{figure}
\centering 
\includegraphics[width=\hsize,clip]{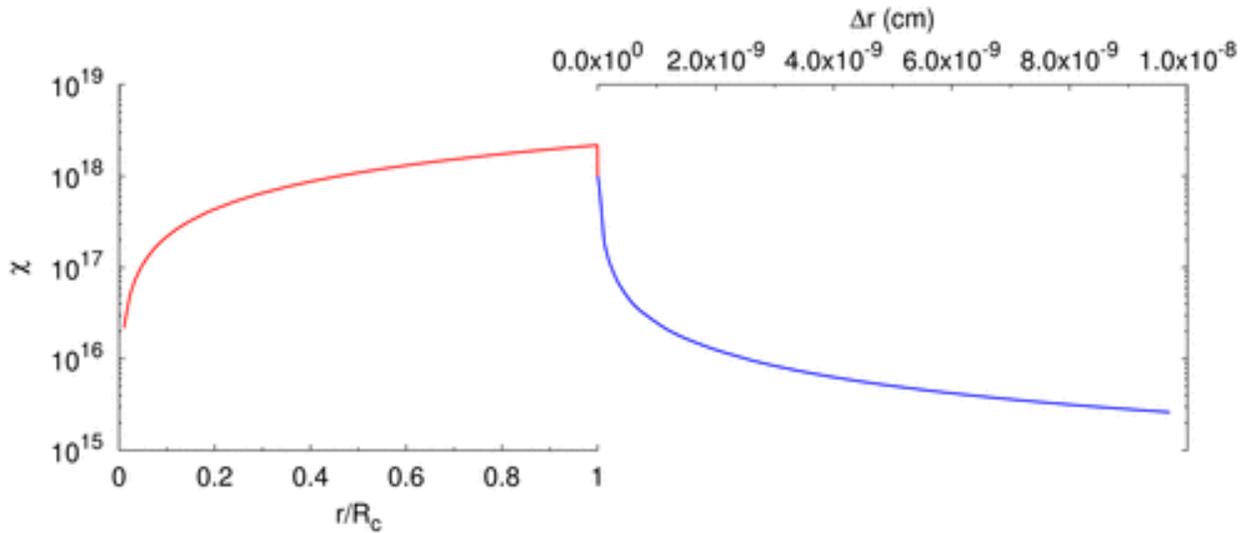}
\caption{The same as Fig.~(\ref{chif}): enlargement around the core radius $R_c$ showing explicitly the continuity of function $\chi$ and its 
derivative $\chi'$ from the internal to the external solution. }%
\label{chircf}%
\end{figure}

In line with the procedure already followed for the heavy atoms  
(\citet{ruffinistella80,ruffinistella81}) we here adopt the relativistic Thomas-Fermi Equation:
\begin{eqnarray}
\frac {1}{x}\frac {d^2\chi(x)}{dx^2}= - 4\pi \alpha\left\{\theta(x-x_c)
- \frac {1}{3\pi^2}\left[\left(\frac {\chi(x)}{x}+\beta\right)^2-\beta^2\right]^{3/2}\right\},
\label{eqless}
\end{eqnarray}
where $\alpha=e^2/(\hbar c)$, $\theta(x-x_c)$ represents the normalized proton density distribution, the variables $x$  and $\chi$  are related to the radial coordinate and the electron Coulomb potential $V_{\rm coul}$ by 
\begin{eqnarray}
x=\frac {r}{R_c}\left(\frac {3N_p}{4\pi}\right)^{1/3};\quad eV_{\rm coul}(r)\equiv \frac {\chi(r)}{r},
\label{dless}
\end{eqnarray} 
and the constants $x_c (r=R_c)$ and $\beta$ are respectively
\begin{eqnarray}
x_c\equiv\left(\frac {3N_p}{4\pi}\right)^{1/3};\quad \beta\equiv  
\frac {mcR_c}{\hbar}\left(\frac{4\pi}{3N_p}\right)^{1/3}.
\label{dbeta}
\end{eqnarray}
The solution has the boundary conditions
\begin{eqnarray}
\chi(0)=0;\quad \chi(\infty)=0,
\label{bchi}
\end{eqnarray} 
with the continuity of the function $\chi$ and its first derivative $\chi'$ at the boundary of the core $R_c$.
The crucial point is the determination of the eigenvalue of the first derivative at the center 
\begin{eqnarray}
\chi'(0)={\rm const}. ,
\label{bchi1}
\end{eqnarray} 
which has to be determined by fulfilling the above boundary conditions (\ref{bchi}) and constraints given by 
Eq.~(\ref{neq}) and Eq.~(\ref{golbalnp}).
The difficulty of the integration of the Thomas-Fermi Equations is certainly one of the most celebrated chapters in theoretical physics and mathematical physics, still challenging a proof of the existence and uniqueness of the solution and strenuously avoiding the occurrence of exact analytic solutions. We recall after the original papers of \citet{Thomas,Fermi}, the works of \citet{scorza28,scorza29,sommerfeld,Miranda} all the way to the many hundredth papers reviewed in the classical articles of \citet{liebsimon,Lieb,Spruch}. The situation here is more difficult since we are working on the special relativistic generalization of the Thomas-Fermi Equation. 
Also in this case, therefore, we have to proceed by numerical integration. The difficulty of this numerical task is further enhanced by a consistency check in order to fulfill all different constraints. It is so that we start the computations by assuming a total number of protons and a value of the core radius $R_c$. We integrate the Thomas-Fermi Equation and we determine the number of neutrons from the Eq.~(\ref{neq}). We iterate the procedure until a value of $A$ is reached consistent 
with our choice of the core radius. The paramount difficulty of the problem is the numerical determination of the 
eigenvalue in Eq.~(\ref{bchi1}) which already for $A \approx 10^{4}$ had presented remarkable numerical difficulties (\citet{ruffinistella80}). In the present context we have been faced for a few months by an  apparently unsurmountable numerical task: the determination of the eigenvalue seemed to necessitate a  significant number of decimals in the first derivative (\ref{bchi1}) comparable to the number of the electrons in the problem! 
We shall discuss elsewhere the way we overcame the difficulty by splitting the problem on the ground of the physical interpretation of the solution (\citet{rrx06}). The solution is given in 
Fig.~(\ref{chif}) and Fig.~(\ref{chircf}).

\begin{figure}
\centering 
\includegraphics[width=\hsize,clip]{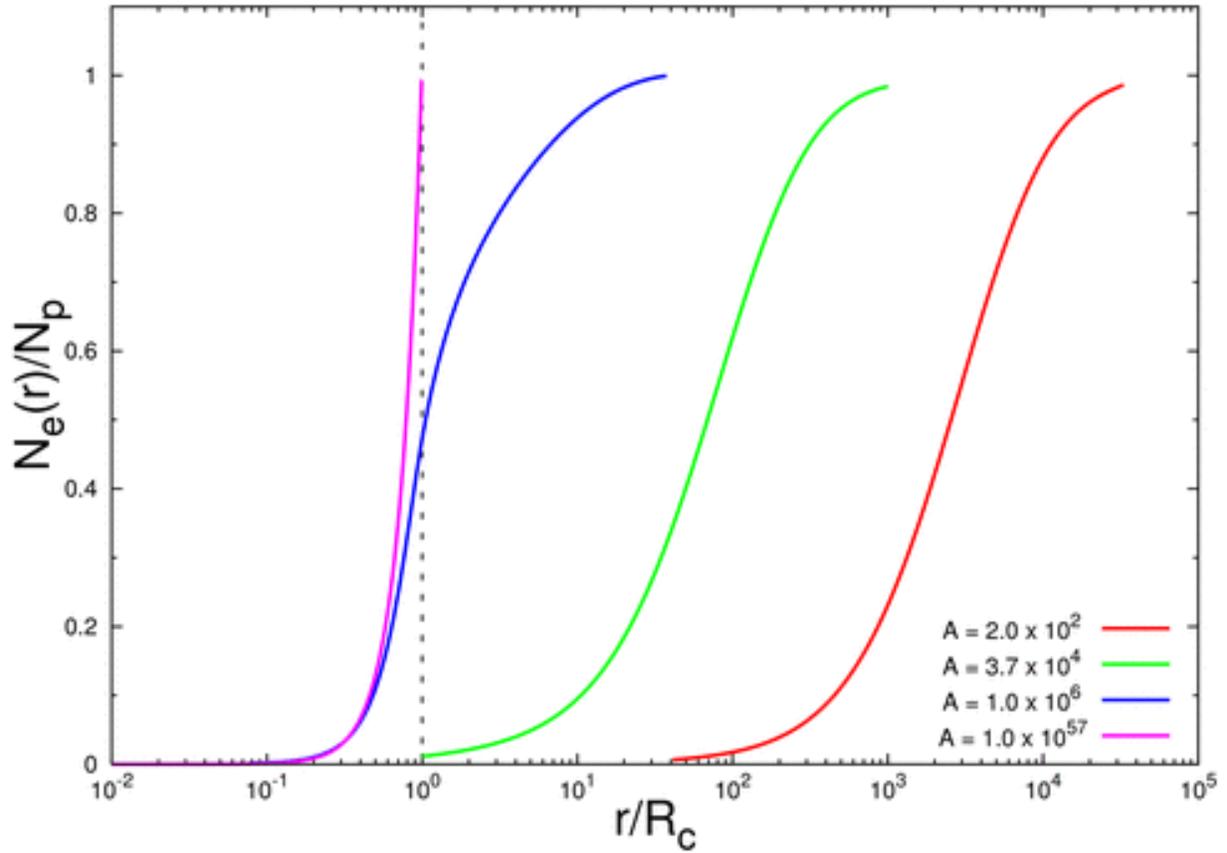}
\caption{The electron number (\ref{tein}) in the unit of the total proton number $N_p$, for selected values of 
$A$, is given as function of radial distance in the unit of the core radius $R_c$, again in logarithmic scale. It is clear how by increasing the value of $A$ the penetration of electrons inside the core increases. The detail shown in Fig.~(\ref{efieldf}) and Fig.~(\ref{enumberf1}) demonstrates how  for $N \simeq (m_{\rm Planck}/m_n)^3$ a relatively small
tail of electron outside the core exists and generates on the baryonic core surface an electric field close to the critical value given in . A significant electron density outside the core is found.
}%
\label{enumberf}%
\end{figure}

\begin{figure}
\centering 
\includegraphics[width=\hsize,clip]{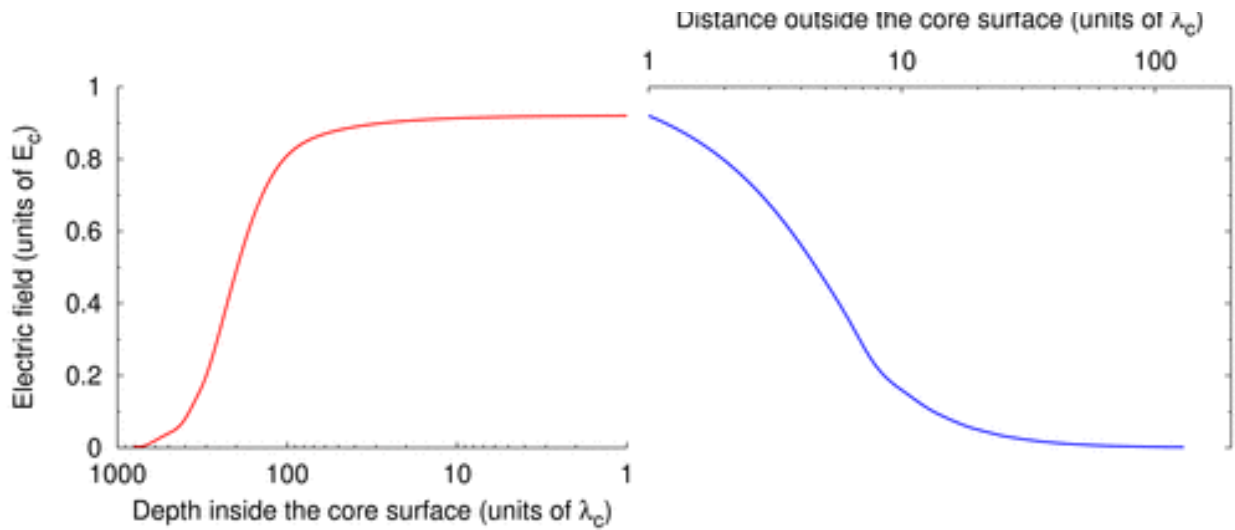}
\caption{The electric field in the unit of the critical field $E_c$ is plotted around the core radius $R_c$. The left (right) diagram in the red (blue) refers the region just inside (outside) the core radius plotted logarithmically. By increasing the density of the star the field approaches the critical field. }%
\label{efieldf}%
\end{figure}

\begin{figure}
\centering 
\includegraphics[width=\hsize,clip]{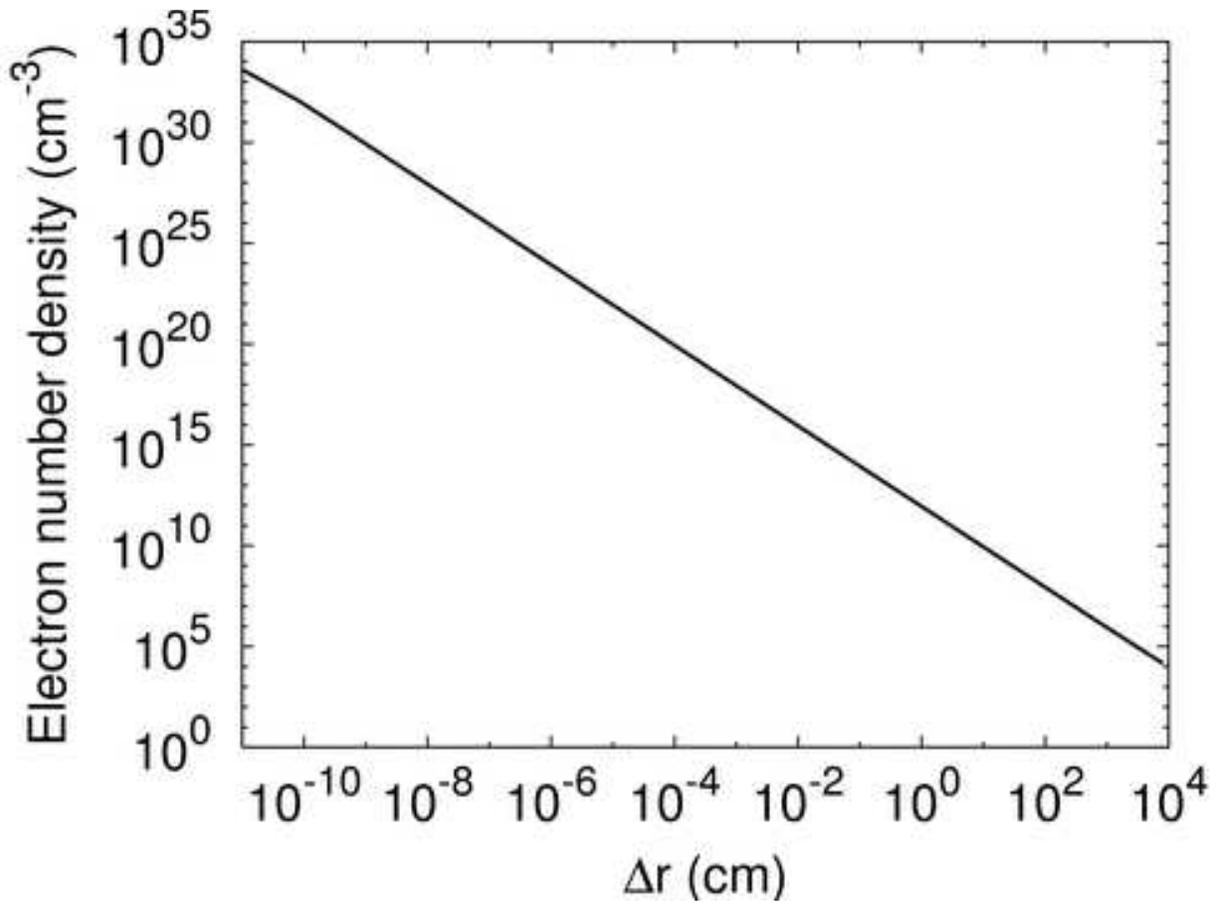}
\caption{ The density of electrons for $A=10^{57}$ in the region outside the core; both scale are logarithmically.
}%
\label{enumberf1}%
\end{figure}

A relevant quantity for exploring the physical significance of the solution is given by the number of electrons within a given radius $r$:
\begin{eqnarray}
N_e(r)=\int_0^{r} 4\pi (r')^2 n_e(r')dr'.
\label{tein}
\end{eqnarray}
This allows to determine, for selected values of the $A$ parameter, the distribution of the electrons within and outside the core and follow the progressive penetration of the electrons in the core at increasing values of $A$ [ see Fig.~(\ref{enumberf})]. 
We can then evaluate, generalizing the results in \citet{ruffinistella80,ruffinistella81}, the net charge inside the core
\begin{eqnarray}
N_{\rm net} = N_p-N_e(R_c) < N_p,
\label{net}
\end{eqnarray} 
and consequently determine of the electric field at the core surface, as well as within and outside the core 
[see Fig.~(\ref{efieldf})] and evaluate as well the  Fermi degenerate electron distribution outside the core 
[see Fig.~(\ref{enumberf1})].
It is interesting to explore the solution of the problem under the same conditions and constraints imposed by the fundamental interactions and the quantum statistics and imposing instead of Eq.~(\ref{localnp}) 
the corresponding Eq.~(\ref{golbalnp}). Indeed a solution exist and is much simpler  
\begin{eqnarray}
n_n(x)=n_p(x)=n_e(x)=0,\quad \chi=0.
\label{trivial}
\end{eqnarray}


\subsection{The energetic stability of solution}

Before concluding as we announce we like to check on the theoretical consistency of the solution. We obtain an overall neutral configuration for the nuclear matter in bulk, with a positively charged baryonic core with  
\begin{equation}
N_{\rm net}= 0.92\left(\frac{m}{m_\pi}\right)^2\left(\frac{e}{m_n\sqrt{G}}\right)^2\left(\frac{1}{\alpha}\right)^2 ,
\label{nete}
\end{equation}
and an electric field on the baryonic core surface (see Fig.~(\ref{efieldf}) )
\begin{equation}
\frac{E}{E_c}=0.92.
\label{esurface}
\end{equation} 
The corresponding Coulomb repulsive energy per nucleon is given by 
\begin{equation}
U^{\rm max}_{\rm coul}= \frac{1}{2\alpha}\left(\frac{m}{m_\pi}\right)^3mc^2\approx 1.78\cdot 10^{-6}({\rm MeV}),
\label{coul1}
\end{equation}
well below the nucleon binding energy per nucleon. It is also important to verify that this charge core is gravitationally stable.
We have in fact 
\begin{equation}
\frac{Q}{\sqrt{G}M}=\alpha^{-1/2}\left(\frac{m}{m_\pi}\right)^2\approx 1.56\cdot 10^{-4}.
\label{nuclb}
\end{equation}
The electric field of the baryonic core is screened to infinity by an electron distribution given in Fig.~(\ref{enumberf1}).
As usual any new solution of Thomas-Fermi systems has relevance and finds its justification in the theoretical physics and mathematical physics domain. We expect that as in the other solutions previously obtained in the literature of the relativistic Thomas-Fermi equations also this one we present in this chapter will find important applications in physics and astrophysics.
There are a variety of new effects that such a generalized approach naturally leads to: (1) the mass-radius relation 
of neutron star may be affected; (2) the electrodynamic aspects of neutron stars and pulsars will be 
different; (3) we expect also important 
consequence in the initial conditions in the physics of gravitational collapse of the baryonic core as soon as the critical mass for gravitational collapse to a black hole is reached. The consequent collapse to a black hole will have very different energetics properties.

\section{Conclusions}

These notes should manifest that GRBs are very likely one of the most complex, but still fully comprehensible, system in physics and astrophysics. In a 9 year activity in our group we have developed the theory and the numerical codes to reach the full description of a GRB. This includes effects of general relativity, special relativity and quantum relativistic field theories.

The basic phenomenon originating the GRBs has been identified, in agreement with other leading works in this field, in a self accelerating optically thick electron-positron plasma. The basic equations have been given to describe the dynamical expansion of such a plasma, the engulfment of a finite amount of baryonic matter, its further acceleration up to reaching the transparency condition and, finally, the interaction of the accelerated baryons with the ISM. Such equations have been compared and contrasted with the ones in the current literature and the differences have been justified. We have created and evolved a complex numerical code to perform all the computations. By this code we have fitted the luminosities in selected energy bands and the spectra of all GRBs we analyzed. The use of this program is not automatic. An attentive analysis lasting a few months is needed for each GRB source in order to interpret the ``tomography'' determining the ISM structure, as well as to identify the two free parameters, $E_{e^\pm}^{tot} = E_{dya}$ and $B$, describing the GRB source.

It is since now clear that GRBs exist on a very large range of isotropic energies, varying from $10^{48}$ to $10^{54}$ ergs. Evidence is also mounting that GRBs originate from the gravitational collapse of very different progenitors. Such progenitors range from of a critical mass neutron star collapsing to the smallest possible black hole all the way to coalescing binary systems formed by a white dwarf and/or a neutron star and/or a black hole in various combinations.

From the observations of GRBs, especially the ones by \emph{Swift}, it is emerging the existence of a canonical GRB structure, quite independent on the nature of the progenitors. This result, which may appear to be surprising and difficult to explain, is indeed a very natural consequence of the uniqueness theorem of the black hole, graphically expressed since 1971 (\citet{rw71}) and object of rigorous mathematical proofs in the intervening years (\citet{RuKerr}). With reference to the ISM, it is also clear that GRBs can only occur in a very finite range of ISM density with $\langle n_{ism} \rangle \le 1$ particle/cm$^3$. For this reason, while we expect GRBs to originating in galactic halos at $\langle n_{ism} \rangle \sim 10^{-3}$ particles/cm$^3$, we do not expect GRBs originating from denser parts of galactic clusters.

The mechanism responsible for the origin and the energetics of the electron-positron plasma originating the GRBs, either in relation to black hole physics or to other physical processes, has often been discussed qualitatively in the GRB scientific literature but never quantitatively with explicit equations. In our model it is identified with the quantum vacuum polarization process in the overcritical field developed during the gravitational collapse to a black hole. The GRB energy source is therefore the blackholic energy. In order to mimic this complex dynamical process for mathematical convenience and definiteness of the computations, we have first used a dyadosphere around a given and already formed Reissner-Nordstr{\o}m black hole. This however does not mean that we need a black hole created and staying there forever waiting to be discharged! The entire process of gravitational collapse we are speaking of lasts less than one second, and the dynamical phases of the existence of the dyadosphere lasts less than $10^{-2}$ s. We are speaking in reality about a very transient phenomenon. Actually GRBs are likely the most energetic and transient phenomenon in the Universe.

In the same spirit, for mathematical convenience and definiteness, we have used some simplified models to probe the dynamical phases of gravitational collapse. The existence of an already overcritical field since the early phases of the gravitational collapse has appeared to be a necessary condition for the creation of an optically thick electron-positron plasma and the extraction of the blackholic energy. Densities on the order of $10^{15}$--$10^{16}$ g/cm$^3$, namely of $10$ times the nuclear density, are necessarily involved in the process of black hole formation from any of the precursors mentioned above. Under such conditions, the physical description of the process of gravitational collapse needs necessarily to take into account strong, weak, electromagnetic and gravitational interactions, as well as quantum statistics.

In the last section of these lecture notes, we have exemplified a system which, although globally neutral, presents an internal charge separation due to the concurrence of strong, weak and electromagnetic interactions, as well as of quantum statistics. What is also very important, this system develops an overcritical electric field. Systems of this kind appear to be a natural initial condition for the gravitational collapse leading to a dyadosphere to occur.

The GRBs are, for the first time, probing a variety of new fundamental physics processes:
\begin{enumerate}
\item the vacuum polarization process and the creation of electron-positron pairs out of the vacuum, which have been for years unsuccessfully attempted in Earth-bound experiments, appears to be here at reach for the first time in an astrophysical setting of unprecedented magnitude;
\item the extraction of blackholic energy is possibly to be observed for the first time;
\item the details of gravitational collapse process can be followed with unprecedented accuracy and lead to possible coincident signals in gravitational waves, neutrino bursts and UHECRs.
\end{enumerate}

\section{Acknowledgments}
We are thankful to Alexei Aksenov, Vladimir Belinski, Donato Bini, Gennadii Bisnovatyi-Kogan, Sergei Blinnikov, Christian Cherubini, Thibault Damour, Gustavo De Barros, Charles Dermer, Andrea Geralico, Vahe Gurzadyan, Roy Kerr, Hagen Kleinert, Iosef Khriplovich, Bahram Mashhoon, Mario Novello, Barbara Patricelli, Santiago Esteban Perez Bergliaffa, Tsvi Piran, Francesca Pompi, Vladimir Popov, Luis Juracy Rangel Lemos, Andreas Ringwald, Jorge Armando Rueda Hernandez, Jay Salmonson, Rashid Sunyaev, Lev Titarchuk, Jim Wilson and Dima Yakovlev for many interesting theoretical discussions, as well as to Lorenzo Amati, Lucio Angelo Antonelli, Enrico Costa, Massimo Della Valle, Filippo Frontera, Luciano Nicastro, Elena Pian, Luigi Piro, Marco Tavani and all the BeppoSAX team and the Italian Swift Team for assistance in the data analysis.

\end{document}